# The genetic code, algebra of projection operators and problems of inherited biological ensembles

Sergey V. Petoukhov

Head of Laboratory of Biomechanical System, Mechanical Engineering Research Institute of the Russian Academy of Sciences, Moscow

spetoukhov@gmail.com, petoukhov@imash.ru, http://petoukhov.com/

**Summary**. This article is devoted to applications of projection operators to simulate phenomenological properties of the molecular-genetic code system. Oblique projection operators are under consideration, which are connected with matrix representations of the genetic coding system in forms of the Rademacher and Hadamard matrices. Evidences are shown that sums of such projectors give abilities for adequate simulations of ensembles of inherited biological phenomena including ensembles of biological cycles, morphogenetic ensembles of phyllotaxis patterns, mirror-symmetric patterns, etc. For such modeling, the author proposes multidimensional vector spaces, whose subspaces are under a selective control (or coding) by means of a set of matrix operators on base of genetic projectors. Development of genetic biomechanics is discussed. The author proposes and describes special systems of multidimensional numbers under names "united-hypercomplex numbers", which attracted his attention when he studied genetic systems and genetic matrices. New rules of long nucleotide sequences are described on the base of the proposed notion of tetra-groups of equivalent oligonucleotides. Described results can be used for developing algebraic biology, biotechnical applications and some other fields of science and technology.

**Content**

## 1. ABOUT THE PARTNERSHIP OF THE GENETIC CODE AND MATHEMATICS

Science has led to a new understanding of life itself: "*Life is a partnership between genes and mathematics*" [Stewart, 1999]. But what kind of mathematics can be a partner for the genetic coding system? This article shows some evidences that algebra of projectors can be one of main parts of such mathematics. Till now the notion of projection operators (or briefly, projectors) was one of important in many fields of non-biological science: physics including quantum mechanics; mathematics; computer science and informatics including theory of digital codes; chemistry; mathematical logic, etc. On basis of materials of this article, the author thinks that projectors can become one of the main notions and effective mathematical tools in mathematical biology. Moreover they will help not only to a development of algebraic biology and a new understanding of living matter but also to a mutual enrichment of different branches of science.

Projectors are expressed by means of square matrices (http://mathworld.wolfram.com/ProjectionMatrix.html, https://en.wikipedia.org/wiki/Projection_(linear_algebra)). A necessary and sufficient condition that a matrix P is a projection operator is the fulfillment of the following condition: $P^2 = P$. A set of projectors is separated into two sub-sets:

- orthogonal projectors, which are expressed by symmetric matrices and theory of which is well developed and has a lot of applications;
- oblique projectors, which are expressed by non-symmetric matrices; their theory and its applications are developed much weaker as the author can judge. Namely oblique projectors will be the main objects of attention in this article.

This article is a continuation and an essential development of the author's article about relations between the genetic system and projection operators [Petoukhov, 2010].

In accordance with Mendel's laws of independent inheritance of traits, information from the micro-world of genetic molecules dictates constructions in the macro-world of living organisms under strong noise and interference. This dictation is realized by means of unknown algorithms of multi-channel noise-immunity coding. For example, in human organism, his skin color, eye color and hair color are inherited genetically independently of each other. It is possible if appropriate kinds of information are conducted via independent informational channels and if a general "phase space" of living organism contains sub-spaces with a possibility of a selective control or a selective coding of processes in them. So, any living organism is an algorithmic machine of multi-channel noise-immunity coding with ability to a selective control and coding of different sub-spaces of its phase space (a model approach to phase spaces with a selective control of their sub-spaces is presented in this article). This machine works in conditions of ontogenetic development of the organism when a multi-dimensionality of its phase space is increased step by step.

To understand such genetic machine, it is appropriate to use the theory of noise-immunity coding and transmission of digital information, taking into account the discrete nature of the genetic code. In this theory, mathematical matrices have the basic importance. The use of matrix representations and analysis in the study of phenomenological features of molecular-genetic ensembles has led to the development of a special scientific direction under a name "Matrix Genetics" [Petoukhov, 2008; Petoukhov, He, 2009]. Namely researches of the "matrix genetics" gave results that are represented in this article.

Concerning the theme of projectors in inherited biological phenomena, one can note that our genetically inherited visual system works on the principle of projection of external objects at the retina. This projection is modeled using projection operators. The author believes that the value of projectors for bioinformatics is not limited to this single fact of biological significance of projection operators, but that the whole system of genetic and sensory informatics is based on their active use. This ubiquitous use of projection operators reflects and ensures (in some degree) the unity of any organism and interrelations of its parts.

The set of projection operators, which are associated with the matrix representation of the genetic code, provides new opportunities for modeling ensembles of inherited cycles; ensembles of phyllotaxis structures; a numeric specificity of reproduction of genetic information in acts of mitosis and meiosis of biological cells, etc. In the frame of the "projector conception" arised here in genetic informatics, some features of evolutionary transformations of variants (or dialects) of the genetic code are clarified.

The main mathematical objects of the article are four matrices $R_4$, $R_8$, $H_4$ and $H_8$ shown on Figure 1. Why these numeric matrices are chosen from infinite set of matrices? The reason is that they are connected with phenomenology of the genetic code system in matrix forms of its representation as it was shown in works [Petoukhov, 2008b, 2011a,b, 2012a,b] and as it will be additionally demonstrated in the end of this article. The matrices $R_4$ and $R_8$ are conditionally termed “Rademacher matrices” because each of their columns represents one of known Rademacher functions. The matrices $H_4$ and $H_8$ belong to a great set of Hadamard matrices, which are widely used for noise-immunity coding in technologies of signals processing and which are connected with complete orthogonal systems of Walsh functions.

$$\mathbf{R_4} = \begin{bmatrix} 1 & 1 & 1 & -1 \\ -1 & 1 & -1 & -1 \\ 1 & -1 & 1 & 1 \\ -1 & -1 & -1 & 1 \end{bmatrix} ; \quad \mathbf{R_8} = \begin{bmatrix} 1 & 1 & 1 & 1 & 1 & 1 & -1 & -1 \\ 1 & 1 & 1 & 1 & 1 & 1 & -1 & -1 \\ -1 & -1 & 1 & 1 & -1 & -1 & -1 & -1 \\ -1 & -1 & 1 & 1 & -1 & -1 & -1 & -1 \\ 1 & 1 & -1 & -1 & 1 & 1 & 1 & 1 \\ 1 & 1 & -1 & -1 & 1 & 1 & 1 & 1 \\ -1 & -1 & -1 & -1 & -1 & -1 & 1 & 1 \\ -1 & -1 & -1 & -1 & -1 & -1 & 1 & 1 \end{bmatrix}$$

$$\mathbf{H_4} = \begin{array}{|cc|cc|} \hline 1 & 1 & -1 & 1 \\ -1 & 1 & 1 & 1 \\ \hline 1 & -1 & 1 & 1 \\ -1 & -1 & -1 & 1 \\ \hline \end{array} ; \qquad \mathbf{H_8} = \begin{array}{|cccc|cccc|} \hline 1 & -1 & 1 & -1 & -1 & 1 & 1 & -1 \\ 1 & 1 & 1 & 1 & -1 & -1 & 1 & 1 \\ -1 & 1 & 1 & -1 & 1 & -1 & 1 & -1 \\ -1 & -1 & 1 & 1 & 1 & 1 & 1 & 1 \\ \hline 1 & -1 & -1 & 1 & 1 & -1 & 1 & -1 \\ 1 & 1 & -1 & -1 & 1 & 1 & 1 & 1 \\ -1 & 1 & -1 & 1 & -1 & 1 & 1 & -1 \\ -1 & -1 & -1 & -1 & -1 & -1 & 1 & 1 \\ \hline \end{array}$$

Figure 1. Numeric matrices $H_4$, $H_8$, $R_4$ and $R_8$ which are connected with phenomenology of the genetic coding system [Petoukhov, 2011, 2012a,b]

Every of these matrices can be decomposed into sum of sparse matrices, each of which contains only one non-zero column. Such decomposition can be conditionally termed a «column decomposition». Every of such matrices can be also decomposed into a sum of sparse matrices, each of which contains only one non-zero row. Such decomposition can be conditionally termed a «row decomposition».

## 2. GENETIC RADEMACHER MATRICES AS SUMS OF PROJECTORS

Let us begin with the column decomposition $R_4=c_0+c_1+c_2+c_3$ and the row decomposition $R_4=r_0+r_1+r_2+r_3$ of the Rademacher (4*4)-matrix $R_4$ (Figure 2).

$$R_4 = c_0+c_1+c_2+c_3 = \begin{vmatrix} 1 & 0 & 0 & 0 \\ -1 & 0 & 0 & 0 \\ 1 & 0 & 0 & 0 \\ -1 & 0 & 0 & 0 \end{vmatrix} + \begin{vmatrix} 0 & 1 & 0 & 0 \\ 0 & 1 & 0 & 0 \\ 0 & -1 & 0 & 0 \\ 0 & -1 & 0 & 0 \end{vmatrix} + \begin{vmatrix} 0 & 0 & 1 & 0 \\ 0 & 0 & -1 & 0 \\ 0 & 0 & 1 & 0 \\ 0 & 0 & -1 & 0 \end{vmatrix} + \begin{vmatrix} 0 & 0 & 0 & -1 \\ 0 & 0 & 0 & -1 \\ 0 & 0 & 0 & 1 \\ 0 & 0 & 0 & 1 \end{vmatrix}$$

$$R_4 = r_0+r_1+r_2+r_3 = \begin{vmatrix} 1 & 1 & 1 & -1 \\ 0 & 0 & 0 & 0 \\ 0 & 0 & 0 & 0 \\ 0 & 0 & 0 & 0 \end{vmatrix} + \begin{vmatrix} 0 & 0 & 0 & 0 \\ -1 & 1 & -1 & -1 \\ 0 & 0 & 0 & 0 \\ 0 & 0 & 0 & 0 \end{vmatrix} + \begin{vmatrix} 0 & 0 & 0 & 0 \\ 0 & 0 & 0 & 0 \\ 1 & -1 & 1 & 1 \\ 0 & 0 & 0 & 0 \end{vmatrix} + \begin{vmatrix} 0 & 0 & 0 & 0 \\ 0 & 0 & 0 & 0 \\ 0 & 0 & 0 & 0 \\ -1 & -1 & -1 & 1 \end{vmatrix}$$

Figure 2. The «column decomposition» (upper layer) and the «row decomposition» (bottom layer) of the Rademacher matrix $R_4$ from Figure 1

Each of these sparse matrices $c_0$, $c_1$, $c_2$, $c_3$ and $r_0$, $r_1$, $r_2$, $r_3$ is a projection operator because it satisfies the criterion of projectors $P^2=P$ (for example, $c_0^2=c_0$, etc). Every of these projectors is an oblique (non-ortogonal) projector because it is expressed by means of a non-symmetrical matrix. Every of sets ($c_0$, $c_1$, $c_2$, $c_3$) and ($r_0$, $r_1$, $r_2$, $r_3$) consists of non-commutative projectors. We will conditionally name projectors $c_0$, $c_1$, $c_2$, $c_3$ as «column projectors» and projectors $r_0$, $r_1$, $r_2$, $r_3$ as «row projectors».

Let us examine all possible variants of sums of pairs of the different column projectors $c_0$, $c_1$, $c_2$ and $c_3$: ($c_0+c_1$), ($c_0+c_2$), ($c_0+c_3$), ($c_1+c_2$), ($c_1+c_3$), ($c_2+c_3$). The result of this examination is the following: matrices ($c_0+c_1$) and ($c_2+c_3$) with a weight coefficient $2^{-0.5}$ lead to cyclic groups with their period 8 in cases of their exponentiation:

$(2^{-0.5}*(c_0+c_1))^n = (2^{-0.5}*(c_0+c_1))^{n+8}$, $(2^{-0.5}*(c_2+c_3))^n = (2^{-0.5}*(c_2+c_3))^{n+8}$, where n = 1, 2, 3,… (Figure 3).

$$(2^{-0.5}*(c_0+c_1))^1 = \begin{vmatrix} 2^{-0.5}, & 2^{-0.5}, & 0, & 0 \\ -2^{-0.5}, & 2^{-0.5}, & 0, & 0 \\ 2^{-0.5}, & -2^{-0.5}, & 0, & 0 \\ -2^{-0.5}, & -2^{-0.5}, & 0, & 0 \end{vmatrix} ; \quad (2^{-0.5}*(c_2+c_3))^1 = \begin{vmatrix} 0, & 0, & 2^{-0.5}, & -2^{-0.5} \\ 0, & 0, & -2^{-0.5}, & -2^{-0.5} \\ 0, & 0, & 2^{-0.5}, & 2^{-0.5} \\ 0, & 0, & -2^{-0.5}, & 2^{-0.5} \end{vmatrix}$$

$$(2^{-0.5}*(c_0+c_1))^2 = \begin{vmatrix} 0, & 1, & 0, & 0 \\ -1, & 0, & 0, & 0 \\ 1, & 0, & 0, & 0 \\ 0, & -1, & 0, & 0 \end{vmatrix} ; \quad (2^{-0.5}*(c_2+c_3))^2 = \begin{vmatrix} 0, & 0, & 1, & 0 \\ 0, & 0, & 0, & -1 \\ 0, & 0, & 0, & 1 \\ 0, & 0, & -1, & 0 \end{vmatrix}$$

$$(2^{-0.5}*(c_0+c_1))^3 = \begin{vmatrix} -2^{-0.5}, & 2^{-0.5}, & 0, & 0 \\ -2^{-0.5}, & -2^{-0.5}, & 0, & 0 \\ 2^{-0.5}, & 2^{-0.5}, & 0, & 0 \\ 2^{-0.5}, & -2^{-0.5}, & 0, & 0 \end{vmatrix} ; \quad (2^{-0.5}*(c_2+c_3))^3 = \begin{vmatrix} 0, & 0, & 2^{-0.5}, & 2^{-0.5} \\ 0, & 0, & 2^{-0.5}, & -2^{-0.5} \\ 0, & 0, & -2^{-0.5}, & 2^{-0.5} \\ 0, & 0, & -2^{-0.5}, & -2^{-0.5} \end{vmatrix}$$

$$(2^{-0.5}*(c_0+c_1))^4 = \begin{vmatrix} -1, & 0, & 0, & 0 \\ 0, & -1, & 0, & 0 \\ 0, & 1, & 0, & 0 \\ 1, & 0, & 0, & 0 \end{vmatrix} ; \quad 2^{-0.5}*(c_2+c_3)^4 = \begin{vmatrix} 0, & 0, & 0, & 1 \\ 0, & 0, & 1, & 0 \\ 0, & 0, & -1, & 0 \\ 0, & 0, & 0, & -1 \end{vmatrix}$$

$$(2^{-0.5}*(c_0+c_1))^5 = \begin{vmatrix} -2^{-0.5}, & -2^{-0.5}, & 0, & 0 \\ 2^{-0.5}, & -2^{-0.5}, & 0, & 0 \\ -2^{-0.5}, & 2^{-0.5}, & 0, & 0 \\ 2^{-0.5}, & 2^{-0.5}, & 0, & 0 \end{vmatrix} ; \quad (2^{-0.5}*(c_2+c_3))^5 = \begin{vmatrix} 0, & 0, & -2^{-0.5}, & 2^{-0.5} \\ 0, & 0, & 2^{-0.5}, & 2^{-0.5} \\ 0, & 0, & -2^{-0.5}, & -2^{-0.5} \\ 0, & 0, & 2^{-0.5}, & -2^{-0.5} \end{vmatrix}$$

$$(2^{-0.5}*(c_0+c_1))^6 = \begin{vmatrix} 0, & -1, & 0, & 0 \\ 1, & 0, & 0, & 0 \\ -1, & 0, & 0, & 0 \\ 0, & 1, & 0, & 0 \end{vmatrix} ; \quad (2^{-0.5}*(c_2+c_3))^6 = \begin{vmatrix} 0, & 0, & -1, & 0 \\ 0, & 0, & 0, & 1 \\ 0, & 0, & 0, & -1 \\ 0, & 0, & 1, & 0 \end{vmatrix}$$

$$(2^{-0.5}*(c_0+c_1))^7 = \begin{vmatrix} 2^{-0.5}, & -2^{-0.5}, & 0, & 0 \\ 2^{-0.5}, & 2^{-0.5}, & 0, & 0 \\ -2^{-0.5}, & -2^{-0.5}, & 0, & 0 \\ -2^{-0.5}, & 2^{-0.5}, & 0, & 0 \end{vmatrix} ; \quad (2^{-0.5}*(c_2+c_3))^7 = \begin{vmatrix} 0, & 0, & -2^{-0.5}, & -2^{-0.5} \\ 0, & 0, & -2^{-0.5}, & 2^{-0.5} \\ 0, & 0, & 2^{-0.5}, & -2^{-0.5} \\ 0, & 0, & 2^{-0.5}, & 2^{-0.5} \end{vmatrix}$$

$$(2^{-0.5}*(c_0+c_1))^8 = \begin{vmatrix} 1, & 0, & 0, & 0 \\ 0, & 1, & 0, & 0 \\ 0, & -1, & 0, & 0 \\ -1, & 0, & 0, & 0 \end{vmatrix} \quad ; \quad (2^{-0.5}*(c_2+c_3))^8 = \begin{vmatrix} 0, & 0, & 0, & -1 \\ 0, & 0, & -1, & 0 \\ 0, & 0, & 1, & 0 \\ 0, & 0, & 0, & 1 \end{vmatrix}$$

$$(2^{-0.5}*(c_0+c_1))^9 = \begin{vmatrix} 2^{-0.5}, & 2^{-0.5}, & 0, & 0 \\ -2^{-0.5}, & 2^{-0.5}, & 0, & 0 \\ 2^{-0.5}, & -2^{-0.5}, & 0, & 0 \\ -2^{-0.5}, & -2^{-0.5}, & 0, & 0 \end{vmatrix} \quad ; \quad (2^{-0.5}*(c_2+c_3))^9 = \begin{vmatrix} 0, & 0, & 2^{-0.5}, & -2^{-0.5} \\ 0, & 0, & -2^{-0.5}, & -2^{-0.5} \\ 0, & 0, & 2^{-0.5}, & 2^{-0.5} \\ 0, & 0, & -2^{-0.5}, & 2^{-0.5} \end{vmatrix}$$

Figure 3. The illustration of cyclic groups of operators on the basis of sums of projection operators $(c_0+c_1)$ and $(c_2+c_3)$ from Figure 2 in cases of their exponentiation.

These two sums of column projectors $(c_0+c_1)$ and $(c_2+c_3)$ are marked by green colour in the left table on Figure 4, where every of cells represents a sum of those projectors, which denote its column and row.

Two other examined sums of column projectors $(c_0+c_2)$ and $(c_1+c_3)$ is doubled when squaring: $(c_0+c_2)^n = 2^{n-1}*(c_0+c_2)$, $(c_1+c_3)^n = 2^{n-1}*(c_1+c_3)$, where n = 1, 2, 3,… (these sums are marked by red colours in the left table on Figure 4) . This doubling reminds dichotomic dividing of biological cells in a result of mitosis when doubling of genetic information occurs.

The last examined sums of the column projectors $(c_0+c_3)$ and $(c_1+c_2)$ possess the following feature. Matrices of their second power is quadrupled in a result of exponentiation in integer powers: $((c_0+c_3)^2)^n = 4^{n-1}*(c_0+c_3)^2$, $((c_1+c_2)^2)^n = 4^{n-1}*(c_1+c_2)^2$, where n = 1, 2, 3… (this feature can be used to simulate a genetic phenomenon of tetra-reproduction of gametes and genetic information in a course of meiosis). The cells with these sums $(c_0+c_3)$ and $(c_1+c_2)$ are marked by yellow colour in the left table on Figure 4.

If we examine the row projectors $r_0$, $r_1$, $r_2$ and $r_3$ from Figure 2, we receive the same tabular structure with a small change: red and yellow cells are swapped (Figure 4, right). Matrices $2^{-0.5}*(r_0+r_1)$ and $2^{-0.5}*(r_2+r_3)$ are basises for cyclic groups with a period 8 in relation to their exponentiation (green cells in the right table on Figure 4). Matrices $(r_0+r_2)$ and $(r_1+r_3)$ is doubled when squaring: $(r_0+r_2)^n = 2^{n-1}*(r_0+r_2)$, $(r_1+r_3)^n = 2^{n-1}*(r_1+r_3)$, where n = 1, 2, 3,… (red cells in the right table on Figure 4). Matrices $(r_0+r_3)$ and $(r_1+r_2)$ possess the «quadruplet» property: $((r_0+r_3)^2)^n = 4^{n-1}*(r_0+r_3)^2$, $((r_1+r_2)^2)^n = 4^{n-1}*(r_1+r_2)^2$, where n = 1, 2, 3… (yellow cells in the right table on Figure 4). The cells on the main diagonal correspond to sum of two projectors themselves: $(c_i+c_i)^n = 2^n*c_i$, $(r_i+r_i)^n = 2^n*r_i$, where i = 0, 1, 2, 3.

| | $c_0$ | $c_1$ | $c_2$ | $c_3$ |
|---|---|---|---|---|
| $c_0$ | - | $c_0+c_1$ | $c_0+c_2$ | $c_0+c_3$ |
| $c_1$ | $c_1+c_0$ | - | $c_1+c_2$ | $c_1+c_3$ |
| $c_2$ | $c_2+c_0$ | $c_2+c_1$ | - | $c_2+c_3$ |
| $c_3$ | $c_3+c_0$ | $c_3+c_1$ | $c_3+c_2$ | - |

| | $r_0$ | $r_1$ | $r_2$ | $r_3$ |
|---|---|---|---|---|
| $r_0$ | - | $r_0+r_1$ | $r_0+r_2$ | $r_0+r_3$ |
| $r_1$ | $r_1+r_0$ | - | $r_1+r_2$ | $r_1+r_3$ |
| $r_2$ | $r_2+r_0$ | $r_2+r_1$ | - | $r_2+r_3$ |
| $r_3$ | $r_3+r_0$ | $r_3+r_1$ | $r_3+r_2$ | - |

Figure 4. Tables of some features of sums of pairs of different «column projectors» $c_0$, $c_1$, $c_2$, $c_3$ and of «row projectors» $r_0$, $r_1$, $r_2$, $r_3$ (from the Rademacher matrix $R_4$ on Figure 2) in relation to their exponentiation. Explanations in text.

One can mention that the structure of this symmetric table is unexpectedly coincided with the structure of the typical (4*4)-matrix of dyadic shifts, which is known in theory of signal processing and which is related with some phenomenologic properties of molecular-genetic systems [Ahmed, Rao, 1975; Petoukhov, 2008, 2012a; Petoukhov, He, 2009].

It should be noted that cyclic features of (4*4)-matrices $2^{-0.5}*(c_0+c_1)$, $2^{-0.5}*(c_2+c_3)$, $2^{-0.5}*(r_0+r_1)$ and $2^{-0.5}*(r_2+r_3)$ in cases of their exponentiation exist due to their connections with matrix representations of 2-parametric complex numbers in 4-dimensional space; these connections are shown by means of a new decomposition of each of these matrices into sum of new sparse matrices $e_0$ and $e_1$ (Figures 5, 6, two upper levels, where the multiplication table for $e_0$ and $e_1$ in the right column is identical to the multiplication table of complex numbers). The dichotomic features of (4*4)-matrices $(c_0+c_2)$, $(c_1+c_3)$, $(r_0+r_3)$, $(r_1+r_2)$ and tetra-reproduction features of (4*4)-matrices $((c_0+c_3)^2)^n = 4^{n-1}*(c_0+c_3)^2$, $((c_1+c_2)^2)^n = 4^{n-1}*(c_1+c_2)^2$, $((r_0+r_2)^2)^n = 4^{n-1}*(r_0+r_2)^2$, $((r_1+r_3)^2)^n = 4^{n-1}*(r_1+r_3)^2$ exist due to their connections with matrix representations of 2-parametric hyperbolic numbers in 4-dimensional space (Figures 5, 6, four bottom levels, where the multiplication table for $e_0$ and $e_1$ in the right column is identical to the multiplication table of hyperbolic numbers). Synonyms of hyperbolic numbers are Lorentz numbers, split-complex numbers, double numbers, perplex numbers, etc. - http://en.wikipedia.org/wiki/Split-complex_number). Hyperbolic numbers are a two-dimensional commutative algebra over the real numbers. Additional details about such (4*4)-matrix representations of complex numbers and hyperbolic numbers see in [Petoukhov, 2012b].

| | | | | | | | |
|---|---|---|---|---|---|---|---|
| $c_0+c_1 =$ | $\begin{matrix}1&1&0&0\\-1&1&0&0\\1&-1&0&0\\-1&-1&0&0\end{matrix}$ | $= e_0+e_1 =$ | $\begin{matrix}1&0&0&0\\0&1&0&0\\0&-1&0&0\\-1&0&0&0\end{matrix}$ | + | $\begin{matrix}0&1&0&0\\-1&0&0&0\\1&0&0&0\\0&-1&0&0\end{matrix}$ | ; | $\begin{matrix}&e_0&e_1\\e_0&e_0&e_1\\e_1&e_1&-e_0\end{matrix}$ |
| $c_2+c_3 =$ | $\begin{matrix}0&0&1&-1\\0&0&-1&-1\\0&0&1&1\\0&0&-1&1\end{matrix}$ | $= e_0+e_1 =$ | $\begin{matrix}0&0&0&-1\\0&0&-1&0\\0&0&1&0\\0&0&0&1\end{matrix}$ | + | $\begin{matrix}0&0&1&0\\0&0&0&-1\\0&0&0&1\\0&0&-1&0\end{matrix}$ | ; | $\begin{matrix}&e_0&e_1\\e_0&e_0&e_1\\e_1&e_1&-e_0\end{matrix}$ |
| $c_0+c_2=$ | $\begin{matrix}1&0&1&0\\-1&0&-1&0\\1&0&1&0\\-1&0&-1&0\end{matrix}$ | $= e_0+e_1 =$ | $\begin{matrix}1&0&0&0\\-1&0&0&0\\0&0&1&0\\0&0&-1&0\end{matrix}$ | + | $\begin{matrix}0&0&1&0\\0&0&-1&0\\1&0&0&0\\-1&0&0&0\end{matrix}$ | ; | $\begin{matrix}&e_0&e_1\\e_0&e_0&e_1\\e_1&e_1&e_0\end{matrix}$ |
| $c_1+c_3 =$ | $\begin{matrix}0&1&0&-1\\0&1&0&-1\\0&-1&0&1\\0&-1&0&1\end{matrix}$ | $= e_0+e_1 =$ | $\begin{matrix}0&1&0&0\\0&1&0&0\\0&0&0&1\\0&0&0&1\end{matrix}$ | + | $\begin{matrix}0&0&0&-1\\0&0&0&-1\\0&-1&0&0\\0&-1&0&0\end{matrix}$ | ; | $\begin{matrix}&e_0&e_1\\e_0&e_0&e_1\\e_1&e_1&e_0\end{matrix}$ |
| $0.5*(c_0+c_3)^2 =$ | $\begin{matrix}1&0&0&-1\\0&0&0&0\\0&0&0&0\\-1&0&0&1\end{matrix}$ | $= e_0+e_1 =$ | $\begin{matrix}1&0&0&0\\0&0&0&0\\0&0&0&0\\0&0&0&1\end{matrix}$ | + | $\begin{matrix}0&0&0&-1\\0&0&0&0\\0&0&0&0\\-1&0&0&0\end{matrix}$ | ; | $\begin{matrix}&e_0&e_1\\e_0&e_0&e_1\\e_1&e_1&e_0\end{matrix}$ |

| | | | | | | |
|---|---|---|---|---|---|---|
| $0.5*(c_1+c_2)^2 =$ | $\begin{matrix} 0 & 0 & 0 & 0 \\ 0 & 1 & -1 & 0 \\ 0 & -1 & 1 & 0 \\ 0 & 0 & 0 & 0 \end{matrix}$ | $= e_0+e_1=$ | $\begin{matrix} 0 & 0 & 0 & 0 \\ 0 & 1 & 0 & 0 \\ 0 & 0 & 1 & 0 \\ 0 & 0 & 0 & 0 \end{matrix}$ | + | $\begin{matrix} 0 & 0 & 0 & 0 \\ 0 & 0 & -1 & 0 \\ 0 & -1 & 0 & 0 \\ 0 & 0 & 0 & 0 \end{matrix}$ | ; $\begin{array}{c\|cc} & e_0 & e_1 \\ \hline e_0 & e_0 & e_1 \\ e_1 & e_1 & e_0 \end{array}$ |

Figure 5. The table represents special decompositions of (4*4)-matrices $(c_0+c_1)$, $(c_2+c_3)$, $(c_0+c_2)$, $(c_1+c_3)$, $0.5*(c_0+c_3)^2$ and $0.5*(c_1+c_2)^2$ into sum of two matrices $e_0+e_1$ (see also Figures 2 and 4). The table shows direct relations of these matrices with matrix representations of 2-parametric complex numbers and hyperbolic numbers. Here $c_0$, $c_1$, $c_2$ and $c_3$ are column projectors from Figure 2. For each set of matrices $e_0$ and $e_1$ at every tabular level, the right column of this table contains its multiplication table: for two upper levels it is a known multiplication table of complex numbers; for other four levels it is a known multiplication table of hyperbolic numbers.

| | | | | | | |
|---|---|---|---|---|---|---|
| $r_0+r_1 =$ | $\begin{matrix} 1 & 1 & 1 & -1 \\ -1 & 1 & -1 & -1 \\ 0 & 0 & 0 & 0 \\ 0 & 0 & 0 & 0 \end{matrix}$ | $= e_0+e_1=$ | $\begin{matrix} 1 & 0 & 1 & 0 \\ 0 & 1 & 0 & -1 \\ 0 & 0 & 0 & 0 \\ 0 & 0 & 0 & 0 \end{matrix}$ | + | $\begin{matrix} 0 & 1 & 0 & -1 \\ -1 & 0 & -1 & 0 \\ 0 & 0 & 0 & 0 \\ 0 & 0 & 0 & 0 \end{matrix}$ | ; $\begin{array}{c\|cc} & e_0 & e_1 \\ \hline e_0 & e_0 & e_1 \\ e_1 & e_1 & -e_0 \end{array}$ |
| $r_2+r_3 =$ | $\begin{matrix} 0 & 0 & 0 & 0 \\ 0 & 0 & 0 & 0 \\ 1 & -1 & 1 & 1 \\ -1 & -1 & -1 & 1 \end{matrix}$ | $= e_0+e_1=$ | $\begin{matrix} 0 & 0 & 0 & 0 \\ 0 & 0 & 0 & 0 \\ 1 & 0 & 1 & 0 \\ 0 & -1 & 0 & 1 \end{matrix}$ | + | $\begin{matrix} 0 & 0 & 0 & 0 \\ 0 & 0 & 0 & 0 \\ 0 & -1 & 0 & 1 \\ -1 & 0 & -1 & 0 \end{matrix}$ | ; $\begin{array}{c\|cc} & e_0 & e_1 \\ \hline e_0 & e_0 & e_1 \\ e_1 & e_1 & -e_0 \end{array}$ |
| $0.5*(r_0+r_2)^2=$ | $\begin{matrix} 1 & 0 & 1 & 0 \\ 0 & 0 & 0 & 0 \\ 1 & 0 & 1 & 0 \\ 0 & 0 & 0 & 0 \end{matrix}$ | $= e_0+e_1=$ | $\begin{matrix} 1 & 0 & 0 & 0 \\ 0 & 0 & 0 & 0 \\ 0 & 0 & 1 & 0 \\ 0 & 0 & 0 & 0 \end{matrix}$ | + | $\begin{matrix} 0 & 0 & 1 & 0 \\ 0 & 0 & 0 & 0 \\ 1 & 0 & 0 & 0 \\ 0 & 0 & 0 & 0 \end{matrix}$ | ; $\begin{array}{c\|cc} & e_0 & e_1 \\ \hline e_0 & e_0 & e_1 \\ e_1 & e_1 & e_0 \end{array}$ |
| $0.5*(r_1+r_3)^2=$ | $\begin{matrix} 0 & 0 & 0 & 0 \\ 0 & 1 & 0 & -1 \\ 0 & 0 & 0 & 0 \\ 0 & -1 & 0 & 1 \end{matrix}$ | $= e_0+e_1=$ | $\begin{matrix} 0 & 0 & 0 & 0 \\ 0 & 1 & 0 & 0 \\ 0 & 0 & 0 & 0 \\ 0 & 0 & 0 & 1 \end{matrix}$ | + | $\begin{matrix} 0 & 0 & 0 & 0 \\ 0 & 0 & 0 & -1 \\ 0 & 0 & 0 & 0 \\ 0 & -1 & 0 & 0 \end{matrix}$ | ; $\begin{array}{c\|cc} & e_0 & e_1 \\ \hline e_0 & e_0 & e_1 \\ e_1 & e_1 & e_0 \end{array}$ |
| $r_0+r_3 =$ | $\begin{matrix} 1 & 1 & 1 & -1 \\ 0 & 0 & 0 & 0 \\ 0 & 0 & 0 & 0 \\ -1 & -1 & -1 & 1 \end{matrix}$ | $= e_0+e_1=$ | $\begin{matrix} 1 & 1 & 0 & 0 \\ 0 & 0 & 0 & 0 \\ 0 & 0 & 0 & 0 \\ 0 & 0 & -1 & 1 \end{matrix}$ | | $\begin{matrix} 0 & 0 & 1 & -1 \\ 0 & 0 & 0 & 0 \\ 0 & 0 & 0 & 0 \\ -1 & -1 & 0 & 0 \end{matrix}$ | ; $\begin{array}{c\|cc} & e_0 & e_1 \\ \hline e_0 & e_0 & e_1 \\ e_1 & e_1 & e_0 \end{array}$ |
| $r_1+r_2 =$ | $\begin{matrix} 0 & 0 & 0 & 0 \\ -1 & 1 & -1 & -1 \\ 1 & -1 & 1 & 1 \\ 0 & 0 & 0 & 0 \end{matrix}$ | $= e_0+e_1=$ | $\begin{matrix} 0 & 0 & 0 & 0 \\ -1 & 1 & 0 & 0 \\ 0 & 0 & 1 & 1 \\ 0 & 0 & 0 & 0 \end{matrix}$ | + | $\begin{matrix} 0 & 0 & 0 & 0 \\ 0 & 0 & -1 & -1 \\ 1 & -1 & 0 & 0 \\ 0 & 0 & 0 & 0 \end{matrix}$ | ; $\begin{array}{c\|cc} & e_0 & e_1 \\ \hline e_0 & e_0 & e_1 \\ e_1 & e_1 & e_0 \end{array}$ |

Figure 6. The table represents special decompositions of (4*4)-matrices $(r_0+r_1)$, $(r_2+r_3)$, $0.5*(r_0+r_2)^2$, $0.5*(r_1+r_3)^2$, $(r_0+r_3)$ and $(r_1+r_2)$ into sum of two matrices $e_0+e_1$. The table shows direct relations of these matrices with matrix representations of 2-parametric complex numbers and hyperbolic numbers. Here $r_0$, $r_1$, $r_2$ and $r_3$ are row projectors from Figure 2. For each set of matrices $e_0$ and $e_1$ at every tabular level, the right column of this table contains its multiplication table: for two upper levels it is a known multiplication table of complex

numbers; for other four levels it is a known multiplication table of hyperbolic numbers.

Now let us turn to the Rademacher (8*8)-matrix $R_8$ (Figure 1) to analyze its column decomposition $R_8 = s_0+s_1+s_2+s_3+s_4+s_5+s_6+s_7$ (Fugure 7) and its row decomposition $R_8 = v_0+v_1+v_2+v_3+v_4+v_5+v_6+v_7$ (Figure 8).

$$\mathbf{R_8 = s_0+s_1+s_2+s_3+ s_4+s_5+s_6+s_7 =} \begin{vmatrix} 1&0&0&0&0&0&0&0\\ 1&0&0&0&0&0&0&0\\ -1&0&0&0&0&0&0&0\\ -1&0&0&0&0&0&0&0\\ 1&0&0&0&0&0&0&0\\ 1&0&0&0&0&0&0&0\\ -1&0&0&0&0&0&0&0\\ -1&0&0&0&0&0&0&0 \end{vmatrix} + \begin{vmatrix} 0&1&0&0&0&0&0&0\\ 0&1&0&0&0&0&0&0\\ 0&-1&0&0&0&0&0&0\\ 0&-1&0&0&0&0&0&0\\ 0&1&0&0&0&0&0&0\\ 0&1&0&0&0&0&0&0\\ 0&-1&0&0&0&0&0&0\\ 0&-1&0&0&0&0&0&0 \end{vmatrix} +$$

$$\begin{vmatrix} 0&0&1&0&0&0&0&0\\ 0&0&1&0&0&0&0&0\\ 0&0&1&0&0&0&0&0\\ 0&0&1&0&0&0&0&0\\ 0&0&-1&0&0&0&0&0\\ 0&0&-1&0&0&0&0&0\\ 0&0&-1&0&0&0&0&0\\ 0&0&-1&0&0&0&0&0 \end{vmatrix} + \begin{vmatrix} 0&0&0&1&0&0&0&0\\ 0&0&0&1&0&0&0&0\\ 0&0&0&1&0&0&0&0\\ 0&0&0&1&0&0&0&0\\ 0&0&0&-1&0&0&0&0\\ 0&0&0&-1&0&0&0&0\\ 0&0&0&-1&0&0&0&0\\ 0&0&0&-1&0&0&0&0 \end{vmatrix} + \begin{vmatrix} 0&0&0&0&1&0&0&0\\ 0&0&0&0&1&0&0&0\\ 0&0&0&0&-1&0&0&0\\ 0&0&0&0&-1&0&0&0\\ 0&0&0&0&1&0&0&0\\ 0&0&0&0&1&0&0&0\\ 0&0&0&0&-1&0&0&0\\ 0&0&0&0&-1&0&0&0 \end{vmatrix} +$$

$$\begin{vmatrix} 0&0&0&0&0&1&0&0\\ 0&0&0&0&0&1&0&0\\ 0&0&0&0&0&-1&0&0\\ 0&0&0&0&0&-1&0&0\\ 0&0&0&0&0&1&0&0\\ 0&0&0&0&0&1&0&0\\ 0&0&0&0&0&-1&0&0\\ 0&0&0&0&0&-1&0&0 \end{vmatrix} + \begin{vmatrix} 0&0&0&0&0&0&-1&0\\ 0&0&0&0&0&0&-1&0\\ 0&0&0&0&0&0&-1&0\\ 0&0&0&0&0&0&-1&0\\ 0&0&0&0&0&0&1&0\\ 0&0&0&0&0&0&1&0\\ 0&0&0&0&0&0&1&0\\ 0&0&0&0&0&0&1&0 \end{vmatrix} + \begin{vmatrix} 0&0&0&0&0&0&0&-1\\ 0&0&0&0&0&0&0&-1\\ 0&0&0&0&0&0&0&-1\\ 0&0&0&0&0&0&0&-1\\ 0&0&0&0&0&0&0&1\\ 0&0&0&0&0&0&0&1\\ 0&0&0&0&0&0&0&1\\ 0&0&0&0&0&0&0&1 \end{vmatrix}$$

Figure 7. The «column decomposition» $R_8 = s_0+s_1+s_2+s_3+s_4+s_5+s_6+s_7$ of the Rademacher (8*8)-matrix $R_8$ (Figure 1) where every of matrices $s_0$, $s_1$, $s_2$, $s_3$, $s_4$, $s_5$, $s_6$, $s_7$ is a projection operator

$$\mathbf{R_8=v_0+v_1+v_2+v_3+v_4+v_5+v_6+v_7=} \begin{vmatrix} 1&1&1&1&1&1&-1&-1\\ 0&0&0&0&0&0&0&0\\ 0&0&0&0&0&0&0&0\\ 0&0&0&0&0&0&0&0\\ 0&0&0&0&0&0&0&0\\ 0&0&0&0&0&0&0&0\\ 0&0&0&0&0&0&0&0\\ 0&0&0&0&0&0&0&0 \end{vmatrix} + \begin{vmatrix} 0&0&0&0&0&0&0&0\\ 1&1&1&1&1&1&-1&-1\\ 0&0&0&0&0&0&0&0\\ 0&0&0&0&0&0&0&0\\ 0&0&0&0&0&0&0&0\\ 0&0&0&0&0&0&0&0\\ 0&0&0&0&0&0&0&0\\ 0&0&0&0&0&0&0&0 \end{vmatrix} +$$

$$\begin{vmatrix} 0&0&0&0&0&0&0&0\\ 0&0&0&0&0&0&0&0\\ -1&-1&1&1&-1&-1&-1&-1\\ 0&0&0&0&0&0&0&0\\ 0&0&0&0&0&0&0&0\\ 0&0&0&0&0&0&0&0\\ 0&0&0&0&0&0&0&0\\ 0&0&0&0&0&0&0&0 \end{vmatrix} + \begin{vmatrix} 0&0&0&0&0&0&0&0\\ 0&0&0&0&0&0&0&0\\ 0&0&0&0&0&0&0&0\\ -1&-1&1&1&-1&-1&-1&-1\\ 0&0&0&0&0&0&0&0\\ 0&0&0&0&0&0&0&0\\ 0&0&0&0&0&0&0&0\\ 0&0&0&0&0&0&0&0 \end{vmatrix} + \begin{vmatrix} 0&0&0&0&0&0&0&0\\ 0&0&0&0&0&0&0&0\\ 0&0&0&0&0&0&0&0\\ 0&0&0&0&0&0&0&0\\ 1&1&-1&-1&1&1&1&1\\ 0&0&0&0&0&0&0&0\\ 0&0&0&0&0&0&0&0\\ 0&0&0&0&0&0&0&0 \end{vmatrix} +$$

$$\begin{vmatrix} 0&0&0&0&0&0&0&0\\ 0&0&0&0&0&0&0&0\\ 0&0&0&0&0&0&0&0\\ 0&0&0&0&0&0&0&0\\ 0&0&0&0&0&0&0&0\\ 1&1&-1&-1&1&1&1&1\\ 0&0&0&0&0&0&0&0\\ 0&0&0&0&0&0&0&0 \end{vmatrix} + \begin{vmatrix} 0&0&0&0&0&0&0&0\\ 0&0&0&0&0&0&0&0\\ 0&0&0&0&0&0&0&0\\ 0&0&0&0&0&0&0&0\\ 0&0&0&0&0&0&0&0\\ 0&0&0&0&0&0&0&0\\ -1&-1&-1&-1&-1&-1&1&1\\ 0&0&0&0&0&0&0&0 \end{vmatrix} + \begin{vmatrix} 0&0&0&0&0&0&0&0\\ 0&0&0&0&0&0&0&0\\ 0&0&0&0&0&0&0&0\\ 0&0&0&0&0&0&0&0\\ 0&0&0&0&0&0&0&0\\ 0&0&0&0&0&0&0&0\\ 0&0&0&0&0&0&0&0\\ -1&-1&-1&-1&-1&-1&1&1 \end{vmatrix}$$

Figure 8. The «row decomposition» $R_8 = v_0+v_1+v_2+v_3+v_4+v_5+v_6+v_7$ of the Rademacher (8*8)-matrix $R_8$ (Figure 1) where every of matrices $v_0$, $v_1$, $v_2$, $v_3$, $v_4$, $v_5$, $v_6$, $v_7$ is a projection operator

By analogy with the described case of the projection (4*4)-operators $c_0$, $c_1$, $c_2$, $c_3$ and $r_0$, $r_1$, $r_2$, $r_3$ (Figures 3-6), one can analyse features of sums of pairs of the column projection (8*8)-operators $s_0$, $s_1$, $s_2$, $s_3$, $s_4$, $s_5$, $s_6$, $s_7$ (Figure 7) and of the row projection (8*8)-operators $v_0$, $v_1$, $v_2$, $v_3$, $v_4$, $v_5$, $v_6$, $v_7$ in relation to their exponentiation. In other words, one can analyze features of matrices $(s_0+s_1)^n$, $(s_0+s_3)^n$,…. and $(v_0+v_1)^n$, $(v_0+v_2)^n$, …. where $n = 1, 2, 3,\ldots$ . Such analysis leads to resulting tables on Figure 9.

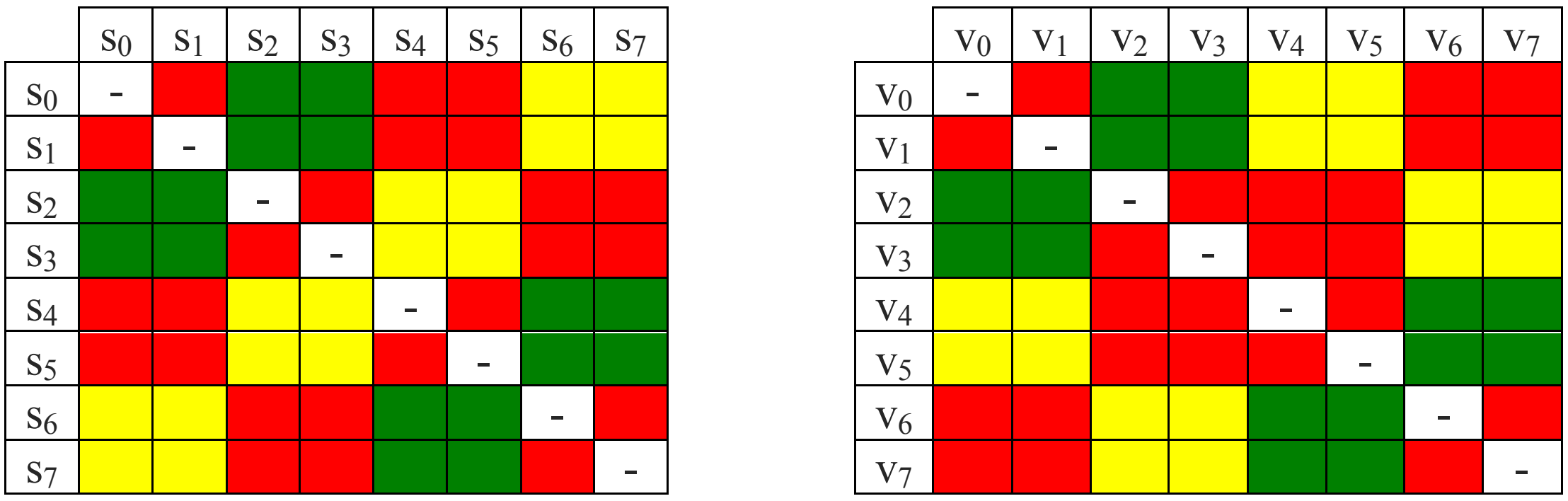

Figure 9. Tables of some features of sums of pairs of different column projectors $s_0, s_1, \ldots, s_7$ (from Figure 7) and of row projectors $v_0, v_1, \ldots, v_7$ (from the Rademacher matrix $R_8$ on Figure 8) in relation to their exponentiation. Explanations in text.

Every of cells in these tables on Figure 9 represents a sum of those projectors which denote its column and row by analogy with Figure 4. Again we have three types of such sums which are marked by green, red and yellow and which possess the similar properties in comparison with the cases on Figure 4.

In each of tables on Figure 9, green cells correspond to those matrices, exponentiations of which generate 8 cyclic groups. The left table contains 16 green cells that correspond to the following cyclic groups with a period 8: $(2^{-0.5}*(s_0+s_2))^n$, $(2^{-0.5}*(s_0+s_3))^n$, $(2^{-0.5}*(s_1+s_2))^n$, $(2^{-0.5}*(s_1+s_3))^n$, $(2^{-0.5}*(s_4+s_6))^n$, $(2^{-0.5}*(s_4+s_7))^n$, $(2^{-0.5}*(s_5+s_6))^n$, $(2^{-0.5}*(s_5+s_7))^n$. The right table contains 16 green cells with the same tabular location that correspond to the following cyclic groups with a period 8: $(2^{-0.5}*(v_0+v_2))^n$, $(2^{-0.5}*(v_0+v_3))^n$, $(2^{-0.5}*(v_1+v_2))^n$, $(2^{-0.5}*(v_1+v_3))^n$, $(2^{-0.5}*(v_4+v_6))^n$, $(2^{-0.5}*(v_4+v_7))^n$, $(2^{-0.5}*(v_5+v_6))^n$, $(2^{-0.5}*(v_5+v_7))^n$.

Red cells in tables on Figure 9 contain those matrices which possess a doubling property in relation to their exponentiation. The left table contains 24 red cells, matrices of which satisfy the following feature: $(s_0+s_1)^n = 2^{n-1}*(s_0+s_1)$, $(s_0+s_4)^n = 2^{n-1}*(s_0+s_4)$, etc. The right table also contains 24 red cells with the same feature but with another tabular location: $(v_0+v_1)^n = 2^{n-1}*(v_0+v_1)$, $(v_0+v_4)^n = 2^{n-1}*(v_0+v_4)$, etc.

16 yellow cells in each of tables on Figure 9 contain those matrices, which have «quadruplet property»: $((s_0+s_6)^2)^n = 4^{n-1}*(s_0+s_6)^2$, $((v_0+v_4)^2)^n = 4^{n-1}*(v_0+v_4)^2$, etc. The cells on the main diagonal correspond to sum of two projectors themselves: $(s_i+s_i)^n = 2^n*s_i$, $(v_i+v_i)^n = 2^n*v_i$, where i = 0, 1, 2, ..., 7.

## 3. GENETIC HADAMARD MATRICES AS SUMS OF PROJECTORS

The genetic Hadamard matrix $H_4$ from Figure 1 can be also decomposed into sum of 4 sparse matrices $H_4 = h_0+h_1+h_2+h_3$ where each of sparse matrices contains only one non-zero column (in a case of the «column decomposition») or only one non-zero row (in a case of the «row decomposition») (Figure 10).

$$H_4 = h_0+h_1+h_2+h_3 = \begin{vmatrix} 1 & 0 & 0 & 0 \\ -1 & 0 & 0 & 0 \\ 1 & 0 & 0 & 0 \\ -1 & 0 & 0 & 0 \end{vmatrix} + \begin{vmatrix} 0 & 1 & 0 & 0 \\ 0 & 1 & 0 & 0 \\ 0 & -1 & 0 & 0 \\ 0 & -1 & 0 & 0 \end{vmatrix} + \begin{vmatrix} 0 & 0 & -1 & 0 \\ 0 & 0 & 1 & 0 \\ 0 & 0 & 1 & 0 \\ 0 & 0 & -1 & 0 \end{vmatrix} + \begin{vmatrix} 0 & 0 & 0 & 1 \\ 0 & 0 & 0 & 1 \\ 0 & 0 & 0 & 1 \\ 0 & 0 & 0 & 1 \end{vmatrix}$$

$$H_4 = g_0+g_1+g_2+g_3 = \begin{vmatrix} 1 & 1 & -1 & 1 \\ 0 & 0 & 0 & 0 \\ 0 & 0 & 0 & 0 \\ 0 & 0 & 0 & 0 \end{vmatrix} + \begin{vmatrix} 0 & 0 & 0 & 0 \\ -1 & 1 & 1 & 1 \\ 0 & 0 & 0 & 0 \\ 0 & 0 & 0 & 0 \end{vmatrix} + \begin{vmatrix} 0 & 0 & 0 & 0 \\ 0 & 0 & 0 & 0 \\ 1 & -1 & 1 & 1 \\ 0 & 0 & 0 & 0 \end{vmatrix} + \begin{vmatrix} 0 & 0 & 0 & 0 \\ 0 & 0 & 0 & 0 \\ 0 & 0 & 0 & 0 \\ -1 & -1 & -1 & 1 \end{vmatrix}$$

Figure 10. The «column decomposition» (upper layer) and the «row decomposition» (bottom layer) of the Hadamard matrix $H_4$ from Figure 1

Each of these sparse matrices $h_0$, $h_1$, $h_2$, $h_3$ and $g_0$, $g_1$, $g_2$, $g_3$ on Fugure 10 is a projector. We will conditionally name projectors $h_0$, $h_1$, $h_2$, $h_3$ again as «column projectors» and projectors $g_0$, $g_1$, $g_2$, $g_3$ as «row projectors».

By analogy with the previous section about the Rademacher matrix $R_4$, one can analyse features of sums of pairs of these column projectors and row projectors in relation to their exponentiation. In other words, one can analyze features of matrices $(h_0+h_1)^n$, $(h_0+h_2)^n$,…. and $(g_0+g_1)^n$, $(g_0+g_2)^n$, …. where n =1, 2, 3,… . Such analysis leads to resulting tables on Figure 11.

| | $h_0$ | $h_1$ | $h_2$ | $h_3$ |
|---|---|---|---|---|
| $h_0$ | - | | | |
| $h_1$ | | - | | |
| $h_2$ | | | - | |
| $h_3$ | | | | - |

| | $g_0$ | $g_1$ | $g_2$ | $g_3$ |
|---|---|---|---|---|
| $g_0$ | - | | | |
| $g_1$ | | - | | |
| $g_2$ | | | - | |
| $g_3$ | | | | - |

Figure 11. Tables of some features of sums of pairs of the different «column projectors» $h_0$, $h_1$, $h_2$, $h_3$ (in the left table) and of the «row projectors» $g_0$, $g_1$, $g_2$, $g_3$ (from the Hadamard matrix $H_4$ on Figure 10) in relation to their exponentiation. Explanations in text.

Both tables on Figure 11 are identical in their mosaics. Every of their cells correspond to a sum of those column projectors (or row projectors), which denote its column and row. 12 green cells correspond to matrices, exponentiations of which lead to cyclic groups with a period 8: $(2^{-0.5}*(h_0+h_1))^n$, $(2^{-0.5}*(h_0+h_2))^n$, … , $(2^{-0.5}*(g_0+g_1))^n$, $(2^{-0.5}*(g_0+g_1))^n$, … , where n = 1, 2, 3,… . The cells on the main diagonal correspond to sum of two projectors themselves: $(h_i+h_i)^n = 2^n*h_i$, $(g_i+g_i)^n = 2^n*g_i$, where i = 0, 1, 2, 3.

Cyclic properties of these (4*4)-matrix operators exist due to a connection of these operators with complex numbers. Figures 12 and 13 show existence of a special decomposition of every of these (4*4)-matrices into such set of two sparse matrices $e_0$ and $e_1$, which is closed relative to multiplication and which defines their multiplication table that coincides with the known multiplication table of complex numbers.

$$\mathbf{h_0+h_1} = \begin{vmatrix} 1 & 1 & 0 & 0 \\ -1 & 1 & 0 & 0 \\ 1 & -1 & 0 & 0 \\ -1 & -1 & 0 & 0 \end{vmatrix} = \mathbf{e_0+e_1} = \begin{vmatrix} 1 & 0 & 0 & 0 \\ 0 & 1 & 0 & 0 \\ 0 & -1 & 0 & 0 \\ -1 & 0 & 0 & 0 \end{vmatrix} + \begin{vmatrix} 0 & 1 & 0 & 0 \\ -1 & 0 & 0 & 0 \\ 1 & 0 & 0 & 0 \\ 0 & -1 & 0 & 0 \end{vmatrix} ;$$

| | $e_0$ | $e_1$ |
|---|---|---|
| $e_0$ | $e_0$ | $e_1$ |
| $e_1$ | $e_1$ | $-e_0$ |

$$h_0+h_2 = \begin{matrix} 1 & 0 & -1 & 0 \\ -1 & 0 & 1 & 0 \\ 1 & 0 & 1 & 0 \\ -1 & 0 & -1 & 0 \end{matrix} = e_0+e_1 = \begin{matrix} 1 & 0 & 0 & 0 \\ -1 & 0 & 0 & 0 \\ 0 & 0 & 1 & 0 \\ 0 & 0 & -1 & 0 \end{matrix} + \begin{matrix} 0 & 0 & -1 & 0 \\ 0 & 0 & 1 & 0 \\ 1 & 0 & 0 & 0 \\ -1 & 0 & 0 & 0 \end{matrix} ;$$

| | $e_0$ | $e_1$ |
|---|---|---|
| $e_0$ | $e_0$ | $e_1$ |
| $e_1$ | $e_1$ | $-e_0$ |

$$h_0+h_3 = \begin{matrix} 1 & 0 & 0 & 1 \\ -1 & 0 & 0 & 1 \\ 1 & 0 & 0 & 1 \\ -1 & 0 & 0 & 1 \end{matrix} = e_0+e_1 = \begin{matrix} 1 & 0 & 0 & 0 \\ 0 & 0 & 0 & 1 \\ 1 & 0 & 0 & 0 \\ 0 & 0 & 0 & 1 \end{matrix} + \begin{matrix} 0 & 0 & 0 & 1 \\ -1 & 0 & 0 & 0 \\ 0 & 0 & 0 & 1 \\ -1 & 0 & 0 & 0 \end{matrix} ;$$

| | $e_0$ | $e_1$ |
|---|---|---|
| $e_0$ | $e_0$ | $e_1$ |
| $e_1$ | $e_1$ | $-e_0$ |

$$h_1+h_2 = \begin{matrix} 0 & 1 & -1 & 0 \\ 0 & 1 & 1 & 0 \\ 0 & -1 & 1 & 0 \\ 0 & -1 & -1 & 0 \end{matrix} = e_0+e_1 = \begin{matrix} 0 & 0 & -1 & 0 \\ 0 & 1 & 0 & 0 \\ 0 & 0 & 1 & 0 \\ 0 & -1 & 0 & 0 \end{matrix} + \begin{matrix} 0 & 1 & 0 & 0 \\ 0 & 0 & 1 & 0 \\ 0 & -1 & 0 & 0 \\ 0 & 0 & -1 & 0 \end{matrix} ;$$

| | $e_0$ | $e_1$ |
|---|---|---|
| $e_0$ | $e_0$ | $e_1$ |
| $e_1$ | $e_1$ | $-e_0$ |

$$h_1+h_3 = \begin{matrix} 0 & 1 & 0 & 1 \\ 0 & 1 & 0 & 1 \\ 0 & -1 & 0 & 1 \\ 0 & -1 & 0 & 1 \end{matrix} = e_0+e_1 = \begin{matrix} 0 & 1 & 0 & 0 \\ 0 & 1 & 0 & 0 \\ 0 & 0 & 0 & 1 \\ 0 & 0 & 0 & 1 \end{matrix} + \begin{matrix} 0 & 0 & 0 & 1 \\ 0 & 0 & 0 & 1 \\ 0 & -1 & 0 & 0 \\ 0 & -1 & 0 & 0 \end{matrix} ;$$

| | $e_0$ | $e_1$ |
|---|---|---|
| $e_0$ | $e_0$ | $e_1$ |
| $e_1$ | $e_1$ | $-e_0$ |

$$h_2+h_3 = \begin{matrix} 0 & 0 & -1 & 1 \\ 0 & 0 & 1 & 1 \\ 0 & 0 & 1 & 1 \\ 0 & 0 & -1 & 1 \end{matrix} = e_0+e_1 = \begin{matrix} 0 & 0 & 0 & 1 \\ 0 & 0 & 1 & 0 \\ 0 & 0 & 1 & 0 \\ 0 & 0 & 0 & 1 \end{matrix} + \begin{matrix} 0 & 0 & -1 & 0 \\ 0 & 0 & 0 & 1 \\ 0 & 0 & 0 & 1 \\ 0 & 0 & -1 & 0 \end{matrix} ;$$

| | $e_0$ | $e_1$ |
|---|---|---|
| $e_0$ | $e_0$ | $e_1$ |
| $e_1$ | $e_1$ | $-e_0$ |

Figure 12. The table represents special decompositions of (4*4)-matrices ($h_0+h_1$), ($h_0+h_2$), ($h_0+h_3$), ($h_1+h_2$), ($h_1+h_3$), ($h_2+h_3$) into sum of two matrices $e_0+e_1$. The table shows direct relations of these matrices with matrix representations of 2-parametric complex numbers. Here $h_0$, $h_1$, $h_2$ and $h_3$ are column projectors of the Hadamard matrix $H_4$ from Figure 10. For each set of matrices $e_0$ and $e_1$ at every tabular level, the right column of the table contains its multiplication table, which coinsides with the multiplication table of complex numbers.

$$g_0+g_1 = \begin{matrix} 1 & 1 & -1 & 1 \\ -1 & 1 & 1 & 1 \\ 0 & 0 & 0 & 0 \\ 0 & 0 & 0 & 0 \end{matrix} = e_0+e_1 = \begin{matrix} 1 & 0 & -1 & 0 \\ 0 & 1 & 0 & 1 \\ 0 & 0 & 0 & 0 \\ 0 & 0 & 0 & 0 \end{matrix} + \begin{matrix} 0 & 1 & 0 & 1 \\ -1 & 0 & 1 & 0 \\ 0 & 0 & 0 & 0 \\ 0 & 0 & 0 & 0 \end{matrix} ;$$

| | $e_0$ | $e_1$ |
|---|---|---|
| $e_0$ | $e_0$ | $e_1$ |
| $e_1$ | $e_1$ | $-e_0$ |

$$g_0+g_2 = \begin{matrix} 1 & 1 & -1 & 1 \\ 0 & 0 & 0 & 0 \\ 1 & -1 & 1 & 1 \\ 0 & 0 & 0 & 0 \end{matrix} = e_0+e_1 = \begin{matrix} 1 & 0 & 0 & 1 \\ 0 & 0 & 0 & 0 \\ 0 & -1 & 1 & 0 \\ 0 & 0 & 0 & 0 \end{matrix} + \begin{matrix} 0 & 1 & -1 & 0 \\ 0 & 0 & 0 & 0 \\ 1 & 0 & 0 & 1 \\ 0 & 0 & 0 & 0 \end{matrix} ;$$

| | $e_0$ | $e_1$ |
|---|---|---|
| $e_0$ | $e_0$ | $e_1$ |
| $e_1$ | $e_1$ | $-e_0$ |

$$g_0+g_3 = \begin{vmatrix} 1 & 1 & -1 & 1 \\ 0 & 0 & 0 & 0 \\ 0 & 0 & 0 & 0 \\ -1 & -1 & -1 & 1 \end{vmatrix} = e_0+e_1 = \begin{vmatrix} 1 & 1 & 0 & 0 \\ 0 & 0 & 0 & 0 \\ 0 & 0 & 0 & 0 \\ 0 & 0 & -1 & 1 \end{vmatrix} + \begin{vmatrix} 0 & 0 & -1 & 1 \\ 0 & 0 & 0 & 0 \\ 0 & 0 & 0 & 0 \\ -1 & -1 & 0 & 0 \end{vmatrix};$$

| | $e_0$ | $e_1$ |
|---|---|---|
| $e_0$ | $e_0$ | $e_1$ |
| $e_1$ | $e_1$ | $-e_0$ |

$$g_1+g_2 = \begin{vmatrix} 0 & 0 & 0 & 0 \\ -1 & 1 & 1 & 1 \\ 1 & -1 & 1 & 1 \\ 0 & 0 & 0 & 0 \end{vmatrix} = e_0+e_1 = \begin{vmatrix} 0 & 0 & 0 & 0 \\ -1 & 1 & 0 & 0 \\ 0 & 0 & 1 & 1 \\ 0 & 0 & 0 & 0 \end{vmatrix} + \begin{vmatrix} 0 & 0 & 0 & 0 \\ 0 & 0 & 1 & 1 \\ 1 & -1 & 0 & 0 \\ 0 & 0 & 0 & 0 \end{vmatrix};$$

| | $e_0$ | $e_1$ |
|---|---|---|
| $e_0$ | $e_0$ | $e_1$ |
| $e_1$ | $e_1$ | $-e_0$ |

$$g_1+g_3 = \begin{vmatrix} 0 & 0 & 0 & 0 \\ -1 & 1 & 1 & 1 \\ 0 & 0 & 0 & 0 \\ -1 & -1 & -1 & 1 \end{vmatrix} = e_0+e_1 = \begin{vmatrix} 0 & 0 & 0 & 0 \\ 0 & 1 & 1 & 0 \\ 0 & 0 & 0 & 0 \\ -1 & 0 & 0 & 1 \end{vmatrix} + \begin{vmatrix} 0 & 0 & 0 & 0 \\ -1 & 0 & 0 & 1 \\ 0 & 0 & 0 & 0 \\ 0 & -1 & -1 & 0 \end{vmatrix};$$

| | $e_0$ | $e_1$ |
|---|---|---|
| $e_0$ | $e_0$ | $e_1$ |
| $e_1$ | $e_1$ | $-e_0$ |

$$g_2+g_3 = \begin{vmatrix} 0 & 0 & 0 & 0 \\ 0 & 0 & 0 & 0 \\ 1 & -1 & 1 & 1 \\ -1 & -1 & -1 & 1 \end{vmatrix} = e_0+e_1 = \begin{vmatrix} 0 & 0 & 0 & 0 \\ 0 & 0 & 0 & 0 \\ 1 & 0 & 1 & 0 \\ 0 & -1 & 0 & 1 \end{vmatrix} + \begin{vmatrix} 0 & 0 & 0 & 0 \\ 0 & 0 & 0 & 0 \\ 0 & -1 & 0 & 1 \\ -1 & 0 & -1 & 0 \end{vmatrix};$$

| | $e_0$ | $e_1$ |
|---|---|---|
| $e_0$ | $e_0$ | $e_1$ |
| $e_1$ | $e_1$ | $-e_0$ |

Fugure 13. The table represents special decompositions of (4*4)-matrices ($g_0$+$g_1$), ($g_0$+$g_2$), ($g_0$+$g_3$), ($g_1$+$g_2$), ($g_1$+$g_3$), ($g_2$+$g_3$) into sum of two matrices $e_0$+$e_1$. The table shows direct relations of these matrices with matrix representations of 2-parametric complex numbers. Here $g_0$, $g_1$, $g_2$ and $g_3$ are row projectors from Figure 10. For each set of matrices $e_0$ and $e_1$ at every tabular level, the right column of the table contains its multiplication table, which coinsides with the multiplication table of complex numbers.

Now let us turn to the genetic Hadamard matrix $H_8$ from Figure 1. It can be also decomposed into sum of 8 sparse matrices $H_8=u_0+u_1+u_2+u_3+u_4+u_5+u_6+u_7$ where each of sparse matrices contains only one non-zero column (in a case of the «column decomposition») or only one non-zero row (in a case of the «row decomposition») (Figure 14).

$$H_8=u_0+u_1+u_2+u_3+u_4+u_5+u_6+u_7 = \begin{vmatrix} 1 & 0 & 0 & 0 & 0 & 0 & 0 & 0 \\ 1 & 0 & 0 & 0 & 0 & 0 & 0 & 0 \\ -1 & 0 & 0 & 0 & 0 & 0 & 0 & 0 \\ -1 & 0 & 0 & 0 & 0 & 0 & 0 & 0 \\ 1 & 0 & 0 & 0 & 0 & 0 & 0 & 0 \\ 1 & 0 & 0 & 0 & 0 & 0 & 0 & 0 \\ -1 & 0 & 0 & 0 & 0 & 0 & 0 & 0 \\ -1 & 0 & 0 & 0 & 0 & 0 & 0 & 0 \end{vmatrix} + \begin{vmatrix} 0 & -1 & 0 & 0 & 0 & 0 & 0 & 0 \\ 0 & 1 & 0 & 0 & 0 & 0 & 0 & 0 \\ 0 & 1 & 0 & 0 & 0 & 0 & 0 & 0 \\ 0 & -1 & 0 & 0 & 0 & 0 & 0 & 0 \\ 0 & -1 & 0 & 0 & 0 & 0 & 0 & 0 \\ 0 & 1 & 0 & 0 & 0 & 0 & 0 & 0 \\ 0 & 1 & 0 & 0 & 0 & 0 & 0 & 0 \\ 0 & -1 & 0 & 0 & 0 & 0 & 0 & 0 \end{vmatrix} +$$

$$
\begin{vmatrix}
0&0&1&0&0&0&0&0\\
0&0&1&0&0&0&0&0\\
0&0&1&0&0&0&0&0\\
0&0&1&0&0&0&0&0\\
0&0&-1&0&0&0&0&0\\
0&0&-1&0&0&0&0&0\\
0&0&-1&0&0&0&0&0\\
0&0&-1&0&0&0&0&0
\end{vmatrix}
+
\begin{vmatrix}
0&0&0&-1&0&0&0&0\\
0&0&0&1&0&0&0&0\\
0&0&0&-1&0&0&0&0\\
0&0&0&1&0&0&0&0\\
0&0&0&1&0&0&0&0\\
0&0&0&-1&0&0&0&0\\
0&0&0&1&0&0&0&0\\
0&0&0&-1&0&0&0&0
\end{vmatrix}
+
\begin{vmatrix}
0&0&0&0&-1&0&0&0\\
0&0&0&0&-1&0&0&0\\
0&0&0&0&1&0&0&0\\
0&0&0&0&1&0&0&0\\
0&0&0&0&1&0&0&0\\
0&0&0&0&1&0&0&0\\
0&0&0&0&-1&0&0&0\\
0&0&0&0&-1&0&0&0
\end{vmatrix}
+
$$

$$
\begin{vmatrix}
0&0&0&0&0&1&0&0\\
0&0&0&0&0&-1&0&0\\
0&0&0&0&0&-1&0&0\\
0&0&0&0&0&1&0&0\\
0&0&0&0&0&-1&0&0\\
0&0&0&0&0&1&0&0\\
0&0&0&0&0&1&0&0\\
0&0&0&0&0&-1&0&0
\end{vmatrix}
+
\begin{vmatrix}
0&0&0&0&0&0&1&0\\
0&0&0&0&0&0&1&0\\
0&0&0&0&0&0&1&0\\
0&0&0&0&0&0&1&0\\
0&0&0&0&0&0&1&0\\
0&0&0&0&0&0&1&0\\
0&0&0&0&0&0&1&0\\
0&0&0&0&0&0&1&0
\end{vmatrix}
+
\begin{vmatrix}
0&0&0&0&0&0&0&-1\\
0&0&0&0&0&0&0&1\\
0&0&0&0&0&0&0&-1\\
0&0&0&0&0&0&0&1\\
0&0&0&0&0&0&0&-1\\
0&0&0&0&0&0&0&1\\
0&0&0&0&0&0&0&-1\\
0&0&0&0&0&0&0&1
\end{vmatrix}
$$

Figure 14. The «column decomposition» $H_8=u_0+u_1+u_2+u_3+u_4+u_5+u_6+u_7$ of the Hadamard (8*8)-matrix $H_8$ (Figure 1) where every of sparse matrices $u_0$, $u_1$, $u_2$, $u_3$, $u_4$, $u_5$, $u_6$, $u_7$ is a projection operator

$$
\mathbf{H_8=d_0+d_1+d_2+d_3+d_4+d_5+d_6+d_7} =
\begin{vmatrix}
1&-1&1&-1&-1&1&1&-1\\
0&0&0&0&0&0&0&0\\
0&0&0&0&0&0&0&0\\
0&0&0&0&0&0&0&0\\
0&0&0&0&0&0&0&0\\
0&0&0&0&0&0&0&0\\
0&0&0&0&0&0&0&0\\
0&0&0&0&0&0&0&0
\end{vmatrix}
+
\begin{vmatrix}
0&0&0&0&0&0&0&0\\
1&1&1&1&-1&-1&1&1\\
0&0&0&0&0&0&0&0\\
0&0&0&0&0&0&0&0\\
0&0&0&0&0&0&0&0\\
0&0&0&0&0&0&0&0\\
0&0&0&0&0&0&0&0\\
0&0&0&0&0&0&0&0
\end{vmatrix}
$$

$$
\begin{vmatrix}
0&0&0&0&0&0&0&0\\
0&0&0&0&0&0&0&0\\
-1&1&1&-1&1&-1&1&-1\\
0&0&0&0&0&0&0&0\\
0&0&0&0&0&0&0&0\\
0&0&0&0&0&0&0&0\\
0&0&0&0&0&0&0&0\\
0&0&0&0&0&0&0&0
\end{vmatrix}
+
\begin{vmatrix}
0&0&0&0&0&0&0&0\\
0&0&0&0&0&0&0&0\\
0&0&0&0&0&0&0&0\\
-1&-1&1&1&1&1&1&1\\
0&0&0&0&0&0&0&0\\
0&0&0&0&0&0&0&0\\
0&0&0&0&0&0&0&0\\
0&0&0&0&0&0&0&0
\end{vmatrix}
+
\begin{vmatrix}
0&0&0&0&0&0&0&0\\
0&0&0&0&0&0&0&0\\
0&0&0&0&0&0&0&0\\
0&0&0&0&0&0&0&0\\
1&-1&-1&1&1&-1&1&-1\\
0&0&0&0&0&0&0&0\\
0&0&0&0&0&0&0&0\\
0&0&0&0&0&0&0&0
\end{vmatrix}
$$

$$
\begin{vmatrix}
0&0&0&0&0&0&0&0\\
0&0&0&0&0&0&0&0\\
0&0&0&0&0&0&0&0\\
0&0&0&0&0&0&0&0\\
0&0&0&0&0&0&0&0\\
1&1&-1&-1&1&1&1&1\\
0&0&0&0&0&0&0&0\\
0&0&0&0&0&0&0&0
\end{vmatrix}
+
\begin{vmatrix}
0&0&0&0&0&0&0&0\\
0&0&0&0&0&0&0&0\\
0&0&0&0&0&0&0&0\\
0&0&0&0&0&0&0&0\\
0&0&0&0&0&0&0&0\\
0&0&0&0&0&0&0&0\\
-1&1&-1&1&-1&1&1&-1\\
0&0&0&0&0&0&0&0
\end{vmatrix}
+
\begin{vmatrix}
0&0&0&0&0&0&0&0\\
0&0&0&0&0&0&0&0\\
0&0&0&0&0&0&0&0\\
0&0&0&0&0&0&0&0\\
0&0&0&0&0&0&0&0\\
0&0&0&0&0&0&0&0\\
0&0&0&0&0&0&0&0\\
-1&-1&-1&-1&-1&-1&1&1
\end{vmatrix}
$$

Figure 15. The «row decomposition» $H_8=d_0+d_1+d_2+d_3+d_4+d_5+d_6+d_7$ of the Hadamard (8*8)-matrix $H_8$ (Figure 1) where every of sparse matrices $d_0$, $d_1$, $d_2$, $d_3$, $d_4$, $d_5$, $d_6$, $d_7$ is a projection operator

Every of these sparse matrices $u_0$, $u_1$, $u_2$, $u_3$, $u_4$, $u_5$, $u_6$, $u_7$ and $d_0$, $d_1$, $d_2$, $d_3$, $d_4$, $d_5$, $d_6$, $d_7$ on Fugures 14, 15 is a projector. We will conditionally name projectors $u_0$, $u_1$, $u_2$, $u_3$, $u_4$, $u_5$, $u_6$, $u_7$ again as «column projectors» and projectors $d_0$, $d_1$, $d_2$, $d_3$, $d_4$, $d_5$, $d_6$, $d_7$ as «row projectors».

By analogy with the previous sections, one can analyse features of sums of pairs of these column projectors and row projectors in relation to their exponentiation. In other words, one can analyze features of matrices $(u_0+u_1)^n$, $(u_0+u_3)^n$,.... and $(d_0+d_1)^n$, $(d_0+d_2)^n$, .... where n =1, 2, 3,.... Such analysis leads to resulting tables on Figure 16.

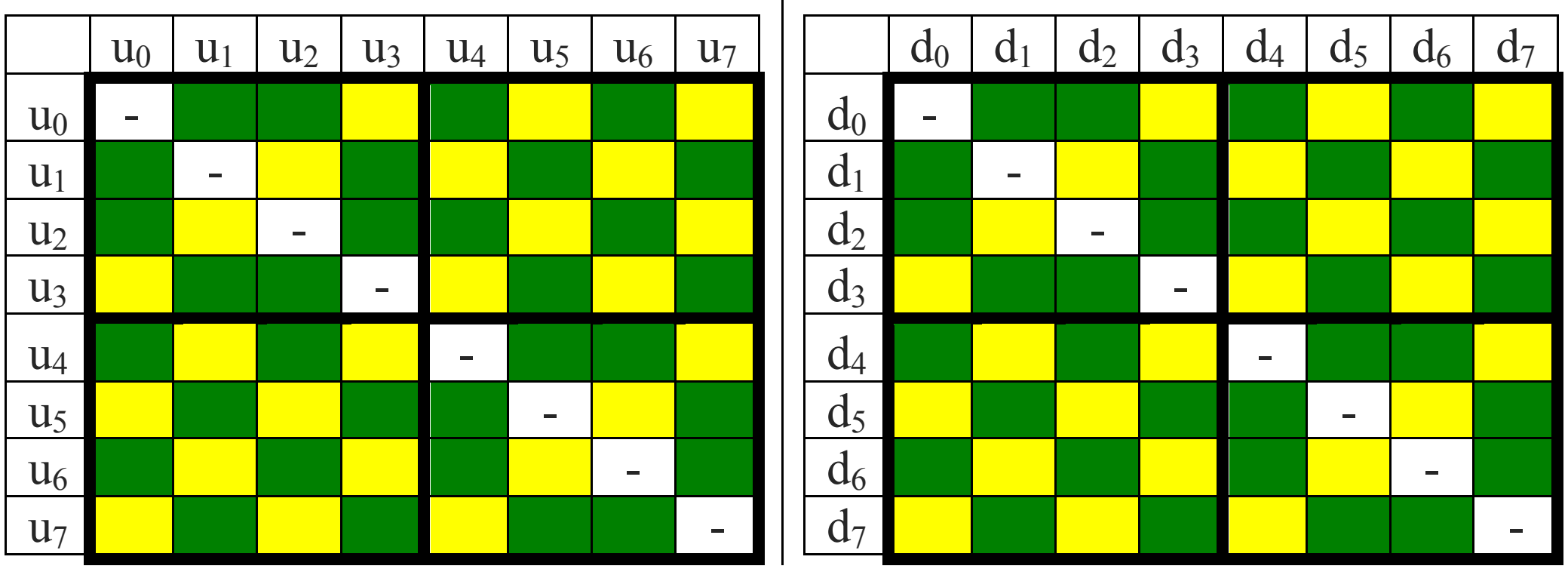

Figure 16. Tables of some features of sums of pairs of the different column projectors $u_0$, $u_1$, …, $u_7$ (from Figure 14) and of the row projectors $d_0$, $d_1$, …, $d_7$ (from Figure 15) in relation to their exponentiation. This is the case of the Hadamard matrix $H_8$ from Figure 1. Explanations in text.

Both tables on Figure 16 have the identical mosaic with 32 green cells and 24 yellow cells. The green cells in these tables correspond to those matrices, exponentiations of which generate cyclic groups with a period 8:

- $(2^{-0.5}*(u_0+u_1))^n$, $(2^{-0.5}*(u_0+u_2))^n$, $(2^{-0.5}*(u_0+u_4))^n$, $(2^{-0.5}*(u_0+u_6))^n$, $(2^{-0.5}*(u_1+u_3))^n$, $(2^{-0.5}*(u_1+u_5))^n$, $(2^{-0.5}*(u_1+u_7))^n$, $(2^{-0.5}*(u_2+u_3))^n$, $(2^{-0.5}*(u_2+u_4))^n$, $(2^{-0.5}*(u_2+u_6))^n$, $(2^{-0.5}*(u_3+u_5))^n$, $(2^{-0.5}*(u_3+u_7))^n$, $(2^{-0.5}*(u_4+u_5))^n$, $(2^{-0.5}*(u_4+u_6))^n$, $(2^{-0.5}*(u_5+u_7))^n$, $(2^{-0.5}*(u_6+u_7))^n$ (in the left table);
- $(2^{-0.5}*(d_0+d_1))^n$, $(2^{-0.5}*(d_0+d_2))^n$, $(2^{-0.5}*(d_0+d_4))^n$, $(2^{-0.5}*(d_0+d_6))^n$, $(2^{-0.5}*(d_1+d_3))^n$, $(2^{-0.5}*(d_1+d_5))^n$, $(2^{-0.5}*(d_1+d_7))^n$, $(2^{-0.5}*(d_2+d_3))^n$, $(2^{-0.5}*(d_2+d_4))^n$, $(2^{-0.5}*(d_2+d_6))^n$, $(2^{-0.5}*(d_3+d_5))^n$, $(2^{-0.5}*(d_3+d_7))^n$, $(2^{-0.5}*(d_4+d_5))^n$, $(2^{-0.5}*(d_4+d_6))^n$, $(2^{-0.5}*(d_5+d_7))^n$, $(2^{-0.5}*(d_6+d_7))^n$ (in the right table).

Cyclic properties of these (8*8)-matrix operators exist due to a connection of these operators with complex numbers. Figure 17 shows some examples of decompositions of the (8*8)-matrices from green cells on Figure 16 into corresponding sets of two sparse matrices, each of which is closed in relation to multiplication and each of which defines the multiplication table of complex numbers (see some additional details about representations of complex numbers by means of $(2^n*2^n)$-matrices in [Petoukhov, 2012b]). It should be noted here that our study in the field of matrix genetics has revealed methods of extension of these (8*8)-genetic matrices $R_4$, $R_8$, $H_4$, $H_8$ (Figure 1) into $(2^n*2^n)$-matrices which are also sums of "column projectors" and "row projectors" and which give by analogy as much cyclic groups as needed to model big ensembles of cyclic processes.

$$u_0+u_4 = \begin{bmatrix} 1&0&0&0&-1&0&0&0 \\ 1&0&0&0&-1&0&0&0 \\ -1&0&0&0&1&0&0&0 \\ -1&0&0&0&1&0&0&0 \\ 1&0&0&0&1&0&0&0 \\ 1&0&0&0&1&0&0&0 \\ -1&0&0&0&-1&0&0&0 \\ -1&0&0&0&-1&0&0&0 \end{bmatrix} = \begin{bmatrix} 1&0&0&0&0&0&0&0 \\ 1&0&0&0&0&0&0&0 \\ -1&0&0&0&0&0&0&0 \\ -1&0&0&0&0&0&0&0 \\ 0&0&0&0&1&0&0&0 \\ 0&0&0&0&1&0&0&0 \\ 0&0&0&0&-1&0&0&0 \\ 0&0&0&0&-1&0&0&0 \end{bmatrix} + \begin{bmatrix} 0&0&0&0&-1&0&0&0 \\ 0&0&0&0&-1&0&0&0 \\ 0&0&0&0&1&0&0&0 \\ 0&0&0&0&1&0&0&0 \\ 1&0&0&0&0&0&0&0 \\ 1&0&0&0&0&0&0&0 \\ -1&0&0&0&0&0&0&0 \\ -1&0&0&0&0&0&0&0 \end{bmatrix} = e_0+e_4;$$

| | $e_0$ | $e_4$ |
|---|---|---|
| $e_0$ | $e_0$ | $e_4$ |
| $e_4$ | $e_4$ | $-e_0$ |

$$u_1+u_5 = \begin{bmatrix} 0&-1&0&0&0&1&0&0 \\ 0&1&0&0&0&-1&0&0 \\ 0&1&0&0&0&-1&0&0 \\ 0&-1&0&0&0&1&0&0 \\ 0&-1&0&0&0&-1&0&0 \\ 0&1&0&0&0&1&0&0 \\ 0&1&0&0&0&1&0&0 \\ 0&-1&0&0&0&-1&0&0 \end{bmatrix} = \begin{bmatrix} 0&-1&0&0&0&0&0&0 \\ 0&1&0&0&0&0&0&0 \\ 0&1&0&0&0&0&0&0 \\ 0&-1&0&0&0&0&0&0 \\ 0&0&0&0&0&-1&0&0 \\ 0&0&0&0&0&1&0&0 \\ 0&0&0&0&0&1&0&0 \\ 0&0&0&0&0&-1&0&0 \end{bmatrix} + \begin{bmatrix} 0&0&0&0&0&1&0&0 \\ 0&0&0&0&0&-1&0&0 \\ 0&0&0&0&0&-1&0&0 \\ 0&0&0&0&0&1&0&0 \\ 0&-1&0&0&0&0&0&0 \\ 0&1&0&0&0&0&0&0 \\ 0&1&0&0&0&0&0&0 \\ 0&-1&0&0&0&0&0&0 \end{bmatrix} = e_1+e_5;$$

| | $e_1$ | $e_5$ |
|---|---|---|
| $e_1$ | $e_1$ | $e_5$ |
| $e_5$ | $e_5$ | $-e_1$ |

$$u_2+u_6 = \begin{bmatrix} 0&0&1&0&0&0&1&0 \\ 0&0&1&0&0&0&1&0 \\ 0&0&1&0&0&0&1&0 \\ 0&0&1&0&0&0&1&0 \\ 0&0&-1&0&0&0&1&0 \\ 0&0&-1&0&0&0&1&0 \\ 0&0&-1&0&0&0&1&0 \\ 0&0&-1&0&0&0&1&0 \end{bmatrix} = \begin{bmatrix} 0&0&1&0&0&0&0&0 \\ 0&0&1&0&0&0&0&0 \\ 0&0&1&0&0&0&0&0 \\ 0&0&1&0&0&0&0&0 \\ 0&0&0&0&0&0&1&0 \\ 0&0&0&0&0&0&1&0 \\ 0&0&0&0&0&0&1&0 \\ 0&0&0&0&0&0&1&0 \end{bmatrix} + \begin{bmatrix} 0&0&0&0&0&0&1&0 \\ 0&0&0&0&0&0&1&0 \\ 0&0&0&0&0&0&1&0 \\ 0&0&0&0&0&0&1&0 \\ 0&0&-1&0&0&0&0&0 \\ 0&0&-1&0&0&0&0&0 \\ 0&0&-1&0&0&0&0&0 \\ 0&0&-1&0&0&0&0&0 \end{bmatrix} = e_2+e_6;$$

| | $e_2$ | $e_6$ |
|---|---|---|
| $e_2$ | $e_2$ | $e_6$ |
| $e_6$ | $e_6$ | $-e_2$ |

$$u_3+u_7 = \begin{bmatrix} 0&0&0&-1&0&0&0&-1 \\ 0&0&0&1&0&0&0&1 \\ 0&0&0&-1&0&0&0&-1 \\ 0&0&0&1&0&0&0&1 \\ 0&0&0&1&0&0&0&-1 \\ 0&0&0&-1&0&0&0&1 \\ 0&0&0&1&0&0&0&-1 \\ 0&0&0&-1&0&0&0&1 \end{bmatrix} = \begin{bmatrix} 0&0&0&-1&0&0&0&0 \\ 0&0&0&1&0&0&0&0 \\ 0&0&0&-1&0&0&0&0 \\ 0&0&0&1&0&0&0&0 \\ 0&0&0&0&0&0&0&-1 \\ 0&0&0&0&0&0&0&1 \\ 0&0&0&0&0&0&0&-1 \\ 0&0&0&0&0&0&0&1 \end{bmatrix} + \begin{bmatrix} 0&0&0&0&0&0&0&-1 \\ 0&0&0&0&0&0&0&1 \\ 0&0&0&0&0&0&0&-1 \\ 0&0&0&0&0&0&0&1 \\ 0&0&0&1&0&0&0&0 \\ 0&0&0&-1&0&0&0&0 \\ 0&0&0&1&0&0&0&0 \\ 0&0&0&-1&0&0&0&0 \end{bmatrix} = e_3+e_7;$$

| | $e_3$ | $e_7$ |
|---|---|---|
| $e_3$ | $e_3$ | $e_7$ |
| $e_7$ | $e_7$ | $-e_3$ |

Figure 17. The decomposition of the (8*8)-matrices $u_0+u_4$, $u_1+u_5$, $u_2+u_6$, $u_3+u_7$, which are examples of (8*8)-matrices from green cells on Figure 16, into corresponding sets of two sparse matrices $e_0$ and $e_4$, $e_1$ and $e_5$, $e_2$ and $e_6$, $e_3$ and $e_7$, each of which is closed in relation to multiplication and each of which defines the multiplication table of complex numbers (on the right)

Figure 17 testifies that the Hadamard (8*8)-matrix $H_8 = (u_0+u_4)+(u_1+u_5)+(u_2+u_6)+(u_3+u_7)$ (Fig. 1) is a sum of 4 complex numbers in 8-dimensional space.

Yellow cells in tables on Figure 16 correspond to matrices with the following property: $((u_i+u_j)^2)^n = 2^{n-1}*(u_i+u_j)$ and $((d_i+d_j)^2)^n = 2^{n-1}*(d_i+d_j)$ where $i \neq j$, $i,j = 0, 1, 2, \ldots, 7$, $n = 1, 2, 3, \ldots$ . Cells on the main diagonal correspond to matrices $(u_i+u_i)^n = 2^n*u_i$.

## 4. INHERITED BIOCYCLES AND A SELECTIVE CONTROL OF CYCLIC CHANGES OF VECTORS IN A MULTIDIMENSIONAL SPACE. PROBLEMS OF GENETIC BIOMECHANICS

Any living organism is an object with a huge ensemble of inherited cyclic processes, which form a hierarchy at different levels. Even every protein is involved in a cycle of the "birth and death," because after a certain time it breaks down into its constituent amino acids and they are then collected into a new protein. According to chronomedicine and biorhythmology, various diseases of the body are associated with disturbances (dys-synchronization) in these cooperative ensembles of biocycles. All inherited physiological subsystems of the body should be agreed with the structural organization of genetic coding for their coding and transmission to descendants; in other words, they bear the stamp of its features. We develop a "genetic biomechanics", which studies deep coherence between inherited physiological systems and molecular-genetic structures.

Our discovery of the described cyclic groups (on basis of genetic projectors), which are connected with phenomenological properties of molecular-genetic systems in their matrix forms of representation, gives a mathematical approach to simulate ensembles of cyclic processes. In this approach an idea of multi-dimensional vector space is used to simulate inherited biological phenomena including cooperative ensembles of cyclic processes. Multidimensional vectors of this bioinformation space can be changed under influence of those matrix operators on the basis of genetic projectors that were decribed in previous section. Due to special properties of these operators a useful possibility exists to provide a selective control (or a selective coding) of cyclic changes (and some other changes) of separate coordinates of multidimensional vectors in this space.

Let us explain this by one example. Let us take, for instance, the cyclic group of operators $Y^n = (2^{-0.5}*(s_0+s_2))^n$ (see Figures 7 and 9, on the left) and an arbitrary 8-dimensional vector $X=[x_0, x_1, x_2, x_3, x_4, x_5, x_6, x_7]$. Then let us analyze an expression $X*Y^n = [x_0, x_1, x_2, x_3, x_4, x_5, x_6, x_7]*(2^{-0.5}*(s_0+s_2))^n = Z_n$ that leads to a new vector $Z_n=[z_0, z_1, z_2, z_3, z_4, z_5, z_6, z_7]$ (here $n = 1, 2, 3, \ldots$). Figure 18 shows a cyclic transformation of coordinates of vectors $Z_n$; vectors $X*Y^1$ and $X*Y^9$ are identical because the period of this cyclic group $Y^n = (2^{-0.5}*(s_0+s_2))^n$ is equal to 8.

| | |
|---|---|
| $X*Y^1 = Z_1 =$ | $2^{-0.5}*[(x_0+x_1-x_2-x_3+x_4+x_5-x_6-x_7),\ 0,\ (x_0+x_1+x_2+x_3-x_4-x_5-x_6-x_7),\ 0, 0, 0, 0, 0]$ |
| $X*Y^2 = Z_2 =$ | $[(x_4-x_3-x_2+x_5),\ 0,\ (x_0+x_1-x_6-x_7),\ 0, 0, 0, 0, 0]$ |
| $X*Y^3 = Z_3 =$ | $2^{-0.5}*[(x_4-x_1-x_2-x_3-x_0+x_5+x_6+x_7),\ 0,\ (x_0+x_1-x_2-x_3+x_4+x_5-x_6-x_7),\ 0, 0, 0, 0, 0]$ |
| $X*Y^4 = Z_4 =$ | $[(x_6-x_1-x_0+x_7),\ 0,\ (x_4-x_3-x_2+x_5),\ 0, 0, 0, 0, 0]$ |
| $X*Y^5 = Z_5 =$ | $2^{-0.5}*[(x_2-x_1-x_0+x_3-x_4-x_5+x_6+x_7),\ 0,\ (x_4-x_1-x_2-x_3-x_0+x_5+x_6+x_7), 0, 0, 0, 0, 0]$ |
| $X*Y^6 = Z_6 =$ | $[(x_2+x_3-x_4-x_5),\ 0,\ (x_6-x_1-x_0+x_7),\ 0, 0, 0, 0, 0]$ |
| $X*Y^7 = Z_7 =$ | $2^{-0.5}*[(x_0+x_1+x_2+x_3-x_4-x_5-x_6-x_7),\ 0,\ (x_2-x_1-x_0+x_3-x_4-x_5+x_6+x_7),\ 0, 0, 0, 0, 0]$ |
| $X*Y^8 = Z_8 =$ | $[(x_0+x_1-x_6-x_7),\ 0,\ (x_2+x_3-x_4-x_5),\ 0, 0, 0, 0, 0]$ |
| $X*Y^9 = Z_9 =$ | $2^{-0.5}*[(x_0+x_1-x_2-x_3+x_4+x_5-x_6-x_7),\ 0,\ (x_0+x_1+x_2+x_3-x_4-x_5-x_6-x_7),\ 0, 0, 0, 0, 0]$ |

Figure 18. The illustration of a selective control (or selective coding) of cyclic changes of coordinates of a vector $X*Y^n = Z_n$ where $Y^n$ is the cyclic group with its period 8.

One can see from Figure 18 that only coordinates $z_0$ and $z_2$ have cyclic changes in this set of new vectors $Z_n$, all other coordinates are equal to zero. In other words, all cycles are realized on a 2-dimensional plane ($z_0$, $z_2$) inside the 8-dimensional space. If one uses another cyclic group of operators, for example, $(2^{-0.5}*(s_1+s_3))^n$ ($s_1$ and $s_2$ are from Figure 7) then the same initial vector $X=[x_0, x_1, x_2, x_3, x_4, x_5, x_6, x_7]$ will be transformed into a cyclic set of vectors in another 2-dimensional plane ($z_1$, $z_3$) of the same 8-dimensional space. One should conclude that, in this model approach, the same initial information in a form of a multidimensional vector X could generate a few cyclic processes in different planes of appropriate multidimensional space by means of using cyclic operators of the described type. In other words, we have here a multi-purpose using of vector information due to such operators (for instance, this informational vector can represent a fragment of a nucleotide sequence that can be used to organize many cyclic processes in different planes or subspaces of a phase space of genetic phenomena).

In the proposed model approach, one more benefit is that different cyclic processes of such cooperative ensemble can be easy coordinated and synchronized including an assignment of their relative phase shifts, starting times and different tempos of their cycles.

One technical remark is needed here. If we use a cyclic operator on the basis of the "column projectors", then a vector X should be multipled by the matrix on the right in accordance with the sample: $[x_0, x_1, x_2, x_3, x_4, x_5, x_6, x_7]*(2^{-0.5}*(s_0+s_2))^n$. But if we use a cyclic operator on the basis of the "row projectors" (for instance, $v_0$ and $v_2$ from Figure 8) then a vector X should be multiplied by the matrix on the left in accordance with the following sample: $(2^{-0.5}*(v_0+v_2))^n*[x_0; x_1; x_2; x_3; x_4; x_5; x_6; x_7]$.

Below we will describe extensions of the genetic (4*)-matrices $R_4$ and $H_4$ (Figure 1) into $2^n*2^n$-matrices every of which consists of $2^n$ "column projectors" (or $2^n$ "row projectors"); summation of projectors from this expanded set leads to new cyclic groups, etc. by analogy with the described cases (see Figures 4, 9, 11, 16). It gives a great number of cyclic groups of operators with similar properties of the selective control (or selective coding) of cyclic changes of coordinates of $2^n$-dimensional vectors. These numerous cyclic groups are useful to simulate big cooperative ensembles of cyclic processes, for instance, an ensemble of cyclic motions of legs, hands and separate muscles during different gaits (walking, running, etc.) simultaneously with heartbeats, breathing cycles, metabolic cycles, etc. Such models and their practical applications are created in the author's laboratory. The problem of inherited ensembles of biological cycles are closely linked to the fundamental problems of the biological clock and time, aging, etc. Taking into account results, which were obtained in "matrix genetics", the author puts forward "a biological concept of projectors" , which interprets the living body as a colony of projection operators.

It should be noted that in a case of a cyclic group of vector transformations with a period 8 (for example, in the case of the cyclic group $(2^{-0.5}*(s_0+s_2))^n$) that has only 8 discrete stages inside one cycle, one can enlarge a quantity of stages in “k” times by changing of the power in a form n/k: the cyclic group $(2^{-0.5}*(s_0+s_2))^{n/k}$ has k*8 stages inside one cycle (here "n" and "k" - integer positive numbers). The more value of "k", the less discretization of the cycle and the more smooth (uninterrupted) type of this cyclic process.

It can be added that many gaits (which are based on cyclic movements of limbs and corresponding muscle actuators) have genetically inherited character. So, newborn turtles and crocodiles, when they hatched from eggs, crawl with quite coordinated movements to water without any training from anybody; a newborn foal, after a bit time, begins to walk and run; centipedes crawl by means of coordinated movements of a great number of their legs (this number sometimes reaches up to 750) on the basis of inherited algorithms of control of legs. One should emphasize that, in the previous history, gaits and locomotion algorithms were studied in biomechanics of movements without any connection with the structures of genetic coding and with inheritance of unified control algorithms. The projection operations are associated with many kinds of movements and planned actions of our body to achieve the goal by the shortest path: for example, sending a billiard ball in the goal, we use a projection operation; directing a finger to the button of computer or piano, we make a projection action, etc. In other words, the concept of projection operators can be additionally used to simulate a broad class of such biomechanical actions.

Subject of genetically inherited ability of coordinating movements of body parts is connected with fundamental problems of congenital knowledge about surrounding space and of physiological foundations of geometry. Various researches have long put forward ideas about the importance of kinematic organization of body and its movements in the genesis of spatial representations of the individual. For example, H. Poincare has put these ideas into the foundation of his teachings about the physiological foundations of geometry and about the origin of spatial representations in individuum.

According to Poincare, the concept of space and geometry arises from an individual on the basis of kinematic organization of his body with using characterizations of positions and movements of body parts relative to each other, ie in the kinematic organization of the body is something that precedes the concept of space [Poincare, 1913]. Evolutionary development of the whole apparatus of kinematic activity of our body has provided a coherence of this apparatus with realities of the physical world. Because of this, each newborn organism receives adequate spatial representations not only through personal contact during ontogeny with the objects of the surrounding world, but also at the expense of achievements of previous generations enshrined in the apparatus of body movements in the phylogenesis. According to Poincare, for organism, which is absolutely immobile, spatial and geometric concepts are excluded. «*To localize an object simply means to represent to oneself the movements that would be necessary to reach it. I will explain myself. It is not a question of representing the movements themselves in space, but solely of representing to oneself the muscular sensations which accompany these movements and which do not presuppose the preexistence of the notion of space*. [Poincare, 1913, p. 247]. «*I have just said that it is to our own body that we naturally refer exterior objects; that we carry about everywhere with us a system of axes to which we refer all the points of space and that this system of axes seems to be invariably bound to our body. It should be noticed that rigorously we could not speak of axes invariably bound to the body unless the different parts of this body were themselves invariably bound to one another. As this is not the case, we ought, before referring exterior objects to these fictitious axes, to suppose our body brought back to the initial attitude"* [Poincare, 1913, p. 247]. «*We should therefore not have been able to construct space if we had not had an instrument to measure it; well, this instrument to which we relate everything, which we use instinctively, it is our own body. It is in relation to our body that we place exterior objects, and the only spatial relations of these objects that we can represent are their relations to our body. It is our body which serves us, so to speak, as system of axes of coordinates*» [Poincare, 1913, p. 418]. In times of Poincare science did not know about the genetic code, but from the modern point of view these thoughts by Poincare testify in favor the importance of the structural organization of the genetic system for physiological foundations of geometry and innate notions of space, which are connected with inherited apparatus and algorithms of body movements. And they are in tune with the results of matrix genetics, which are presented in our paper.

Modern physiology makes a significant addition to the teachings of the Poincare about an innate relationship of body and spatial representations, claiming an existence of a priori notions about our body shell. This statement is due to the study of the so-called phantom sensations in disabled: a special sense of the presence of natural parts of the body, which are absent in reality. It was found [Vetter, Weinstein, 1967; Weinstein, Sersen, 1961] that phantom sensations occur not only in cases of disabled with amputees, but also in people with congenital absence of limbs. Hence, the notion of the individual scheme of our body is not conditioned by our experiences, but has an innate character.

Additional materials relating to innate spatial representations, including the concept of B. Russell [Russel, 1956] about an innate character of ideas of projective geometry for each person, as well as an overview of works E. Schroedinger and other researchers about the geometry of spaces of visual perception, can be found in the book [Petoukhov, 1981].

We note here that although the concept of space is the primary concept for most physical theories, one can develop a meaningful theory in theoretical physics, in which it serves as only one of secondary notions, which are deduced from primary bases of a numeric system of a discrete character. We mean the "binary geometrophysics" [Vladimirov, 2008], ideas of which generate some associations with the ability of animal organisms (initially endowed with discrete molecular genetic information) to receive spatial representations and to create spatial movements on the basis of this primary information of discrete character.

## 5. ABOUT A DIRECTION OF ROTATION OF VECTORS UNDER INFLUENCE OF THE CYCLIC GROUPS OF THE OPERATORS

In configurations and functions of biological objects frequently one direction of rotation is preferable (it concerns the famous problem of biological dissymmetry). Taking this into account, it is interesting what one can say about directions of rotation of 4-dimensional and 8-dimensional vectors under influence of the cyclic groups described in previous sections. Figure 19 gives answer and shows directions of cyclic rotation of vectors $[x_0, x_1, x_2, x_3]*(2^{-0.5}*(c_i+c_j))^n$, $(2^{-0.5}*(r_i+r_j))^n*[x_0, x_1, x_2, x_4]$, $[x_0, x_1, x_2, x_3, x_4, x_5, x_6, x_7]*(2^{-0.5}*(s_i+s_j))^n$, $(2^{-0.5}*(v_i+v_j))^n*[x_0, x_1, x_2, x_3, x_4, x_5, x_6, x_7]$ under enlarging "n" ($i \neq j$; all these cyclic operators correspond to green cells in tables on Figure 19, they are based on summation of pairs of the projectors of the Rademacher matrices $R_4$ and $R_8$ from Figure 1).

| | $c_0$ | $c_1$ | $c_2$ | $c_3$ |
|---|---|---|---|---|
| $c_0$ | | ↺ | | |
| $c_1$ | ↺ | - | | |
| $c_2$ | | | - | ↺ |
| $c_3$ | | | ↺ | - |

| | $r_0$ | $r_1$ | $r_2$ | $r_3$ |
|---|---|---|---|---|
| $r_0$ | | ↻ | | |
| $r_1$ | ↻ | - | | |
| $r_2$ | | | - | ↻ |
| $r_3$ | | | ↻ | - |

| | $s_0$ | $s_1$ | $s_2$ | $s_3$ | $s_4$ | $s_5$ | $s_6$ | $s_7$ |
|---|---|---|---|---|---|---|---|---|
| $s_0$ | - | | ↺ | ↺ | | | | |
| $s_1$ | | - | ↺ | ↺ | | | | |
| $s_2$ | ↺ | ↺ | - | | | | | |
| $s_3$ | ↺ | ↺ | | - | | | | |
| $s_4$ | | | | | - | | ↺ | ↺ |
| $s_5$ | | | | | | - | ↺ | ↺ |
| $s_6$ | | | | | ↺ | ↺ | - | |
| $s_7$ | | | | | ↺ | ↺ | | - |

| | $v_0$ | $v_1$ | $v_2$ | $v_3$ | $v_4$ | $v_5$ | $v_6$ | $v_7$ |
|---|---|---|---|---|---|---|---|---|
| $v_0$ | - | | ↻ | ↻ | | | | |
| $v_1$ | | - | ↻ | ↻ | | | | |
| $v_2$ | ↻ | ↻ | - | | | | | |
| $v_3$ | ↻ | ↻ | | - | | | | |
| $v_4$ | | | | | - | | ↻ | ↻ |
| $v_5$ | | | | | | - | ↻ | ↻ |
| $v_6$ | | | | | ↻ | ↻ | - | |
| $v_7$ | | | | | ↻ | ↻ | | - |

Figure 19. In addition to Figures 4 and 9, the tables show directions of rotations of 4-dimensional and 8-dimensional vectors under influence of the cyclic groups of operators, which correspond to green cells and which are based on summation of pairs of the "column projectors" (on the left, see Figures 2, 4, 7, 9) and of the "row projectors" (on the right, see Figures 2, 4, 8, 9) of the Rademacher matrices $R_4$ and $R_8$ (Figure 1). The symbol ↺ means counter-clockwise rotation, the symbol ↻ means clockwise rotation.

Figure 20 shows directions of cyclic rotation of vectors $[x_0, x_1, x_2, x_3]*(2^{-0.5}*(h_i+h_j))^n$, $(2^{-0.5}*(g_i+g_j))^n*[x_0, x_1, x_2, x_4]$, $[x_0, x_1, x_2, x_3, x_4, x_5, x_6, x_7]*(2^{-0.5}*(u_i+u_j))^n$, $(2^{-0.5}*(d_i+d_j))^n*[x_0, x_1, x_2, x_3, x_4, x_5, x_6, x_7]$ under enlarging "n" ($i \neq j$; all these operators correspond to green cells in tables on Figure 19, they are based on summation of pairs of projectors of the Rademacher matrices $R_4$ and $R_8$ from Figure 1).).

| | $h_0$ | $h_1$ | $h_2$ | $h_3$ |
|---|---|---|---|---|
| $h_0$ | - | ↺ | ↻ | ↺ |
| $h_1$ | ↺ | - | ↺ | ↺ |
| $h_2$ | ↻ | ↺ | - | ↺ |
| $h_3$ | ↺ | ↺ | ↺ | - |

| | $g_0$ | $g_1$ | $g_2$ | $g_3$ |
|---|---|---|---|---|
| $g_0$ | - | ↻ | ↺ | ↻ |
| $g_1$ | ↻ | - | ↻ | ↻ |
| $g_2$ | ↺ | ↻ | - | ↻ |
| $g_3$ | ↻ | ↻ | ↻ | - |

| | $u_0$ | $u_1$ | $u_2$ | $u_3$ | $u_4$ | $u_5$ | $u_6$ | $u_7$ |
|---|---|---|---|---|---|---|---|---|
| $u_0$ | - | ↻ | ↺ | | ↻ | | ↺ | |
| $u_1$ | ↻ | - | | ↺ | | ↻ | | ↺ |
| $u_2$ | ↺ | | - | ↻ | ↺ | | ↺ | |
| $u_3$ | | ↺ | ↻ | - | | ↺ | | ↺ |
| $u_4$ | ↻ | | ↺ | | - | ↻ | ↺ | |
| $u_5$ | | ↻ | | ↺ | ↻ | - | | ↺ |
| $u_6$ | ↺ | | ↺ | | ↺ | | - | ↻ |
| $u_7$ | | ↺ | | ↺ | | ↺ | ↻ | - |

| | $d_0$ | $d_1$ | $d_2$ | $d_3$ | $d_4$ | $d_5$ | $d_6$ | $d_7$ |
|---|---|---|---|---|---|---|---|---|
| $d_0$ | - | ↺ | ↻ | | ↺ | | ↻ | |
| $d_1$ | ↺ | - | | ↻ | | ↺ | | ↻ |
| $d_2$ | ↻ | | - | ↺ | ↻ | | ↻ | |
| $d_3$ | | ↻ | ↺ | - | | ↻ | | ↻ |
| $d_4$ | ↺ | | ↻ | | - | ↺ | ↻ | |
| $d_5$ | | ↺ | | ↻ | ↺ | - | | ↻ |
| $d_6$ | ↻ | | ↻ | | ↻ | | - | ↺ |
| $d_7$ | | ↻ | | ↻ | | ↻ | ↺ | - |

Figure 20. In addition to Figures 11 and 16, the tables show directions of rotations of 4-dimensional and 8-dimensional vectors under influence of the cyclic groups of operators, which correspond to green cells and which are based on summation of pairs of the "column projectors" (on the left, see Figures 10, 11, 14, 16) and of the "row projectors" (on the right, see Figures 10, 11, 15, 16) of the Hadamard matrices $H_4$ and $H_8$ (Figure 1). The symbol ↺ means counter-clockwise rotation, the symbol ↻ means clockwise rotation.

Each of tables on Figures 19 and 20 contains completely different (asymmetrical) number of rotations in the clockwise and counterclockwise. These facts give evidences in favor of an idea that living matter at its basic level of genetic information has certain informational reasons to provide dissymmetry of inherited biological structures and processes. Taking this into account, the author thinks about a possibility of informational reasons for biological dis-symmetry. Here one can remind for a comparison that usually scientists look for reasons of biological dis-symmetry in physical or chemical sciences but not in informatic science.

## 6. HAMILTON'S QUATERNIONS, COCKLE'S SPLIT-QUATERNIONS, THEIR EXTENSIONS AND PROJECTION OPERATORS

In previous sections we described cases of summation of pairs of the oblique projectors. Now let us consider cases of summation of 4 of these projectors and cases of summation of 8 of these projectors.

The matrix $H_4$ (Figure 1) is sum of the four "column projectors" or the four "row projector" (Figure 10). But $H_4$ has also another decomposition in a form of four sparse matrices $H_{40}$, $H_{41}$, $H_{42}$ and $H_{43}$ (Figure 21). This set is closed in relation to multiplication and it defines their multiplication table (Figure 21, bottom level) that is identical to the known multiplication table of quaternions by Hamilton. From this point of view, the matrix $H_4$ is the quaternion by Hamilton with unit coordinates. (Such type of decompositions is termed a dyadic-shift decomposition because it corresponds to structures of matrices of dyadic shifts, well known in technology of signals processing [Ahmed, Rao, 1975]).

$H_4 = H_{40} + H_{41} + H_{42} + H_{43} =$

| | | | | | | | | | | | | | | | | | | | | | |
|---|---|---|---|---|---|---|---|---|---|---|---|---|---|---|---|---|---|---|---|---|---|
| 1 | 0 | 0 | 0 | | 0 | 1 | 0 | 0 | | 0 | 0 | -1 | 0 | | 0 | 0 | 0 | 1 |
| 0 | 1 | 0 | 0 | + | -1 | 0 | 0 | 0 | + | 0 | 0 | 0 | 1 | + | 0 | 0 | 1 | 0 |
| 0 | 0 | 1 | 0 | | 0 | 0 | 0 | 1 | | 1 | 0 | 0 | 0 | | 0 | -1 | 0 | 0 |
| 0 | 0 | 0 | 1 | | 0 | 0 | -1 | 0 | | 0 | -1 | 0 | 0 | | -1 | 0 | 0 | 0 |

| | 1 | $H_{41}$ | $H_{42}$ | $H_{43}$ |
|---|---|---|---|---|
| 1 | 1 | $H_{41}$ | $H_{42}$ | $H_{43}$ |
| $H_{41}$ | $H_{41}$ | - 1 | $H_{43}$ | - $H_{42}$ |
| $H_{42}$ | $H_{42}$ | - $H_{43}$ | - 1 | $H_{41}$ |
| $H_{43}$ | $H_{43}$ | $H_{42}$ | - $H_{41}$ | - 1 |

Figure 21. The dyadic-shift decomposition of the (4*4)-matrix $H_4$ (from Figure 1) gives the set of 4 sparse matrices $H_{40}$, $H_{41}$, $H_{42}$ and $H_{43}$, which corresponds to the multiplication table of quatrnions by Hamilton (bottom row). The matrix $H_{40}$ is identity matrix

Here one can mention that Hamilton quaternions are closely related to the Pauli matrices, the theory of the electromagnetic field (Maxwell wrote his equation on the language of quaternions Hamilton), the special theory of relativity, the theory of spins, quantum theory of chemical valency, etc. In the twentieth century thousands of works were devotes to quaternions in physics [http://arxiv.org/abs/math-ph/0511092]. Now Hamilton quaternions are manifested in the genetic code system. Our scientific direction - "matrix genetics" - has led to the discovery of an important bridge among physics, biology and computer science for their mutual enrichment. In our studies, we have received a new example of the effectiveness of mathematics: abstract mathematical structures, which have been derived by mathematicians at the tip of the pen 160 years ago, are embodied long ago in the information basis of living matter - the system of genetic coding. The mathematical structures, which are discovered by mathematicians in a result of painful reflections (like Hamilton, who has wasted 10 years of continuous thought to reveal his quaternions), are already represented in the genetic coding system.

Let us turn now to the (8*8)-matrix $H_8$ (Figure 1) that can be represented as sum of two matrices $HL_8 = u_0+u_2+u_4+u_6$ and $HR_8 = u_1+u_3+u_5+u_7$ (Figure 22). Here $u_0, u_1, \ldots, u_7$ are the «column projectors» from Figure 14.

$H_8 = HL_8+HR_8 =$

| | | | | | | | |
|---|---|---|---|---|---|---|---|
| 1 | 0 | 1 | 0 | -1 | 0 | 1 | 0 |
| 1 | 0 | 1 | 0 | -1 | 0 | 1 | 0 |
| -1 | 0 | 1 | 0 | 1 | 0 | 1 | 0 |
| -1 | 0 | 1 | 0 | 1 | 0 | 1 | 0 |
| 1 | 0 | -1 | 0 | 1 | 0 | 1 | 0 |
| 1 | 0 | -1 | 0 | 1 | 0 | 1 | 0 |
| -1 | 0 | -1 | 0 | -1 | 0 | 1 | 0 |
| -1 | 0 | -1 | 0 | -1 | 0 | 1 | 0 |

+

| | | | | | | | |
|---|---|---|---|---|---|---|---|
| 0 | -1 | 0 | -1 | 0 | 1 | 0 | -1 |
| 0 | 1 | 0 | 1 | 0 | -1 | 0 | 1 |
| 0 | 1 | 0 | -1 | 0 | -1 | 0 | -1 |
| 0 | -1 | 0 | 1 | 0 | 1 | 0 | 1 |
| 0 | -1 | 0 | 1 | 0 | -1 | 0 | -1 |
| 0 | 1 | 0 | -1 | 0 | 1 | 0 | 1 |
| 0 | 1 | 0 | 1 | 0 | 1 | 0 | -1 |
| 0 | -1 | 0 | -1 | 0 | -1 | 0 | 1 |

**Figure 22**. The representation of the Hadamard matrix $H_8$ (from Figure 1) as sum of two sparse matrices $HL_8$ and $HR_8$

Figure 23 shows a decomposition of the matrix $HL_8$ (from Figure 22) as a sum of 4 matrices: $HL_8 = HL_{80} + HL_{81} + HL_{82} + HL_{83}$. The set of matrices $HL_{80}$, $HL_{81}$, $HL_{82}$ and $HL_{83}$ is closed in relation to multiplication and it defines the multiplication table that is identical to the multiplication table of quaternions by Hamilton. General expression for quaternions in this case can be written as $Q_L = a_0*HL_{80} + a_1*HL_{81} + a_2*HL_{82} + a_3*HL_{83}$, where $a_0$, $a_1$, $a_2$, $a_3$ are real numbers. From this point of view, the (8*8)-genomatrix $HL_8$ is the 4-parametric quaternion by Hamilton with unit coordinates.

$$HL_8 = HL_{80} + HL_{81} + HL_{82} + HL_{83} =$$

$$\begin{pmatrix} 1&0&1&0&-1&0&1&0 \\ 1&0&1&0&-1&0&1&0 \\ -1&0&1&0&1&0&1&0 \\ -1&0&1&0&1&0&1&0 \\ 1&0&-1&0&1&0&1&0 \\ 1&0&-1&0&1&0&1&0 \\ -1&0&-1&0&-1&0&1&0 \\ -1&0&-1&0&-1&0&1&0 \end{pmatrix} = \begin{pmatrix} 1&0&0&0&0&0&0&0 \\ 1&0&0&0&0&0&0&0 \\ 0&0&1&0&0&0&0&0 \\ 0&0&1&0&0&0&0&0 \\ 0&0&0&0&1&0&0&0 \\ 0&0&0&0&1&0&0&0 \\ 0&0&0&0&0&0&1&0 \\ 0&0&0&0&0&0&1&0 \end{pmatrix} + \begin{pmatrix} 0&0&1&0&0&0&0&0 \\ 0&0&1&0&0&0&0&0 \\ -1&0&0&0&0&0&0&0 \\ -1&0&0&0&0&0&0&0 \\ 0&0&0&0&0&0&1&0 \\ 0&0&0&0&0&0&1&0 \\ 0&0&0&0&-1&0&0&0 \\ 0&0&0&0&-1&0&0&0 \end{pmatrix}$$

$$+ \begin{pmatrix} 0&0&0&0&-1&0&0&0 \\ 0&0&0&0&-1&0&0&0 \\ 0&0&0&0&0&0&1&0 \\ 0&0&0&0&0&0&1&0 \\ 1&0&0&0&0&0&0&0 \\ 1&0&0&0&0&0&0&0 \\ 0&0&-1&0&0&0&0&0 \\ 0&0&-1&0&0&0&0&0 \end{pmatrix} + \begin{pmatrix} 0&0&0&0&0&0&1&0 \\ 0&0&0&0&0&0&1&0 \\ 0&0&0&0&1&0&0&0 \\ 0&0&0&0&1&0&0&0 \\ 0&0&-1&0&0&0&0&0 \\ 0&0&-1&0&0&0&0&0 \\ -1&0&0&0&0&0&0&0 \\ -1&0&0&0&0&0&0&0 \end{pmatrix}$$

| | $HL_{80}$ | $HL_{81}$ | $HL_{82}$ | $HL_{83}$ |
|---|---|---|---|---|
| $HL_{80}$ | $HL_{80}$ | $HL_{81}$ | $HL_{82}$ | $HL_{83}$ |
| $HL_{81}$ | $HL_{81}$ | $-HL_{80}$ | $HL_{83}$ | $-HL_{82}$ |
| $HL_{82}$ | $HL_{82}$ | $-HL_{83}$ | $-HL_{80}$ | $HL_{81}$ |
| $HL_{83}$ | $HL_{83}$ | $HL_{82}$ | $-HL_{81}$ | $-HL_{80}$ |

Figure 23. Upper rows: the decomposition of the matrix $HL_8$ (from Figure 22) as sum of 4 matrices: $HL_8 = HL_{80} + HL_{81} + HL_{82} + HL_{83}$. Bottom row: the multiplication table of these 4 matrices $HL_{80}$, $HL_{81}$, $HL_{82}$ and $HL_{83}$, which is identical to the multiplication table of quaternions by Hamilton. The matrix $HL_{80}$ represents the real unit for this matrix set.

The similar situation holds true for the matrix $HR_8$ (from Figure 22). Figure 24 shows a decomposition of the matrix $HR_8$ as a sum of 4 matrices: $HR_8 = HR_{80} + HR_{81} + HR_{82} + HR_{83}$. The set of matrices $HR_{80}$, $HR_{81}$, $HR_{82}$ and $HR_{83}$ is closed in relation to multiplication and it defines the multiplication table that is identical to the same multiplication table of quaternions by Hamilton. General expression of quaternions in this case can be written as $Q_R = a_0*HR_{80} + a_1*HR_{81} + a_2*HR_{82} + a_3*HR_{83}$, where $a_0$, $a_1$, $a_2$, $a_3$ are real numbers. From this point of view, the (8*8)-genomatrix $HR_8$ is the quaternion by Hamilton with unit coordinates.

$$HR_8 = HR_{80} + HR_{81} + HR_{82} + HR_{83} =$$

| | | | | | | | |
|---|---|---|---|---|---|---|---|
| 0 | -1 | 0 | -1 | 0 | 1 | 0 | -1 |
| 0 | 1 | 0 | 1 | 0 | -1 | 0 | 1 |
| 0 | 1 | 0 | -1 | 0 | -1 | 0 | -1 |
| 0 | -1 | 0 | 1 | 0 | 1 | 0 | 1 |
| 0 | -1 | 0 | 1 | 0 | -1 | 0 | -1 |
| 0 | 1 | 0 | -1 | 0 | 1 | 0 | 1 |
| 0 | 1 | 0 | 1 | 0 | 1 | 0 | -1 |
| 0 | -1 | 0 | -1 | 0 | -1 | 0 | 1 |

=

| | | | | | | | |
|---|---|---|---|---|---|---|---|
| 0 | -1 | 0 | 0 | 0 | 0 | 0 | 0 |
| 0 | 1 | 0 | 0 | 0 | 0 | 0 | 0 |
| 0 | 0 | 0 | -1 | 0 | 0 | 0 | 0 |
| 0 | 0 | 0 | 1 | 0 | 0 | 0 | 0 |
| 0 | 0 | 0 | 0 | 0 | -1 | 0 | 0 |
| 0 | 0 | 0 | 0 | 0 | 1 | 0 | 0 |
| 0 | 0 | 0 | 0 | 0 | 0 | 0 | -1 |
| 0 | 0 | 0 | 0 | 0 | 0 | 0 | 1 |

+

| | | | | | | | |
|---|---|---|---|---|---|---|---|
| 0 | 0 | 0 | -1 | 0 | 0 | 0 | 0 |
| 0 | 0 | 0 | 1 | 0 | 0 | 0 | 0 |
| 0 | 1 | 0 | 0 | 0 | 0 | 0 | 0 |
| 0 | -1 | 0 | 0 | 0 | 0 | 0 | 0 |
| 0 | 0 | 0 | 0 | 0 | 0 | 0 | -1 |
| 0 | 0 | 0 | 0 | 0 | 0 | 0 | 1 |
| 0 | 0 | 0 | 0 | 0 | 1 | 0 | 0 |
| 0 | 0 | 0 | 0 | 0 | -1 | 0 | 0 |

+

| | | | | | | | |
|---|---|---|---|---|---|---|---|
| 0 | 0 | 0 | 0 | 0 | 1 | 0 | 0 |
| 0 | 0 | 0 | 0 | 0 | -1 | 0 | 0 |
| 0 | 0 | 0 | 0 | 0 | 0 | 0 | -1 |
| 0 | 0 | 0 | 0 | 0 | 0 | 0 | 1 |
| 0 | -1 | 0 | 0 | 0 | 0 | 0 | 0 |
| 0 | 1 | 0 | 0 | 0 | 0 | 0 | 0 |
| 0 | 0 | 0 | 1 | 0 | 0 | 0 | 0 |
| 0 | 0 | 0 | -1 | 0 | 0 | 0 | 0 |

+

| | | | | | | | |
|---|---|---|---|---|---|---|---|
| 0 | 0 | 0 | 0 | 0 | 0 | 0 | -1 |
| 0 | 0 | 0 | 0 | 0 | 0 | 0 | 1 |
| 0 | 0 | 0 | 0 | 0 | -1 | 0 | 0 |
| 0 | 0 | 0 | 0 | 0 | 1 | 0 | 0 |
| 0 | 0 | 0 | 1 | 0 | 0 | 0 | 0 |
| 0 | 0 | 0 | -1 | 0 | 0 | 0 | 0 |
| 0 | 1 | 0 | 0 | 0 | 0 | 0 | 0 |
| 0 | -1 | 0 | 0 | 0 | 0 | 0 | 0 |

| | $HR_{80}$ | $HR_{81}$ | $HR_{82}$ | $HR_{83}$ |
|---|---|---|---|---|
| $HR_{80}$ | $HR_{80}$ | $HR_{81}$ | $HR_{82}$ | $HR_{83}$ |
| $HR_{81}$ | $HR_{81}$ | $-HR_{80}$ | $HR_{83}$ | $-HR_{82}$ |
| $HR_{82}$ | $HR_{82}$ | $-HR_{83}$ | $-HR_{80}$ | $HR_{81}$ |
| $HR_{83}$ | $HR_{83}$ | $HR_{82}$ | $-HR_{81}$ | $-HR_{80}$ |

Figure 24. upper rows: the decomposition of the matrix $HR_8$ (from Figure 22) as sum of 4 matrices: $H_{8R} = H0_{8R} + H1_{8R} + H2_{8R} + H3_{8R}$. Bottom row: the multiplication table of these 4 matrices $HR_{80}$, $HR_{81}$, $HR_{82}$ and $HR_{83}$, which is identical to the multiplication table of quaternions by Hamilton. $HR_{80}$ represents the real unit for this matrix set

The initial (8*8)-matrix $H_8$ (Figure 1) can be also decomposed in another way on the base of dyadic-shift decomposition. Figure 25 shows such dyadic-shift decomposition $H_8 = H_{80}+H_{81}+H_{82}+H_{83}+H_{84}+H_{85}+H_{86}+H_{87}$, when 8 sparse matrices $H_{80}$, $H_{81}$, $H_{82}$, $H_{83}$, $H_{84}$, $H_{85}$, $H_{86}$, $H_{87}$ arise ($H_{80}$ is identity matrix). The set $H_{80}$, $H_{81}$, $H_{82}$, $H_{83}$, $H_{84}$, $H_{85}$, $H_{86}$, $H_{87}$ is closed in relation to multiplication and it defines the multiplication table on Figure 25. This multiplication table is identical to the multiplication table of 8-dimensional hypercomplex numbers that are termed as biquaternions by Hamilton (or Hamiltons' quaternions over the field of complex numbers). General expression for biquaternions in this case can be written as $Q_8 = a_0*H_{80}+a_1*H_{81}+a_2*H_{82}+a_3*H_{83}+ a_4*H_{84} +a_5*H_{85}+a_6*H_{86}+a_7*H_{87}$, where $a_0$, $a_1$, $a_2$, $a_3$, $a_4$, $a_5$, $a_6$, $a_7$ are real numbers. From this point of view, the (8*8)-genomatrix $H_8$ is Hamilton's biquaternion with unit coordinates.

$H_8 = H_{80}+H_{81}+H_{82}+H_{83}+H_{84}+H_{85}+H_{86}+H_{87} =$

| | | | | | | | |
|---|---|---|---|---|---|---|---|
| 1 | 0 | 0 | 0 | 0 | 0 | 0 | 0 |
| 0 | 1 | 0 | 0 | 0 | 0 | 0 | 0 |
| 0 | 0 | 1 | 0 | 0 | 0 | 0 | 0 |
| 0 | 0 | 0 | 1 | 0 | 0 | 0 | 0 |
| 0 | 0 | 0 | 0 | 1 | 0 | 0 | 0 |
| 0 | 0 | 0 | 0 | 0 | 1 | 0 | 0 |
| 0 | 0 | 0 | 0 | 0 | 0 | 1 | 0 |
| 0 | 0 | 0 | 0 | 0 | 0 | 0 | 1 |

+

| | | | | | | | |
|---|---|---|---|---|---|---|---|
| 0 | -1 | 0 | 0 | 0 | 0 | 0 | 0 |
| 1 | 0 | 0 | 0 | 0 | 0 | 0 | 0 |
| 0 | 0 | 0 | -1 | 0 | 0 | 0 | 0 |
| 0 | 0 | 1 | 0 | 0 | 0 | 0 | 0 |
| 0 | 0 | 0 | 0 | 0 | -1 | 0 | 0 |
| 0 | 0 | 0 | 0 | 1 | 0 | 0 | 0 |
| 0 | 0 | 0 | 0 | 0 | 0 | 0 | -1 |
| 0 | 0 | 0 | 0 | 0 | 0 | 1 | 0 |

+

| | | | | | | | |
|---|---|---|---|---|---|---|---|
| 0 | 0 | 1 | 0 | 0 | 0 | 0 | 0 |
| 0 | 0 | 0 | 1 | 0 | 0 | 0 | 0 |
| -1 | 0 | 0 | 0 | 0 | 0 | 0 | 0 |
| 0 | -1 | 0 | 0 | 0 | 0 | 0 | 0 |
| 0 | 0 | 0 | 0 | 0 | 0 | 1 | 0 |
| 0 | 0 | 0 | 0 | 0 | 0 | 0 | 1 |
| 0 | 0 | 0 | 0 | -1 | 0 | 0 | 0 |
| 0 | 0 | 0 | 0 | 0 | -1 | 0 | 0 |

+

| | | | | | | | |
|---|---|---|---|---|---|---|---|
| 0 | 0 | 0 | -1 | 0 | 0 | 0 | 0 |
| 0 | 0 | 1 | 0 | 0 | 0 | 0 | 0 |
| 0 | 1 | 0 | 0 | 0 | 0 | 0 | 0 |
| -1 | 0 | 0 | 0 | 0 | 0 | 0 | 0 |
| 0 | 0 | 0 | 0 | 0 | 0 | 0 | -1 |
| 0 | 0 | 0 | 0 | 0 | 0 | 1 | 0 |
| 0 | 0 | 0 | 0 | 0 | 1 | 0 | 0 |
| 0 | 0 | 0 | 0 | -1 | 0 | 0 | 0 |

+

$$
\begin{bmatrix} 0&0&0&0&-1&0&0&0\\ 0&0&0&0&0&-1&0&0\\ 0&0&0&0&0&0&1&0\\ 0&0&0&0&0&0&0&1\\ 1&0&0&0&0&0&0&0\\ 0&1&0&0&0&0&0&0\\ 0&0&-1&0&0&0&0&0\\ 0&0&0&-1&0&0&0&0 \end{bmatrix}
+
\begin{bmatrix} 0&0&0&0&0&1&0&0\\ 0&0&0&0&-1&0&0&0\\ 0&0&0&0&0&0&0&-1\\ 0&0&0&0&0&0&1&0\\ 0&-1&0&0&0&0&0&0\\ 1&0&0&0&0&0&0&0\\ 0&0&0&1&0&0&0&0\\ 0&0&-1&0&0&0&0&0 \end{bmatrix}
+
\begin{bmatrix} 0&0&0&0&0&0&1&0\\ 0&0&0&0&0&0&0&1\\ 0&0&0&0&1&0&0&0\\ 0&0&0&0&0&1&0&0\\ 0&0&-1&0&0&0&0&0\\ 0&0&0&-1&0&0&0&0\\ -1&0&0&0&0&0&0&0\\ 0&-1&0&0&0&0&0&0 \end{bmatrix}
+
\begin{bmatrix} 0&0&0&0&0&0&0&-1\\ 0&0&0&0&0&0&1&0\\ 0&0&0&0&0&-1&0&0\\ 0&0&0&0&1&0&0&0\\ 0&0&0&1&0&0&0&0\\ 0&0&-1&0&0&0&0&0\\ 0&1&0&0&0&0&0&0\\ -1&0&0&0&0&0&0&0 \end{bmatrix}
$$

| | **1** | $H_{81}$ | $H_{82}$ | $H_{83}$ | $H_{84}$ | $H_{85}$ | $H_{86}$ | $H_{87}$ |
|---|---|---|---|---|---|---|---|---|
| **1** | **1** | $H_{81}$ | $H_{82}$ | $H_{83}$ | $H_{84}$ | $H_{85}$ | $H_{86}$ | $H_{87}$ |
| $H_{81}$ | $H_{81}$ | **-1** | $H_{83}$ | $-H_{82}$ | $H_{85}$ | $-H_{84}$ | $H_{87}$ | $-H_{86}$ |
| $H_{82}$ | $H_{82}$ | $H_{83}$ | **-1** | $-H_{81}$ | $-H_{86}$ | $-H_{87}$ | $H_{84}$ | $H_{85}$ |
| $H_{83}$ | $H_{83}$ | $-H_{82}$ | $-H_{81}$ | **1** | $-H_{87}$ | $H_{86}$ | $H_{85}$ | $-H_{84}$ |
| $H_{84}$ | $H_{84}$ | $H_{85}$ | $H_{86}$ | $H_{87}$ | **-1** | $-H_{81}$ | $-H_{82}$ | $-H_{83}$ |
| $H_{85}$ | $H_{85}$ | $-H_{84}$ | $H_{87}$ | $-H_{86}$ | $-H_{81}$ | **1** | $-H_{83}$ | $H_{82}$ |
| $H_{86}$ | $H_{86}$ | $H_{87}$ | $-H_{84}$ | $-H_{85}$ | $H_{82}$ | $H_{83}$ | **-1** | $-H_{81}$ |
| $H_{87}$ | $H_{87}$ | $-H_{86}$ | $-H_{85}$ | $H_{84}$ | $H_{83}$ | $-H_{82}$ | $-H_{81}$ | **1** |

Figure 25. Upper rows: the decomposition of the matrix $H_8$ (from Figure 1) as sum of 8 matrices: $H_8 = H_{80}+H_{81}+H_{82}+H_{83}+H_{84}+H_{85}+H_{86}+H_{87}$. Bottom row: the multiplication table of these 8 matrices $H_{80}$, $H_{81}$, $H_{82}$, $H_{83}$, $H_{84}$, $H_{85}$, $H_{86}$, $H_{87}$, which is identical to the multiplication table of biquaternions by Hamilton (or Hamiltons' quaternions over the field of complex numbers). $H_{80}$ is identity matrix

One can analyze the Rademacher genomatrices $R_4$ and $R_8$ (From Figure 1) by a similar way [Petoukhov, 2012b]). In particular, in this case the following results arise:

- The Rademacher (4*4)-matrix $R_4$ represents split-quaternion by J.Cockle with unit coordinates (http://en.wikipedia.org/wiki/Split-quaternion) in the case of its dyadic-shift decomposition;
- The Rademacher (8*8)-matrix $R_8$ represents bisplit-quaternion by J.Cockle with unit coordinates in the case of its dyadic-shift decomposition;
- If the Rademacher (4*4)-matrix $R_4$ is represented as sum of two sparse matrices $(c_0+c_2) + (c_1+c_3)$ (here $c_0$, $c_1$, $c_2$, $c_3$ are the column projectors from Figure 2), then the matrix $R_4$ is sum of two hyperbolic numbers with unit coordunates because each of summands $(c_0+c_2)$ and $(c_1+c_3)$ is hyperbolic number with unit coordinates. A similar is true for the case of the "row projectors" $r_0$, $r_1$, $r_2$, $r_3$ from Figure 2.

Now let us pay a special attention to the Rademacher (8*8)-matrix $R_8$ as a sum of the following two sparse matrices $RL_8$ and $RR_8$, the first of which is a sum of 4 projectors with even indexes $s_0$, $s_2$, $s_4$, $s_6$ and the second of which is a sum of 4 projectors with odd indexes $s_1$, $s_3$, $s_5$, $s_7$: $R_8 = (s_0+s_2+s_4+s_6) + (s_1+s_3+s_5+s_7) = RL_8 + RR_8$ (here $s_0$, $s_1$, …, $s_7$ are the column projectors from Figure 7). Below this decomposition will be useful for analysis of a correspondence between 64 triplets and 20 amino acids with stop-codons.

$$R_8 = RL_8 + RR_8 = \begin{bmatrix} 1&0&1&0&1&0&-1&0 \\ 1&0&1&0&1&0&-1&0 \\ -1&0&1&0&-1&0&-1&0 \\ -1&0&1&0&-1&0&-1&0 \\ 1&0&-1&0&1&0&1&0 \\ 1&0&-1&0&1&0&1&0 \\ -1&0&-1&0&-1&0&1&0 \\ -1&0&-1&0&-1&0&1&0 \end{bmatrix} + \begin{bmatrix} 0&1&0&1&0&1&0&-1 \\ 0&1&0&1&0&1&0&-1 \\ 0&-1&0&1&0&-1&0&-1 \\ 0&-1&0&1&0&-1&0&-1 \\ 0&1&0&-1&0&1&0&1 \\ 0&1&0&-1&0&1&0&1 \\ 0&-1&0&-1&0&-1&0&1 \\ 0&-1&0&-1&0&-1&0&1 \end{bmatrix}$$

Figure 26. The decomposition of the Rademacher (8*8)-genomatrix $R_8$ from Figure 1

Each of these sparse matrices $RL_8$ and $RR_8$ can be decomposed into a set of 4 sparse matrices: $RL_8 = RL_{80} + RL_{81} + RL_{82} + RL_{83}$ and $RR_8 = RR_{80} + RR_{81} + RR_{82} + RR_{83}$ (Figures 27 and 28). The first set of matrices $RL_{80}$, $RL_{81}$, $RL_{82}$, $RL_{83}$ is closed relative to multiplication and it defines a known multiplication table of split-quaternions by J. Cockle (http://en.wikipedia.org/wiki/Split-quaternion) on Figure 27. The second set of matrices $RR_{80}$, $RR_{81}$, $RR_{82}$, $RR_{83}$ is also closed relative to multiplication and it defines the same multiplication table of split-quaternions by J. Cockle (Figure 28). Consequently, each of matrices $RL_8$ and $RR_8$ is split-quaternion by Cockle with unit coordinates.

$$RL_8 = RL_{80} + RL_{81} + RL_{82} + RL_{83} = \begin{bmatrix} 1&0&0&0&0&0&0&0 \\ 1&0&0&0&0&0&0&0 \\ 0&0&1&0&0&0&0&0 \\ 0&0&1&0&0&0&0&0 \\ 0&0&0&0&1&0&0&0 \\ 0&0&0&0&1&0&0&0 \\ 0&0&0&0&0&0&1&0 \\ 0&0&0&0&0&0&1&0 \end{bmatrix} + \begin{bmatrix} 0&0&1&0&0&0&0&0 \\ 0&0&1&0&0&0&0&0 \\ -1&0&0&0&0&0&0&0 \\ -1&0&0&0&0&0&0&0 \\ 0&0&0&0&0&0&1&0 \\ 0&0&0&0&0&0&1&0 \\ 0&0&0&0&-1&0&0&0 \\ 0&0&0&0&-1&0&0&0 \end{bmatrix} +$$

$$+ \begin{bmatrix} 0&0&0&0&1&0&0&0 \\ 0&0&0&0&1&0&0&0 \\ 0&0&0&0&0&0&-1&0 \\ 0&0&0&0&0&0&-1&0 \\ 1&0&0&0&0&0&0&0 \\ 1&0&0&0&0&0&0&0 \\ 0&0&-1&0&0&0&0&0 \\ 0&0&-1&0&0&0&0&0 \end{bmatrix} + \begin{bmatrix} 0&0&0&0&0&0&-1&0 \\ 0&0&0&0&0&0&-1&0 \\ 0&0&0&0&-1&0&0&0 \\ 0&0&0&0&-1&0&0&0 \\ 0&0&-1&0&0&0&0&0 \\ 0&0&-1&0&0&0&0&0 \\ -1&0&0&0&0&0&0&0 \\ -1&0&0&0&0&0&0&0 \end{bmatrix} ;$$

| | $RL_{80}$ | $RL_{81}$ | $RL_{82}$ | $RL_{83}$ |
|---|---|---|---|---|
| $RL_{80}$ | $RL_{80}$ | $RL_{81}$ | $RL_{83}$ | $RL_{83}$ |
| $RL_{81}$ | $RL_{81}$ | $- RL_{80}$ | $RL_{83}$ | $- RL_{82}$ |
| $RL_{82}$ | $RL_{82}$ | $- RL_{83}$ | $RL_{80}$ | $- RL_{81}$ |
| $RL_{83}$ | $RL_{83}$ | $RL_{82}$ | $RL_{81}$ | $RL_{80}$ |

Figure 27. The decomposition of the matrix $RL_8$ from Figure 26 into the set of 4 matrices $RL_{80}$, $RL_{81}$, $RL_{82}$, $RL_{83}$, which defines the known multiplication table of split-quaternions by J. Cockle (http://en.wikipedia.org/wiki/Split-quaternion)

$$RR_8 = RR_{80} + RR_{81} + RR_{82} + RR_{83} = \begin{bmatrix} 0&1&0&0&0&0&0&0 \\ 0&1&0&0&0&0&0&0 \\ 0&0&0&1&0&0&0&0 \\ 0&0&0&1&0&0&0&0 \\ 0&0&0&0&0&1&0&0 \\ 0&0&0&0&0&1&0&0 \\ 0&0&0&0&0&0&0&1 \\ 0&0&0&0&0&0&0&1 \end{bmatrix} + \begin{bmatrix} 0&0&0&1&0&0&0&0 \\ 0&0&0&1&0&0&0&0 \\ 0&-1&0&0&0&0&0&0 \\ 0&-1&0&0&0&0&0&0 \\ 0&0&0&0&0&0&0&1 \\ 0&0&0&0&0&0&0&1 \\ 0&0&0&0&0&-1&0&0 \\ 0&0&0&0&0&-1&0&0 \end{bmatrix} +$$

$$+\begin{pmatrix} 0&0&0&0&0&1&0&0\\ 0&0&0&0&0&1&0&0\\ 0&0&0&0&0&0&0&-1\\ 0&0&0&0&0&0&0&-1\\ 0&1&0&0&0&0&0&0\\ 0&1&0&0&0&0&0&0\\ 0&0&0&-1&0&0&0&0\\ 0&0&0&-1&0&0&0&0 \end{pmatrix} + \begin{pmatrix} 0&0&0&0&0&0&0&-1\\ 0&0&0&0&0&0&0&-1\\ 0&0&0&0&0&-1&0&0\\ 0&0&0&0&0&-1&0&0\\ 0&0&0&-1&0&0&0&0\\ 0&0&0&-1&0&0&0&0\\ 0&-1&0&0&0&0&0&0\\ 0&-1&0&0&0&0&0&0 \end{pmatrix};$$

| | $RR_{80}$ | $RR_{81}$ | $RR_{82}$ | $RR_{83}$ |
|---|---|---|---|---|
| $RR_{80}$ | $RR_{80}$ | $RR_{81}$ | $RR_{83}$ | $RR_{83}$ |
| $RR_{81}$ | $RR_{81}$ | $-RR_{80}$ | $RR_{83}$ | $-RR_{82}$ |
| $RR_{82}$ | $RR_{82}$ | $-RR_{83}$ | $RR_{80}$ | $-RR_{81}$ |
| $RR_{83}$ | $RR_{83}$ | $RR_{82}$ | $RR_{81}$ | $RR_{80}$ |

Figure 28. The decomposition of the matrix $RR_8$ from Figure 26 into the set of 4 matrices $RR_{80}$, $RR_{81}$, $RR_{82}$, $RR_{83}$, which defines the same multiplication table of split-quaternions by J. Cockle (http://en.wikipedia.org/wiki/Split-quaternion)

But each of the same (8*8)-matrices $RL_8$ and $RR_8$ (Figure 26)) can be decomposed in another way, which leads to its representation in a form of sum of two hyperbolic numbers: $RL_8= (s_0+s_4)+(s_2+s_6)$ and $RR_8=(s_1+s_5)+(s_3+s_7)$, where is $s_0$, $s_1$, …, $s_7$ are column projectors from the decomposition of $R_8$ on Figure 7. Figure 29 shows that each of sums $(s_0+s_4)$, $(s_2+s_6)$, $(s_1+s_5)$, $(s_3+s_7)$ in these decompositions for $RL_8$ and $RR_8$ is a (8*8)-matrix representation of hyperbolic number, whose coordinates are equal to 1. It means also that the whole (8*8)-matrix $R_8$ is sum of 4 hyperbolic numbers, whose coordinates are equal to 1, in an 8-dimensional space. These decompositions are useful for analyzing the degeneration of the genetic code in next Sections.

$$s_0+s_4=\begin{pmatrix} 1&0&0&0&1&0&0&0\\ 1&0&0&0&1&0&0&0\\ -1&0&0&0&-1&0&0&0\\ -1&0&0&0&-1&0&0&0\\ 1&0&0&0&1&0&0&0\\ 1&0&0&0&1&0&0&0\\ -1&0&0&0&-1&0&0&0\\ -1&0&0&0&-1&0&0&0 \end{pmatrix} = \begin{pmatrix} 1&0&0&0&0&0&0&0\\ 1&0&0&0&0&0&0&0\\ -1&0&0&0&0&0&0&0\\ -1&0&0&0&0&0&0&0\\ 0&0&0&0&1&0&0&0\\ 0&0&0&0&1&0&0&0\\ 0&0&0&0&-1&0&0&0\\ 0&0&0&0&-1&0&0&0 \end{pmatrix} + \begin{pmatrix} 0&0&0&0&1&0&0&0\\ 0&0&0&0&1&0&0&0\\ 0&0&0&0&-1&0&0&0\\ 0&0&0&0&-1&0&0&0\\ 1&0&0&0&0&0&0&0\\ 1&0&0&0&0&0&0&0\\ -1&0&0&0&0&0&0&0\\ -1&0&0&0&0&0&0&0 \end{pmatrix} = e_0+e_4;$$

| | $e_0$ | $e_4$ |
|---|---|---|
| $e_0$ | $e_0$ | $e_4$ |
| $e_4$ | $e_4$ | $e_0$ |

$$s_2+s_6=\begin{pmatrix} 0&0&1&0&0&0&-1&0\\ 0&0&1&0&0&0&-1&0\\ 0&0&1&0&0&0&-1&0\\ 0&0&1&0&0&0&-1&0\\ 0&0&-1&0&0&0&1&0\\ 0&0&-1&0&0&0&1&0\\ 0&0&-1&0&0&0&1&0\\ 0&0&-1&0&0&0&1&0 \end{pmatrix} = \begin{pmatrix} 0&0&1&0&0&0&0&0\\ 0&0&1&0&0&0&0&0\\ 0&0&1&0&0&0&0&0\\ 0&0&1&0&0&0&0&0\\ 0&0&0&0&0&0&1&0\\ 0&0&0&0&0&0&1&0\\ 0&0&0&0&0&0&1&0\\ 0&0&0&0&0&0&1&0 \end{pmatrix} + \begin{pmatrix} 0&0&0&0&0&0&-1&0\\ 0&0&0&0&0&0&-1&0\\ 0&0&0&0&0&0&-1&0\\ 0&0&0&0&0&0&-1&0\\ 0&0&-1&0&0&0&0&0\\ 0&0&-1&0&0&0&0&0\\ 0&0&-1&0&0&0&0&0\\ 0&0&-1&0&0&0&0&0 \end{pmatrix} = e_2+e_6;$$

| | $e_2$ | $e_6$ |
|---|---|---|
| $e_2$ | $e_2$ | $e_6$ |
| $e_6$ | $e_6$ | $e_2$ |

$$s_1+s_5=\begin{pmatrix} 0&1&0&0&0&1&0&0\\ 0&1&0&0&0&1&0&0\\ 0&-1&0&0&0&-1&0&0\\ 0&-1&0&0&0&-1&0&0\\ 0&1&0&0&0&1&0&0\\ 0&1&0&0&0&1&0&0\\ 0&-1&0&0&0&-1&0&0\\ 0&-1&0&0&0&-1&0&0 \end{pmatrix} = \begin{pmatrix} 0&1&0&0&0&0&0&0\\ 0&1&0&0&0&0&0&0\\ 0&-1&0&0&0&0&0&0\\ 0&-1&0&0&0&0&0&0\\ 0&0&0&0&0&1&0&0\\ 0&0&0&0&0&1&0&0\\ 0&0&0&0&0&-1&0&0\\ 0&0&0&0&0&-1&0&0 \end{pmatrix} + \begin{pmatrix} 0&0&0&0&0&1&0&0\\ 0&0&0&0&0&1&0&0\\ 0&0&0&0&0&-1&0&0\\ 0&0&0&0&0&-1&0&0\\ 0&1&0&0&0&0&0&0\\ 0&1&0&0&0&0&0&0\\ 0&-1&0&0&0&0&0&0\\ 0&-1&0&0&0&0&0&0 \end{pmatrix} = e_1+e_5;$$

| | $e_1$ | $e_5$ |
|---|---|---|
| $e_1$ | $e_1$ | $e_5$ |
| $e_5$ | $e_5$ | $e_1$ |

$$s_3+s_7 = \begin{pmatrix} 0&0&0&1&0&0&0&-1\\ 0&0&0&1&0&0&0&-1\\ 0&0&0&1&0&0&0&-1\\ 0&0&0&1&0&0&0&-1\\ 0&0&0&-1&0&0&0&1\\ 0&0&0&-1&0&0&0&1\\ 0&0&0&-1&0&0&0&1\\ 0&0&0&-1&0&0&0&1 \end{pmatrix} = \begin{pmatrix} 0&0&0&1&0&0&0&0\\ 0&0&0&1&0&0&0&0\\ 0&0&0&1&0&0&0&0\\ 0&0&0&1&0&0&0&0\\ 0&0&0&0&0&0&0&1\\ 0&0&0&0&0&0&0&1\\ 0&0&0&0&0&0&0&1\\ 0&0&0&0&0&0&0&1 \end{pmatrix} + \begin{pmatrix} 0&0&0&0&0&0&0&-1\\ 0&0&0&0&0&0&0&-1\\ 0&0&0&0&0&0&0&-1\\ 0&0&0&0&0&0&0&-1\\ 0&0&0&-1&0&0&0&0\\ 0&0&0&-1&0&0&0&0\\ 0&0&0&-1&0&0&0&0\\ 0&0&0&-1&0&0&0&0 \end{pmatrix} = e_3+e_7;$$

| | $e_3$ | $e_7$ |
|---|---|---|
| $e_3$ | $e_3$ | $e_7$ |
| $e_7$ | $e_7$ | $e_3$ |

Figure 29. Decompositions of sums of projectors ($s_0+s_4$), ($s_2+s_6$), ($s_1+s_5$), ($s_3+s_7$), which show that each of these sums is an (8*8)-matrix representation of hyperbolic numbers, whose coordinates are equal to 1, in an 8-dimensional space.

## 7. GENETIC MATRICES AS SUMS OF TENSOR PRODUCTS OF OBLIQUE (2*2)-PROJECTORS. EXTENSIONS OF GENETIC MATRICES INTO ($2^N*2^N$)-MATRICES

The Rademacher matrices $R_4$ and $R_8$ and also Hadamard matrices $H_4$ and $H_8$ (Figure 1) are interconnected by means of the following expressions:

$$R_4 \otimes [1\ 1;\ 1\ 1] = R_8, \quad H_4 \otimes [1\ -1;\ 1\ 1] = H_8 \qquad (1)$$

where ⊗ means tensor multiplication; the matrix [1 1; 1 1] is a traditional (2*2)-matrix representation of hyperbolic number with unit coordinates; the matrix [1 -1; 1 1] is a traditional (2*2)-matrix representation of complex number with unit coordinates.

The following extensions of the expressions (1) lead to ($2^n*2^n$)-matrices $R_K$ and $H_K$ (where $K=2^n$, n = 4, 5, 6,…; (n-2) means a tensor power):

$$R_4 \otimes [1\ 1;\ 1\ 1]^{(n-2)} = R_K, \quad H_4 \otimes [1\ -1;\ 1\ 1]^{(n-2)} = H_K \qquad (2)$$

In this algorithmic way we get a great set of ($2^n*2^n$)-matrices $R_K$ and $H_K$, each of which can be represented as a sum of $2^n$ «column projectors» (or $2^n$ «row projectors») by analogy with cases described above. Summations of these new «column projectors» (and also «row projectors») in different combinations (in pairs, in fours, in eights, etc.) give many new operators, exponentiation of which generates a great number of cyclic groups and other kind of operators. They also give many new representations of complex numbers, hyperbolic numbers, Hamilton's quaternion, split-quaternions and their extensions in a form of ($2^n*2^n$)-matrices that correspond to $2^n$-dimensional spaces. These new operators possess many similar properties, including a selective control (or coding) of different subspaces in $2^n$-dimensional space, in analogy with operators described in previous sections.

Why one can declare that each of matrices in the «column decomposition» (or in the «row decomposition») of any of matrices $R_k$ and $H_k$ in the expressions (2) is a projection operator? It can be declared on basis of the following simple theorem, taking into account that main diagonals of all matrices $R_k$ and $H_k$ contains only entries +1.

**Theorem**: any sparse square matrix P, which contains only a single non-zero column or a single non-zero row and which has its entry +1 on the main diagonal, is a projection operator (it satisfies the criterion $P^2=P$).

**Proof**. When multiplying two matrices $|A_{ik}|$ and $|B_{kj}|$, the elements of the rows in the first matrix are multiplied with corresponding columns in the second matrix to receive the resulting matrix $(AB)_{ij}| = \sum_{k=1}^{m} A_{ik}*B_{kj}$ (http://en.wikipedia.org/wiki/Matrix_multiplication ). Let us consider a case of a square matrix $|P_{ij}|$ (here i,j = 1, 2,…, m) with only a single non-zero column $P_{is} \neq 0$, which is numerated by an index “s” and which contains +1 in its cell on the main diagonal of this matrix: $P_{ss} = 1$. It means that all entries $P_{ik} = 0$ if $k \neq s$. The second degree of this matrix gives a square matrix:

$$\sum_{k=1}^{m} P_{ik}*P_{kj} = \sum_{k=1}^{m} P_{ik}*P_{ks} \qquad (3)$$

But among all $P_{ik}$ only one column differs from zero: $P_{is} \neq 0$. In the equation (3), $P_{is}$ corresponds to the second factor $P_{ss} = 1$. By these reasons we have $P_{is}*P_{ss} = P_{is}$ for the equation (3). So the sparse square matrix $|P_{ij}|^2$ contains only the same single non-zero column $P_{is}$ like as the matrix $|P_{ij}|$. Consequently $|P_{ij}|^2=|P_{ij}|$; in other words, this matrix $|P_{ij}|$ is a projection operator, Q.E.D. The case of similar representations of such $2^n*2^n$-matrices on basis of a sum of “row projectors” has its proof by analogy.

This theorem allows making the following conclusion about any variant of matrix presentations of complex numbers, hyperbolic numbers and their extensions into $2^n$-dimensional numerical systems (including Hamilton's quaternions and biquaternions, split-quaternions and bisplit-quaternions by Cockle, etc.): if the real part of such $2^n$-dimensional number is equal to +1, then its matrix presentation is a sum of $2^n$ «column projectors» (and «row projectors»). It is provided by the fact that real parts of such multidimensional numerical systems are represented by matrix diagonal that contains only entries +1. Figure 30 shows an example of one of matrix presentations of Hamilton quaternions in a case when their real parts are equal to +1.

| | | | | | | | | | | |
|---|---|---|---|---|---|---|---|---|---|---|
| 1 | b | c | d | | 1 0 0 0 | | 0 b 0 0 | | 0 0 c 0 | | 0 0 0 d |
| -b | 1 | -d | c | = | -b 0 0 0 | + | 0 1 0 0 | + | 0 0 -d 0 | + | 0 0 0 c |
| -c | d | 1 | -b | | -c 0 0 0 | | 0 d 0 0 | | 0 0 1 0 | | 0 0 0 -b |
| -d | -c | b | 1 | | -d 0 0 0 | | 0 -c 0 0 | | 0 0 b 0 | | 0 0 0 1 |

Figure 30. The "column decomposition" of the classical (4*4)-matrix representation of Hamilton's quaternions (http://en.wikipedia.org/wiki/Quaternion#Matrix_representations) in cases when their real parts are equal to +1. Each of 4 matrices (on the right) is a projection operator in accordance with the described theorem. Here b, c, d are real numbers.

So, many kinds of hypercomplex numbers are based on sums of projectors. In this sense the notion "projectors" can be considered as more fundemental than the notion "hypercomplex numbers" of the mentioned types. Many of these hypercomplex numbers are applied widely in different fields of science: physics, chemistry, informatics, etc. Awareness of the fact that these systems of hypercomplex numbers are based on sums of projectors may help in a rethinking of existing theories and in developing new theories in the field of mathematical natural science. In particularly, it concerns Hamilton's quaternions. For example, Maxwell has used them in creation of his equations of electro-magnetic field. Could one develop an alternative description and development of the theory of electro-magnetic field on basis of sums of appropriate projectors? It is one of many open questions in theoretical aplications of projectors.

Now we show that each of genetic Rademacher and Hadamard matrices (including $R_2$, $R_8$, $H_4$, $H_8$ from Figure 1 and their extensions into $2^n*2^n$-matrices $R_K$ and $H_K$ in expressions (2)) can be expressed as sums and tensor multiplications of four (2*2)-matrices of «column projectors» (or of analogical «row projectors»). Figure 31 shows these 4 basic (2*2)-projectors, which are marked by 4 different colours for visibility, and some examples of expressions of a few Rademacher and Hadamard matrices by means of their using.

$$\begin{vmatrix} 1 & 0 \\ 1 & 0 \end{vmatrix} ; \begin{vmatrix} 1 & 0 \\ -1 & 0 \end{vmatrix} ; \begin{vmatrix} 0 & -1 \\ 0 & 1 \end{vmatrix} ; \begin{vmatrix} 0 & 1 \\ 0 & 1 \end{vmatrix}$$

$$\begin{vmatrix} 1 & 1 \\ 1 & 1 \end{vmatrix} = \begin{vmatrix} 1 & 0 \\ 1 & 0 \end{vmatrix} + \begin{vmatrix} 0 & 1 \\ 0 & 1 \end{vmatrix} ; \qquad \begin{vmatrix} 1 & -1 \\ 1 & 1 \end{vmatrix} = \begin{vmatrix} 1 & 0 \\ 1 & 0 \end{vmatrix} + \begin{vmatrix} 0 & -1 \\ 0 & 1 \end{vmatrix}$$

$$R_4 = \begin{vmatrix} 1 & 0 \\ 1 & 0 \end{vmatrix} \otimes \begin{vmatrix} 1 & 0 \\ -1 & 0 \end{vmatrix} + \begin{vmatrix} 0 & 1 \\ 0 & 1 \end{vmatrix} \otimes \begin{vmatrix} 1 & 0 \\ -1 & 0 \end{vmatrix} + \begin{vmatrix} 1 & 0 \\ -1 & 0 \end{vmatrix} \otimes \begin{vmatrix} 0 & 1 \\ 0 & 1 \end{vmatrix} + \begin{vmatrix} 0 & -1 \\ 0 & 1 \end{vmatrix} \otimes \begin{vmatrix} 0 & 1 \\ 0 & 1 \end{vmatrix}$$

$$H_4 = \begin{vmatrix} 1 & 0 \\ 1 & 0 \end{vmatrix} \otimes \begin{vmatrix} 1 & 0 \\ -1 & 0 \end{vmatrix} + \begin{vmatrix} 0 & -1 \\ 0 & 1 \end{vmatrix} \otimes \begin{vmatrix} 1 & 0 \\ -1 & 0 \end{vmatrix} + \begin{vmatrix} 1 & 0 \\ -1 & 0 \end{vmatrix} \otimes \begin{vmatrix} 0 & 1 \\ 0 & 1 \end{vmatrix} + \begin{vmatrix} 0 & 1 \\ 0 & 1 \end{vmatrix} \otimes \begin{vmatrix} 0 & 1 \\ 0 & 1 \end{vmatrix}$$

Figure 31. Examples of using the 4 basic (2*2)-projectors (upper level) to express (2*2)-matrix representations of hyperbolic number and of complex number with unit coordinates (the second level) and to express the Rademacher (4*4)-matrix $R_4$ and the Hadamard (4*4)-matrix $H_4$ from Figure 1 (two lower levels)

One can also note that every of the genetic "column $(2^n*2^n)$-projectors" and the "row $(2^n*2^n)$-projectors" can be expressed by means of tensor multiplications of appropriate (2*2)-projectors from their basic set of the 4 projectors (Figure 31, upper level). It means that cases of 2-dimensional spaces can be considered as basic in this model approach. It is interesting because of the known fact that namely 2-dimensional sub-spaces play a fundamental role in morphological organization and development of living bodies (see for example about a fundamental role of primary tissue layers or primary germ layers in http://en.wikipedia.org/wiki/Germ_layer; in accordance with germ layer theory, for example, all different organs of human bodies develop from one of the 3 germ layers).

## 8. AN APPLICATION OF OBLIQUE PROJECTORS TO SIMULATE ENSEMBLES OF PHYLLOTAXIS PATTERNS IN LIVING BODIES

In the field of mathematical biology, phyllotaxis phenomena are one of the most known [Adler, Barabe, Jean, 1997; Jean, 1995; http://www.goldenmuseum.com/0604Phillotaxis_engl.html]. Usually phyllotaxis laws are described as those inherited spiral-like dislocations of leaves and some other parts of plants, which are connected with Fibonacci numbers. But the similar phyllotaxis laws dictate also inherited configurations of some biological molecules, parts of animal bodies, etc. (see, for example, a review in [Jean, 1995]). In other words, phyllotaxis laws appear in inherited morphological structures at very different levels and branches of biological evolution. Figure 32 shows a few examples of phyllotaxis spirals.

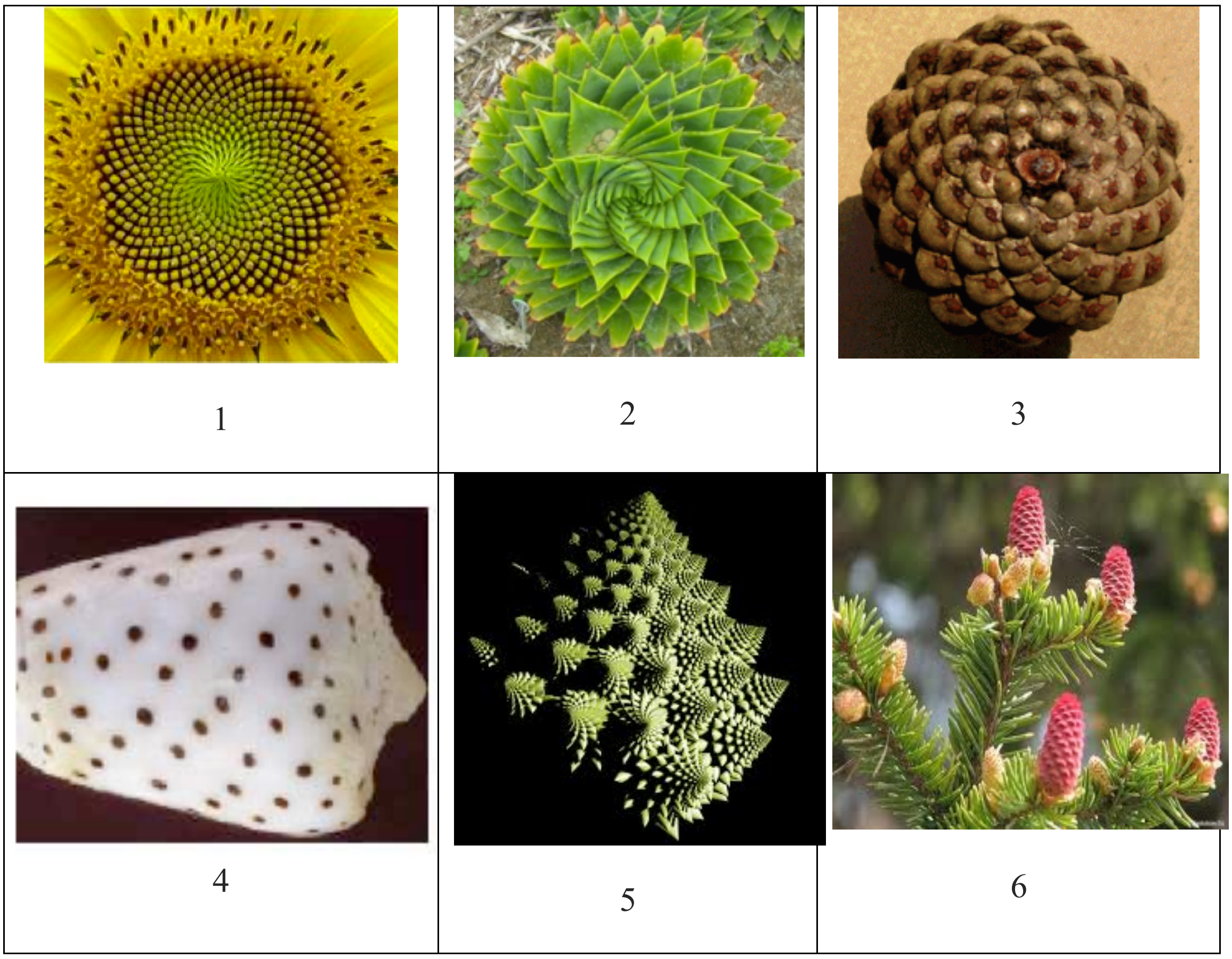

Figure 32. Examples of phyllotaxis patterns: 1) sunflower; 2) alloe (from http://radiolus.com/index.php/nepoznannoe/255-paradoksy-posledovatelnosti-chisel-fibonachchi ); 3) a pine cone (from http://www.maths.surrey.ac.uk/hosted-sites/R.Knott/Fibonacci/fibnat.html); 4) a seashell (from http://www.eb.tuebingen.mpg.de/?id=476 ); 5) a fractal vegetable Romanesco Broccoli (Brassica oleracea) (from http://egregores.blogspot.ru/2010/12/extremely-cool-natural-fractals.html ); 6) a spruce with cones (from http://foto.rambler.ru/users/nadezhda-rodnaja/albums/53780408/photo/4e9c302c-1a1a-3e9d-c61e-db70c258d714/ )

On Figure 32, images 5 and 6 illustrate that a whole organism can contain many parts with similar phyllotaxis patterns in each (like as a spruce with many phyllotaxis cones). For physicist or mathematician is natural to think that such organism can be modelled as a multidimensional phase space (or a configuration space) with an appropriate number of similar subspaces, each of which receives the same phyllotaxis pattern due to a selective control of subspaces or their selective genetic coding. Such selective control or coding in this phase space should be based on an appropriate system of operators (about a phase space see http://en.wikipedia.org/wiki/Phase_space).

Our model approach allows such modelling due to a discovered system of operators based on described sums of oblique projectors (including $2^n*2^n$-dimensional matrices $R_k$ and $H_k$ in expressions (2)), which have such properties of a selective control (or coding) of subspaces of $2^n$-dimensional space. In other words, we propose an approach to model ensembles of phyllotaxis patterns (or other patterns and processes) inside a multidmensional phase space that represents a whole organism. The described system of operators for a selective control or coding can be conditionally named briefly as a «genetic system of operators» (or more briefly, «G-system of operators). As we can judge, in the field of phyllotaxis study, other authors didn't simulate such ensembles of phyllotaxis patterns though many different models of separate patterns (without their ensembles in a joint phase space) exist. In addition, known models of inherited phyllotaxis patterns don't associate them with structural properties of the genetic coding system in contrast to our genetic approach. Let us explain our model approach to phyllotaxis phenomena in more details.

It is known that classical phyllotaxis patterns arise in the result of iterative rotations of initial object approximately on an angle $137^0$ with a simultaneous increase of its distance from the center of the phyllotaxis pattern. On a complex plane such iterative operations can be simulated by means of iterative multiplication of an initial vector (or a point) with appropriate complex number $z = x+i*y$, which provides such angle of rotation (due to its appropriate argument) and increase (due to its appropriate modulus) in accordance with known properties of complex numbers. The described G-system of operators, which contains many variants of sparse $2^n*2^n$-matrix presentations of complex numbers on basis of sums of some genetic projectors (from "column decompositions" of $2^n*2^n$-matrices $R_K$ and $H_K$ in expressions (2)), allows generating many phyllotaxis patterns, each of which belong to its own 2-dimensional plane inside a whole phase space. In these cases, each of phyllotaxis patterns in a separate phase plane can have its own degree of maturation (or development) and its own type of a phyllotaxis picture; it depends on a kind of complex numbers, which is chosen for its iterative generating.

This model approach does not pretend to a new explanation for existence of Fibonacci numbers in phyllotaxis patterns. But this model approach give abilities to simulate bunches of phyllotaxis patterns in separate organisms. Concerning to Fibonacci or Luca numbers in phyllotaxis laws, one should remind here that "*the phyllotaxis rules ... cannot be taken as applying to all circumstances, like a law of nature. Rather, in the words of the famous Canadian mathematician Coxeter, they are 'only a fascinatingly prevalent tendency*" (http://goldenratiomyth.weebly.com/phyllotaxis-the-fibonacci-sequence-in-nature.html ). One can think that the role of iterative operations in living nature is much more important than particular realizations of Fibonacci or Luca numbers.

To simulate an ensemble of phyllotaxis 3d-patterns (an ensemble of many cones of a spruce, etc.), each of which belongs to a separate subspace of a whole $2^n*2^n$-dimensional phase space, an iterative application of Hamilton's quaternions can be used in their $2^n*2^n$-matrix forms of presentation in the described G-system of operators.

In addition the author reminds here about cyclic groups on basis of Hamilton's quaternion and biquaternion with unit coordinates: these cyclic groups allow simulating some heritable biological phenomena including color perception, properties of which correspond to the Newton's color circle (see [Petoukhov, 2011b] and Section 17 in [Petoukhov, 2012a]). Using Hamilton's quaternions and biquaternions as $2^n*2^n$-operators from the described G-system allows simulating some inherited ensembles of biological patterns including some inherited ensembles of color patterns and color changes of biological bodies.

## 9. HYPERBOLIC NUMBERS, GENETIC PROJECTORS AND THE WEBER-FECHNER LAW OF PSYCHOPHYSICS

This Section is devoted to hyperbolic numbers in separate planes of $2^N$-dimensional spaces, their connections with some genetic projectors (from Figures 2, 4-6, 8 and 9) and with inherited psychophysical phenomena. The name «hyperbolic» for this kind of 2-dimensional numbers is traditionally used because of their connection with hyperbolas and hyperbolic rotations (http://en.wikipedia.org/wiki/Split-complex_number, http://www.physicsinsights.org/hyperbolic_rotations.html ). Hyperbolic numbers have properties corresponding to Lorentz group of two-dimensional Special Relativity (www.springer.com/.../9783642179761-c1.pdf?, , http://garretstar.com/secciones/publications/docs/HYP2.PDF ).

$$G_{xy} = x*1+y*i = \begin{bmatrix} x, & y \\ y, & x \end{bmatrix} = x* \begin{bmatrix} 1 & 0 \\ 0 & 1 \end{bmatrix} + y* \begin{bmatrix} 0 & 1 \\ 1 & 0 \end{bmatrix} ;$$

| | 1 | i |
|---|---|---|
| 1 | 1 | i |
| i | i | 1 |

$$J = x*1+(x^2-a)^{0.5}*i = \begin{bmatrix} x, & (x^2-a)^{0.5} \\ (x^2-a)^{0.5}, & x \end{bmatrix} = x* \begin{bmatrix} 1 & 0 \\ 0 & 1 \end{bmatrix} + (x^2-a)^{0.5}* \begin{bmatrix} 0 & 1 \\ 1 & 0 \end{bmatrix}$$

Figure 33. Upper level: the matrix representation of hyperbolic numbers, where two sparse (2*2)-matrices represent real and imaginary units (1 and i) of these numbers. The multiplication table of these basic elements 1 and i is shown on the right side. Bottom level: the special case of the set of hyperbolic numbers J (with the fixed value "a") describes a hyperbola $x^2+y^2=a$, where "x" is a variable.

Hyperbolic numbers $G_{xy} = x*1+y*i$ (where «i» is the imaginary unit of hyperbolic numbers with its property $i^2=+1$) have a known (2*2)-matrix form of their representation shown on Figure 33. A special case of hyperbolic numbers

$$J=[x, (x^2-a)^{0.5}; (x^2-a)^{0.5}, x] \quad (4)$$

(where "x" is a variable and the parameter "a" is fixed) describes a hyperbola, which corresponds to the equation $x^2+y^2=a$. This equation describes the hyperbola in the coordinate system (x,y), axes of which coincide with axes of symmetry of the hyperbole. In another coordinate system, axes of which coincide with asymptotes of the hyperbola, the same hyperbola is described by the equation y=a/x (Figure 34). Any point of any hyperbola can be transformed into a new point of the same hyperbola by means of so called hyperbolic rotation, which is described by the same matrix representation of hyperbolic numbers [x, y; y, x] if its determinant is equal to 1.

It is known that hyperbolic numbers and hyperbolic operators are closely connected with natural logarithms, which can be defined on the base of hyperbolic rotations because of their relations with values of area under hyperbolas (http://mathworld.wolfram.com/NaturalLogarithm.html). By this reason natural logarithm «was formerly called hyperbolic logarithm as it corresponds to the area under a hyperbola» (http://en.wikipedia.org/wiki/Natural_logarithm ). Area s of a curvilinear trapezoid inside boundaries, which are created by the hyperbola x*y = a, the x-axis and the lines $x = x_0$ and $x = x_1$, is equal to

$$s = a*\ln(x_1/x_0) = a*\{\ln(x_1)-\ln(x_0)\}, \quad (5)$$

where ln – natural (or hyperbolic) logarithm (Figure 34). History of hyperbolic logarithms is described for example in the book [Klein, 2009].

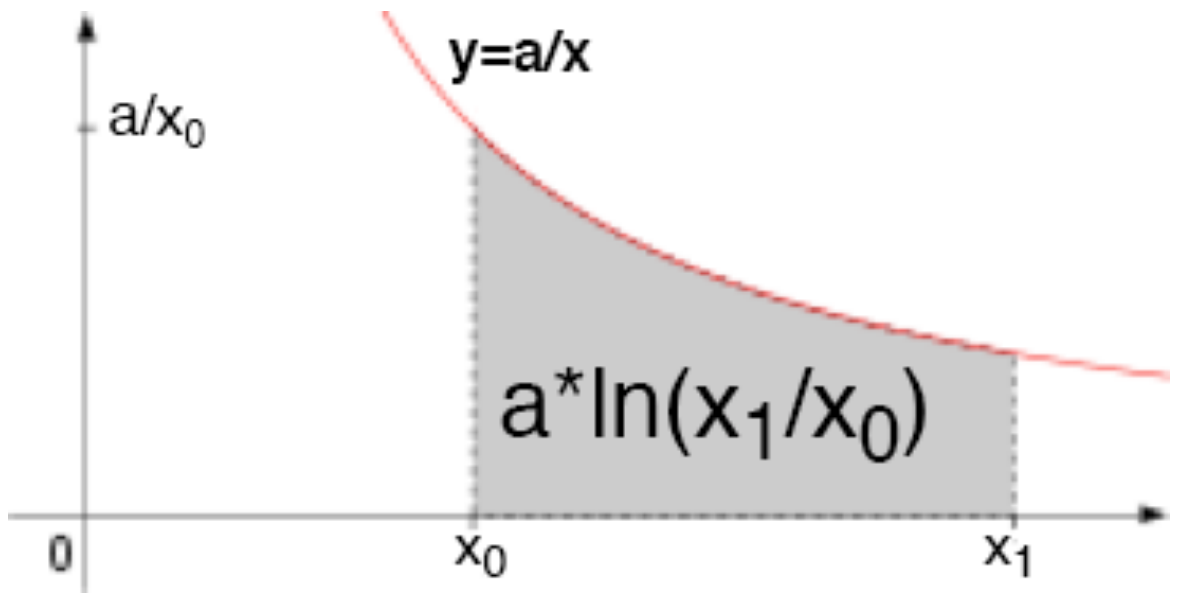

Figure 34. The function $a*\ln(x_1/x_0)$ is illustrated as the area under the hyperbola $y = a/x$ from $x_0$ to $x_1$.

But a wide class of genetically inherited physiological phenomena is organized by the nature by means of the same logarithmic law (5) and it can be described mathematically on the base of hyperbolic numbers. We mean here the main psychophysiologic law by Weber-Fechner (http://en.wikipedia.org/wiki/Weber–Fechner_law): the intensity of the perception is proportional to the logarithm of stimulus intensity; it is expressed by the equation

$$p = k*\ln(V/V_0) = k*\{\ln(V) - \ln(V_0)\}, \quad (6)$$

where p - the intensity of perception, V - stimulus, $V_0$ - threshold stimulus, ln - natural logarithm. Proportionality factor k in the expression (6) is different for different channels of sensory perception (vision, hearing, etc.); this difference of values «k» is associated with different values "a" in the equation of the hyperbola: $x*y = a$. The threshold stimulus $V_0$ in (5) is also different for different channels of sensory perception.

The identity of expressions (5) and (6) allows to propose a geometric model of the Weber-Fechner law on the base of the described matrix representation of hyperbolic numbers in connection with a phenomenology of the molecular-genetic system. In this model the threshold stimulus $V_0$ in the expression (6) is interpreted as the value $x_0$ from the expression (5); the proportionality factor "k" in (6) is interpreted as the value "a" from (5); the stimulus V(t), which is varied in time, is interpreted as the variable x(t) from (5) and the perception "p" in (6) is interpreted as the area "s" from (5). Taking into account the type of hyperbolic numbers J from the expression (4), it is obvious that the hyperbolic number $J_0=[x_0, (x_0^2-a)^{0.5}; (x_0^2-a)^{0.5}, x_0]$ corresponds to the fixed area $s_0=a*\ln(x_0)$ of the curvilinear trapezoid, points of which have their x-coordinates inside the interval from 1 till $x_0$; another hyperbolic number $J=[x, (x^2-a)^{0.5}; (x^2-a)^{0.5}, x_0]$ corresponds to the area $s_x=a*\ln(x)$ of the curvilinear trapezoid with its x-coordinates inside the bigger interval from 1 till x. The total area s of the third curvilinear trapezoid, points of which have their x-coordinates inside the interval from $x_0$ till x, is equal to difference $s_x$-$s_0$, that is

$$s = s_x-s_0 = a*\ln(x) - a*\ln(x_0) = a*\ln(x_1/x_0) \quad (7)$$

This curvilinear trapezoid with its area s corresponds to the hyperbolic number, which is equal to the following:

$$J - J_0 = [(x-x_0), ((x-x_0)^2-a)^{0.5}; ((x-x_0)^2-a)^{0.5}, x_0] \quad (8)$$

We have described the geometric model for the case of one sensory channel. Let us generalize now this geometric model for a case of a multi-dimensional space with different 2-dimensional planes inside it, each of which contains the described hyperbolic model of the Weber-Fechner law for one of many sensory channels of an organism. It is known that different types of inherited sensory perception are subordinated to this law: sight, hearing, smell, touch, taste, etc. One can suppose that the Weber–Fechner law is the law specially for nervous system. But it is not true since its meaning is much more wider because it is realized in many kinds of lower organisms without a nervous system in them: "*this law is applicable to chemo-tropical, helio-tropical and geo-tropical movements of bacteria, fungi and antherozoids of ferns, mosses and phanerogams .... The Weber-Fechner law, therefore, is not the law of the nervous system and its centers, but the law of protoplasm in general and its ability to respond to stimuli*" [Shultz, 1916, p. 126]. So the whole system of perception of an organism demonstrates itself as a multi-parametric system with many sub-systems of perception, which are subordinated to the logarithmic Weber-Fechner law. A generalized model of multi-dimensional space with appropriate sub-spaces is needed to describe this inherited multi-channel organization of the logarithmic perception in the case of the whole living organism. Let us show that our approach on the base of sums of genetic projectors allows to create such generalized model.

Hyperbolic numbers $G_{xy}=x*1+y*i$ have the multiplication table of their real and imaginary units shown on Figure 33. But the same multiplication table of hyperbolic numbers we have already received above:

- on Figures 5 and 6 for sums $(c_0+c_2)$, $(c_1+c_3)$, $(r_0+r_3)$, $(r_1+r_2)$ of genetic (4*4)-projectors $c_0$, $c_1$, $c_2$, $c_3$, $r_0$, $r_1$, $r_2$, $r_3$ from Figure 2;
- on Figure 29 for sums $(s_0+s_4)$, $(s_1+s_5)$, $(s_2+s_6)$, $(s_3+s_7)$ of genetic (8*8)-projectors $s_0$, $s_1$,…, $s_7$ from Figure 7.

It means that these sums of genetic (4*4)- and (8*8)-projectors are matrix representations of hyperbolic numbers with unit coordinates in corresponding 2-dimensional planes of 4-dimensional and 8-dimensional spaces. On the basis of each of these sums, one can construct a general representation of hyperbolic numbers (with an arbitrary values of their coordinates) in an appropriate hyperbolic plane of such multi-dimensional space. For example, if $e_0$ and $e_4$ are (8*8)-matrices taken from the decomposition of the (8*8)-matrix $s_0+s_4$ on Figure 29, then the expression $G_{04}=a_0*e_0+b_4*e_4$ represents the hyperbolic number with real coordinates «$a_0$» and «$b_4$» in the plane $(x_0, x_4)$ inside an 8-dimensional space with its coordinate system $(x_0, x_1, …, x_7)$. Similar situations holds true for expressions $G_{15}=a_1*e_1+b_5*e_5$, $G_{26}=a_2*e_2+b_6*e_6$, $G_{37}=a_3*e_3+b_7*e_7$, each of which represents hyperbolic number in an appropriate plane inside (8*8)-dimensional space, if $e_1$, $e_2$, $e_3$, $e_5$, $e_6$, $e_7$ are (8*8)-matrices taken from decompositions on Figure 29. The (8*8)-matrix $G_{04}+G_{15}+G_{26}+G_{37}$ represents an operator of an 8-dimensional space with a set of 4 hyperbolic planes $(x_0, x_4)$, $(x_1, x_5)$, $(x_2, x_6)$, $(x_3, x_7)$ inside it.

For each of these four 2-dimensional planes $(x_0, x_4)$, $(x_1, x_5)$, $(x_2, x_6)$, $(x_3, x_7)$ of an 8-dimensional space one can choose an individual hyperbola $x^2-y^2=a$ and an individual threshold stimulus $x_0$ to create the hyperbolic model of the Weber-Fechner law in the united set of four perception channels of the organism. In this case we have the following expression of the model for each of the planes, if we take (8*8)-matrices $e_0$, $e_1$, …, $e_7$ from Figure 29 to express real and imaginary units of hyperbolic numbers (here $V_0$, $V_1$, $V_2$, $V_3$ mean stimulus variables on $x_0$-, $x_1$-, $x_2$-, $x_3$-axis correspondingly; $V_{00}$, $V_{01}$, $V_{02}$, $V_{03}$ mean threshold stimuluses on the same $x_0$-, $x_1$-, $x_2$-, $x_3$-axis correspondingly; parameters $a_0$, $a_1$, $a_2$, $a_3$ for corresponding planes mean the parameter "a" from the general equation $x^2$-$y^2$=a of hyperbola; $S_{04}$, $S_{15}$, $S_{26}$, $S_{37}$ mean area of corresponding trapezoids defined by the hyperbolic numbers in these four planes; parameters $k_0$, $k_1$, $k_2$, $k_3$ for corresponding planes mean the parameter k from (6)):

- for the plane ($x_0$, $x_4$): $G_{04}=(V_0-V_{00})*e_0+\{(V_0-V_{00})^2-a_0)^{0.5}\}*e_4$; this hyperbolic number $G_{04}$ corresponds to the area $S_{04}= k_0*\ln(V_0/V_{00})$, which simulates the Weber-Fechner law in the plane ($x_0$, $x_4$);
- for the plane ($x_1$, $x_5$): $G_{15}=(V_1-V_{01})*e_1+\{(V_1-V_{01})^2-a_1)^{0.5}\}*e_5$; this hyperbolic number $G_{15}$ corresponds to the area $S_{15}= k_1*\ln(V_1/V_{01})$, which simulates the Weber-Fechner law in the plane ($x_1$, $x_5$);
- for the plane ($x_2$, $x_6$): $G_{26}=(V_2-V_{02})*e_2+\{(V_2-V_{02})^2-a_2)^{0.5}\}*e_6$; this hyperbolic number $G_{26}$ corresponds to the area $S_{26}= k_2*\ln(V_2/V_{02})$, which simulates the Weber-Fechner law in the plane ($x_2$, $x_6$);
- for the plane ($x_3$, $x_7$): $G_{37}=(V_3-V_{03})*e_3+\{(V_3-V_{03})^2-a_0)^{0.5}\}*e_7$; this hyperbolic number $G_{37}$ corresponds to the area $S_{37}= k_3*\ln(V_3/V_{03})$, which simulates the Weber-Fechner law in the plane ($x_3$, $x_7$).

Figure 35 shows the (8*8)-matrix W, which represents the sum of the (8*8)-matrix representations of these hyperbolic numbers: $W=G_{04}+G_{15}+G_{26}+G_{37}$. One can see that the (8*8)-matrix W on Figure 35 has a special structure: its both (4*4)-quadrants along each diagonals are identical to each other. This fact allows expressing the matrix W in the following form:

$$W = [1\ 0;\ 0\ 1]\otimes M_0 + [0\ 1;\ 1\ 0]\otimes M_1 \ , \qquad (9)$$

where [1 0; 0 1] and [0 1; 1 0] are matrix representations of real and imaginary units of hyperbolic numbers; $M_0$ is the (4*4)-matrix, which reproduces each of (4*4)-quadrants along the main diagonal; $M_1$ is the (4*4)-matrix, which reproduces each of (4*4)-quadrants along the second diagonal. The expression (9) means that the matrix W belongs to so called «tensornumbers» (more precisely, to a category of «tensorhyperbolic numbers»), which will be introduced below in a special Section.

Our approach, which was described above on the base of genetic matrices, allows natural modeling such kind of $2^N$-parametric systems with its 2-parametric hyperbolic sub-systems by means of the described type of a $(2^N*2^N)$-matrix operator of a $2^N$-dimensional space with appropriate quantity of hyperbolic planes inside it (in this case each of hyperbolic planes corresponds to an individual channel of perception with its own coefficient k and the threshold value $V_0$ in the expression (6)). This $(2^N*2^N)$-matrix operator also belongs to the category of tensorhyperbolic numbers, because it can be expressed by means of the expression (9), where $M_0$ and $M_1$ are $(2^{N-1}*2^{N-1})$-matrices.

| | | | | | | | |
|---|---|---|---|---|---|---|---|
| $(V_0-V_{00})$ | $(V_1-V_{01})$ | $(V_2-V_{02})$ | $(V_3-V_{03})$ | $\{(V_0-V_{00})^2-a_0)^{0.5}\}$ | $\{(V_1-V_{01})^2-a_1)^{0.5}\}$ | $-\{(V_2-V_{02})^2-a_2)^{0.5}\}$ | $-\{(V_3-V_{03})^2-a_0)^{0.5}\}$ |
| $(V_0-V_{00})$ | $(V_1-V_{01})$ | $(V_2-V_{02})$ | $(V_3-V_{03})$ | $\{(V_0-V_{00})^2-a_0)^{0.5}\}$ | $\{(V_1-V_{01})^2-a_1)^{0.5}\}$ | $-\{(V_2-V_{02})^2-a_2)^{0.5}\}$ | $-\{(V_3-V_{03})^2-a_0)^{0.5}\}$ |
| $-(V_0-V_{00})$ | $-(V_1-V_{01})$ | $(V_2-V_{02})$ | $(V_3-V_{03})$ | $-\{(V_0-V_{00})^2-a_0)^{0.5}\}$ | $-\{(V_1-V_{01})^2-a_1)^{0.5}\}$ | $-\{(V_2-V_{02})^2-a_2)^{0.5}\}$ | $-\{(V_3-V_{03})^2-a_0)^{0.5}\}$ |
| $-(V_0-V_{00})$ | $-(V_1-V_{01})$ | $(V_2-V_{02})$ | $(V_3-V_{03})$ | $-\{(V_0-V_{00})^2-a_0)^{0.5}\}$ | $-\{(V_1-V_{01})^2-a_1)^{0.5}\}$ | $-\{(V_2-V_{02})^2-a_2)^{0.5}\}$ | $-\{(V_3-V_{03})^2-a_0)^{0.5}\}$ |
| $\{(V_0-V_{00})^2-a_0)^{0.5}\}$ | $\{(V_1-V_{01})^2-a_1)^{0.5}\}$ | $-\{(V_2-V_{02})^2-a_2)^{0.5}\}$ | $-\{(V_3-V_{03})^2-a_0)^{0.5}\}$ | $(V_0-V_{00})$ | $(V_1-V_{01})$ | $(V_2-V_{02})$ | $(V_3-V_{03})$ |
| $\{(V_0-V_{00})^2-a_0)^{0.5}\}$ | $\{(V_1-V_{01})^2-a_1)^{0.5}\}$ | $-\{(V_2-V_{02})^2-a_2)^{0.5}\}$ | $-\{(V_3-V_{03})^2-a_0)^{0.5}\}$ | $(V_0-V_{00})$ | $(V_1-V_{01})$ | $(V_2-V_{02})$ | $(V_3-V_{03})$ |
| $-\{(V_0-V_{00})^2-a_0)^{0.5}\}$ | $-\{(V_1-V_{01})^2-a_1)^{0.5}\}$ | $-\{(V_2-V_{02})^2-a_2)^{0.5}\}$ | $-\{(V_3-V_{03})^2-a_0)^{0.5}\}$ | $-(V_0-V_{00})$ | $-(V_1-V_{01})$ | $(V_2-V_{02})$ | $(V_3-V_{03})$ |
| $-\{(V_0-V_{00})^2-a_0)^{0.5}\}$ | $-\{(V_1-V_{01})^2-a_1)^{0.5}\}$ | $-\{(V_2-V_{02})^2-a_2)^{0.5}\}$ | $-\{(V_3-V_{03})^2-a_0)^{0.5}\}$ | $-(V_0-V_{00})$ | $-(V_1-V_{01})$ | $(V_2-V_{02})$ | $(V_3-V_{03})$ |

Figure 35. The matrix representation of the sum $W=G_{04}+G_{15}+G_{26}+G_{37}$ of the hyperbolic numbers $G_{04}$, $G_{15}$, $G_{26}$, $G_{37}$, which belong to four planes $(x_0, x_4)$, $(x_1, x_5)$, $(x_2, x_6)$, $(x_3, x_7)$ inside the 8-dimensional space (explanation in the text).

So we have two important facts:

- the logarithmic Weber-Fechner law has a total meaning for different sub-systems of perception inside the whole perception system of organism;
- this unity of all subsystems of perception inside the whole organism, which are subordinated to the Weber-Fechner law, can be expressed by means of $2^N$-dimensional tensorhyperbolic numbers by analogy with the expression (9).

These facts allow the author to put forward the following statement (or hypothesis): a living organism percepts an external world as a multi-parametric system, which belongs to a tensorhyperbolic category. In other words, for the whole perception system of a living organism, the external world is a life of tensorhyperbolic numbers in time. Correspondingly, interrelations of a living organism with the external world are realized on the base of processing of percepted tensorhyperbolic numbers, which are systematically changed over time in accordance with changes of external stimuluses (in addition, the author believes that a living organism can be regarded as a life of tensornumbers over time; in this approach tensorcomplex numbers and their extensions deserve special attention). This mathematical approach to phenomenology of perception of the world is closely connected with the multi-parametric system of the genetic coding in its matrix form of representation; such connection allows a genetical transfer of this general biological property along a chain of generations.

## 10. **REFLECTION OPERATORS AND GENETIC PROJECTORS**.

By definition, a linear operator L is the reflection operator (or briefly, a "reflection") if and only if it satisfies the following criterion: $L^2 = E$, where E is the identity operator (it is also denoted as «1»), that is the real unit (see for example [Vinberg, 2003, Chapter 6]). The imaginary unit «i» of hyperbolic numbers satisfies this criterion and consequently it is the reflection operator: $i^2=+1$ (see Figure 33). Hyperbolic number with unit coordinates (1+i) is sum of the identity operator «1» and the reflection operator «i». The well-known (2*2)-matrix represenation of the imaginary unit is the following: [0 1; 1 0] (Figure 33). The acion of this reflection operator on a voluntary 2-dimensional vector generates a new vector, which is a mirror-symmetrical analogue of the initial vector relative to the bisector of the angle between the x-axis and y-axis of the coordinate system (x, y). For example, [3, 5]*[0 1; 1 0] = [5, 3]. A reflection is an involution: when applied twice in succession, every point returns to its original location, and every geometrical object is restored to its original state.

But we have shown above that many of sums of genetic projectors are the ($2^N*2^N$)-matrix representation of hyperbolic numbers, which have their own real and imaginary units in respective planes inside $2^N$-dimensional space (see for example Figures 5, 6 and 29). Correspondingly, imaginary units of these hyperbolic numbers are ($2^N*2^N$)-operators of reflections in these planes inside $2^N$-dimensional space. It gives evidences in favor of that the system of genetic coding actively uses also reflection operators. It is interesting because mirror reflections exist in many genetically inherited biological structures, including left and right halves of human and animal bodies. In the author's laboratory, the genetic ($2^N*2^N$)-matrices of reflection operators are used to analyze mirror symmetries in molecular-genetic systems including long genetic sequences, genetic palindromes, chromosomal inversion, etc.

## 11. THE SYMBOLIC MATRICES OF GENETIC DUPLETS AND TRIPLETS

In Section 1 the author promised to explain a relation of numeric matrices $R_4$, $R_8$, $H_4$, $H_8$ (Figure 1), which were the initial matrices in this article, with a phenomenology of the genetic coding system in matrix forms of its representation. This Section is devoted to the explanation. Theory of noise-immunity coding is based on matrix methods. For example, matrix methods allow transferring high-quality photos of Mar's surface via millions of kilometers of strong interference. In particularly, tensor families of Hadamard matrices are used for this aim. Tensor multiplication of matrices is the well-known operation in fields of signals processing technology, theoretical physics, etc. It is used for transition from spaces with a smaller dimension to associated spaces of higher dimension.

By analogy with theory of noise-immunity coding, the 4-letter alphabet of RNA (adenine A, cytosine C, guanine G and uracil U) can be represented in a form of the symbolic (2*2)-matrix [C U; A G] (Figure 36) as a kernel of the tensor family of symbolic matrices $[C\ U;\ A\ G]^{(n)}$, where (n) means a tensor power (Figure 36). Inside this family, this 4-letter alphabet of monoplets is connected with the alphabet of 16 duplets and 64 triplets by means of the second and third tensor powers of the kernel matrix: $[C\ U;\ A\ G]^{(2)}$ and $[C\ U;\ A\ G]^{(3)}$, where all duplets and triplets are disposed in a strict order (Figure 36). We begin with the alphabet A, C, G, U of RNA here because of mRNA-sequences of triplets define protein sequences of amino acids in a course of its reading in ribosomes.

Figure 36 contains not only 64 triplets but also amino acids and stop-codons encoded by the triplets in the case of the Vertebrate mitochondrial genetic code that is the most symmetrical among known variants of the genetic code (http://www.ncbi.nlm.nih.gov/Taxonomy/Utils/wprintgc.cgi). Let us explain the black-and-white mosaics of $[C\ U;\ A\ G]^{(2)}$ and $[C\ U;\ A\ G]^{(3)}$ (Figure 36) which reflect important features of the genetic code. These features are connected with a specificity of reading of mRNA-sequences in ribosomes to define protein sequences of amino acids (this is the reason, why we use the alphabet A, C, G, U of RNA in matrices on Figure 36; below we will consider the case of DNA-sequences separately).

| | |
|---|---|
| C | U |
| A | G |

| | | | |
|---|---|---|---|
| CC | CU | UC | UU |
| CA | CG | UA | UG |
| AC | AU | GC | GU |
| AA | AG | GA | GG |

| | | | | | | | |
|---|---|---|---|---|---|---|---|
| CCC Pro | CCU Pro | CUC Leu | CUU Leu | UCC Ser | UCU Ser | UUC Phe | UUU Phe |
| CCA Pro | CCG Pro | CUA Leu | CUG Leu | UCA Ser | UCG Ser | UUA Leu | UUG Leu |
| CAC His | CAU His | CGC Arg | CGU Arg | UAC Tyr | UAU Tyr | UGC Cys | UGU Cys |
| CAA Gln | CAG Gln | CGA Arg | CGG Arg | UAA Stop | UAG Stop | UGA Trp | UGG trp |
| ACC Thr | ACU Thr | AUC Ile | AUU Ile | GCC Ala | GCU Ala | GUC Val | GUU Val |
| ACA Thr | ACG Thr | AUA Met | AUG Met | GCA Ala | GCG Ala | GUA Val | GUG Val |
| AAC Asn | AAU Asn | AGC Ser | AGU Ser | GAC Asp | GAU Asp | GGC Gly | GGU Gly |
| AAA Lys | AAG Lys | AGA Stop | AGG Stop | GAA Glu | GAG Glu | GGA Gly | GGG Gly |

Figure 36. The first three representatives of the tensor family of RNA-alphabetic matrices [C U; A G]$^{(n)}$. Black color marks 8 strong duplets in the matrix [C U; A G]$^{(2)}$ (at the top) and 32 triplets with strong roots in the matrix [C U; A G]$^{(3)}$ (bottom). 20 amino acids and stop-codons, which correspond to the triplets, are also shown in the matrix [C U; A G]$^{(3)}$ for the case of the Vertebrate mitochondrial genetic code

A combination of letters on the two first positions of each triplet is ususally termed as a "root" of this triplet [Konopelchenko, Rumer, 1975a,b; Rumer, 1968]. Modern science recognizes many variants (or dialects) of the genetic code, data about which are shown on the NCBI's website http://www.ncbi.nlm.nih.gov/Taxonomy/Utils/wprintgc.cgi. 19 variants (or dialects) of the genetic code exist that differ one from another by some details of correspondences between triplets and objects encoded by them (these dialects are known at July 10, 2013, but perhaps later their list be increased). Most of these dialects (including the so called Standard Code and the Vertebrate Mitochondrial Code) have the symmetric general scheme of these correspondences, where 32 "black" triplets with "strong roots" and 32 "white" triplets with "weak" roots exist (the next Section shows all of these 19 dialects in details). In this basic scheme, the set of 64 triplets contains 16 subfamilies of triplets, every one of which contains 4 triplets with the same two letters on the first positions (an example of such subsets is the case of four triplets CAC, CAA, CAT, CAG with the same two letters CA on their first positions). In the described basic scheme, the set of these 16 subfamilies of *NN*-triplets is divided into two equal subsets. The first subset contains 8 subfamilies of so called "two-position" *NN*-triplets, a coding value of which is independent on a letter on their third position: (CCC, CCT, CCA, CCG), (CTC, CTT, CTA, CTG), (CGC, CGT, CGA, CGG), (TCC, TCT, TCA, TCG), (ACC, ACT, ACA, ACG), (GCC, GCT, GCA, GCG), (GTC, GTT, GTA, GTG), (GGC, GGT, GGA, GGG). An example of such subfamilies is the four triplets CGC, CGA, CGT, CGC, all of which encode the same amino acid Arg, though they have different letters on their third position. The 32 triplets of the first subset are termed as "triplets with strong roots" [Konopelchenko, Rumer, 1975a,b; Rumer, 1968]. The following duplets are appropriate 8 strong roots for them: CC, CT, CG, AC, TC, GC, GT, GG (strong duplets). All members of these 32 *NN*-triplets and 8 strong duplets are marked by black color in the matrices [C U; A G]$^{(3)}$ and [C U; A G]$^{(2)}$ on Figure 36.

The second subset contains 8 subfamilies of “three-position” NN-triplets, the coding value of which depends on a letter on their third position: (CAC, CAT, CAA, CAG), (TTC, TTT, TTA, TTG), (TAC, TAT, TAA, TAG), (TGC, TGT, TGA, TGG), (AAC, AAT, AAA, AAG), (ATC, ATT, ATA, ATG), (AGC, AGT, AGA, AGG), (GAA, GAT, GAA, GAG). An example of such subfamilies is the four triplets CAC, CAA, CAT, CAC, two of which (CAC, CAT) encode the amino acid His and the other two (CAA, CAG) encode another amino acid Gln. The 32 triplets of the second subset are termed as “triplets with weak roots” [Konopelchenko, Rumer, 1975a,b; Rumer, 1968]. The following duplets are appropriate 8 weak roots for them: CA, AA, AT, AG, TA, TT, TG, GA (weak duplets). All members of these 32 *NN*-triplets and 8 weak duplets are marked by white color in the matrices [C U; A G]$^{(3)}$ and [C U; A G]$^{(2)}$ on Figure 36.

From the point of view of its black-and-white mosaic, each of columns of genetic matrices [C U; A G]$^{(2)}$ and [C U; A G]$^{(3)}$ has a meander-like character and coincides with one of Rademacher functions that form orthogonal systems and well known in discrete signals processing. These functions contain elements “+1” and “-1” only. Due to this fact, one can construct Rademacher representations of the symbolic genomatrices [C U; A G]$^{(2)}$ and [C U; A G]$^{(3)}$ (Figure 36) by means of the following operation: each of black duplets and of black triplets is replaced by number “+1” and each of white duplets and white triplets is replaced by number “-1”. This operation leads immediately to the matrices $R_4$ and $R_8$ from Figure 1, that are the Rademacher representations of the phenomenological genomatrices [C U; A G]$^{(2)}$ and [C U; A G]$^{(3)}$.

If columns of the matrix [C U; A G]$^{(3)}$ on Figure 36 are numerated from left to right by indexes 0, 1, 2, …, 7, one can see that 4 columns with even indexes 0, 2, 4, 6 contain 32 triplets, each of which has nitrogenous bases C or A on its third position, that is in its suffix (these C and A are usually termed “amino bases”). Other 4 columns with odd indexes 1, 3, 5, 7 contain other 32 triplets, each of which has nitrogenous bases T or G on its third position (these T and G are usually termed “keto bases”). The following important phenomen is connected with this separation of the matrix [C U; A G]$^{(3)}$ in columns with even and odd indexes: adjacent columns with indexes “0 and 1”, “2 and 3”, “4 and 5” and “6 and 7” contain identical list of amino acids and stop-codons (these adjacent columns are twins from this point of view). Consequently the symbolic matrix [C U; A G]$^{(3)}$ can be represented as a sum of two sparse matrices with identical lists of amino acids and stop-codons: the first of these two matrices coincides with the matrix [C U; A G]$^{(3)}$ in columns with even indexes and has zero columns with odd indexes; the second one coincides with the matrix [C U; A G]$^{(3)}$ in columns with odd indexes and has zero columns with even indexes.

By analogy the Rademacher representation $R_8$ of this symbolic matrix [C U; A G]$^{(3)}$ can be also decomposed into two sparse matrices $RL_8$ and $RR_8$ (Figure 26), the first of which has all non-zero columns with even indexes and the second one has all non-zero columns with odd indexes. As it was shown above, each of these numeric (8*8)-matrices $RL_8$ and $RR_8$ represents split-quaternion by Cockle, whose coordinates are equal to 1, in an 8-dimensional space. It means that the system of correspondences between the set of 64 triplets (with their internal separation into subsets of triplets with strong and weak roots) and the set of 20 amino acids and stop-codon is created by the nature in accordance with the layout of these two split-quaternions $RL_8$ and $RR_8$ in an 8-dimensional space. In some extend this double numeric construction resembles double helix of DNA.

But each of these (8*8)-matrices $RL_8$ and $RR_8$ consists of two hyperbolic numbers (Figure 29): $RL_8 = (e_0+e_4)+(e_2+e_6)$ and $RR_8=(e_1+e_5)+(e_3+e_7)$. In other words the system of correspondences between the set of triplets and the set of amino acids and stop-codons is based on the mentioned 4 hyperbolic numbers in an 8-dimensional space. One can mention that here we are meeting again with a set of 4 elements in some analogy with the sets of 4 elements in the genetic alphabets of nitrogenous bases in DNA and RNA - A, C, G, T/U (and also with the Ancient set of Pythagorean Tetraktys - http://en.wikipedia.org/wiki/Tetractys). Whether any symmetry exists between structures of these 4 hyperbolic numbers and the phenomenological disposition of amino acids and stop-codons in the genomatrix $[C\ U;\ A\ G]^{(3)}$ on Figure 36? To receive an answer on this question, let us compare a content of corresponding cells of the symbolic genomatrix $[C\ U;\ A\ G]^{(3)}$ (Figure 36) with non-zero cells of matrices $(e_0+e_4)$, $(e_2+e_6)$, $(e_1+e_5)$ and $(e_3+e_7)$, which represent these 4 hyperbolic numbers (Figure 29). Figure 37 shows results of such comparison for the hyperbolic numbers $(e_0+e_4)$ and $(e_2+e_6)$ with even indexes of their column projectors; results of such comparison for the hyperbolic numbers $(e_1+e_5)$ and $(e_3+e_7)$ with odd indexes are identical because in the matrix $[C\ U;\ A\ G]^{(3)}$ (Figure 36) adjacent columns with indexes "0 and 1", "2 and 3", "4 and 5" and "6 and 7" contain identical lists of amino acids and stop-codons. Those amino acids, which belong to matrix cells with triplets of strong roots, are marked by bold letters on Figure 37. One can see here the following symmetrical feature: each of real and imagine parts of these hyperbolic numbers $(e_0+e_4)$ and $(e_2+e_6)$ contains an equal quantity of amino acids marked by bold letters and also an equal quantity of amino acids of another type. By such way we receive a special separation of the set of 20 amino acids into a few groups, which belong to real parts or to imagine parts of these hyperbolic numbers and which should be analized in future more attentively.

$$e_0+e_4 = \begin{bmatrix} 1&0&0&0&0&0&0&0\\ 1&0&0&0&0&0&0&0\\ -1&0&0&0&0&0&0&0\\ -1&0&0&0&0&0&0&0\\ 0&0&0&0&1&0&0&0\\ 0&0&0&0&1&0&0&0\\ 0&0&0&0&-1&0&0&0\\ 0&0&0&0&-1&0&0&0 \end{bmatrix} + \begin{bmatrix} 0&0&0&0&1&0&0&0\\ 0&0&0&0&1&0&0&0\\ 0&0&0&0&-1&0&0&0\\ 0&0&0&0&-1&0&0&0\\ 1&0&0&0&0&0&0&0\\ 1&0&0&0&0&0&0&0\\ -1&0&0&0&0&0&0&0\\ -1&0&0&0&0&0&0&0 \end{bmatrix} \Rightarrow \begin{bmatrix} \mathbf{Pro}&0&0&0&0&0&0&0\\ \mathbf{Pro}&0&0&0&0&0&0&0\\ \text{His}&0&0&0&0&0&0&0\\ \text{Gln}&0&0&0&0&0&0&0\\ 0&0&0&0&\mathbf{Ala}&0&0&0\\ 0&0&0&0&\mathbf{Ala}&0&0&0\\ 0&0&0&0&\text{Asp}&0&0&0\\ 0&0&0&0&\text{Glu}&0&0&0 \end{bmatrix} + \begin{bmatrix} 0&0&0&0&\mathbf{Ser}&0&0&0\\ 0&0&0&0&\mathbf{Ser}&0&0&0\\ 0&0&0&0&\text{Tyr}&0&0&0\\ 0&0&0&0&\text{Stop}&0&0&0\\ \mathbf{Thr}&0&0&0&0&0&0&0\\ \mathbf{Thr}&0&0&0&0&0&0&0\\ \text{Asn}&0&0&0&0&0&0&0\\ \text{Lys}&0&0&0&0&0&0&0 \end{bmatrix}$$

$$e_2+e_6 = \begin{bmatrix} 0&0&1&0&0&0&0&0\\ 0&0&1&0&0&0&0&0\\ 0&0&1&0&0&0&0&0\\ 0&0&1&0&0&0&0&0\\ 0&0&0&0&0&0&1&0\\ 0&0&0&0&0&0&1&0\\ 0&0&0&0&0&0&1&0\\ 0&0&0&0&0&0&1&0 \end{bmatrix} + \begin{bmatrix} 0&0&0&0&0&0&-1&0\\ 0&0&0&0&0&0&-1&0\\ 0&0&0&0&0&0&-1&0\\ 0&0&0&0&0&0&-1&0\\ 0&0&-1&0&0&0&0&0\\ 0&0&-1&0&0&0&0&0\\ 0&0&-1&0&0&0&0&0\\ 0&0&-1&0&0&0&0&0 \end{bmatrix} \Rightarrow \begin{bmatrix} 0&0&\mathbf{Leu}&0&0&0&0&0\\ 0&0&\mathbf{Leu}&0&0&0&0&0\\ 0&0&\mathbf{Arg}&0&0&0&0&0\\ 0&0&\mathbf{Arg}&0&0&0&0&0\\ 0&0&0&0&0&0&\mathbf{Val}&0\\ 0&0&0&0&0&0&\mathbf{Val}&0\\ 0&0&0&0&0&0&\mathbf{Gly}&0\\ 0&0&0&0&0&0&\mathbf{Gly}&0 \end{bmatrix} + \begin{bmatrix} 0&0&0&0&0&0&\text{Phe}&0\\ 0&0&0&0&0&0&\text{Leu}&0\\ 0&0&0&0&0&0&\text{Cys}&0\\ 0&0&0&0&0&0&\text{Trp}&0\\ 0&0&\text{Ile}&0&0&0&0&0\\ 0&0&\text{Met}&0&0&0&0&0\\ 0&0&\text{Ser}&0&0&0&0&0\\ 0&0&\text{Stop}&0&0&0&0&0 \end{bmatrix}$$

Figure 37. The separation of 20 amino acids and stop-codons in accordance with real and imagine parts of hyperbolic numbers of $e_0+e_4$ and $e_2+e_6$. Those amino acids, which belong to matrix cells with triplets of strong roots, are marked by bold letters.

Now let us pay attention to Figure 38, where beginnings of appropriate tensor family of matrices [C T; A G]$^{(n)}$ for the case of the DNA alphabet (adenine A, cytosine C, guanine G and thymine T) are shown. What kind of black-and-white mosaics (or a disposition of elements “+1” and “-1” in numeric representations of these symbolic matrices) can be appropriate in the case of the DNA alphabet for the basic matrix [C T; A G] and [C T; A G]$^{(2)}$?

The important phenomenological fact is that the thymine T is a single nitrogenous base in DNA which is replaced in RNA by another nitrogenous base U (uracil) for unknown reason (this is one of the mysteries of the genetic system). In other words, in this system the letter T is the opposition in relation to the letter U, and so the letter T can be symbolized by number “-1” (instead of number “+1” for U). Taking this into account, a simple algorithm exists, which transforms the black-and-white mosaics of matrices [C U; A G]$^{(2)}$ and [C U; A G]$^{(3)}$ into other mosaics of matrices [C T; A G]$^{(2)}$ and [C T; A G]$^{(3)}$ that are shown on Figure 38. Concerning to their mosaics, the matrices [C T; A G]$^{(2)}$ and [C T; A G]$^{(3)}$ coincide with mosaics of the Hadamard matrices $H_4$ and $H_8$ (Figure 1), which are their Hadamard represenations (here one should remind that Hadamard matrices contain only entries +1 and -1).

The mentioned algorithm was desribed in a few author's works (see for example [Petoukhov, 2012a,b]). The Appendix 3 describes another way to construct Hadamard (4*4)- and (8*8)-matrices on the base of the unique status of the letter T in the genetic alphabet A, C, G and T in DNA.

| | |
|---|---|
| C | T |
| A | G |

| | | | |
|---|---|---|---|
| CC | CT | TC | TT |
| CA | CG | TA | TG |
| AC | AT | GC | GT |
| AA | AG | GA | GG |

;

| | | | | | | | |
|---|---|---|---|---|---|---|---|
| CCC Pro | CCT Pro | CTC Leu | CTT Leu | TCC Ser | TCT Ser | TTC Phe | TTT Phe |
| CCA Pro | CCG Pro | CTA Leu | CTG Leu | TCA Ser | TCG Ser | TTA Leu | TTG Leu |
| CAC His | CAT His | CGC Arg | CGT Arg | TAC Tyr | TAT Tyr | TGC Cys | TGT Cys |
| CAA Gln | CAG Gln | CGA Arg | CGG Arg | TAA Stop | TAG Stop | TGA Trp | TGG Trp |
| ACC Thr | ACT Thr | ATC Ile | ATT Ile | GCC Ala | GCT Ala | GTC Val | GTT Val |
| ACA Thr | ACG Thr | ATA Met | ATG Met | GCA Ala | GCG Ala | GTA Val | GTG Val |
| AAC Asn | AAT Asn | AGC Ser | AGT Ser | GAC Asp | GAT Asp | GGC Gly | GGT Gly |
| AAA Lys | AAG Lys | AGA Stop | AGG Stop | GAA Glu | GAG Glu | GGA Gly | GGG Gly |

Figure 38. The first three representatives [C T; A G], [C T; A G]$^{(2)}$ and [C T; A G]$^{(3)}$ of the tensor family of DNA-alphabetic matrices [C T; A G]$^{(n)}$. These symbolic matrices [C T; A G]$^{(2)}$ and [C T; A G]$^{(3)}$ have mosaics, which coincide with the mosaics of their Hadamard representations $H_4$ and $H_8$ on Figure 1. All amino acids and stop-codons are shown for the case of the Vertebrate mitochondrial genetic code by analogy with Figure 36.

## 12. GENETIC PROJECTORS AND THE EXCLUSION PRINCIPLE FOR EVOLUTIONARY CHANGES OF DIALECTS OF THE GENETIC CODE

This Section describes an exclusion principle of evolution of dialects of the genetic code. This principle, which was discovered by the author, shows that evolutionary changes of dialects of the genetic code are related with the genetic projectors. One should note that discovering of exclusion principles of nature is a significant task of mathematical natural science (the exclusion principle by Pauli in quantum mechanics is one of examples).

The following list contains all known 19 dialects of the genetic code presented at July 10, 2013 on the NCBI's website http://www.ncbi.nlm.nih.gov/Taxonomy/Utils/wprintgc.cgi:

1 The Standard Code
2 The Vertebrate Mitochondrial Code
3 The Yeast Mitochondrial Code
4 The Mold, Protozoan, and Coelenterate Mitochondrial Code and the Mycoplasma/Spiroplasma Code
5 The Invertebrate Mitochondrial Code
6 The Ciliate, Dasycladacean and Hexamita Nuclear Code
7 The Echinoderm and Flatworm Mitochondrial Code
8 The Euplotid Nuclear Code
9 The Bacterial, Archaeal and Plant Plastid Code
10 The Alternative Yeast Nuclear Code
11 The Ascidian Mitochondrial Code
12 The Alternative Flatworm Mitochondrial Code
13 Blepharisma Nuclear Code
14 Chlorophycean Mitochondrial Code
15 Trematode Mitochondrial Code
16 Scenedesmus Obliquus Mitochondrial Code
17 Thraustochytrium Mitochondrial Code
18 Pterobranchia Mitochondrial Code
19 Candidate Division SR1 and Gracilibacteria Code

Figure 39 shows these dialects in typical forms of black-and-white genetic matrices $[C\ U;\ A\ G]^{(3)}$, where black cells correspond to triplets with strong roots and white cells correspond to triplets with weak roots in cases of each individual dialect. One can see that the vast majority of dialects (13 dialect from the set of 19 dialects) possesses the identical black-and-white mosaics though some triplets have different code meanings in different dialects (they encode different amino acid or stop-signal in different dialects) in comparison with their meanings in the vertebrate mitochondria genetic code (the author takes the case of the vertebrate mitochondria genetic code as the basic case because this dialect is the most symmetrical). All such triplets, which change their code meaning, are marked by red letters on Figure 39. From the list of 19 dialects only the following 6 dialects have their matrix $[C\ U;\ A\ G]^{(3)}$ with atypical black-and-white mosaics (see Figure 39): 5) The Invertebrate Mitochondrial Code; 7) The Echinoderm and Flatworm Mitochondrial Code; 10) The Alternative Yeast Nuclear Code; 12) The Alternative Flatworm Mitochondrial Code; 15) Trematode Mitochondrial Code; 16) Scenedesmus Obliquus Mitochondrial Code.

By analogy with the Rademacher presentation $R_8$ (see above Figures 1, 36, and Section 10), one can again replace black triplets by elements «+1» and white triplets by elements «-1» to receive numeric representations of these genetic matrices $[C\ U;\ A\ G]^{(3)}$ of all dialects. Such numeric representations of genetic matrices can be conditionally called as

«±1-represenations». The result is the following: this numeric ±1-representation of matrices [C U; A G]$^{(3)}$ of every of 19 dialects is decomposed into a sum of 8 sparse (8*8)-matrices of «column projectors» (or «row projectors») (see Figure 39). It is connected with the fact that all cells on main diagonals of these numeric matrices contain only «+1» (see the theorem in Section 7). This general feature of all dialects is a consequence of the following phenomenologic fact: biological evolution never changes code meaning of 16 black triplets, which occupies (2*2)-sub-quadrants along the main diagonal of these matrices (CCC, CCU, CCA, CCG, CGC, CGU, CGA, CGG, GCC, GCU, GCA, GCG, GGC, GGU, GGA, GGG).

From the point of view of algebra of projection operators, the described facts mean that biologic evolution of dialects of the genetic code is connected with a condition of conservation of the numeric ±1-representation of the genetic matrix [C U; A G]$^{(3)}$ as a sum of 8 column projectors (or 8 row projectors). In other words, algebra of projectors shows an existence of an algebraic invariant of biological evolution.

One can formulate here the phenomenologic **exclusion principle for evolutionary changes of dialects of the genetic code**: it is forbidden for biological evolution to violate a separation of the set of 64 triplets into two subsets of triplets with strong and weak roots (black and white triplets) in a such way that a black-and-white mosaic of the genetic matrix [C U; A G]$^{(3)}$ in its «±1-representation» ceases to be a sum of 8 column projectors (or 8 row projectors).

The Vertebrate Mitochondrial Code:

| | | | | | | | |
|---|---|---|---|---|---|---|---|
| CCC Pro | CCU Pro | CUC Leu | CUU Leu | UCC Ser | UCU Ser | UUC Phe | UUU Phe |
| CCA Pro | CCG Pro | CUA Leu | CUG Leu | UCA Ser | UCG Ser | UUA Leu | UUG Leu |
| CAC His | CAU His | CGC Arg | CGU Arg | UAC Tyr | UAU Tyr | UGC Cys | UGU Cys |
| CAA Gln | CAG Gln | CGA Arg | CGG Arg | UAA Stop | UAG Stop | UGA Trp | UGG trp |
| ACC Thr | ACU Thr | AUC Ile | AUU Ile | GCC Ala | GCU Ala | GUC Val | GUU Val |
| ACA Thr | ACG Thr | AUA Met | AUG Met | GCA Ala | GCG Ala | GUA Val | GUG Val |
| AAC Asn | AAU Asn | AGC Ser | AGU Ser | GAC Asp | GAU Asp | GGC Gly | GGU Gly |
| AAA Lys | AAG Lys | AGA Stop | AGG Stop | GAA Glu | GAG Glu | GGA Gly | GGG Gly |

The Standard Code:

| | | | | | | | |
|---|---|---|---|---|---|---|---|
| CCC Pro | CCU Pro | CUC Leu | CUU Leu | UCC Ser | UCU Ser | UUC Phe | UUU Phe |
| CCA Pro | CCG Pro | CUA Leu | CUG Leu | UCA Ser | UCG Ser | UUA Leu | UUG Leu |
| CAC His | CAU His | CGC Arg | CGU Arg | UAC Tyr | UAU Tyr | UGC Cys | UGU Cys |
| CAA Gln | CAG Gln | CGA Arg | CGG Arg | UAA Stop | UAG Stop | UGA Stop | UGG trp |
| ACC Thr | ACU Thr | AUC Ile | AUU Ile | GCC Ala | GCU Ala | GUC Val | GUU Val |
| ACA Thr | ACG Thr | AUA Ile | AUG Met | GCA Ala | GCG Ala | GUA Val | GUG Val |
| AAC Asn | AAU Asn | AGC Ser | AGU Ser | GAC Asp | GAU Asp | GGC Gly | GGU Gly |
| AAA Lys | AAG Lys | AGA Arg | AGG Arg | GAA Glu | GAG Glu | GGA Gly | GGG Gly |

The Yeast Mitochondrial Code:

| | | | | | | | |
|---|---|---|---|---|---|---|---|
| CCC<br>Pro | CCU<br>Pro | CUC<br>Thr | CUA<br>Thr | UCC<br>Ser | UCU<br>Ser | UUC<br>Phe | UUU<br>Phe |
| CCA<br>Pro | CCG<br>Pro | CUU<br>Thr | CUG<br>Thr | UCA<br>Ser | UCG<br>Ser | UUA<br>Leu | UUG<br>Leu |
| CAC<br>His | CAU<br>His | CGC<br>Arg | CGU<br>Arg | UAC<br>Tyr | UAU<br>Tyr | UGC<br>Cys | UGU<br>Cys |
| CAA<br>Gln | CAG<br>Gln | CGA<br>Arg | CGG<br>Arg | UAA<br>Stop | UAG<br>Stop | UGA<br>Trp | UGG<br>trp |
| ACC<br>Thr | ACU<br>Thr | AUC<br>Ile | AUU<br>Ile | GCC<br>Ala | GCU<br>Ala | GUC<br>Val | GUU<br>Val |
| ACA<br>Thr | ACG<br>Thr | AUA<br>Met | AUG<br>Met | GCA<br>Ala | GCG<br>Ala | GUA<br>Val | GUG<br>Val |
| AAC<br>Asn | AAU<br>Asn | AGC<br>Ser | AGU<br>Ser | GAC<br>Asp | GAU<br>Asp | GGC<br>Gly | GGU<br>Gly |
| AAA<br>Lys | AAG<br>Lys | AGA<br>Arg | AGG<br>Arg | GAA<br>Glu | GAG<br>Glu | GGA<br>Gly | GGG<br>Gly |

The Mold, Protozoan, and Coelenterate Mitochondrial Code and the Mycoplasma/Spiroplasma Code:

| | | | | | | | |
|---|---|---|---|---|---|---|---|
| CCC<br>Pro | CCU<br>Pro | CUC<br>Leu | CUU<br>Leu | UCC<br>Ser | UCU<br>Ser | UUC<br>Phe | UUU<br>Phe |
| CCA<br>Pro | CCG<br>Pro | CUA<br>Leu | CUG<br>Leu | UCA<br>Ser | UCG<br>Ser | UUA<br>Leu | UUG<br>Leu |
| CAC<br>His | CAU<br>His | CGC<br>Arg | CGU<br>Arg | UAC<br>Tyr | UAU<br>Tyr | UGC<br>Cys | UGU<br>Cys |
| CAA<br>Gln | CAG<br>Gln | CGA<br>Arg | CGG<br>Arg | UAA<br>Stop | UAG<br>Stop | UGA<br>Trp | UGG<br>trp |
| ACC<br>Thr | ACU<br>Thr | AUC<br>Ile | AUU<br>Ile | GCC<br>Ala | GCU<br>Ala | GUC<br>Val | GUU<br>Val |
| ACA<br>Thr | ACG<br>Thr | AUA<br>Ile | AUG<br>Met | GCA<br>Ala | GCG<br>Ala | GUA<br>Val | GUG<br>Val |
| AAC<br>Asn | AAU<br>Asn | AGC<br>Ser | AGU<br>Ser | GAC<br>Asp | GAU<br>Asp | GGC<br>Gly | GGU<br>Gly |
| AAA<br>Lys | AAG<br>Lys | AGA<br>Arg | AGG<br>Arg | GAA<br>Glu | GAG<br>Glu | GGA<br>Gly | GGG<br>Gly |

The Invertebrate Mitochondrial Code:

| | | | | | | | |
|---|---|---|---|---|---|---|---|
| CCC<br>Pro | CCU<br>Pro | CUC<br>Leu | CUU<br>Leu | UCC<br>Ser | UCU<br>Ser | UUC<br>Phe | UUU<br>Phe |
| CCA<br>Pro | CCG<br>Pro | CUA<br>Leu | CUG<br>Leu | UCA<br>Ser | UCG<br>Ser | UUA<br>Leu | UUG<br>Leu |
| CAC<br>His | CAU<br>His | CGC<br>Arg | CGU<br>Arg | UAC<br>Tyr | UAU<br>Tyr | UGC<br>Cys | UGU<br>Cys |
| CAA<br>Gln | CAG<br>Gln | CGA<br>Arg | CGG<br>Arg | UAA<br>Stop | UAG<br>Stop | UGA<br>Trp | UGG<br>trp |
| ACC<br>Thr | ACU<br>Thr | AUC<br>Ile | AUU<br>Ile | GCC<br>Ala | GCU<br>Ala | GUC<br>Val | GUU<br>Val |
| ACA<br>Thr | ACG<br>Thr | AUA<br>Met | AUG<br>Met | GCA<br>Ala | GCG<br>Ala | GUA<br>Val | GUG<br>Val |
| AAC<br>Asn | AAU<br>Asn | AGC<br>Ser | AGU<br>Ser | GAC<br>Asp | GAU<br>Asp | GGC<br>Gly | GGU<br>Gly |
| AAA<br>Lys | AAG<br>Lys | AGA<br>Ser | AGG<br>Ser | GAA<br>Glu | GAG<br>Glu | GGA<br>Gly | GGG<br>Gly |

The Ciliate, Dasycladacean and Hexamita Nuclear Code:

| | | | | | | | |
|---|---|---|---|---|---|---|---|
| CCC Pro | CCU Pro | CUC Leu | CUU Leu | UCC Ser | UCU Ser | UUC Phe | UUU Phe |
| CCA Pro | CCG Pro | CUA Leu | CUG Leu | UCA Ser | UCG Ser | UUA Leu | UUG Leu |
| CAC His | CAU His | CGC Arg | CGU Arg | UAC Tyr | UAU Tyr | UGC Cys | UGU Cys |
| CAA Gln | CAG Gln | CGA Arg | CGG Arg | UAA Gln | UAG Gln | UGA Stop | UGG trp |
| ACC Thr | ACU Thr | AUC Ile | AUU Ile | GCC Ala | GCU Ala | GUC Val | GUU Val |
| ACA Thr | ACG Thr | AUA Ile | AUG Met | GCA Ala | GCG Ala | GUA Val | GUG Val |
| AAC Asn | AAU Asn | AGC Ser | AGU Ser | GAC Asp | GAU Asp | GGC Gly | GGU Gly |
| AAA Lys | AAG Lys | AGA Arg | AGG Arg | GAA Glu | GAG Glu | GGA Gly | GGG Gly |

The Echinoderm and Flatworm Mitochondrial Code:

| | | | | | | | |
|---|---|---|---|---|---|---|---|
| CCC Pro | CCU Pro | CUC Leu | CUU Leu | UCC Ser | UCU Ser | UUC Phe | UUU Phe |
| CCA Pro | CCG Pro | CUA Leu | CUG Leu | UCA Ser | UCG Ser | UUA Leu | UUG Leu |
| CAC His | CAU His | CGC Arg | CGU Arg | UAC Tyr | UAU Tyr | UGC Cys | UGU Cys |
| CAA Gln | CAG Gln | CGA Arg | CGG Arg | UAA Stop | UAG Stop | UGA Trp | UGG trp |
| ACC Thr | ACU Thr | AUC Ile | AUU Ile | GCC Ala | GCU Ala | GUC Val | GUU Val |
| ACA Thr | ACG Thr | AUA Ile | AUG Met | GCA Ala | GCG Ala | GUA Val | GUG Val |
| AAC Asn | AAU Asn | AGC Ser | AGU Ser | GAC Asp | GAU Asp | GGC Gly | GGU Gly |
| AAA Asn | AAG Lys | AGA Ser | AGG Ser | GAA Glu | GAG Glu | GGA Gly | GGG Gly |

The Euplotid Nuclear Code:

| | | | | | | | |
|---|---|---|---|---|---|---|---|
| CCC Pro | CCU Pro | CUC Leu | CUU Leu | UCC Ser | UCU Ser | UUC Phe | UUU Phe |
| CCA Pro | CCG Pro | CUA Leu | CUG Leu | UCA Ser | UCG Ser | UUA Leu | UUG Leu |
| CAC His | CAU His | CGC Arg | CGU Arg | UAC Tyr | UAU Tyr | UGC Cys | UGU Cys |
| CAA Gln | CAG Gln | CGA Arg | CGG Arg | UAA Stop | UAG Stop | UGA Cys | UGG trp |
| ACC Thr | ACU Thr | AUC Ile | AUU Ile | GCC Ala | GCU Ala | GUC Val | GUU Val |
| ACA Thr | ACG Thr | AUA Ile | AUG Met | GCA Ala | GCG Ala | GUA Val | GUG Val |
| AAC Asn | AAU Asn | AGC Ser | AGU Ser | GAC Asp | GAU Asp | GGC Gly | GGU Gly |
| AAA Lys | AAG Lys | AGA Arg | AGG Arg | GAA Glu | GAG Glu | GGA Gly | GGG Gly |

The Bacterial, Archaeal and Plant Plastid Code:

| | | | | | | | |
|---|---|---|---|---|---|---|---|
| CCC<br>Pro | CCU<br>Pro | CUC<br>Leu | CUU<br>Leu | UCC<br>Ser | UCU<br>Ser | UUC<br>Phe | UUU<br>Phe |
| CCA<br>Pro | CCG<br>Pro | CUA<br>Leu | CUG<br>Leu | UCA<br>Ser | UCG<br>Ser | UUA<br>Leu | UUG<br>Leu |
| CAC<br>His | CAU<br>His | CGC<br>Arg | CGU<br>Arg | UAC<br>Tyr | UAU<br>Tyr | UGC<br>Cys | UGU<br>Cys |
| CAA<br>Gln | CAG<br>Gln | CGA<br>Arg | CGG<br>Arg | UAA<br>Stop | UAG<br>Stop | UGA<br>Stop | UGG<br>trp |
| ACC<br>Thr | ACU<br>Thr | AUC<br>Ile | AUU<br>Ile | GCC<br>Ala | GCU<br>Ala | GUC<br>Val | GUU<br>Val |
| ACA<br>Thr | ACG<br>Thr | AUA<br>Ile | AUG<br>Met | GCA<br>Ala | GCG<br>Ala | GUA<br>Val | GUG<br>Val |
| AAC<br>Asn | AAU<br>Asn | AGC<br>Ser | AGU<br>Ser | GAC<br>Asp | GAU<br>Asp | GGC<br>Gly | GGU<br>Gly |
| AAA<br>Lys | AAG<br>Lys | AGA<br>Arg | AGG<br>Arg | GAA<br>Glu | GAG<br>Glu | GGA<br>Gly | GGG<br>Gly |

The Alternative Yeast Nuclear Code:

| | | | | | | | |
|---|---|---|---|---|---|---|---|
| CCC<br>Pro | CCU<br>Pro | CUC<br>Leu | CUU<br>Leu | UCC<br>Ser | UCU<br>Ser | UUC<br>Phe | UUU<br>Phe |
| CCA<br>Pro | CCG<br>Pro | CUA<br>Leu | CUG<br>Ser | UCA<br>Ser | UCG<br>Ser | UUA<br>Leu | UUG<br>Leu |
| CAC<br>His | CAU<br>His | CGC<br>Arg | CGU<br>Arg | UAC<br>Tyr | UAU<br>Tyr | UGC<br>Cys | UGU<br>Cys |
| CAA<br>Gln | CAG<br>Gln | CGA<br>Arg | CGG<br>Arg | UAA<br>Stop | UAG<br>Stop | UGA<br>Stop | UGG<br>trp |
| ACC<br>Thr | ACU<br>Thr | AUC<br>Ile | AUU<br>Ile | GCC<br>Ala | GCU<br>Ala | GUC<br>Val | GUU<br>Val |
| ACA<br>Thr | ACG<br>Thr | AUA<br>Ile | AUG<br>Met | GCA<br>Ala | GCG<br>Ala | GUA<br>Val | GUG<br>Val |
| AAC<br>Asn | AAU<br>Asn | AGC<br>Ser | AGU<br>Ser | GAC<br>Asp | GAU<br>Asp | GGC<br>Gly | GGU<br>Gly |
| AAA<br>Lys | AAG<br>Lys | AGA<br>Arg | AGG<br>Arg | GAA<br>Glu | GAG<br>Glu | GGA<br>Gly | GGG<br>Gly |

The Ascidian Mitochondrial Code:

| | | | | | | | |
|---|---|---|---|---|---|---|---|
| CCC<br>Pro | CCU<br>Pro | CUC<br>Leu | CUU<br>Leu | UCC<br>Ser | UCU<br>Ser | UUC<br>Phe | UUU<br>Phe |
| CCA<br>Pro | CCG<br>Pro | CUA<br>Leu | CUG<br>Leu | UCA<br>Ser | UCG<br>Ser | UUA<br>Leu | UUG<br>Leu |
| CAC<br>His | CAU<br>His | CGC<br>Arg | CGU<br>Arg | UAC<br>Tyr | UAU<br>Tyr | UGC<br>Cys | UGU<br>Cys |
| CAA<br>Gln | CAG<br>Gln | CGA<br>Arg | CGG<br>Arg | UAA<br>Stop | UAG<br>Stop | UGA<br>Trp | UGG<br>trp |
| ACC<br>Thr | ACU<br>Thr | AUC<br>Ile | AUU<br>Ile | GCC<br>Ala | GCU<br>Ala | GUC<br>Val | GUU<br>Val |
| ACA<br>Thr | ACG<br>Thr | AUA<br>Met | AUG<br>Met | GCA<br>Ala | GCG<br>Ala | GUA<br>Val | GUG<br>Val |
| AAC<br>Asn | AAU<br>Asn | AGC<br>Ser | AGU<br>Ser | GAC<br>Asp | GAU<br>Asp | GGC<br>Gly | GGU<br>Gly |
| AAA<br>Lys | AAG<br>Lys | AGA<br>Gly | AGG<br>Gly | GAA<br>Glu | GAG<br>Glu | GGA<br>Gly | GGG<br>Gly |

The Alternative Flatworm Mitochondrial Code:

| | | | | | | | |
|---|---|---|---|---|---|---|---|
| CCC Pro | CCU Pro | CUC Leu | CUU Leu | UCC Ser | UCU Ser | UUC Phe | UUU Phe |
| CCA Pro | CCG Pro | CUA Leu | CUG Leu | UCA Ser | UCG Ser | UUA Leu | UUG Leu |
| CAC His | CAU His | CGC Arg | CGU Arg | UAC Tyr | UAU Tyr | UGC Cys | UGU Cys |
| CAA Gln | CAG Gln | CGA Arg | CGG Arg | UAA Stop | UAG Stop | UGA Trp | UGG trp |
| ACC Thr | ACU Thr | AUC Ile | AUU Ile | GCC Ala | GCU Ala | GUC Val | GUU Val |
| ACA Thr | ACG Thr | AUA Ile | AUG Met | GCA Ala | GCG Ala | GUA Val | GUG Val |
| AAC Asn | AAU Asn | AGC Ser | AGU Ser | GAC Asp | GAU Asp | GGC Gly | GGU Gly |
| AAA Asn | AAG Lys | AGA Ser | AGG Ser | GAA Glu | GAG Glu | GGA Gly | GGG Gly |

Blepharisma Nuclear Code:

| | | | | | | | |
|---|---|---|---|---|---|---|---|
| CCC Pro | CCU Pro | CUC Leu | CUU Leu | UCC Ser | UCU Ser | UUC Phe | UUU Phe |
| CCA Pro | CCG Pro | CUA Leu | CUG Leu | UCA Ser | UCG Ser | UUA Leu | UUG Leu |
| CAC His | CAU His | CGC Arg | CGU Arg | UAC Tyr | UAU Tyr | UGC Cys | UGU Cys |
| CAA Gln | CAG Gln | CGA Arg | CGG Arg | UAA Stop | UAG Gln | UGA Stop | UGG trp |
| ACC Thr | ACU Thr | AUC Ile | AUU Ile | GCC Ala | GCU Ala | GUC Val | GUU Val |
| ACA Thr | ACG Thr | AUA Ile | AUG Met | GCA Ala | GCG Ala | GUA Val | GUG Val |
| AAC Asn | AAU Asn | AGC Ser | AGU Ser | GAC Asp | GAU Asp | GGC Gly | GGU Gly |
| AAA Lys | AAG Lys | AGA Arg | AGG Arg | GAA Glu | GAG Glu | GGA Gly | GGG Gly |

Chlorophycean Mitochondrial Code:

| | | | | | | | |
|---|---|---|---|---|---|---|---|
| CCC Pro | CCU Pro | CUC Leu | CUU Leu | UCC Ser | UCU Ser | UUC Phe | UUU Phe |
| CCA Pro | CCG Pro | CUA Leu | CUG Leu | UCA Ser | UCG Ser | UUA Leu | UUG Leu |
| CAC His | CAU His | CGC Arg | CGU Arg | UAC Tyr | UAU Tyr | UGC Cys | UGU Cys |
| CAA Gln | CAG Gln | CGA Arg | CGG Arg | UAA Stop | UAG Leu | UGA Stop | UGG trp |
| ACC Thr | ACU Thr | AUC Ile | AUU Ile | GCC Ala | GCU Ala | GUC Val | GUU Val |
| ACA Thr | ACG Thr | AUA Ile | AUG Met | GCA Ala | GCG Ala | GUA Val | GUG Val |
| AAC Asn | AAU Asn | AGC Ser | AGU Ser | GAC Asp | GAU Asp | GGC Gly | GGU Gly |
| AAA Lys | AAG Lys | AGA Arg | AGG Arg | GAA Glu | GAG Glu | GGA Gly | GGG Gly |

Trematode Mitochondrial Code:

| | | | | | | | |
|---|---|---|---|---|---|---|---|
| CCC Pro | CCU Pro | CUC Leu | CUU Leu | UCC Ser | UCU Ser | UUC Phe | UUU Phe |
| CCA Pro | CCG Pro | CUA Leu | CUG Leu | UCA Ser | UCG Ser | UUA Leu | UUG Leu |
| CAC His | CAU His | CGC Arg | CGU Arg | UAC Tyr | UAU Tyr | UGC Cys | UGU Cys |
| CAA Gln | CAG Gln | CGA Arg | CGG Arg | UAA Stop | UAG Stop | UGA Trp | UGG trp |
| ACC Thr | ACU Thr | AUC Ile | AUU Ile | GCC Ala | GCU Ala | GUC Val | GUU Val |
| ACA Thr | ACG Thr | AUA Met | AUG Met | GCA Ala | GCG Ala | GUA Val | GUG Val |
| AAC Asn | AAU Asn | AGC Ser | AGU Ser | GAC Asp | GAU Asp | GGC Gly | GGU Gly |
| AAA Asn | AAG Lys | AGA Ser | AGG Ser | GAA Glu | GAG Glu | GGA Gly | GGG Gly |

Scenedesmus Obliquus Mitochondrial Code:

| | | | | | | | |
|---|---|---|---|---|---|---|---|
| CCC Pro | CCU Pro | CUC Leu | CUU Leu | UCC Ser | UCU Ser | UUC Phe | UUU Phe |
| CCA Pro | CCG Pro | CUA Leu | CUG Leu | UCA Stop | UCG Ser | UUA Leu | UUG Leu |
| CAC His | CAU His | CGC Arg | CGU Arg | UAC Tyr | UAU Tyr | UGC Cys | UGU Cys |
| CAA Gln | CAG Gln | CGA Arg | CGG Arg | UAA Stop | UAG Leu | UGA Stop | UGG trp |
| ACC Thr | ACU Thr | AUC Ile | AUU Ile | GCC Ala | GCU Ala | GUC Val | GUU Val |
| ACA Thr | ACG Thr | AUA Ile | AUG Met | GCA Ala | GCG Ala | GUA Val | GUG Val |
| AAC Asn | AAU Asn | AGC Ser | AGU Ser | GAC Asp | GAU Asp | GGC Gly | GGU Gly |
| AAA Lys | AAG Lys | AGA Arg | AGG Arg | GAA Glu | GAG Glu | GGA Gly | GGG Gly |

Thraustochytrium Mitochondrial Code:

| | | | | | | | |
|---|---|---|---|---|---|---|---|
| CCC Pro | CCU Pro | CUC Leu | CUU Leu | UCC Ser | UCU Ser | UUC Phe | UUU Phe |
| CCA Pro | CCG Pro | CUA Leu | CUG Leu | UCA Ser | UCG Ser | UUA Stop | UUG Leu |
| CAC His | CAU His | CGC Arg | CGU Arg | UAC Tyr | UAU Tyr | UGC Cys | UGU Cys |
| CAA Gln | CAG Gln | CGA Arg | CGG Arg | UAA Stop | UAG Stop | UGA Stop | UGG trp |
| ACC Thr | ACU Thr | AUC Ile | AUU Ile | GCC Ala | GCU Ala | GUC Val | GUU Val |
| ACA Thr | ACG Thr | AUA Ile | AUG Met | GCA Ala | GCG Ala | GUA Val | GUG Val |
| AAC Asn | AAU Asn | AGC Ser | AGU Ser | GAC Asp | GAU Asp | GGC Gly | GGU Gly |
| AAA Lys | AAG Lys | AGA Arg | AGG Arg | GAA Glu | GAG Glu | GGA Gly | GGG Gly |

Pterobranchia Mitochondrial Code:

| | | | | | | | |
|---|---|---|---|---|---|---|---|
| CCC Pro | CCU Pro | CUC Leu | CUU Leu | UCC Ser | UCU Ser | UUC Phe | UUU Phe |
| CCA Pro | CCG Pro | CUA Leu | CUG Leu | UCA Ser | UCG Ser | UUA Leu | UUG Leu |
| CAC His | CAU His | CGC Arg | CGU Arg | UAC Tyr | UAU Tyr | UGC Cys | UGU Cys |
| CAA Gln | CAG Gln | CGA Arg | CGG Arg | UAA Stop | UAG Stop | UGA Trp | UGG trp |
| ACC Thr | ACU Thr | AUC Ile | AUU Ile | GCC Ala | GCU Ala | GUC Val | GUU Val |
| ACA Thr | ACG Thr | AUA Ile | AUG Met | GCA Ala | GCG Ala | GUA Val | GUG Val |
| AAC Asn | AAU Asn | AGC Ser | AGU Ser | GAC Asp | GAU Asp | GGC Gly | GGU Gly |
| AAA Lys | AAG Lys | AGA Ser | AGG Lys | GAA Glu | GAG Glu | GGA Gly | GGG Gly |

Candidate Division SR1 and Gracilibacteria Code:

| | | | | | | | |
|---|---|---|---|---|---|---|---|
| CCC Pro | CCU Pro | CUC Leu | CUU Leu | UCC Ser | UCU Ser | UUC Phe | UUU Phe |
| CCA Pro | CCG Pro | CUA Leu | CUG Leu | UCA Ser | UCG Ser | UUA Leu | UUG Leu |
| CAC His | CAU His | CGC Arg | CGU Arg | UAC Tyr | UAU Tyr | UGC Cys | UGU Cys |
| CAA Gln | CAG Gln | CGA Arg | CGG Arg | UAA Stop | UAG Stop | UGA Gly | UGG trp |
| ACC Thr | ACU Thr | AUC Ile | AUU Ile | GCC Ala | GCU Ala | GUC Val | GUU Val |
| ACA Thr | ACG Thr | AUA Ile | AUG Met | GCA Ala | GCG Ala | GUA Val | GUG Val |
| AAC Asn | AAU Asn | AGC Ser | AGU Ser | GAC Asp | GAU Asp | GGC Gly | GGU Gly |
| AAA Lys | AAG Lys | AGA Arg | AGG Arg | GAA Glu | GAG Glu | GGA Gly | GGG Gly |

Figure 39. The matrices $[C\ U;\ A\ G]^{(3)}$ show 19 known dialects of the genetic code. Black (white) cells contain triplets with strong (weak) roots. Red color shows triplets, which have different code meanings in a considered dialect in comparison with their code meanings in the Vertebrate Mitochondrial Code, which is the most symmetrical among all dialects.

## 13. ABOUT «TENSORCOMPLEX» NUMBERS

This Section describes a system of multidimensional numbers, which seem to be a new one for mathematical natural sciences and which are constructed on the basis of sums of genetic projectors described above. Here the author will take some data from his work [Petoukhov, 2012b].

First of all, let us return to two sums of genetic projectors $h_0+h_2$ and $h_1+h_3$ (from Figure 10), which were received from the Hadamard matrix $H_4$ (Figure 1). The first sum $h_0+h_2$ is decomposed into two basic matrices $e_0$ = [1 0 0 0; -1 0 0 0; 0 0 1 0; 0 0 -1 0] and $e_2$ = [0 0 -1 0; 0 0 1 0; 1 0 0 0; -1 0 0 0], the set of which is closed relative to multiplication and defines the multiplication table of complex numbers. The second sum $h_1+h_3$ is decomposed into other basic matrices $e_1$ = [0 1 0 0; 0 1 0 0; 0 0 0 1; 0 0 0 1] and $e_3$ = [0 0 0 1; 0 0 0 1;

0 -1 0 0; 0 -1 0 0], the set of which is closed relative to multiplication and also defines the multiplication table of complex numbers (see Figure 10). Now one can consider linear compositions $C_L$ and $G_R$ on basis of these basic matrices (Figure 37, upper level). The work [Petoukhov, 2012b] shows that each of $C_L$ and $C_R$ is a (4*4)-matrix representation of 2-parametric complex numbers over field of real numbers but these representations concern different 2-dimensional planes $(x_0, x_2)$ and $(x_1, x_3)$ of a 4-dimensional vector space.

Many people know that the sum of two complex numbers gives a new complex number and that the product of two complex numbers is commutative. This is true when these complex numbers belong to the same complex plane. But the sum of these (4x4)-representations of two complex numbers $C_L$ and $C_R$, which belong to different planes of 4-dimensional space, is not equal to a new complex number, and their product is not commutative: $C_L*C_R \neq C_R*C_L$. Each of these products $C_L*C_R$ and $C_R*C_L$ gives a new complex number. Figure 40 (two middle levels) show expressions $C_L*C_R$ and $C_R*C_L$ with their decompositions into sets of two matrices, which correspond to the multiplication table of complex numbers. Figure 40 (bottom level) also shows an expression of a corresponding commutator. One should note here that the expression of the commutator $C_L*C_R-C_R*C_L$ on Figure 40 belongs to so called “tensorcomplex numbers”, which will be introduced below (Figures 41 and 42) with marks of their quadrants by means of yellow and green colors to emphasise a special cross-like structure of this type of numbers.

$$C_L=a_0*e_0+a_2*e_2=\begin{bmatrix} a_0 & 0 & -a_2 & 0 \\ -a_0 & 0 & a_2 & 0 \\ a_2 & 0 & a_0 & 0 \\ -a_2 & 0 & -a_0 & 0 \end{bmatrix}; \quad C_R=a_1*e_1+a_3*e_3=\begin{bmatrix} 0 & a_1 & 0 & a_3 \\ 0 & a_1 & 0 & a_3 \\ 0 & -a_3 & 0 & a_1 \\ 0 & -a_3 & 0 & a_1 \end{bmatrix}$$

| | | | | | |
|---|---|---|---|---|---|
| $C_L*C_R=$ | $0,\ a_0*a_1+a_2*a_3,\ 0,\ a_0*a_3-a_1*a_2$<br>$0,\ -a_0*a_1-a_2*a_3,\ 0,\ a_1*a_2-a_0*a_3$<br>$0,\ a_1*a_2-a_0*a_3,\ 0,\ a_0*a_1+a_2*a_3$<br>$0,\ a_0*a_3-a_1*a_2,\ 0,\ -a_0*a_1-a_2*a_3$ | $= -(a_0*a_1+a_2*a_3)*$ | 0 -1 0 0<br>0 1 0 0<br>0 0 0 -1<br>0 0 0 1 | $+(a_1*a_2-a_0*a_3)*$ | 0 0 0 -1<br>0 0 0 1<br>0 1 0 0<br>0 -1 0 0 |
| $C_R*C_L=$ | $-a_0*a_1-a_2*a_3,\ 0,\ a_1*a_2-a_0*a_3,\ 0$<br>$-a_0*a_1-a_2*a_3,\ 0,\ a_1*a_2-a_0*a_3,\ 0$<br>$a_0*a_3-a_1*a_2,\ 0,\ -a_0*a_1-a_2*a_3,\ 0$<br>$a_0*a_3-a_1*a_2,\ 0,\ -a_0*a_1-a_2*a_3,\ 0$ | $=-(a_0*a_1+a_2*a_3)*$ | 1 0 0 0<br>1 0 0 0<br>0 0 1 0<br>0 0 1 0 | $-(a_1*a_2-a_0*a_3)*$ | 0 0 -1 0<br>0 0 -1 0<br>1 0 0 0<br>1 0 0 0 |

| | | |
|---|---|---|
| $C_L*C_R-C_R*C_L=$ | $a_0*a_1+a_2*a_3,\quad a_0*a_1+a_2*a_3$<br>$a_0*a_1+a_2*a_3,\quad -a_0*a_1-a_2*a_3$ | $a_0*a_3-a_1*a_2,\quad a_0*a_3-a_1*a_2$<br>$a_0*a_3-a_1*a_2,\quad a1*a_2-a_0*a_3$ |
| | $a_1*a_2-a_0*a_3,\quad a_1*a_2-a_0*a_3$<br>$a_1*a_2-a_0*a_3,\quad a_0*a_3-a_1*a_2$ | $a_0*a_1+a_2*a_3,\quad a_0*a_1+a_2*a_3$<br>$a_0*a_1+a_2*a_3,\quad -a_0*a_1-a_2*a_3$ |

Figure 40. Upper level: $C_L$ is a (4*4)-matrix representation of complex numbers $z = a_0 + a_2*i$ on a 2-dimensional plane $(x_0, x_2)$ in a 4-dimensional space $(x_0, x_1, x_2, x_3)$; here i is imaginary unit of complex numbers ($i^2 = -1$). $C_R$ is a (4*4)-matrix representation of complex numbers $z = a_1 + a_3*i$ on another 2-dimensional plane $(x_1, x_3)$ in the same 4-dimensional space $(x_0, x_1, x_2, x_3)$. Here $a_0$, $a_1$, $a_2$ and $a_3$ are real numbers. Two middle levels: expressions of products $C_L*C_R$ and $C_R*C_L$; in both cases two basic matrices define the multiplication table of complex numbers. Bottom level: the expression of the commutator $C_L*C_R-C_R*C_L$.

Now let us take a sum $V = C_L + C_R$ of these (4*4)-matrix representations of complex numbers, which are related with different planes of the 4-dimensional space (Figure 41).

$$V = C_L + C_R = \begin{bmatrix} a_0 & a_1 & -a_2 & a_3 \\ -a_0 & a_1 & a_2 & a_3 \\ a_2 & -a_3 & a_0 & a_1 \\ -a_2 & -a_3 & -a_0 & a_1 \end{bmatrix} = [1\ 0;\ 0\ 1]\otimes M + [0\ 1;\ -1\ 0]\otimes P$$

$$V^{-1} = (2*(a_0^2 + a_2^2))^{-1} * \begin{bmatrix} a_0 & -a_0 & a_2 & -a_2 \\ 0 & 0 & 0 & 0 \\ -a_2 & a_2 & a_0 & -a_0 \\ 0 & 0 & 0 & 0 \end{bmatrix} + (2*(a_1^2 + a_3^2))^{-1} * \begin{bmatrix} 0 & 0 & 0 & 0 \\ a_1 & a_1 & -a_3 & -a_3 \\ 0 & 0 & 0 & 0 \\ a_3 & a_3 & a_1 & a_1 \end{bmatrix}$$

Figure 41. Upper level: the sum $V = C_L + C_R$ (see Figure 40). Here $M=[a_0\ a_1;\ -a_0\ a_1]$ and $P=[-a_2\ a_3;\ a_2\ a_3]$; ⊗ is a symbol of tensor (or Kronecker) multiplication. Bottom level: the expression of the inverse matrix $V^{-1}$.

Figure 41 shows that $V=C_L+C_R$ is a (4*4)-matrix of a special type, where both quadrants (marked by yellow color) along the main diagonal are identical each other, and two other quadrants (marked by green color) differ each from other only by inversion of sign in their entries. This matrix can be written in a form $V = [1\ 0;\ 0\ 1]\otimes M + [0\ 1;\ -1\ 0]\otimes P$, where $M=[a_0\ a_1;\ -a_0\ a_1]$ and $P=[-a_2\ a_3;\ a_2\ a_3]$; ⊗ is a symbol of tensor multiplication; [1 0; 0 1] – a matrix representation of real unit; [0 1; -1 0] – a matrix representation of imaginary unit; $a_0$, $a_1$, $a_2$ and $a_3$ are real numbers. Such form of denotation resembles a well-known matrix representation of usual complex numbers: z = [1 0; 0 1]*a + [0 1; -1 0]*b, where a and b are real numbers. But it differs in the following aspects:

1) the expression V = [1 0; 0 1]⊗M + [0 1; -1 0]⊗P includes tensor multiplication ⊗ instead of usual multiplication in the matrix representation of complex numbers;
2) in the case of V, multipliers of the basic elements are square matrices M and P instead of real numbers «a» and «b» in the case of complex numbers.
3) The order of the factors inside V is essential since tensor multiplication is not commutative.

Let us consider a set of matrices, which includes all matrices of this kind V together with their inverse matrices $V^{-1}$ and together with all products of matrices of this kind. This set of matrices has properties of multi-dimensional numeric system as it is described below.

What one can say about algebraic properties of matrices of such type V (Figure 41)? Matrices of this type can be added and subtracted. The matrix V has its inverse matrix $V^{-1}$ (Figure 42), which is defined on the basis of the condition $V*V^{-1}=V^{-1}*V=E_4$, where $E_4$ is identity matrix [1 0 0 0; 0 1 0 0; 0 0 1 0; 0 0 0 1]. Product of two different matrices of this type (for example G and S on Figure 42) generates a new (4*4)-matrix W=G*S (Figure 42), where both (2*2)-quadrants (marked by yellow color) along the main diagonal are identical each other, and two other (2*2)-quadrants (marked by green color) differ each from other only by inversion of sign in their entries. This matrix can be written in a form W = [1 0; 0 1]⊗Q + [0 1; -1 0]⊗K, where Q=[ a*k-c*k-b*n-d*n, a*m+c*m+b*p-d*p; b*n-c*k-a*k-d*n, c*m-a*m-b*p-d*p] and K=[c*n-a*n-d*k-b*k, a*p-b*m+d*m+c*p; b*k+a*n-d*k+c*n,

b*m-a*p+d*m+c*p]; ⊗ is a symbol of tensor multiplication; [1 0; 0 1] – a matrix representation of real unit; [0 1; -1 0] – a matrix representation of imaginary unit; a, b, c, d, k, m, n and p are real numbers. Such form of denotation resembles the known matrix representation of usual complex numbers: z = [1 0; 0 1]*a + [0 1; -1 0]*b, where a and b are real numbers. But it differs again in the following aspects:

1) the expression W = [1 0; 0 1]⊗Q + [0 1; -1 0]⊗K includes tensor multiplication ⊗ instead of usual multiplication in the matrix representation of complex numbers;
2) in the case of W, multipliers of the basic elements are square matrices Q and K instead of real numbers «a» and «b» in the case of complex numbers.
3) The order of the factors inside W is essential since tensor multiplication is not commutative.

Taking into account a significant role of tensor multiplication ⊗ in W, the author names algebraic constructions in a form W as "tensorcomplex numbers" because such matrices W have the following algebraic properties in relation to usual operations of addition, subtraction, multiplication and division:

- Addition and subtraction of two different matrices of this type W create a new matrix of the same type. Multiplication of different matrices of this type with each other is noncommutative and it gives a new matrix of the same type (Figure 42).
- Each non-zero matrix W=G*S has an inverse matrix $W^{-1} = G^{-1}*S^{-1}$ (expressions for $G^{-1}$ and $S^{-1}$ were shown on Figure 41). It allows a definition of operation of division of two matrices of this type as a multiplication with an inverse matrix.

Such properties of tensorcomplex numbers resemble algebraic properties of quaternions by Hamilton, which represent noncommutative division algebra (http://en.wikipedia.org/wiki/Quaternion). Here one should emphasize that tensorcomplex numbers cardinally differ from hypercomplex numbers $x_0+x_1*i_1+\ldots+x_n*i_n$, where $x_0, x_1,\ldots, x_n$ are real numbers, because, in the case of tensorcomplex numbers, multipliers of the basic elements are square matrices but not real numbers. By this reason, the famous Frobenius theorem (http://en.wikipedia.org/wiki/Frobenius_theorem_(real_division_algebras)) for hypercomplex numbers is not related to tensorcomplex numbers. This theorem says that any finite-dimensional associative division algebra is isomorphic to one of the following algebras: the real numbers, the complex numbers, the quaternions by Hamilton.

G=

| | | | |
|---|---|---|---|
| a | c | -b | d |
| -a | c | b | d |
| b | -d | a | c |
| -b | -d | -a | c |

; S =

| | | | |
|---|---|---|---|
| k | m | -n | p |
| -k | m | n | p |
| n | -p | k | m |
| -n | -p | -k | m |

G*S=

| | | | |
|---|---|---|---|
| a*k-c*k-b*n-d*n | a*m+c*m+b*p-d*p | c*n-a*n-d*k-b*k | a*p-b*m+d*m+c*p |
| b*n-c*k-a*k-d*n | c*m-a*m-b*p-d*p | b*k+a*n-d*k +c*n | b*m-a*p+d*m+c*p |
| b*k+a*n+d*k-c*n | b*m-a*p-d*m-c*p | a*k-c*k-b*n-d*n | a*m+c*m+b*p-d*p |
| d*k-a*n-b*k-c*n | a*p-b*m-d*m-c*p | b*n-c*k-a*k-d*n | c*m-a*m-b*p-d*p |

Figure 42. Multiplication of two matrices G and S gives a new matrix W = G*S, which belong to so called «tensorcomplex numbers». Here a, b, c, d, k, m, n and p are real numbers.

This set of tensorcomplex numbers is one of examples of numeric systems, where real numbers exist only inside matrices in a form of whole ensembles but not as individual multipliers (or as individual personages) inside such numeric systems. It is one of differences of tensorcomplex numbers from hypercomplex systems. One can mention that the commutator $C_L*C_R-C_R*C_L$ (Figure 40) belongs to tensorcomplex numbers.

Till now in this Section we considered the case of tensorcomplex numbers, which were represented by means of the expression W = [1 0; 0 1]⊗Q + [0 1; -1 0]⊗K, where Q and K are (2*2)-matrices. But our work [Petoukhov, 2012b,….] describes that complex numbers can be also represented by means of sparse $(2^N*2^N)$-matrices on the basis of sums of projectors (here N = 2, 3, 4, …). One can take sum of two complex numbers, which belong to different planes of the same $2^N$-dimensional space and which are represented by means of appropriate $(2^N*2^N)$-matrices. In this case new types of tensorcomplex numbers arise. Theory and expressions for such tensorcomplex numbers in spaces of higher dimensions are developed now for a publication in the nearest future.

The author hopes that tensorcomplex numbers, which seem to be a new type of multidimensional numbers for mathematical natural sciences, will be useful not only in bioinformatics, but also in physics, theory of communication, logic and other fields.

## 14. ABOUT «TENSORHYPERBOLIC» NUMBERS

Now let us return to the Rademacher (4*4)-matrix $R_4$, which is a sum of 4 column projectors $c_0$, $c_1$, $c_2$, $c_3$: $R_4 = c_0+c_1+c_2+c_3$ (Figures 1 and 2). Our work [Petoukhov, 2012b, Figures 11-13] shows that sum $c_0+c_2$ is decomposed into basic matrices $e_0$ = [1 0 0 0; -1 0 0 0; 0 0 1 0; 0 0 -1 0] and $e_2$ = [0 0 1 0; 0 0 -1 0; 1 0 0 0; -1 0 0 0], the set of which is closed relative to multiplication and defines the multiplication table of hyperbolic numbers; their known synonyms are "*split-complex numbers*", "*double numbers*" or "*Lorentz numbers*" (http://en.wikipedia.org/wiki/Split-complex_number). In this paper we will prefer using the name «hyperbolic numbers» for such type of 2-dimensional numbers. Another sum $c_1+c_3$ is decomposed into basic matrices $e_1$ = [0 1 0 0; 0 1 0 0; 0 0 0 1; 0 0 0 1] and $e_3$ = [0 0 0 -1; 0 0 0 -1; 0 -1 0 0; 0 -1 0 0]. Now one can consider linear compositions $D_L$ and $D_R$ on basis of these basic matrices (Figure 43). The work [Petoukhov, 2012b, Figures 12, 13] shows that each of $D_L$ and $D_R$ is a (4*4)-matrix representation of 2-parametric hyperbolic numbers over field of real numbers but these representations concern different 2-dimensional planes ($x_0$, $x_2$) and ($x_1$, $x_3$) of a 4-dimensional vector space.

$D_L=a_0*e_0+a_2*e_2=$

| | | | |
|---|---|---|---|
| $a_0$ | 0 | $a_2$ | 0 |
| $-a_0$ | 0 | $-a_2$ | 0 |
| $a_2$ | 0 | $a_0$ | 0 |
| $-a_2$ | 0 | $-a_0$ | 0 |

; $D_R=a_1*e_1+a_3*e_3=$

| | | | |
|---|---|---|---|
| 0 | $a_1$ | 0 | $-a_3$ |
| 0 | $a_1$ | 0 | $-a_3$ |
| 0 | $-a_3$ | 0 | $a_1$ |
| 0 | $-a_3$ | 0 | $a_1$ |

Figure 43. Left side: $D_L$ is a (4*4)-matrix representation of hyperbolic (or split-complex) numbers $z = a_0 + a_2*j$ on a 2-dimensional plane ($x_0$, $x_2$) in a 4-dimensional space ($x_0$, $x_1$, $x_2$, $x_3$); here j is imaginary unit of hyperbolic numbers ($j^2=1$). Right side: $D_R$ is a (4*4)-matrix representation of hyperbolic numbers $z = a_1 + a_3*j$ on another 2-dimensional plane ($x_1$, $x_3$) in the same 4-dimensional space ($x_0$, $x_1$, $x_2$, $x_3$). Here $a_0$, $a_1$, $a_2$ and $a_3$ are real numbers.

Now let us take a sum DL+DR of these (4*4)-matrix representations of hyperbolic numbers, which are related with different planes of the 4-dimensional space (Figure 44).

$$D = D_L + D_R = \begin{bmatrix} a_0 & a_1 & a_2 & -a_3 \\ -a_0 & a_1 & -a_2 & -a_3 \\ a_2 & -a_3 & a_0 & a_1 \\ -a_2 & -a_3 & -a_0 & a_1 \end{bmatrix} = [1\ 0;\ 0\ 1] \otimes M + [0\ 1;\ 1\ 0] \otimes K$$

$$\begin{bmatrix} a & 0 & a & 0 \\ -a & 0 & a & 0 \\ a & 0 & a & 0 \\ -a & 0 & -a & 0 \end{bmatrix} * \begin{bmatrix} 0 & c & 0 & -c \\ 0 & c & 0 & -c \\ 0 & -c & 0 & c \\ 0 & -c & 0 & c \end{bmatrix} = \begin{bmatrix} 0 & 0 & 0 & 0 \\ 0 & 0 & 0 & 0 \\ 0 & 0 & 0 & 0 \\ 0 & 0 & 0 & 0 \end{bmatrix}$$

$$D^{-1} = (2*(a_0^2 - a_2^2))^{-1} * \begin{bmatrix} a_0 & -a_0 & -a_2 & a_2 \\ 0 & 0 & 0 & 0 \\ -a_2 & a_2 & a_0 & -a_0 \\ 0 & 0 & 0 & 0 \end{bmatrix} + (2*(a_1^2 - a_3^2))^{-1} * \begin{bmatrix} 0 & 0 & 0 & 0 \\ a_1 & a_1 & -a_3 & -a_3 \\ 0 & 0 & 0 & 0 \\ -a_3 & -a_3 & a_1 & a_1 \end{bmatrix}$$

Figure 44. Upper level: the sum $D = D_L + D_R$ (see Figure 43). Here $M=[a_0\ a_1;\ -a_0\ a_1]$ and $K=[a_2\ -a_3;\ -a_2\ -a_3]$; $\otimes$ is a symbol of tensor multiplication. Middle level: the example of zero divizors in this type of matrices. Bottom level: the inverse matix $D^{-1}$.

Figure 44 shows that $D = D_L+D_R$ is a (4*4)-matrix of a special type, where both quadrants along each of diagonals are identical (they are marked by yellow and blue colors). This matrix can be written in a form $D = [1\ 0;\ 0\ 1] \otimes M + [0\ 1;\ 1\ 0] \otimes K$, where $M=[a_0\ a_1;\ -a_0\ a_1]$ and $K=[a_2\ -a_3;\ -a_2\ -a_3]$, where [1 0; 0 1] – a matrix representation of real unit; [0 1; 1 0] – a matrix representation of imaginary unit j of hyperbolic numbers ($j^2=1$); $a_0$, $a_1$, $a_2$ and $a_3$ are real numbers. Such form of denotation resembles a well-known matrix representation of usual hyperbolic numbers: y = [1 0; 0 1]*a + [0 1; 1 0]*b, where «a» and «b» are real numbers. But it differs again in the following aspects:

1) the expression $D = [1\ 0;\ 0\ 1] \otimes M + [0\ 1;\ 1\ 0] \otimes K$ includes tensor multiplication $\otimes$ instead of usual multiplication in the matrix representation of hyperbolic numbers;
2) in the case of D, multipliers of the basic elements are square matrices M and K instead of real numbers «a» and «b» in the case of hyperbolic numbers;
3) The order of factors inside D is essential since tensor multiplication is not commutative.

The set of matrix D has zero divisors, examples of which are shown on Figure 44. Figure 44 also shows a general expression of the inverse matrix $D^{-1}$ for the matrix D.

Multiplication of two matrices $D_0$ and $D_1$ of this type gives a new matrix L (Figure 45), where both quadrants along each of diagonals are identical (they are marked by yellow and blue colors). This matrix can be represented in the following form: $L = [1\ 0;\ 0\ 1] \otimes Q + [0\ 1;\ 1\ 0] \otimes K$, where Q = [a*k-c*k+b*n+d*n, a*m+c*m-b*p+d*p; d*n-c*k-b*n-a*k, c*m-a*m +b*p +d*p], K=[a*n+b*k-c*n+d*k, b*m-a*p-c*p-d*m; d*k-b*k-c*n-a*n, a*p-b*m-c*p-d*m].

Taking into account a significant role of tensor multiplication $\otimes$ in L, the author names algebraic constructions in a form L (Figure 45) as "tensorhyperbolic numbers" because such matrices L have the following algebraic properties in relation to usual operations of addition, subtraction, multiplication and division:

- Addition, subtraction and multiplication of two different matrices of this type create a new matrix of the same type. Multiplication of different matrices of this type with each other is noncommutative. The set of matrices L has zero divisors.
- Each non-zero matrix L=D0*D1, if it is not a zero divisor, has an inverse matrix $L^{-1} = D_0^{-1}*D_1^{-1}$ (expressions for $D^{-1}$ was shown on Figure 44). It allows a definition of operation of division of two matrices of this type as a multiplication with an inverse matrix.

$$D_0=\begin{array}{|cc|cc|} a & c & b & -d \\ -a & c & -b & -d \\ \hline b & -d & a & c \\ -b & -d & -a & c \end{array} \;;\quad D_1=\begin{array}{|cc|cc|} k & m & n & -p \\ -k & m & -n & -p \\ \hline n & -p & k & m \\ -n & -p & -k & m \end{array}$$

$$L=D_0*D_1=\begin{array}{|cc|cc|} a*k-c*k+b*n+d*n, & a*m+c*m-b*p+d*p & a*n+b*k-c*n+d*k, & b*m-a*p-c*p-d*m \\ d*n-c*k-b*n-a*k, & c*m-a*m+b*p+d*p & d*k-b*k-c*n-a*n, & a*p-b*m-c*p-d*m \\ \hline a*n+b*k-c*n+d*k, & b*m-a*p-c*p-d*m & a*k-c*k+b*n+d*n, & a*m+c*m-b*p+d*p \\ d*k-b*k-c*n-a*n, & a*p-b*m-c*p-d*m & d*n-c*k-b*n-a*k, & c*m-a*m+b*p+d*p \end{array}$$

Figure 45. Multiplication of two matrices $D_0$ and $D_1$ (of the type D from Figure 44) gives a new matrix $L = D_0*D_1$, which belongs to so called «tensorhyperbolic numbers». Here a, b, c, d, k, m, n and p are real numbers.

This Section has described the case of tensorhyperbolic numbers in the form of (4*4)-matrices for 4-dimensional spaces. But tensorhyperbolic numbers and their generalization can be expressed in forms of $(2^N*2^N)$-matrices for $2^N$-dimensional spaces. These materials will be published later together with data about «tensordual» numbers, «tensorquaternions», etc.

Different types of such multidimensional numbers can be combined under a brief name «tensornumbers». Tensornumbers $[1]\otimes M_0+[i_1]\otimes M_1+\ldots+[i_n]\otimes M_n$ (here $M_n$ are square matrices) are a generalization of hypercomplex numbers in the case when the following changes are made in the usual denotation of hypercomplex numbers $1*x_0+i_1*x_1+\ldots+i_n*x_n$:

- Real multipliers $x_0, x_1, \ldots, x_n$ are replaced by square matrices $M_0, M_1,\ldots, M_n$ of special kinds;
- Usual multiplication is replaced by tensor multiplication.

It is obvious that hypercomplex numbers $1*x_0+i_1*x_1+\ldots+i_n*x_n$ are a degenerate case of tensornumbers when their matrices $M_n$ have the first order: 1) (1x1)-matrices $M_n = [x_n]$ are real numbers $x_n$; 2) tensor multiplication $[i_n]\otimes[x_n]$ of (1x1)-matrix $[x_n]$ with the matrix representation $[i_n]$ of any of the basic elements is commutative and it coincides with usual multiplication. By these reasons the following equation is true: $[1]\otimes[x_0]+[i_1]\otimes[x_1]+\ldots+[i_n]\otimes[x_n] = [1]*x_0+[i_1]*x_1+\ldots+[i_n]*x_n$.

What one can say at this initial stage about a future of tensornumbers in mathematical natural sciences and technologies? Two extreme points of view are possible here: 1) tensorcomplex numbers will have no applications; 2) tensorcomplex numbers will have a great significance for mathematical natural sciences including a creation of new theories of physical fields and their generalization, new laws of conservation, generalization of many physical and other rules and knowledge, new approaches in engineering and biological informatics, mathematical logics, etc. The author believes that the second point of view will coincide with a real future in a higher extent.

## 15. UNITED-HYPERCOMPLEX NUMBERS AND GENETIC MATRICES

As known, hypercomplex numbers have a single real unit and additional quantity of imaginary units, all of which in the total are called basic elements of hypercomplex numbers. The expression (10) shows their linear form of writing in general case:

$$a_0*E + a_1*e_1 + a_2*e_2 + ....... + a_n*e_n \qquad (10)$$

where E denotes the real unit (or the identity matrix E in the case of the matrix representation of hypercomplex numbers), $e_1$, $e_2$, …, $e_n$ denote imaginary units, $a_0$, $a_1$, …., $a_n$ denote real coefficients. The set of these basic elements (E, $e_1$, $e_2$,…, $e_n$) should be closed under multiplication. This means that multiplication of any two basic elements with each other always gives again one of these elements or their linear superposition [Kantor, Solodovnikov, 1989].

In the result of our study of systems of structured alphabets of DNA and RNA in matrix forms of their Walsh-representations $R_4$, $R_8$, $H_4$, $H_8$ (Fig. 1) we have revealed that these alphabetic structures are connected with special $(2^n*2^n)$-matrix operators. Various decompositions of these operators show their connection with special $2^n$-dimensional numeric systems with the following features: 1) they are not hypercomplex numbers in the whole because of absence of the identity matrix E in their decompositions (that is the general real unit is absent among basic elements of such systems); 2) they contain two or more blocks, each of whose represents complex numbers or hypercomplex numbers with its individual set of basic elements, which play role of its individual real unit and imaginary units but only inside a separate block; by this reason we call them as local-real units and local-imaginary units. In other words, two or more different systems of complex or hypercomplex numbers exist in this case not separately but are united together into a single whole in a form of special block matrices. Multiplication tables (in cases of their existence) of basic elements of such systems have a character of plexus structures. The author calls such multi-dimensional numeric systems as "united-hypercomplex numeric systems" (or briefly, U-hypercomplex numbers or simply U-numbers). These united-hypercomplex numbers possess interesting properties to model biological phenomena and multi-parametrical systems in general. One can think that U-complex numbers and U-hypercomplex numbers play an important role in bioinformation phenomena since they are connected with the phenomenology of alphabets of DNA and RNA.

Let us consider the Walsh-representaion $R_4$ (Fig. 1) of the matrix of 16 doublets $[C, T; A, G]^{(2)}$. The matrix $R_4$ can be decomposed into sum of 4 sparse matrices $e_0$, $e_1$, $e_2$, $e_3$, for example, in the way shown on Fig. 46. It is interesting that this variant of the decomposition consists of two sets of matrices, each of which is closed in relation to multiplication unexpectedly: the first set consists of matrices $e_0$ and $e_1$; the second set – matrices $e_2$ and $e_3$. Fig. 46 shows multiplication tables for both sets, in each of which multiplication between its elements always gives an element from the same set. These multiplication tables coincide with the multiplication table of basic elements of complex numbers.

| | | | | | | | | | | |
|---|---|---|---|---|---|---|---|---|---|---|
| 1 1 1 -1<br>-1 1 -1 -1<br>1 -1 1 1<br>-1 -1 -1 1 | = | 1 0 0 0<br>0 1 0 0<br>0 -1 0 0<br>-1 0 0 0 | + | 0 1 0 0<br>-1 0 0 0<br>1 0 0 0<br>0 -1 0 0 | + | 0 0 0 -1<br>0 0 -1 0<br>0 0 1 0<br>0 0 0 1 | + | 0 0 1 0<br>0 0 0 -1<br>0 0 0 1<br>0 0 -1 0 | = $e_0+e_1+e_2+e_3$ |

| * | **$e_0$** | **$e_1$** |
|---|---|---|
| **$e_0$** | $e_0$ | $e_1$ |
| **$e_1$** | $e_1$ | $-e_0$ |

| * | **$e_2$** | **$e_3$** |
|---|---|---|
| **$e_2$** | $e_2$ | $e_3$ |
| **$e_3$** | $e_3$ | $-e_2$ |

Fig. 46. Top: the decomposition of the matrix $R_4$ (Fig. 1) into 4 sparse matrices $e_0$, $e_1$, $e_2$, $e_3$. Bottom: multiplication tables of pairs of matrices $e_0$ and $e_1$ (left) and $e_2$ and $e_3$ (right) coincide with the multiplication table of basic elements of complex numbers.

It means that each of expressions $ZL=a_0*e_0+a_1*e_1$ and $ZR=a_2*e_2+a_3*e_3$ , where $e_0$, $e_1$, $e_2$, $e_3$ are taken from Fig. 46, represents its own system of complex numbers in the unusual form of the sparse (4*4)-matrix with their 2 independent parameters $a_0$, $a_1$ and $a_2$, $a_3$ correspondingly (Fig. 47). It can be formulated also in the following way: each of the systems ZL and ZR is isomorphic to the classical system of complex numbers. Inside the system ZL, the matrix $e_0+e_1$ represents the complex number with unit coordinates ($a_0=a_1=1$). Inside the system ZR, the matrix $e_2+e_3$ represents the complex number with unit coordinates ($a_2=a_3=1$).

$$ZL = \begin{bmatrix} a_0, & a_1, & 0, & 0 \\ -a_1, & a_0, & 0, & 0 \\ a_1, & -a_0, & 0, & 0 \\ -a_0, & -a_1, & 0, & 0 \end{bmatrix} ; \qquad ZL^{-1} = (a_0^2+a_1^2)^{-1} * \begin{bmatrix} a_0, & -a_1, & 0, & 0 \\ a_1, & a_0, & 0, & 0 \\ -a_1, & -a_0, & 0, & 0 \\ -a_0, & a_1, & 0, & 0 \end{bmatrix}$$

$$ZR = \begin{bmatrix} 0, & 0, & a_3, & -a_2 \\ 0, & 0, & -a_2, & -a_3 \\ 0, & 0, & a_2, & a_3 \\ 0, & 0, & -a_3, & a_2 \end{bmatrix} ; \qquad ZR^{-1} = (a_2^2+a_3^2)^{-1} * \begin{bmatrix} 0, & 0, & -a_3, & -a_2 \\ 0, & 0, & -a_2, & a_3 \\ 0, & 0, & a_2, & -a_3 \\ 0, & 0, & a_3, & a_2 \end{bmatrix}$$

Fig. 47. Left: two systems of complex numbers $ZL=a_0*e_0+a_1*e_1$ and $ZR=a_2*e_2+a_3*e_3$ in their form of (4*4)-matrices. Right: inverse matrices $ZL^{-1}$ and $ZR^{-1}$ in relation to matrices $e_0$ and $e_2$, which play the role of identity matrices in the systems ZL and ZR correspondingly.

The classical identity matrix E=[1 0 0 0; 0 1 0 0; 0 0 1 0; 0 0 0 1] is absent in the set of matrices ZL and ZR, where - beside this - each matrix has zero determinant. Consequently the usual notion of the inverse matrix $ZL^{-1}$ or $ZR^{-1}$ (as $ZL*ZL^{-1}=E$ or $ZR*ZR^{-1}=E$) can't be defined in relation to the identity matrix E in accordance with the famous theorem about inverse matrices for matrices with zero determinant in the case of the complete set of matrices (Bellman, 1960, Chapter 6, § 4). But we analyze not the complete set of (4*4)-matrices but very limited special sets of matrices ZL and ZR. The system ZL has the matrix $e_0$ (Fig. 46), which possesses all properties of the identity matrix for any matrix ZL since $e_0*ZL = ZL*e_0 = ZL$ and $e_0^2 = e_0$. In the frame of the system of matrices ZL, where locally the matrix $e_0$ plays the role of the identity matrix (the local-identity matrix), one can define - for any non-zero matrix ZL - its inverse matrix $ZL^{-1}$ in relation to the matrix $e_0$ on the basis of equations: $ZL*ZL^{-1} = ZL^{-1}*ZL = e_0$. (Fig. 47).

By analogy, the system of matrices ZR has the matrix $e_2$ (Fig. 46), which possesses all properties of the identity matrix for any matrix ZR since $e_2*ZR = ZR*e_2 = ZR$ and $e_2^2 = e_2$. In the frame of the system of matrices ZR, where locally the matrix $e_2$ plays the role of the identity matrix (the local-identity matrix), one can define - for any non-zero matrix ZR - its inverse matrix $ZR^{-1}$ in relation to the matrix $e_2$ on the base of equations: $ZR*ZR^{-1} = ZR^{-1}*ZR = e_2$ (Fig. 47).

Multiplication of two members from the same system ZL (or ZR) is commutative as it is true for complex numbers. But multiplication of two members from different systems (one member from ZL and one member from ZR) is not commutative and gives a new matrix from one of these systems.

Multiplication of a complex number $ZL=a_0*e_0+a_1*e_1$ with its conjugate complex number $a_0*e_0-a_1*e_1$ gives the norm of this complex number: $(a_0^2+a_1^2)*e_0$. Multiplication of a complex number $ZR=a_2*e_2+a_3*e_3$ with its conjugate complex number $a_2*e_2-a_3*e_3$ gives the norm of this complex number: $(a_2^2+a_3^2)*e_2$. Each of these norms can be called as a local-norm inside the set of matrices $Z_2 = ZL+ZR = a_0*e_0+a_1*e_1+a_2*e_2+a_3*e_3$.

One should emphasized that the total set of sparse matrices $e_0$, $e_1$, $e_2$, $e_3$ does not contain the identity matrix E=[1,0,0,0; 0,1,0,0; 0,0,1,0; 0,0,0,1] and therefore the total sum $Z_2=a_0*e_0+a_1*e_1+a_2*e_2+a_3*e_3$ does not represent a system of hypercomplex numbers. It contains two blocks of complex numbers ($a_0*e_0+a_1*e_1$) and ($a_2*e_2+a_3*e_3$) and by this reason it is one of many types of U-complex numbers. It can be called as the 2-block U-complex numeric system (other U-complex systems can include more quantity of blocks as shown below). The set of its basic elements is closed under multiplication and defines the multiplication table in Fig. 48. In the system of U-complex numbers $Z_2$, the matrices $e_0$ and $e_1$ are conditionally called as the local-real unit and the local-imaginary unit correspondingly in the block $ZL=a_0*e_0+a_1*e_1$; the matrices $e_2$ and $e_3$ are called as the local-real unit and the local-imaginary unit in the block $ZR=a_2*e_2+a_3*e_3$.

$$Z_2 = ZL+ZR = a_0*e_0+a_1*e_1+a_2*e_2+a_3*e_3.$$

| * | $e_0$ | $e_1$ | $e_2$ | $e_3$ |
|---|---|---|---|---|
| $e_0$ | $e_0$ | $e_1$ | $e_2$ | $e_3$ |
| $e_1$ | $e_1$ | $-e_0$ | $-e_3$ | $e_2$ |
| $e_2$ | $e_0$ | $e_1$ | $e_2$ | $e_3$ |
| $e_3$ | $-e_1$ | $e_0$ | $e_3$ | $-e_2$ |

Fig. 48. Left: the system of U-complex numbers $Z_2$ of the 2-block type, where basic elements $e_0$, $e_1$, $e_2$, $e_3$ are taken from Fig. 46. Right: the multiplication table of the basic elements of this 4-parametric numeric system.

This multiplication table has a cruciform character with an element of a fractal plexus in relative locations of sets of basic units:

- the structure of both (2*2)-quadrants along the main diagonal coincides with the multiplication table of basic elements of complex numbers; in opposite to this, the structure of both (2*2)-quadrants along the second diagonal coincides with the known matrix representation of complex numbers [a, b; -b, a];
- all cells of the main diagonal of the table contain only local-real units $e_0$ and $e_2$; all cells of the second diagonal contain only local-imaginary units $e_1$ and $e_3$;

• each of the four (2*2)-sub-quadrants of the table also has a cruciform character since its main diagonal contains only local-real units $e_0$ or $e_2$ and its second diagonal contains only local-imaginary units $e_1$ and $e_3$.

Ordinary U-hypercomplex systems can be written in the following linear form (11) in the case of using ordinary multiplication in definition of their separate blocks (below we will also represent «tensor U-hypercomplex numbers», where tensor multiplication participates in definition of separate blocks of U-hypercomplex numbers):

$$(a_0*E_1 + a_1*e_1 + \dots + a_k*e_k) + (b_0*E_2 + b_1*q_1 + \dots + b_m*q_m) + \\ + \dots + (d_0*E_n + d_1*j_1 + \dots + d_p*j_p) \qquad (11)$$

where each of expressions represents its own complex or hypercomplex system; $E_1, E_2,\dots,E_n$ are local-real units; $e_1,\dots, e_k$, $q_1,\dots,q_m$, $j_1,\dots,j_p$ are local-imaginary units; $a_0,..a_k$, $b_0,\dots,b_m$, $d_0,\dots,d_p$ – real coefficients. It is obvious that U-hypercomplex systems (11) are a generalization of hypercomplex systems. The known Frobenius theorem for hypercomplex systems [Kantor, Solodovnikov, 1989] does not apply to U-hypercomplex systems in the whole.

The system of U-hypercomplex numbers $Z_2$=ZL+ZR has operations of addition, subtraction and non-commutative multiplication: applications of any of these operations for two U-complex numbers $Z_2$ gives a new U-complex number of the same type. But this system has no operation of division since the determinant of $Z_2$ is equal to zero (below we will show systems of U-complex numbers with zero divisors, where each of non-zero numbers has its inverse U-complex number).

Results of actions of the matrix operators ZL (or ZR) on a 4-dimensional vector $X=[x_0,x_1,x_2,x_3]$ depend on the side of the action: results of multiplication from right side X*ZL and from left side ZL*X differ and they have important properties concerning 2-dimensional subspaces of configurational 4-dimensional spaces (Fig. 49, 50).

| | | | | |
|---|---|---|---|---|
| $[x_0,x_1,x_2,x_3]$ * ZL = $[a_0*x_0-a_1*x_1-a_0*x_3+a_1*x_2,$ | $a_0*x_1+a_1*x_0-a_0*x_2-a_1*x_3,$ | 0, | 0] | |
| $[x_0,x_1,x_2,x_3]$ * ZR = [0, | 0, | $a_3*x_0-a_2*x_1+a_2*x_2- a_3*x_3,$ | $a_2*x_3-a_3*x_1-a_2*x_0+a_3*x_2]$ | |

Fig. 49. Multiplication of matrix operators ZL and ZR (Fig. 47) with a vector $X=[x_0,x_1,x_2,x_3]$ from the right side leads to a selective manage of subspaces.

In the case of multiplication from the right side (Fig. 49), a possibility is revealed for a selective manage of vectors of 2-dimensional subspaces of configurational 4-dimensional spaces. The action $[x_0,x_1,x_2,x_3]$*ZL gives the result that non-zero vectors exist only in the subspace $(x_0,x_1)$, where they can be managed arbitrary by means of parameters $a_0$ and $a_1$ of the operator ZL. If the parameters $a_0$ and $a_1$ are functions of time, then the vector $[x_0,x_1,x_2,x_3]$*ZL moves along different trajectories in the plane $(x_0,x_1)$ of the 4-dimensional space $[x_0,x_1,x_2,x_3]$. In other words, the corresponding 2-parametrical subsystem of the 4-parametrical system is running under the selective manage of the operator ZL. It leads to valuable capabilities to model multi-parametrical biological systems and bioinformational spaces, where different subsystems and their configurational subspaces are managed independently. One can tell more about such capabilities in connection with a generalization of such U-hypercomplex operators for $2^n$-dimensional spaces and $2^n$-parametrical systems.

A similar situation of the selective manage is true for results of the action $[x_0,x_1,x_2,x_3]$*ZR (Fig. 49): in this case non-zero vectors exist only in the subspace $(x_2,x_3)$, where they can be managed arbitrary by means of parameters $a_2$ and $a_3$ of the operator ZR. If the parameters $a_2$ and $a_3$ are functions of time, then the vector $[x_0,x_1,x_2,x_3]*ZR$ moves along different trajectories in the plane $(x_2,x_3)$ of the 4-dimensional space $[x_0,x_1,x_2,x_3]$.

In the case of multiplication from the left side, the interesting effects of twinning appears (Fig. 50).

| | | | | |
|---|---|---|---|---|
| $ZL*[x_0,x_1,x_2,x_3]^T$ = | $[a_0*x_0+a_1*x_1;$ | $a_0*x_1-a_1*x_0;$ | $-(a_0*x_1-a_1*x_0);$ | $-(a_0*x_0+a_1*x_1)]$ |
| $ZR*[x_0,x_1,x_2,x_3]^T$ = | $[a_3*x_2-a_2*x_3;$ | $-(a_2*x_2+a_3*x_3);$ | $a_2*x_2+a_3*x_3;$ | $-(a_3*x_2-a_2*x_3)]$ |

Fig. 50. Multiplication of the matrix operators ZL and ZR (Fig. 47) with a vector $X=[x_0,x_1,x_2,x_3]$ from the left side leads to effects of twinning.

Fig. 50 shows that multiplication from the left side $ZL*[x_0,x_1,x_2,x_3]^T$ generates a new vector $[y_0,y_1,y_2,y_3]$ whose components $y_0$ and $y_3$ are identical to each other up to sign ($y_0=(a_0*x_0+a_1*x_1)$ and $y_3=-(a_0*x_0+a_1*x_1)$) and whose components $y_1$ and $y_2$ are also identical to each other up to sign ($y_1 = a_0*x_1-a_1*x_0$ and $y_2 = -(a_0*x_1-a_1*x_0)$). Such mathematical peculiarities of the multiplication from the left side $ZL*[x_0,x_1,x_2,x_3]^T$ can be called as effects of twinning. All components of the resulting vector $[y_0,y_1,y_2,y_3]$ in this case are independent on components $x_2$ and $x_3$ of the initial vector.

Analogically, multiplication from the left side $ZR*[x_0,x_1,x_2,x_3]^T$ generates a new vector $[y_0,y_1,y_2,y_3]$ whose components $y_0$ and $y_3$ are identical to each other up to sign ($y_0 = (a_3*x_2-a_2*x_3)$ and $y_3 = -(a_3*x_2-a_2*x_3)$) and whose components $y_1$ and $y_2$ are also identical to each other up to sign ($y_1 = -(a_2*x_2+a_3*x_3)$ and $y_2 = a_2*x_2+a_3*x_3$). In this case all components of the resulting vector $[y_0,y_1,y_2,y_3]$ are independent on components $x_0$ and $x_1$ of the initial vector.

Results of actions of the matrix operators $Z_2$ = ZL+ZR on a 4-dimensional vector $X=[x_0,x_1,x_2,x_3]$ depend on the side of the action (Fig. 51) and they unite results of actions of separate operators ZL and ZR together (Fig. 49, 50). The above described property of the selective manage of 2-dimensional subspaces of the 4-dimensional space is conserved in the case of multiplication by $Z_2$ from the right side. The described effect of twinning is also conserved in the case of multiplication by $Z_2$ from the left side.

| |
|---|
| $[x_0,x_1,x_2,x_3]*Z_2 = [a_0*x_0-a_1*x_1-a_0*x_3+a_1*x_2,\quad a_0*x_1+a_1*x_0-a_0*x_2-a_1*x_3,$<br>$a_3*x_0-a_2*x_1+a_2*x_2-a_3*x_3,\quad a_2*x_3-a_3*x_1-a_2*x_0+a_3*x_2]$ |
| $Z_2*[x_0,x_1,x_2,x_3]^T = [a_0*x_0+a_1*x_1-a_2*x_3+a_3*x_2;\quad a_0*x_1-a_1*x_0-a_2*x_2-a_3*x_3;$<br>$a_1*x_0-a_0*x_1+a_2*x_2+a_3*x_3;\quad a_2*x_3-a_1*x_1-a_0*x_0-a_3*x_2]$ |

Fig. 51. Multiplication of the matrix operators $Z_2$=ZL+ZR with a vector $X=[x_0,x_1,x_2,x_3]$ from the right side and from the left side.

One should note the existence of another decomposition of the Walsh-representation $R_4$ (Fig. 1) of the matrix of 16 doublets $[C, A; T, G]^{(2)}$ into sum of another set of 4 sparse matrices $q_0, q_1, q_2, q_3$ (Fig. 52). Each of its two sub-sets ($q_0$ and $q_1$) and ($q_2$ and $q_3$) is closed in relation to multiplication and corresponds again to the multiplication table of complex numbers (Fig. 52).

$$\begin{matrix}1&0&1&0\\0&1&0&-1\\0&0&0&0\\0&0&0&0\end{matrix} + \begin{matrix}0&1&0&-1\\-1&0&-1&0\\0&0&0&0\\0&0&0&0\end{matrix} + \begin{matrix}0&0&0&0\\0&0&0&0\\1&0&1&0\\0&-1&0&1\end{matrix} + \begin{matrix}0&0&0&0\\0&0&0&0\\0&-1&0&1\\-1&0&-1&0\end{matrix} = q_0+q_1 +q_2+q_3.$$

| * | $q_0$ | $q_1$ |
|---|---|---|
| $q_0$ | $q_0$ | $q_1$ |
| $q_1$ | $q_1$ | $-q_0$ |

| * | $q_2$ | $q_3$ |
|---|---|---|
| $q_2$ | $q_2$ | $q_3$ |
| $q_3$ | $q_3$ | $-q_2$ |

Fig. 52. Another decomposition of the same matrix $R_4$ (Fig. 1) into 4 sparse matrices $q_0$, $q_1$, $q_2$, $q_3$. Right: multiplication tables of pairs of matrices $q_0$ and $q_1$ (left) and $q_2$ and $q_3$ (right) coincide with the multiplication table of basic elements of complex numbers.

It means that the expression $B_2 = a_0*q_0+a_1*q_1+a_2*q_2+a_3*q_3$ defines a new variant of 4-parametrical U-complex numbers of the 2-block character, where two different complex numbers $a_0*q_0+a_1*q_1$ and $a_2*q_2+a_3*q_3$ are united. This new U-complex numbers $B_2$ have analogical properties with the described U-complex numbers $Z_2$ (Fig. 46-51): their system has non-commutative multiplication and has no operation of division.

It is interesting that the same matrix $R_4$ can be decomposed by other ways into sets of 4 sparse matrices, each of which corresponds to a union of two different split-complex numbers into a single whole; this union can be called as «2-block united-split-complex numbers» or «2-block U-split-complex numbers» (Fig. 53, 54). One should remind that split-complex numbers are known also as «double numbers» or «hyperbolic numbers» or «Lorentz numbers (http://en.wikipedia.org/wiki/Split-complex_number). By this reason «2-block U-split-complex numbers» can be also called as «2-block U-double numbers» or «2-block U-hyperbolic numbers» or «2-block U-Lorentz numbers».

| | | | | | | | | | | |
|---|---|---|---|---|---|---|---|---|---|---|
| 1 1 1 -1 | | 1 0 0 0 | | 0 0 1 0 | | 0 1 0 0 | | 0 0 0 -1 | |
| -1 1 -1 -1 | = | -1 0 0 0 | + | 0 0 -1 0 | + | 0 1 0 0 | + | 0 0 0 -1 | = $k_0+k_1+k_2+k_3$ |
| 1 -1 1 1 | | 0 0 1 0 | | 1 0 0 0 | | 0 0 0 1 | | 0 -1 0 0 | |
| -1 -1 -1 1 | | 0 0 -1 0 | | -1 0 0 0 | | 0 0 0 1 | | 0 -1 0 0 | |

| * | **$k_0$** | **$k_1$** |
|---|---|---|
| **$k_0$** | $k_0$ | $k_1$ |
| **$k_1$** | $k_1$ | $k_0$ |

| * | **$k_2$** | **$k_3$** |
|---|---|---|
| **$k_2$** | $k_2$ | $k_3$ |
| **$k_3$** | $k_3$ | $k_2$ |

Fig. 53. Top: the decomposition of the matrix $R_4$ (Fig. 1) into 4 sparse matrices $k_0$, $k_1$, $k_2$, $k_3$. Bottom: multiplication tables of pairs of matrices $k_0$ and $k_1$ (left) and $k_2$ and $k_3$ (right) coincide with the multiplication table of basic elements of split-complex numbers.

| | | | | | | | | | | |
|---|---|---|---|---|---|---|---|---|---|---|
| 1 1 1 -1 | | 1 1 0 0 | | 0 0 1 -1 | | 0 0 0 0 | | 0 0 0 0 | |
| -1 1 -1 -1 | = | 0 0 0 0 | + | 0 0 0 0 | + | -1 1 0 0 | + | 0 0 -1 -1 | = $j_0+j_1+j_2+j_3$ |
| 1 -1 1 1 | | 0 0 0 0 | | 0 0 0 0 | | 0 0 1 1 | | 1 -1 0 0 | |
| -1 -1 -1 1 | | 0 0 -1 1 | | -1 -1 0 0 | | 0 0 0 0 | | 0 0 0 0 | |

| * | **$j_0$** | **$j_1$** |
|---|---|---|
| **$j_0$** | $j_0$ | $j_1$ |
| **$j_1$** | $j_1$ | $j_0$ |

| * | **$j_2$** | **$j_3$** |
|---|---|---|
| **$j_2$** | $j_2$ | $j_3$ |
| **$j_3$** | $j_3$ | $j_2$ |

Fig. 54. Top: the decomposition of the matrix $R_4$ (Fig. 1) into 4 sparse matrices $j_0$, $j_1$, $j_2$, $j_3$. Bottom: multiplication tables of pairs of matrices $j_0$ and $j_1$ (left) and $j_2$ and $j_3$ (right) coincide with the multiplication table of basic elements of split-complex numbers.

Correspondingly the expressions $KL=a_0*k_0+a_1*k_1$ and $KR=a_2*k_2+a_3*k_3$ present two different systems of 2-parametrical split-complex numbers (here $k_0$, $k_1$, $k_2$ and $k_3$ are taken from Fig. 53). Their union $K = KL+KR = a_0*k_0+a_1*k_1+ a_2*k_2+a_3*k_3$ presents 2-block U-split-complex numbers, where $k_0$ and $k_2$ play their roles of local-identity units and $k_1$ and $k_3$ play their roles of local-imaginary units. Each of non-zero numbers KL, if $a_0 \neq a_1$, has its inverse number $KL^{-1} = (a_0^2-a_1^2)^{-1}*(a_0*k_0-a_1*k_1)$ in relation to the local-identity unit $k_0$: $KL*KL^{-1}= KL^{-1}*KL= k_0$. Each of non-zero numbers KL has its inverse number $KL^{-1} = (a_0^2-a_1^2)^{-1}*(a_0*k_0-a_1*k_1)$ in relation to the local-identity unit $k_0$: $KL*KL^{-1}=KL^{-1}*KL= k_0$. Each of non-zero numbers KR, if $a_2 \neq a_3$, has its inverse number $KR^{-1} = (a_2^2-a_3^2)^{-1}*(a_2*k_2-a_3*k_3)$ in relation to the local-identity unit $k_2$: $KR*KR^{-1}=KR^{-1}*KR= k_2$. The system of 4-parametrical split-complex numbers $K = KL+KR = a_0*k_0+a_1*k_1+a_2*k_2+a_3*k_3$ is not so interesting from the algevraic standpoint since it has only operations of addition and subtraction.

Analogically the expressions $JL=a_0*j_0+a_1*j_1$ and $JR=a_2*j_2+a_3*j_3$ present two different systems of 2-parametrical split-complex numbers (here $j_0$, $j_1$, $j_2$ and $j_3$ are taken from Fig. 54). Their union $J = JL+JR = a_0*j_0+a_1*j_1+a_2*j_2+a_3*j_3$ presents 2-block U-split-complex numbers, where $j_0$ and $j_2$ play their roles of local-identity units and $j_1$ and $j_3$ play their roles of local-imaginary units. Each of non-zero numbers JL, where $a_0 \neq a_1$, has its inverse number $JL^{-1} = (a_0^2-a_1^2)^{-1}*(a_0*j_0-a_1*j_1)$ in relation to the local-identity unit $j_0$: $JL*JL^{-1}=JL^{-1}*JL= j_0$. Each of non-zero numbers JL has its inverse number $JL^{-1} = (a_0^2-a_1^2)^{-1}*(a_0*j_0-a_1*j_1)$ in relation to the local-identity unit $j_0$: $JL*JL^{-1}=JL^{-1}*JL= j_0$. Each of non-zero numbers JR, where $a_2 \neq a_3$, has its inverse number $JR^{-1} = (a_2^2-a_3^2)^{-1}*(a_2*j_2-a_3*j_3)$ in relation to the local-identity unit $j_2$: $JR*JR^{-1}=JR^{-1}*JR= j_2$. The system of 4-parametrical split-complex numbers $J = JL+JR = a_0*j_0+a_1*j_1+a_2*j_2+a_3*j_3$ is not so interesting from the algevraic standpoint since it has only operations of addition and subtraction.

For a comparison Fig. 55 shows the described four systems of 2-block U-numbers (two systems of U-complex numbers and two systems of U-split-complex numbers), which are constructed on different decompositions (Fig. 46, 52, 53, 54 ) of the same matrix $R_4$.

| | 00 | 01 | 10 | 11 |
|---|---|---|---|---|
| 00 | CC 00 | CT 01 | TC 10 | TT 11 |
| 01 | CA 01 | CG 00 | TA 11 | TG 10 |
| 10 | AC 10 | AT 11 | GC 00 | GT 01 |
| 11 | AA 11 | AG 10 | GA 01 | GG 00 |

| | | | |
|---|---|---|---|
| $a_0$ | $a_1$ | $a_3$ | $-a_2$ |
| $-a_1$ | $a_0$ | $-a_2$ | $-a_3$ |
| $a_1$ | $-a_0$ | $a_2$ | $a_3$ |
| $-a_0$ | $-a_1$ | $-a_3$ | $a_2$ |

a)

| | | | |
|---|---|---|---|
| $a_0$ | $a_1$ | $a_0$ | $-a_1$ |
| $-a_1$ | $a_0$ | $-a_1$ | $-a_0$ |
| $a_2$ | $-a_3$ | $a_2$ | $a_3$ |
| $-a_3$ | $-a_2$ | $-a_3$ | $a_2$ |

b)

| | | | |
|---|---|---|---|
| $a_0$ | $a_2$ | $a_1$ | $-a_3$ |
| $-a_0$ | $a_2$ | $-a_1$ | $-a_3$ |
| $a_1$ | $-a_3$ | $a_0$ | $a_2$ |
| $-a_1$ | $-a_3$ | $-a_0$ | $a_2$ |

c)

| | | | |
|---|---|---|---|
| $a_0$ | $a_0$ | $a_1$ | $-a_1$ |
| $-a_2$ | $a_2$ | $-a_3$ | $-a_3$ |
| $a_3$ | $-a_3$ | $a_2$ | $a_2$ |
| $-a_1$ | $-a_1$ | $-a_0$ | $a_0$ |

d)

Fig. 55. Left: the matrix $[C, T; A, G]^{(2)}$ from Fig. 38 as the dyadic-shift matrix. Upper row: two 4-parametrical systems of 2-block U-complex numbers connected with decompositions in Fig. 46 and 52. Bottom row: two 4-parametrical systems of 2-block U-split-complex numbers connected with decompositions in Fig. 53 and 54.

One can see from Fig. 55 that all these 4 systems of 2-block U-numbers have the identical mosaic of locations of signs + and – (marked by black and white colors). They differ only by the distribution of parameters $a_0$, $a_1$, $a_2$, $a_3$, which reflects structures of basic elements of corresponding 2-block U-numbers «a», «b», «c» and «d» in Fig. 55. This structures are not chaotic but they are regular from the standpoint of the representation of genetic matrices $[C, T; A, G]^{(n)}$ as the dyadic-shift matrix (the notion of dyadic-shift matrices is used in discrete signals processing and noise-immunity coding [Ahmed, Rao, 1975]). This representation has been described in our works [Petoukhov, 2008b, 2011a; Petoukhov, He, 2010] on the basis of 3 pairs of binary-oppositional traits of nitrogenous bases of DNA and RNA. Fig. 55,left shows the representation of genetic matrices $[C, T; A, G]^{(2)}$ as the dyadic-shift matrix, where each of cells has its binary numeration. For example, in the matrix «a», which correspond to the set of basic matrices $e_0$, $e_1$, $e_2$, $e_3$ in Fig. 46, non-zero entries of the matrices $e_0$, $e_1$, $e_2$, $e_3$ are located in cells with the following binary numerations in the dyadic-shift matrix:

- for $e_0$ - in cells 00, 00, 11, 11 (the parameter $a_0$ is located in these cells);
- for $e_1$ - in cells 01, 01, 10, 10 (the parameter $a_1$ is located in these cells);
- for $e_2$ - in cells 00, 00, 11, 11 (the parameter $a_2$ is located in these cells);
- for $e_3$ – in cells 01, 01, 10, 10 (the parameter $a_3$ is located in these cells).

Here the matrices $e_0$ and $e_2$, which are local-real units in U-numbers "a", have the identical set of binary numerations 00, 00, 11, 11 for their non-zero entries; but non-zero entries of the matrix $e_0$ correspond to doublets with the first letter C and A in contrast to non-zero entries of the marix $e_2$, which correspond to doublets with the first letter G and T. The matrices $e_1$ and $e_3$, which are local-imaginary units in U-numbers "a", also have the identical set of binary numerations 01, 01, 10, 10 for their non-zero entries; but non-zero entries of the matrix $e_1$ correspond to doublets with the first letter C and A in contrast to non-zero entries of the marix $e_3$, which correspond to doublets with the first letter G and T.

The similar situation is realized for the systems of U-numbers «b», «c» and «d» in Fig. 55 and also for 4-block U-complex numbers, which are based on appropriate decompositions of the matrix $R_8$ (Fig. 1).

As known the set of 16 doubles is divided into 2 equal subsets of 8 strong doublets and 8 weak doublets (Fig. 36). The connection of the genetic alphabets of DNA and RNA with the described systems of 4-parametrical U-numbers (the systems «a», «b», «c» and «d» in Fig. 55) allows to suppose the following: the set of 16 doublets can be divided by nature into other separate subsets of equivalency, where 4 subsets exist with 4 doublets in each on the basis of appropriate traits of equivalency connected with the dyadic-shift matrices and with binary-oppositional traits of nitrogenous bases. For example, in the case of U-complex numbers represented by the matrix «a» in Fig. 55, the following division can be supposed:

- the subset for non-zero entries of the basic matrix $e_0$ includes doublets CC, CG, AT, AA;

- the subset for non-zero entries of the basic matrix $e_1$ includes doublets CT, CA, AC, AG;

- the subset for non-zero entries of the basic matrix $e_2$ includes doublets TT, TA, GC, GG;

- the subset for non-zero entries of the basic matrix $e_3$ includes doublets TC, TG, GT, GA.

For systems of U-numbers represented by matrices «b», «c» and «d» (Fig. 55), other approprite combinations of doublets exist to provide their 4 sets with 4 doublets in each on the basis of relevant traits of equivalency. It was described in publications [Petoukhov, 2008b, 2011a, 2012b; Petoukhov, He, 2009] that each of 16 doublets and each of 64 triplets has its own binary numeration on the basis of 3 kinds of objective binary-oppositional traits of nitrogenous bases of DNA and RNA in connection with dyadic groups of binary numbers, the tensor family of genetic matrices $[C, A; T, G]^{(n)}$ and with binary numeration in dyadic-shift matrices. It gives an opportunity for genetic system to mark each of cases of 4 subsets of doublets in matrices «a», «b», «c» and «d» (Fig. 55) by means of its individual token. Using appropriate systems of such tokens, bioinformational system of living organism can parallely apply different kinds of U-hypercomplex numbers in providing multi-channel communication among its different physiological parts and also communication with the external world. From the standpoint of such approach, a living organism has no a single kind of numeric system but many different kinds of systems U-hypercomplex numbers (an organism is a poly-linguistic object). The author believes that many secrets of genetic systems and genetic phenomena are connected with such multi-linguistics of genetic systems.

Let us turn now to the matrix $R_8$, which is the Walsh-representation of the genetic matrix $[C, T; A, G]^{(3)}$. We show here only two of several examples of its possible decompositions (Fig. 56 and 57), which lead to 4-block U-complex numbers and 4-block U-split-complex numbers.

| | | | | | | | |
|---|---|---|---|---|---|---|---|
| $R_8$ = | 1 0 0 0 0 0 0 0<br>1 0 0 0 0 0 0 0<br>0 0 1 0 0 0 0 0<br>0 0 1 0 0 0 0 0<br>0 0 -1 0 0 0 0 0<br>0 0 -1 0 0 0 0 0<br>-1 0 0 0 0 0 0 0<br>-1 0 0 0 0 0 0 0 | + | 0 0 1 0 0 0 0 0<br>0 0 1 0 0 0 0 0<br>-1 0 0 0 0 0 0 0<br>-1 0 0 0 0 0 0 0<br>1 0 0 0 0 0 0 0<br>1 0 0 0 0 0 0 0<br>0 0 -1 0 0 0 0 0<br>0 0 -1 0 0 0 0 0 | + | 0 1 0 0 0 0 0 0<br>0 1 0 0 0 0 0 0<br>0 0 0 1 0 0 0 0<br>0 0 0 1 0 0 0 0<br>0 0 0 -1 0 0 0 0<br>0 0 0 -1 0 0 0 0<br>0 -1 0 0 0 0 0 0<br>0 -1 0 0 0 0 0 0 | + | 0 0 0 1 0 0 0 0<br>0 0 0 1 0 0 0 0<br>0 -1 0 0 0 0 0 0<br>0 -1 0 0 0 0 0 0<br>0 1 0 0 0 0 0 0<br>0 1 0 0 0 0 0 0<br>0 0 0 -1 0 0 0 0<br>0 0 0 -1 0 0 0 0 |
| + | 0 0 0 0 0 0 -1 0<br>0 0 0 0 0 0 -1 0<br>0 0 0 0 -1 0 0 0<br>0 0 0 0 -1 0 0 0<br>0 0 0 0 1 0 0 0<br>0 0 0 0 1 0 0 0<br>0 0 0 0 0 0 1 0<br>0 0 0 0 0 0 1 0 | + | 0 0 0 0 1 0 0 0<br>0 0 0 0 1 0 0 0<br>0 0 0 0 0 0 -1 0<br>0 0 0 0 0 0 -1 0<br>0 0 0 0 0 0 1 0<br>0 0 0 0 0 0 1 0<br>0 0 0 0 -1 0 0 0<br>0 0 0 0 -1 0 0 0 | + | 0 0 0 0 0 0 0 -1<br>0 0 0 0 0 0 0 -1<br>0 0 0 0 0 -1 0 0<br>0 0 0 0 0 -1 0 0<br>0 0 0 0 0 1 0 0<br>0 0 0 0 0 1 0 0<br>0 0 0 0 0 0 0 1<br>0 0 0 0 0 0 0 1 | + | 0 0 0 0 0 1 0 0<br>0 0 0 0 0 1 0 0<br>0 0 0 0 0 0 0 -1<br>0 0 0 0 0 0 0 -1<br>0 0 0 0 0 0 0 1<br>0 0 0 0 0 0 0 1<br>0 0 0 0 0 -1 0 0<br>0 0 0 0 0 -1 0 0 |

| * | $v_0$ | $v_1$ |
|---|---|---|
| $v_0$ | $v_0$ | $v_1$ |
| $v_1$ | $v_1$ | $-v_0$ |

| * | $v_2$ | $v_3$ |
|---|---|---|
| $v_2$ | $v_2$ | $v_3$ |
| $v_3$ | $v_3$ | $-v_2$ |

| * | $v_4$ | $v_5$ |
|---|---|---|
| $v_4$ | $v_4$ | $v_5$ |
| $v_5$ | $v_5$ | $-v_4$ |

| * | $v_6$ | $v_7$ |
|---|---|---|
| $v_6$ | $v_6$ | $v_7$ |
| $v_7$ | $v_7$ | $-v_6$ |

Fig. 56. The decomposition of the matrix $R_8$ (Fig. 1) into the set of 8 sparse matrices:

$R_8=v_0+v_1+v_2+v_3+v_4+v_5+v_6+v_7$. Each of the following pairs of these 8 matrices ($v_0$ and $v_2$; $v_2$ and $v_3$; $v_4$ and $v_5$; $v_6$ and $v_7$) forms a set, which is closed under multiplication. These sets define multiplication tables, which coincide with the multiplication table of the system of complex numbers (the lower row).

Each of the following expressions means a separate system of complex numbers represented in the form of (8*8)-matrices: $a_0*v_0+a_1*v_1$, $a_2*v_2+a_3*v_3$, $a_4*v_4+a_5*v_5$, $a_6*v_6+a_7*v_7$. Their sum V = $a_0*v_0+a_1*v_1$, $a_2*v_2+a_3*v_3$, $a_4*v_4+a_5*v_5$, $a_6*v_6+a_7*v_7$ represents the system of 4-block U-complex numbers.

Fig. 57 shows another decomposition of the same matrix $R_8$ into 8 sparse matrices: $R_8 = p_0+p_1+p_2+p_3+p_4+p_5+p_6+p_7$:

| | | | | | | | |
|---|---|---|---|---|---|---|---|
| | 1 0 0 0 0 0 0 0 | | 0 1 0 0 0 0 0 0 | | 0 0 1 0 0 0 0 0 | | 0 0 0 1 0 0 0 0 |
| | 0 1 0 0 0 0 0 0 | | 1 0 0 0 0 0 0 0 | | 0 0 0 1 0 0 0 0 | | 0 0 1 0 0 0 0 0 |
| | -1 0 0 0 0 0 0 0 | | 0 -1 0 0 0 0 0 0 | | 0 0 1 0 0 0 0 0 | | 0 0 0 1 0 0 0 0 |
| $R_8 =$ | 0 -1 0 0 0 0 0 0 | + | -1 0 0 0 0 0 0 0 | + | 0 0 0 1 0 0 0 0 | + | 0 0 1 0 0 0 0 0 |
| | 1 0 0 0 0 0 0 0 | | 0 1 0 0 0 0 0 0 | | 0 0 -1 0 0 0 0 0 | | 0 0 0 -1 0 0 0 0 |
| | 0 1 0 0 0 0 0 0 | | 1 0 0 0 0 0 0 0 | | 0 0 0 -1 0 0 0 0 | | 0 0 -1 0 0 0 0 0 |
| | -1 0 0 0 0 0 0 0 | | 0 -1 0 0 0 0 0 0 | | 0 0 -1 0 0 0 0 0 | | 0 0 0 -1 0 0 0 0 |
| | 0 -1 0 0 0 0 0 0 | | -1 0 0 0 0 0 0 0 | | 0 0 0 -1 0 0 0 0 | | 0 0 -1 0 0 0 0 0 |
| | 0 0 0 0 1 0 0 0 | | 0 0 0 0 0 1 0 0 | | 0 0 0 0 0 0 -1 0 | | 0 0 0 0 0 0 0 -1 |
| | 0 0 0 0 0 1 0 0 | | 0 0 0 0 1 0 0 0 | | 0 0 0 0 0 0 0 -1 | | 0 0 0 0 0 0 -1 0 |
| | 0 0 0 0 -1 0 0 0 | | 0 0 0 0 0 -1 0 0 | | 0 0 0 0 0 0 -1 0 | | 0 0 0 0 0 0 0 -1 |
| + | 0 0 0 0 0 -1 0 0 | + | 0 0 0 0 -1 0 0 0 | + | 0 0 0 0 0 0 0 -1 | + | 0 0 0 0 0 0 -1 0 |
| | 0 0 0 0 1 0 0 0 | | 0 0 0 0 0 1 0 0 | | 0 0 0 0 0 0 1 0 | | 0 0 0 0 0 0 0 1 |
| | 0 0 0 0 0 1 0 0 | | 0 0 0 0 1 0 0 0 | | 0 0 0 0 0 0 0 1 | | 0 0 0 0 0 0 1 0 |
| | 0 0 0 0 -1 0 0 0 | | 0 0 0 0 0 -1 0 0 | | 0 0 0 0 0 0 1 0 | | 0 0 0 0 0 0 0 1 |
| | 0 0 0 0 0 -1 0 0 | | 0 0 0 0 -1 0 0 0 | | 0 0 0 0 0 0 0 1 | | 0 0 0 0 0 0 1 0 |

| * | $p_0$ | $p_1$ |
|---|---|---|
| $p_0$ | $p_0$ | $p_1$ |
| $p_1$ | $p_1$ | $p_0$ |

| * | $p_2$ | $p_3$ |
|---|---|---|
| $p_2$ | $p_2$ | $p_3$ |
| $p_3$ | $p_3$ | $p_2$ |

| * | $p_4$ | $p_5$ |
|---|---|---|
| $p_4$ | $p_4$ | $p_5$ |
| $p_5$ | $p_5$ | $p_4$ |

| * | $p_6$ | $p_7$ |
|---|---|---|
| $p_6$ | $p_6$ | $p_7$ |
| $p_7$ | $p_7$ | $p_6$ |

Fig. 57. The decomposition of the matrix $R_8$ (Fig. 1) into the set of 8 sparse matrices: $R_8=p_0+p_1+p_2+p_3+p_4+p_5+p_6+p_7$. Each of the following pairs of these 8 matrices ($v_0$ and $v_2$; $v_2$ and $v_3$; $v_4$ and $v_5$; $v_6$ and $v_7$) forms a set, which is closed under multiplication. These sets define multiplication tables, which coincide with the multiplication table of the system of split-complex numbers (the lower row).

Each of the following expressions means a separate system of split-complex numbers represented in the form of (8*8)-matrices: $a_0*p_0+a_1*p_1$, $a_2*p_2+a_3*p_3$, $a_4*p_4+a_5*p_5$, $a_6*p_6+a_7*p_7$. Their sum P = $a_0*p_0+a_1*p_1+a_2*p_2+a_3*p_3+a_4*p_4+a_5*p_5+a_6*p_6+a_7*p_7$ represents the system of 4-block U-split-complex numbers.

Decompositions of the matrix $R_8$ can lead not only to systems of U-numbers with 2-parametrical blocks but also to systems of U-numbers with 4 parametrical blocks as the following example shows (Fig. 58).

| | | | | | | | |
|---|---|---|---|---|---|---|---|
| | 1 0 0 0 0 0 0 0 | | 0 0 1 0 0 0 0 0 | | 0 0 0 0 1 0 0 0 | | 0 0 0 0 0 0 -1 0 |
| | 1 0 0 0 0 0 0 0 | | 0 0 1 0 0 0 0 0 | | 0 0 0 0 1 0 0 0 | | 0 0 0 0 0 0 -1 0 |
| | 0 0 1 0 0 0 0 0 | | -1 0 0 0 0 0 0 0 | | 0 0 0 0 0 0 -1 0 | | 0 0 0 0 -1 0 0 0 |
| $R_8 =$ | 0 0 1 0 0 0 0 0 | + | -1 0 0 0 0 0 0 0 | + | 0 0 0 0 0 0 -1 0 | + | 0 0 0 0 -1 0 0 0 |
| | 0 0 0 0 1 0 0 0 | | 0 0 0 0 0 0 1 0 | | 1 0 0 0 0 0 0 0 | | 0 0 -1 0 0 0 0 0 |
| | 0 0 0 0 1 0 0 0 | | 0 0 0 0 0 0 1 0 | | 1 0 0 0 0 0 0 0 | | 0 0 -1 0 0 0 0 0 |
| | 0 0 0 0 0 0 1 0 | | 0 0 0 0 -1 0 0 0 | | 0 0 -1 0 0 0 0 0 | | -1 0 0 0 0 0 0 0 |
| | 0 0 0 0 0 0 1 0 | | 0 0 0 0 -1 0 0 0 | | 0 0 -1 0 0 0 0 0 | | -1 0 0 0 0 0 0 0 |
| | 0 1 0 0 0 0 0 0 | | 0 0 0 1 0 0 0 0 | | 0 0 0 0 0 1 0 0 | | 0 0 0 0 0 0 0 -1 |
| | 0 1 0 0 0 0 0 0 | | 0 0 0 1 0 0 0 0 | | 0 0 0 0 0 1 0 0 | | 0 0 0 0 0 0 0 -1 |
| | 0 0 0 1 0 0 0 0 | | 0 -1 0 0 0 0 0 0 | | 0 0 0 0 0 0 0 -1 | | 0 0 0 0 0 -1 0 0 |
| + | 0 0 0 1 0 0 0 0 | + | 0 -1 0 0 0 0 0 0 | + | 0 0 0 0 0 0 0 -1 | + | 0 0 0 0 0 -1 0 0 |
| | 0 0 0 0 0 1 0 0 | | 0 0 0 0 0 0 0 1 | | 0 1 0 0 0 0 0 0 | | 0 0 0 -1 0 0 0 0 |
| | 0 0 0 0 0 1 0 0 | | 0 0 0 0 0 0 0 1 | | 0 1 0 0 0 0 0 0 | | 0 0 0 -1 0 0 0 0 |
| | 0 0 0 0 0 0 0 1 | | 0 0 0 0 0 -1 0 0 | | 0 0 0 -1 0 0 0 0 | | 0 -1 0 0 0 0 0 0 |
| | 0 0 0 0 0 0 0 1 | | 0 0 0 0 0 -1 0 0 | | 0 0 0 -1 0 0 0 0 | | 0 -1 0 0 0 0 0 0 |

| * | $w_0$ | $w_1$ | $w_2$ | $w_3$ |
|---|---|---|---|---|
| $w_0$ | $w_0$ | $w_1$ | $w_2$ | $w_3$ |
| $w_1$ | $w_1$ | $-w_0$ | $w_3$ | $-w_2$ |
| $w_2$ | $w_2$ | $-w_3$ | $w_0$ | $-w_1$ |
| $w_3$ | $w_3$ | $w_2$ | $w_1$ | $w_0$ |

| * | $w_4$ | $w_5$ | $w_6$ | $w_7$ |
|---|---|---|---|---|
| $w_4$ | $w_4$ | $w_5$ | $w_6$ | $w_7$ |
| $w_5$ | $w_5$ | $-w_4$ | $w_7$ | $-w_6$ |
| $w_6$ | $w_6$ | $-w_7$ | $w_4$ | $-w_5$ |
| $w_7$ | $w_7$ | $w_6$ | $w_5$ | $w_4$ |

Fig. 58. The decomposition of the matrix $R_8$ (Fig. 1) into the set of 8 sparse matrices: $R_8=w_0+w_1+w_2+w_3+w_4+w_5+w_6+w_7$. Each of the following Each of the following quadruples of matrices ($w_0$, $w_1$, $w_2$ and $w_3$; $w_4$, $w_5$, $w_6$ and $w_7$) forms a set, which is closed under multiplication. These sets define multiplication tables, which coincide with the multiplication table of the 4-parametrical system of split-quaternions of J.Cockle (https://en.wikipedia.org/wiki/Split-quaternion).

Each of the following expressions means a separate system of split-quaternions represented in the form of (8*8)-matrices: $a_0*w_0+a_1*w_1+a_2*w_2+a_3*w_3$ and $a_4*w_4+a_5*w_5+a_6*w_6+a_7*w_7$. Their sum $W = a_0*w_0+a_1*w_1+a_2*w_2+a_3*w_3+a_4*w_4+a_5*w_5 +a_6*w_6+a_7*w_7$ represents the system of 2-block U-split-quaternions. The system of 2-block U-split-quaternions has operations of addition, subtraction and non-commutative multiplication.

## 16. TENSOR FAMILIES OF DIAGONAL UNITED-HYPERCOMPLEX NUMBERS

The following important question arises: whether such systems of U-complex numbers exist, which possess operations of commutative multiplication and division? Below we give positive answer on this question and show systems with these operations and addiionally with divisors of zero. These systems are connected with the genetic matrices and their Walsh-representations $R_4$ and $R_8$ (Fig. 1).

In the matrix $R_4$ (Fig. 1) the mosaic of two quadrants along the main diagonal is mirror-antisymmetric to the mosaic of two quadrants along the second diagonal in relation to the middle horizontal line. The operation of the mirror-antisymmetric transformation is given by the matrix J in the expression (12), the square of which $J^2$ is equal to the identical (4*4)-matrix:

$$J = [0,0,0,-1;\ 0,0,-1,0;\ 0,-1,0,0;\ -1,\ 0,0,0],\quad J^2 = [1,0,0,0;\ 0,1,0,0;\ 0,0,1,0;\ 0,0,0,1] \qquad (12)$$

Fig. 59 illustrates that the matrix $R_4$ is the sum of 2 sparse matrices $R_{40}$ and $R_{41}$ of diagonal types, non-zero parts of which coincide with two quadrants along the main diagonal and along the second diagonal correspondingly.

$$\mathbf{R_4 = R_{40}+R_{41} =} \begin{bmatrix} 1 & 1 & 1 & -1 \\ -1 & 1 & -1 & -1 \\ 1 & -1 & 1 & 1 \\ -1 & -1 & -1 & 1 \end{bmatrix} = \begin{bmatrix} 1 & 1 & 0 & 0 \\ -1 & 1 & 0 & 0 \\ 0 & 0 & 1 & 1 \\ 0 & 0 & -1 & 1 \end{bmatrix} + \begin{bmatrix} 0 & 0 & 1 & -1 \\ 0 & 0 & -1 & -1 \\ 1 & -1 & 0 & 0 \\ -1 & -1 & 0 & 0 \end{bmatrix}$$

Fig. 59. The matrix $R_4$ is the sum of two matrices $R_{40}$ and $R_{41}$.

The mirror-antisymmetric transformation J from the expression (12) transforms the matrix $R_{41}$ into the matrix $R_{40}$. Taking this into account, the initial matrix $R_4 = R_{40}+R_{41} = R_{40} + J*R_{40} = R_{40}*(E+J)$, where E is the identity (4*4)-matrix and E+J is the split-complex number with unit coordinates since the set of matrices E and J is closed under multiplication and its multiplication table coincides with the muliplication table of split-complex numbers (the mirror-antisymmetric transformation J represents the imaginary unit in this set). In other word, the matrix $R_4$ is the product of the matrix $R_{40}$ with this split-complex number. Correspondingly a special attention should be devoted to the sparse matrix $R_{40}$ of the diagonal type, non-zero parts of which represents numeric mosaic of the quadrants along the main diagonal in the genetic matrix $R_4$.

Fig.60 shows a decomposition of this matrix $R_{40} = d_0+d_1+d_2+d_3$, from the standpoint of which $R_{40}$ is diagonal 2-block U-complex number with unit coordinates.

$$R_{40} = \begin{bmatrix} 1 & 1 & 0 & 0 \\ -1 & 1 & 0 & 0 \\ 0 & 0 & 1 & 1 \\ 0 & 0 & -1 & 1 \end{bmatrix} = \begin{bmatrix} 1 & 0 & 0 & 0 \\ 0 & 1 & 0 & 0 \\ 0 & 0 & 0 & 0 \\ 0 & 0 & 0 & 0 \end{bmatrix} + \begin{bmatrix} 0 & 1 & 0 & 0 \\ -1 & 0 & 0 & 0 \\ 0 & 0 & 0 & 0 \\ 0 & 0 & 0 & 0 \end{bmatrix} + \begin{bmatrix} 0 & 0 & 0 & 0 \\ 0 & 0 & 0 & 0 \\ 0 & 0 & 1 & 0 \\ 0 & 0 & 0 & 1 \end{bmatrix} + \begin{bmatrix} 0 & 0 & 0 & 0 \\ 0 & 0 & 0 & 0 \\ 0 & 0 & 0 & 1 \\ 0 & 0 & -1 & 0 \end{bmatrix} = d_0+d_1+d_2+d_3$$

| * | $\mathbf{d_0}$ | $\mathbf{d_1}$ |
|---|---|---|
| $\mathbf{d_0}$ | $d_0$ | $d_1$ |
| $\mathbf{d_1}$ | $d_1$ | $-d_0$ |

| * | $\mathbf{d_2}$ | $\mathbf{d_3}$ |
|---|---|---|
| $\mathbf{d_2}$ | $d_2$ | $d_3$ |
| $\mathbf{d_3}$ | $d_3$ | $-d_2$ |

Fig. 59a. Top: the decomposition of the block-diagonal matrix $R_{40}$ into 4 sparse matrices $d_0$, $d_1$, $d_2$, $d_3$. Bottom: multiplication tables for the pair $d_0$ and $d_2$ and for the pair $d_2$ and $d_3$ coincide with the multiplication table of complex numbers.

The pair of matrices $d_0$ and $d_1$ forms the set, which is closed under multiplication and which defines the multiplication table of complex numbers (Fig. 59a, bottom). The same is true for the pair of matrices $d_2$ and $d_3$. It means that the matrix $R_{40}$ is the 2-block united-complex number of the diagonal type with unit coordinates. These basis elements $d_0$, $d_1$, $d_2$, $d_3$ of two systems of complex numbers define the following general view of these complex numbers SL and SR: $SL = a_0*d_0+a_1*d_1$ and $SR = a_2*d_2+a_3*d_3$. Fig. 60 shows the corresponding general view of the system of 2-block diagonal U-complex numbers $S = a_0*d_0+a_1*d_1+a_2*d_2+a_3*d_3$ having 4 parameters ($a_0$, $a_1$, $a_2$, $a_3$). One should mention that the whole set of matrices $d_0$, $d_1$, $d_2$, $d_3$ is not closed under multiplication and it does not represent any of hypercomplex numbers.

| | | |
|---|---|---|
| $SL = a_0*d_0+a_1*d_1 =$<br>$[a_0, a_1, 0, 0$<br>$-a_1, a_0, 0, 0$<br>$0, 0, 0, 0$<br>$0, 0, 0, 0]$ | $SR = a_2*d_2+a_3*d_3 =$<br>$[0, 0, 0, 0$<br>$0, 0, 0, 0$<br>$0, 0, a_2, a_3$<br>$0, 0, -a_3, a_2]$ | $S = SL+SR = a_0*d_0+a_1*d_1 +a_2*d_2+a_3*d_3 =$<br>$[a_0, a_1, 0, 0$<br>$-a_1, a_0, 0, 0$<br>$0, 0, a_2, a_3$<br>$0, 0, -a_3, a_2]$ |
| $SL^{-1} = (a_0^2 + a_1^2)^{-1} *$<br>$(a_0*d_0-a_1*d_1)$;<br>$SL*SL^{-1}=SL^{-1}*SL=d_0$,<br>where $d_0$ – the local-identity matrix for SL. | $SR^{-1} = (a_2^2 + a_3^2)^{-1}*$<br>$(a_2*d_2-a_3*d_3)$;<br>$SR*SR^{-1}=SR^{-1}*SR=d_2$<br>where $d_2$ – the local-identity matrix for SR. | $S^{-1} = (a_0^2+a_1^2)^{-1}*(a_0*d_0-a_1*d_1) +$<br>$(a_2^2+a_3^2)^{-1}*(a_2*d_2-a_3*d_3) = SL^{-1} + SR^{-1}$.<br>$S*S^{-1} = S^{-1}*S = E$,<br>where E = [1,0,0,0; 0,1,0,0; 0,0,1,0; 0,0,0,1] – the identity matrix |

Fig. 60. Upper row: diagonal U-complex numbers S of 2-block type, which unite complex numbers SL and SR. Lower row: expressions for inverse matrices $SL^{-1}$, $SR^{-1}$ and $S^{-1}$.

The classical identity matrix E=[1 0 0 0; 0 1 0 0; 0 0 1 0; 0 0 0 1] is absent in the set of matrices SL and SR, where - beside this - each matrix has zero determinant (Fig. 60). Consequently the notion of the inverse matrix $SL^{-1}$ or $SR^{-1}$ (as $SL*SL^{-1}=E$ or $SR*SR^{-1}=E$) can't be defined in relation to the identity matrix E in accordance with the famous theorem about inverse matrices for matrices with zero determinant in the case of the complete set of matrices [Bellman, 1960, Chapter 6, § 4]. But we analyze not the complete set of (4*4)-matrices but very limited special sets of matrices SL and SR. The system SL has the matrix $d_0$ (Fig. 59), which possesses all properties of its identity matrix for any matrix SL since $d_0*SL = SL*d_0 = SL$ and $d_0^2 = d_0$. In the frame of the system of matrices SL, where locally the matrix $d_0$ plays the role of the identity matrix (the local-identity matrix), one can define - for any non-zero matrix SL - the inverse matrix $SL^{-1}$ in relation to the matrix $d_0$ on the basis of equations: $SL*SL^{-1} = SL^{-1}*SL = d_0$. One can check that such inverse matrix is defined by the following expression: $SL^{-1} = (a_0^2 + a_1^2)^{-1}*(a_0*d_0-a_1*d_1)$ (Fig. 60).

By analogy, the system of matrices SR has the matrix $d_2$ (Fig. 59a), which possesses all properties of the identity matrix for any matrix SR since $d_2*SR = SR*d_2 = SR$ and $d_2^2 = d_2$. In the frame of the system of matrices SR, where the matrix $d_2$ plays locally the role of the identity matrix (the local-identity matrix), one can define - for any non-zero matrix SR - the inverse matrix $SR^{-1}$ in relation to the matrix $d_2$ on the base of equations: $SR*SR^{-1} = SR^{-1}*SR = d_2$ .

Any of nonzero 2-block diagonal U-complex numbers S = SL+SR = $a_0*d_0+a_1*d_1$ $+a_2*d_2+a_3*d_3$ has its inverse number $S^{-1}$ of the same type in relation to the ordinary identity (4*4)-matrix E=[1,0,0,0; 0,1,0,0; 0,0,1,0; 0,0,0,1]: $S*S^{-1} = S^{-1}*S=E$, where $S^{-1} = (a_0^2+a_1^2)^{-1}*(a_0*d_0-a_1*d_1) + (a_2^2+a_3^2)^{-1}*(a_2*d_2-a_3*d_3) = SL^{-1} + SR^{-1}$. In other words, in this system the inverse U-number $S^{-1}$ is equal to the sum of inverse numbers $SL^{-1}$ and $SR^{-1}$.

Fig. 61 shows that the multiplication between any two members of the system of 2-block diagonal U-complex numbers is commutative and it gives a new member on the same system.

| |
|---|
| $(a_0*d_0+a_1*d_1+a_2*d_2+a_3*d_3)*(b_0*d_0+b_1*d_1+b_2*d_2+b_3*d_3) =$<br>$(b_0*d_0+b_1*d_1+b_2*d_2+b_3*d_3)*(a_0*d_0+a_1*d_1+a_2*d_2+a_3*d_3) = (c_0*d_0+c_1*d_1+c_2*d_2+c_3*d_3)$,<br>where $c_0 = a_0*b_0-a_1*b_1$, $c_1 = a_0*b_1+a_1*b_0$, $c_2= a_2*b_2-a_3*b_3$, $c_3 = a_2*b_3+a_3*b_2$. |

Fig. 61. Commutative multiplication of two members of the system of 2-block diagonal U-complex numbers from Fig. 60.

The system of 2-block diagonal U-complex numbers S has operations of addition, subtraction, commutative multiplication and division (division is defined by multiplication with the inverse diagonal U-numbers $S^{-1}$, which is equal to the sum of inverse complex numbers $SL^{-1}$ and $SR^{-1}$). This system has zero divisors; it means that the system has non-zero matrices, the product of which gives zero. Indeed, the product of non-zero matrices SL and SR is equal to zero matrix: SL*SR=SR*SL=0. (If split-complex numbers are used in both diagonal blocks of the matrix S in Fig. 60 instead of complex numbers, the system of diagonal 2-block U-split-complex numbers arises, which also possesses operations of addition, subtraction, commutative multiplication and division and which has divisors of zero).

Now let us consider questions about norms and the scalar product in this system of 2-block diagonal U-complex numbers. As known a conjugate number $\hat{Z}$ for complex number Z=a+b*i is defined by inversion of sign of its imaginary part: $\hat{Z}$ = a-b*i. The norm |Z| of complex number Z is defined by the expression $|Z| = Z*\hat{Z} = a^2+b^2$.

Correspondingly, conjugate numbers $\widehat{SL}$ and $\widehat{SR}$ for complex numbers $SL = a_0*d_0+a_1*d_1$ and $SR = a_2*d_2+a_3*d_3$ are created by inversion of sign in their local-imaginary parts: $\widehat{SL} = a_0*d_0-a_1*d_1$ and $\widehat{SR} = a_2*d_2-a_3*d_3$. In this case the norm |SL| of complex numbers SL is equal to: $|SL| = SL*\widehat{SL} = (a_0^2+a_1^2)*d_0$. The norm |SR| of complex numbers SR is equal to: $|SR| = SR*\widehat{SR} = (a_2^2+a_3^2)*d_2$.

By definition, the conjugate number $\hat{S}$ for any of DU-complex numbers $S = SL+SR = a_0*d_0+a_1*d_1+a_2*d_2+a_3*d_3$ is also created by means of inversion of signs of its local-imaginary parts $a_1*d_1$ and $a_3*d_3$: $\hat{S} = a_0*d_0-a_1*d_1+a_2*d_2-a_3*d_3$. The norm |S| is defined by the expression $|S| = S*\hat{S} = (a_0*d_0+a_1*d_1+a_2*d_2+a_3*d_3) * (a_0*d_0-a_1*d_1+a_2*d_2-a_3*d_3) = (a_0^2+a_1^2)*d_0 + (a_2^2+a_3^2)*d_2 = |SL|+|SR|$. In other words,the norm |S| of any of these diagonal U-complex numbers is equal to the sum of norms |SL| and |SR| in its two separate blocks of complex numbers.

As known, the scalar product $<Z_1,Z_2>$ for two vectors of the complex plane $Z_1 = a_0+a_1*i$ and $Z_2 = b_0+b_1*i$, which are usually represented by (2*2)-matrices $[a_0, a_1; -a_1, a_0]$ and $[b_0, b_1; -b_1, b_0]$, is defined by means the expression $<Z_1,Z_2> = Z_1*\hat{Z}_2 = (a_0+a_1*i)*(b_0-b_1*i) =$ (a0*b0+a1*b1)+(a1*b0-a0*b1)*i. By analogy one can define the scalar product $<SL_1,SL_2>$ for two vectors $SL_1 = a_0*d_0+a_1*d_1$ and $SL_2 = b_0*d_0+b_1*d_1$ from the set of (4*4)-matrices SL (Fig. 60) by means of the similar expression: $<SL_1,SL_2> = SL_1*\widehat{SL_2} = (a_0*d_0+a_1*d_1) * (b_0*d_0-b_1*d_1) = (a_0*b_0+a_1*b_1)*d_0 + (a_1*b_0-a_0*b_1)*d_1$. So the result of scalar product $<SL_1,SL_2>$ coincides with the scalar product $<Z_1,Z_2>$ but - instead of the real unit "1" and the imaginary unit "i" in $<Z_1,Z_2>$ - the expression $<SL_1,SL_2>$ uses the local-real unit $d_0$ and the local-imaginary unit $d_1$. Analogically the scalar product $<SR_1,SR_2>$ for two vectors $SR_1 = a_2*d_2+a_3*d_3$ and $SR_2 = b_2*d_2+b_3*d_3$ from the set of (4*4)-matrices SR (Fig. 60) can be defined by means of the expression $<SR_1,SR_2> = SR_1*\widehat{SR}_2 = (a_2*d_2+a_3*d_3)*(b_2*d_2-b_3*d_3) = (a_2*b_2+a_3*b_3)*d_2 + (a_3*b_2-a_2*b_3)*d_3$, where the local-real unit $d_2$ and the local-imaginary unit $d_3$ are used.

The scalar product $<S_1,S_2>$ of two 2-block diagonal U-complex numbers $S_1=SL_1+SR_1=a_0*d_0+a_1*d_1+a_2*d_2+a_3*d_3$ and $S_2= SL_2+SR_2= b_0*d_0+b_1*d_1+b_2*d_2+b_3*d_3$ is defined by the analogical expression: $<S_1,S_2> = S_1*\hat{S}_2 = (a_0*d_0+a_1*d_1+a_2*d_2+a_3*d_3) * (b_0*d_0-b_1*d_1+b_2*d_2-b_3*d_3) = (a_0*b_0+a_1*b_1)*d_0 + (a_1*b_0-a_0*b_1)*d_1 + (a_2*b_2+a_3*b_3)*d_2 + (a_3*b_2-a_2*b_3)*d_3 = <SL_1,SL_2> + <SR_1,SR_2>$. In other words, the scalar product of two 2-block diagonal U-complex numbers $S_1$ and $S_2$ is equal to the sum of scalar products of their corresponding blocks of complex numbers.

| | | | | |
|---|---|---|---|---|
| $[x_0,x_1,x_2,x_3]$ * SL = [ | $a_0*x_0-a_1*x_1$, | $a_0*x_1+a_1*x_0$, | 0, | 0 ] |
| SL * $[x_0,x_1,x_2,x_3]$.' = [ | $a_0*x_0+a_1*x_1$; | $a_0*x_1-a_1*x_0$; | 0; | 0 ] |
| $[x_0,x_1,x_2,x_3]$ * SR = [ | 0, | 0, | $a_2*x_2-a_3*x_3$, | $a_2*x_3+a_3*x_2$] |
| SR * $[x_0,x_1,x_2,x_3]$.' = [ | 0; | 0; | $a_2*x_2+a_3*x_3$; | $a_2*x_3-a_3*x_2$ ] |

Fig. 62. Multiplication of matrix operators SL and SR (Fig. 54) with a vector $X=[x_0,x_1,x_2,x_3]$ from the right side and from the left side leads to a possibility of a selective manage of subspaces.

Fig. 62 shows results of actions of the matrix operators SL and SR on a 4-dimensional vector $X=[x_0,x_1,x_2,x_3]$ from the right side and from the left side. In both cases the action of matrix operators SL leads to a possibility of a selective manage of vectors of 2-dimensional subspaces inside configurational 4-dimensional spaces $[x_0,x_1,x_2,x_3]$: the resulting non-zero vectors exist only in the subspace $(x_0,x_1)$, where they can be managed arbitrary by means of parameters $a_0$ and $a_1$ of the operator SL. If the parameters $a_0$ and $a_1$ are functions of time, then vectors $[x_0,x_1,x_2,x_3]$*SL and SL*$[x_0,x_1,x_2,x_3]$ move along corresponding trajectories only in the plane $(x_0,x_1)$.

The action of matrix operators SR in both cases also leads to a possibility of a selective manage of vectors of 2-dimensional subspaces inside configurational 4-dimensional spaces $[x_0,x_1,x_2,x_3]$. In this case the resulting non-zero vectors exist only in another subspace $(x_2,x_3)$, where they can be managed by means of parameters $a_2$ and $a_3$ of the operator SR. If the parameters $a_2$ and $a_3$ are functions of time, then vectors $[x_0,x_1,x_2,x_3]$*SR and SR*$[x_0,x_1,x_2,x_3]$ move along corresponding trajectories only in the plane $(x_2,x_3)$.

It leads to valuable capabilities to model multi-parametrical biological systems and bioinformational spaces, where different subsystems and their configurational subspaces are managed independently. Similar capabilities of selective managements of 2-dimensional subspaces exist also in the case of a generalization of these 2-block diagonal U-complex operators into $2^n$-block diagonal U-complex operators for operations with $2^n$-dimensional spaces and $2^n$-parametrical systems.

How one can make such generalization to get the system of $2^n$-block diagonal U-complex numbers? Let us return to the sparse matrix $DR_4$ (Fig. 59), which contains two identical non-zero quadrants along the main diagonal. Each of these (2*2)-quadrants is the matrix representation [1, 1; -1, 1] of complex number with unit coordinates: 1+i. This matrix $DR_4$ can be written in the form $DR_4=[1, 0; 0, 1]\otimes[1,1; -1,1]$ by means of the tensor multiplication of the identity matrix E=[1, 0; 0, 1] with the matrix [1, 1; -1, 1] (Fig. 63, left).

On the second step - by the recursive application of tensor multiplication of the identity matrix [1, 0; 0, 1] with the block-diagonal (4*4)-matrix $DR_4$ - we receive a new block-diagonal (8*8)-matrix $DR_8$=[1, 0; 0, 1]⊗([1, 0; 0, 1]⊗[1,1; -1,1]) = $[1, 0; 0, 1]^{(2)}$⊗[1,1; -1,1] (Fig. 63, right). This recursive operation of the tensor multiplication of the identity matrix [1, 0; 0, 1] with the block-diagonal matrix, which was received on the previous step of the recursive algorithm, can be repeated again and again to get $2^n$-block diagonal U-complex numbers with unit coordinates: $[1, 0; 0, 1]^{(n)}$⊗[1,1; -1,1], where (n) means tensor power, n = 1, 2, 3,… .

By analogy with the decomposition of the matrix $DR_4$ (Fig. 53), each of members of the tensor family $[1, 0; 0, 1]^{(n)}$⊗[1,1; -1,1] can be decomposed into sum of sparse matrices $m_0$, $m_1$, $m_2$,…, appropriate pairs of which form sets, which are closed under multiplication and the multiplication tables of which are identical to the multiplication table of complex numbers. When each of the sparse matrices of such decomposition is multiplied with one of real coefficients $a_0$, $a_1$, $a_2$,..., a system of $2^n$-block diagonal U-complex numbers is realized. Each of these systems of 2n-block diagonal U-complex numbers has operations of addition, subtraction and commutative multiplication. These systems of $2^n$-block diagonal U-complex numbers have zero divisors (with the exception of the case n=0 when only one block of complex numbers is realized).

Let us illustrate this recursive algorithm of the generalization by means of the additional example of a creation of the system of $2^2$-block diagonal U-complex numbers with 8 parameters using the expression: $DR_8$ = $[1, 0; 0, 1]^{(2)}$⊗[1,1; -1,1] (Fig. 63).

$DR_4$=[1, 0; 0, 1]⊗[1,1; -1,1] =

| | | | |
|---|---|---|---|
| 1 | 1 | 0 | 0 |
| -1 | 1 | 0 | 0 |
| 0 | 0 | 1 | 1 |
| 0 | 0 | -1 | 1 |

$DR_8$=$[1, 0; 0, 1]^{(2)}$⊗[1,1; -1,1] =

| | | | | | | | |
|---|---|---|---|---|---|---|---|
| 1 | 1 | 0 | 0 | 0 | 0 | 0 | 0 |
| -1 | 1 | 0 | 0 | 0 | 0 | 0 | 0 |
| 0 | 0 | 1 | 1 | 0 | 0 | 0 | 0 |
| 0 | 0 | -1 | 1 | 0 | 0 | 0 | 0 |
| 0 | 0 | 0 | 0 | 1 | 1 | 0 | 0 |
| 0 | 0 | 0 | 0 | -1 | 1 | 0 | 0 |
| 0 | 0 | 0 | 0 | 0 | 0 | 1 | 1 |
| 0 | 0 | 0 | 0 | 0 | 0 | -1 | 1 |

Fig. 63. The first members of the tensor family of systems of $2^n$-block diagonal U-complex numbers with unit coordinates: $[1, 0; 0, 1]^{(n)}$⊗[1,1; -1,1], n = 1,2,3,… Each of diagonal (2*2)-blocks contains the matrix representation [1,1;-1,1] of complex number 1+i with unit coordinates.

Fig. 64 shows the decomposition of the matrix $DR_8$=$[1, 0; 0, 1]^{(2)}$⊗[1,1; -1,1] into 8 sparse matrices $m_0$, $m_1$, $m_2$, $m_3$, $m_4$, $m_5$, $m_6$, $m_7$. Each of the pairs $m_0$ and $m_1$, $m_2$ and $m_3$, $m_4$ and $m_5$, $m_6$ and $m_7$ is represented a set, which is closed under multiplication and the multiplication table of which coincides with the multiplication table of complex number (Fig. 64, bottom).

| DR8=[1, 0; 0, 1]$^{(2)}$⊗[1,1; -1,1]= | | | | | | | | | = |
|---|---|---|---|---|---|---|---|---|---|
| | 1 | 1 | 0 | 0 | 0 | 0 | 0 | 0 | |
| | -1 | 1 | 0 | 0 | 0 | 0 | 0 | 0 | |
| | 0 | | 1 | 1 | 0 | 0 | 0 | 0 | |
| | 0 | 0 | -1 | 1 | 0 | 0 | 0 | 0 | |
| | 0 | 0 | 0 | 0 | 1 | 1 | 0 | 0 | |
| | 0 | 0 | 0 | 0 | -1 | 1 | 0 | 0 | |
| | 0 | 0 | 0 | 0 | 0 | 0 | 1 | 1 | |
| | 0 | 0 | 0 | 0 | 0 | 0 | -1 | 1 | |

| | | | | | | | | |
|---|---|---|---|---|---|---|---|---|
| | 1 0 0 0 0 0 0 0 | | 0 1 0 0 0 0 0 0 | | 0 0 0 0 0 0 0 0 | | 0 0 0 0 0 0 0 0 | |
| | 0 1 0 0 0 0 0 0 | | -1 0 0 0 0 0 0 0 | | 0 0 0 0 0 0 0 0 | | 0 0 0 0 0 0 0 0 | |
| | 0 0 0 0 0 0 0 0 | | 0 0 0 0 0 0 0 0 | | 0 0 1 0 0 0 0 0 | | 0 0 0 1 0 0 0 0 | |
| | 0 0 0 0 0 0 0 0 | | 0 0 0 0 0 0 0 0 | | 0 0 0 1 0 0 0 0 | | 0 0 -1 0 0 0 0 0 | |
| = | 0 0 0 0 0 0 0 0 | + | 0 0 0 0 0 0 0 0 | + | 0 0 0 0 0 0 0 0 | + | 0 0 0 0 0 0 0 0 | + |
| | 0 0 0 0 0 0 0 0 | | 0 0 0 0 0 0 0 0 | | 0 0 0 0 0 0 0 0 | | 0 0 0 0 0 0 0 0 | |
| | 0 0 0 0 0 0 0 0 | | 0 0 0 0 0 0 0 0 | | 0 0 0 0 0 0 0 0 | | 0 0 0 0 0 0 0 0 | |
| | 0 0 0 0 0 0 0 0 | | 0 0 0 0 0 0 0 0 | | 0 0 0 0 0 0 0 0 | | 0 0 0 0 0 0 0 0 | |

| | | | | | | | |
|---|---|---|---|---|---|---|---|
| 0 0 0 0 0 0 0 0 | | 0 0 0 0 0 0 0 0 | | 0 0 0 0 0 0 0 0 | | 0 0 0 0 0 0 0 0 | |
| 0 0 0 0 0 0 0 0 | | 0 0 0 0 0 0 0 0 | | 0 0 0 0 0 0 0 0 | | 0 0 0 0 0 0 0 0 | |
| 0 0 0 0 0 0 0 0 | | 0 0 0 0 0 0 0 0 | | 0 0 0 0 0 0 0 0 | | 0 0 0 0 0 0 0 0 | |
| 0 0 0 0 0 0 0 0 | | 0 0 0 0 0 0 0 0 | | 0 0 0 0 0 0 0 0 | | 0 0 0 0 0 0 0 0 | = $m_0+m_1$ |
| 0 0 0 0 1 0 0 0 | + | 0 0 0 0 0 1 0 0 | + | 0 0 0 0 0 0 0 0 | + | 0 0 0 0 0 0 0 0 | $+m_2+m_3$ |
| 0 0 0 0 0 1 0 0 | | 0 0 0 0 -1 0 0 0 | | 0 0 0 0 0 0 0 0 | | 0 0 0 0 0 0 0 0 | $+m_4+m_5$ |
| 0 0 0 0 0 0 0 0 | | 0 0 0 0 0 0 0 0 | | 0 0 0 0 0 0 1 0 | | 0 0 0 0 0 0 0 1 | $+m_6+m_7$ |
| 0 0 0 0 0 0 0 0 | | 0 0 0 0 0 0 0 0 | | 0 0 0 0 0 0 0 1 | | 0 0 0 0 0 0 -1 0 | |

| * | **$m_0$** | **$m_1$** |
|---|---|---|
| **$m_0$** | $m_0$ | $m_1$ |
| **$m_1$** | $m_1$ | $-m_0$ |

| * | **$m_2$** | **$m_3$** |
|---|---|---|
| **$m_2$** | $m_2$ | $m_3$ |
| **$m_3$** | $m_3$ | $-m_2$ |

| * | **$m_4$** | **$m_5$** |
|---|---|---|
| **$m_4$** | $m_4$ | $m_5$ |
| **$m_5$** | $m_5$ | $-m_4$ |

| * | **$m_6$** | **$m_7$** |
|---|---|---|
| **$m_6$** | $m_6$ | $m_7$ |
| **$m_7$** | $m_7$ | $-m_6$ |

Fig. 64. Top: the decomposition of the matrix $DR_8$=[1, 0; 0, 1]$^{(2)}$⊗[1,1; -1,1] into 8 sparse matrices $m_0$, $m_1$, $m_2$, $m_3$, $m_4$, $m_5$, $m_6$, $m_7$. Bottom: multiplication tables for the following pairs: $m_0$ and $m_1$; $m_2$ and $m_3$; $m_4$ and $m_5$; $m_6$ and $m_7$.

Taking each of sparse matrices $m_0$, $m_1$, $m_2$, $m_3$, $m_4$, $m_5$, $m_6$, $m_7$ with its own real coefficients $a_0$, $a_1$, $a_2$,..., we receive the system of $2^2$-block diagonal U-complex numbers: $a_0*m_0+a_1*m_1+a_2*m_2+a_3*m_3+a_4*m_4+a_5*m_5+a_6*m_6+a_7*m_7$ (Fig. 65). This system has operations of addition, subtraction and commutative multiplication and it has zero divisors. Any non-zero member of this system has its inverse number of the same type. Due to this, the operation of division can be defined here by usual manner as multiplication with inverse numbers.

| | | | | | | | |
|---|---|---|---|---|---|---|---|
| $a_0$ | $a_1$ | 0 | 0 | 0 | 0 | 0 | 0 |
| $-a_1$ | $a_0$ | 0 | 0 | 0 | 0 | 0 | 0 |
| 0 | 0 | $a_2$ | $a_3$ | 0 | 0 | 0 | 0 |
| 0 | 0 | $-a_3$ | $a_2$ | 0 | 0 | 0 | 0 |
| 0 | 0 | 0 | 0 | $a_4$ | $a_5$ | 0 | 0 |
| 0 | 0 | 0 | 0 | $-a_5$ | $a_4$ | 0 | 0 |
| 0 | 0 | 0 | 0 | 0 | 0 | $a_6$ | $a_7$ |
| 0 | 0 | 0 | 0 | 0 | 0 | $-a_7$ | $a_6$ |

Fig. 65. The matrix system of $2^2$-block diagonal U-complex numbers

In general, by means of such recursive algorithm using the expression $[1, 0; 0, 1]^{(n)}\otimes[1,1; -1,1]$, we have constructed the tensor family of $2^n$-block diagonal U-complex numbers. Parameters $a_0$, $a_1$, $a_2$,... in separate blocks of each of these systems can be functions of time and can be interrelated with each other to model multi-parametrical systems with various interrelations between their parts. Special attention we pay to applications of $2^n$-block diagonal U-complex numbers for modeling interrelated cyclic processes in living organisms:

- cycles of "life-and-death" of proteins, cells and other parts of organisms;
- interpretation of various diseases (including oncological and cardio-vascular) as consequences of disturbances in interrelations of physiological cycles in accordance with the doctrine of the ancient chronomedicine;
- cycles of activity and rest of physiological sub-systems with applications in medical diagnostics, pharmacology, ergonomics of professions with the change of time zones (pilots, flight attendants, machinists, etc..), acupuncture and eastern chronomedicine in general, sports of highest achievements;
- approaches to study molecular-genetic systems from the standpoint of the chrono-cyclic conception [Petoukhov, 2001];
- theory of hypercycles of M.Eigen and P.Shuster;
- study of cycles in cardiac system and respiratory system,
- study of the biological clock inside living bodies;
- study of cyclic metamorphoses of butterflies and other organisms;
- ensembles of cyclic processes in the field of Artificial Life (ALife), etc.

Beside this, $2^n$-block diagonal U-complex numbers can be applied also to model interrelated cyclic processes in growing multi-parametrical systems in digital communications, devices of artificial intelligence, physics, economics, finances, engineering, biotechnology, etc.

The system of complex numbers plays significant role in different branches of modern science and technology. One can hope that systems of $2^n$-block diagonal U-complex numbers will be also useful not only in algebraic biology but also in other fields of science and technology. It seems interesting to develop the theory of functions of several U-complex variables by analogy with the existing theory of functions of several complex variables.

One can mention else the following additional aspect. Till now we focused our attention on the matrix $R_{40}$ (Fig. 59) of two quadrants along the main diagonal of the matrix $R_4$ from Fig. 1. Now let us pay our attention to the matrix $R_{41}$ (Fig. 59) of two quadrants along the second diagonal of the matrix $R_4$. As it was mentioned above, both quadrants along the second diagonal are mirror-antisymmetric concerning both quadrants along the main diagonal in relation of the middle horizontal line of $R_4$. Taking this into account, let us consider the mirror-antisymmetric analog W of 2-block diagonal U-complex numbers S from Fig. 60. Fig. 66 shows that the square of this matrix $W^2$ is again 2-block diagonal U-complex numbers, where each of non-zero coordinates is a combination of all 4 parameters $a_0$, $a_1$, $a_2$ and $a_3$. (It is interesting that in the block-diagonal matrix $W^2$ the upper diagonal block represents complex number $(a_0*a_2+a_1*a_3)+i*(a_1*a_2-a_0*a_3)$ and the lower diagonal block represents the conjugate complex number $(a_0*a_2+a_1*a_3)-i*(a_1*a_2-a_0*a_3)$). The similar relation is true for $2^n$-block diagonal U-complex numbers and their mirror-antisymmetric analogues.

| W = | $W^2$ = |
|---|---|
| [0, 0, $a_3$, $-a_2$ | [ $a_0*a_2+a_1*a_3$, $a_1*a_2-a_0*a_3$, 0, 0 |
| 0, 0, $-a_2$, $-a_3$ | $-(a_1*a_2-a_0*a_3)$, $a_0*a_2+a_1*a_3$, 0, 0 |
| $a_1$, $-a_0$, 0, 0 | 0, 0, $a_0*a_2+a_1*a_3$, $-(a_1*a_2-a_0*a_3)$ |
| $-a_0$, $-a_1$, 0, 0] | 0, 0, $(a_1*a_2-a_0*a_3)$, $a_0*a_2 + a_1*a_3$] |

Fig. 66. The mirror-antisymmetric analogue W of 2-block diagonal U-complex numbers S from Fig. 60. Its square $W^2$ is again 2-block diagonal U-complex numbers.

Let us return to the recursive algorithm with the expression $[1, 0; 0, 1]^{(n)}\otimes[1,1; -1,1]$. This algorithm can be extended to cases of expressions $[1, 0; 0, 1]^{(n)}\otimes M$, where the square matrix M represents not the complex number with unit coordinates [1,1; -1,1] but other kinds of complex and hypercomplex numbers with unit coordinates: split-complex numbers (their synonyms are double numbers and hyperbolic numbers), dual numbers, quaternions of Hamilton, split-quaternions of Cockle, octonions, etc. In these cases we meet with the following kinds of $2^n$-block diagonal U-numbers:

- $2^n$-block diagonal U-double numbers;
- $2^n$-block diagonal U-dual numbers;
- $2^n$-block diagonal U-quaternions;
- $2^n$-block diagonal U-split-quaternions, etc.

Algebraic properties of these kinds of $2^n$-block diagonal U-numbers will be published later.

The introduced systems of multi-block U-numbers define new – for mathematical natural sciences - types of multi-dimensional spaces (U-numerical spaces), some of which are vector spaces with special interrelations of their vectors.

What one can say about difference between systems of multi-block U-complex numbers and a corresponding set of separate complex numbers, which can be also applied to model a behavior of multi-parametrical systems? Firstly, multi-block U-complex numbers operate inside appropriate multi-dimensional spaces, where - in cases of vector spaces – Descarte's systems of coordinates can be introduced with certain orders of coordinate axes, appropriate orders of components of multi-dimensional vectors, appropriate numerations of sub-spaces and appropriate interrelations between multi-dimensional vectors. In a simple set of separate complex numbers such order of its members is absent. Secondly, the difference exists in relation to the operation of tensor multiplication, which is very essential to model multi-parametrical systems with increasing quantities of parameters: a result of tensor multiplication of two multi-block U-complex numbers differs from a result of tensor multiplication of two separate complex numbers. Similar differences are true for systems of multi-block U-split-complex numbers, etc.

In our approach we intensively used tensor multiplication of matrices taking into account, in particular, the following facts: DNA-molecules belong to the quantum mechanical level, but in quantum mechanics the tensor multiplication is the important operation: when considering a quantum system consisting of two subsystems, the whole space of states is constructed in the form of tensor product of their states.

Of separate interest are multi-block U-numbers with different types of multidimensional numbers in separate blocks (symbiotic U-numbers). In these cases corresponding multi-dimensional spaces can be called as symbiotic spaces.

Our results of researches in the field of matrix genetics additionally attract an attention to matrix operators in a form of $2^n$-block diagonal united operators, for example, $2^n$-block diagonal united Hadamard matrices for spectral analysis in different sub-spaces of multi-dimensional space.

## 17. TETRA-GROUPS OF OLIGONUCLEOTIDES AND RULES OF LONG NUCLEOTIDE SEQUENCES

By unknown reasons, the Nature has chosen the DNA- and RNA-alphabets, each of which contains only 4 letters (A, T, C, G in DNA and A, U, C, G in RNA) with complementary relations among them. In other words, in these basic cases of DNA and RNA, the Nature uses the tetra-groups of letters. In this Section we study a role of special tetra-groups of oligonucleotides in long nucleotide sequences. We use separations of each of complete sets of oligonucleotides ($4^2$ doublets, $4^3$ triplets, …, $4^n$ n-plets) into tetra-groups of equivalency, criterium of which are based on these 4 letters, to study hidden symmetries in long nucleotide sequences. The tetra-group of mononucleotides A, T, C, G is the limiting degenerate case of the introduced set of tetra-groups of oligonucleotides.

One can remind that many authors devoted their works from the middle of XX century to searching of hidden regularities or symmetry principles in long nucleotide sequences. The Chargaff's first parity rule speaks that in any double-stranded DNA segment, the number of occurrences (or frequencies) of adenine A and thymine T are equal, and so are the frequencies of cytosine C and guanine G [Chargaff, 1951, 1971]. This rule was used by Watson and Crick to support their famous DNA double-helix structure model. Chargaff also perceived that the parity rule approximately holds in the single-stranded DNA segment for long nucleotide sequences. This last rule is known as Chargaff's second parity rule (CSPR), and it has been confirmed in several organisms [Mitchell & Bride, 2006]. Originally, CSPR is meant to be valid only to mononucleotide frequencies (that is quantities of monoplets) in the single-stranded DNA. "*But, it occurs that oligonucleotide frequencies follow a generalized Chargaff's second parity rule (GCSPR) where the frequency of an oligonucleotide is approximately equal to its complement reverse oligonucleotide frequency [Prahbu, 1993]. This is known in the literature as the Symmetry Principle*" [Yamagishi, Herai, 2011, p. 2]. The work [Prahbu, 1993] shows the implementation of the Symmetry Principle in long DNA-sequences for cases of complementary reverse n-plets with n = 2, 3, 4, 5 at least. (In literature, a few synonimes of the term "n-plets" are used: n-tuples, n-words or n-mers). These parity rules for long nucleotide sequences are analysed and discussed in many works [Bell, Forsdyke, 1999; Chargaff, 1971, 1975; Dong, Cuticchia, 2001; Forsdyke, 2002; Forsdyke, Bell, 2004; Mitchell, Bridge, 2006; Prabhu, 1993; Yamagishi, Herai, 2011].

These parity rules (or rules of equality), including generalized Chargaff's second parity rule, concerns the equality of frequencies of two separate mononucleotides or two separate oligonucleotides: the equality of frequencies of adenine and thymine, the equality of frequences of the doublet CA and its complement reverse doublet TG, etc. By contrast to this, to study hidden symmetries in long nucleotide sequences, we apply a comparative analysis of equalities not for frequencies of separate oligonucleotides but for aggregated frequencies (or collective frequences) of oligonucleotides from each of 4 equal subsets, which are formed by means of a certain criterion of equivalency. More precisely, first of all, we divide each of complete alphabets of n-plets into 4 subsets with $4^{n-1}$ n-plets in each by the criterion of the same first letter in each n-plet. The complect of these four subsets is called the tetra-group of this equivalence. The sum of the frequencies of all oligonucleotides of each subset is called the collective frequency of a given subset in the considered long sequence of nucleotides.

For example, in the case of the alphabet of 16 doublets, the tetra-group of this equivalence is formed with 4 equivalent doublets in each of its 4 collective members:

- the first collective member: AA=AC=AG=AT;
- the second collective member: TC=TA=TT=TG;
- the third collective member: CC=CA=CT=CG; (13)
- the fourth collective member: GC=GA=GT=GG.

Taking into account the complementarities of bases A-T and C-G in DNA, two subsets of oligonuclotides with the complementary first letters are called "complementary subsets by the first letter". For example, in the described tetra-group of doublets (13) the first and the second collective members are complementary each other by the first letter (A and T are complemetary); similar is true for the third and fourth collective members (their first letters C and G are complementary).

In long nucleotide sequences, each of oligonucleotides has its individual frequency F of existence, which can be denoted, for example, for doublets as F(CC), F(CA), etc. We denote collective frequencies for each of 4 collective members of the tetra-group as $F_n(C)$, $F_n(G)$, $F_n(A)$ and $F_n(T)$, where the index n = 1, 2, 3, 4,… means the length of considered oligonucleotides (that is n-plets). For example, in the case of the described tetra-group of 16 doublets (13), we study collective frequencies $F_2(C)$, $F_2(G)$, $F_2(A)$ and $F_2(T)$:

- $F_2(A) = F(AA) + F(AC) + F(AG) + F(AT)$,
- $F_2(T) = F(TC) + F(TA) + F(TT) + F(TG)$.
- $F_2(C) = F(CC) + F(CA) + F(CT) + F(CG)$, (14)
- $F_2(G) = F(GC) + F(GA) + F(GT) + F(GG)$,

Simultaneously we study another kind of tetra-groups of equivalence by the criterion of the same last letter in n-plets. For example in the case of 16 doublets, the tetra-group of this «reverse» equivalence is the following:

- the first collective member: AA=CA=GA=TA;
- the second collective member: CT=AT=TT=GT;
- the third collective member: CC=AC=TC=GC; (15)
- the fourth collective member: CG=AG=TG=GG.

Two subsets of oligonuclotides with the complementary last letters are called "complementary subsets by the last letter". For example, in the described tetra-group of doublets (15) the first and the second collective members are complementary each other by the last letter (A and T); similar is true for the third and fourth collective members (their last letters C and G are complementary).

We denote collective frequencies for each of 4 collective members of this "reverse" tetra-group as $FR_n(C)$, $FR_n(G)$, $FR_n(A)$ and $FR_n(T)$, where the index n = 1, 2, 3, 4,… means the length of considered oligonucleotides (that is n-plets). For example, in the case of the described tetra-group of 16 doublets, we study collective frequences $FR_2(C)$, $FR_2(G)$, $FR_2(A)$ and $FR_2(T)$:

- $FR_2(A) = F(AA) + F(CA) + F(GA) + (TA)$,
- $FR_2(T) = F(CT) + F(AT) + F(TT) + F(GT)$.
- $FR_2(C) = F(CC) + F(AC) + F(TC) + F(GC)$,
- $FR_2(G) = F(CG) + F(AG) + F(TG) + F(GG)$,

Below we describe the first interesting results of our comparative study of 8 collective frequencies $F_n(A)$, $F_n(T)$, $F_n(C)$, $F_n(G)$, $FR_n(A)$, $FR_n(T)$, $FR_n(C)$ and $FR_n(G)$ of n-plets of such tetra-groups in long nucleotide sequences. These results give evidences in favour of existence of new universal rules of equality for long nucleotide sequences in addition to the known Chargaff's rules of equality. By analogy with the generalized Chargaff's second rule, in cases of these new rules it is assumed that the value n of considered n-plets is much smaller than the length of the studied sequence.

Fig. 67 shows some results of our studies of the mentioned 8 collective frequencies of n-plets (n=1, 2, 3, 4, 5) for both types of tetra-groups in 17 different kinds of long nucleotide sequences taken from GenBank. Individual lengths of these long sequences are more than 90000 nucleotides. Kinds of studied sequences here are taken from the article [Prahbu, 1993] to avoid a suspicion about a special choice of sequences. For each of sequences, there are shown its title and accession data in Genbank, its length and also quantities of mononucleotides, doublets, triplets, 4-plets and 5-plets denoted by $\Sigma_1$, $\Sigma_2$, $\Sigma_3$, $\Sigma_4$ and $\Sigma_5$ correspondingly. Together with each of collective frequencies $F_n(A)$, $F_n(T)$, $F_n(C)$, $F_n(G)$, $FR_n(A)$, $FR_n(T)$, $FR_n(C)$ and $FR_n(G)$, Fig. 67 shows the percentage of each of these collective frequencies in the total quantity $\Sigma_1$ of n-plets taken as 100%. For example, in the case of $F_2(A)$=24800 and $\Sigma_2$=114677, the percentage of the collective frequency $F_2(A)$ is calculated by the formula: $(F_2(A)/ \Sigma_2)*100\% = (24800/114677)*100\% = 21,63\%$.

| № | NAMES OF SEQUENCES | NUCLEOTIDES | DOUBLETS | TRIPLETS | 4-PLETS | 5-PLETS |
|---|---|---|---|---|---|---|
| 1 | Human cytomegalovirus strain AD169 complete genome. 229354 bp. Accession X17403.1 | $F_1(A)$=49475<br>21,57% | $F_2(A)$=24800<br>21,63% | $F_3(A)$=16825<br>22,01% | $F_4(A)$=12423<br>21,67% | $F_5(A)$=9981<br>21,76% |
| | | $F_1(T)$=48776<br>21,27% | $F_2(T)$=24340<br>21,22% | $F_3(T)$=16601<br>21,71% | $F_4(T)$=12266<br>21,39% | $F_5(T)$=9660<br>21,06% |
| | | $F_1(C)$=64911<br>28,30% | $F_2(C)$=32407<br>28,26% | $F_3(C)$=21303<br>27,86% | $F_4(C)$=16132<br>28,13% | $F_5(C)$=12992<br>28,32% |
| | | $F_1(G)$=66192<br>28,86% | $F_2(G)$=33130<br>28,89% | $F_3(G)$=21722<br>28,41% | $F_4(G)$=16517<br>28,81% | $F_5(G)$=13237<br>28,86% |
| | | $\Sigma_1$ = 229354 | $\Sigma_2$ =114677 | $\Sigma_3$ =76451 | $\Sigma_4$ =57338 | $\Sigma_5$ =45870 |
| | | | $FR_2(A)$=24675<br>21,52% | $FR_3(A)$=16648<br>21,78% | $FR_4(A)$=12279<br>21,42% | $FR_5(A)$=10057<br>21,93% |
| | | | $FR_2(T)$=24436<br>21,31% | $FR_3(T)$=16211<br>21,20% | $FR_4(T)$=12320<br>21,49% | $FR_5(T)$=9789<br>21,34% |
| | | | $FR_2(C)$=32504<br>28,34% | $FR_3(C)$=21491<br>28,11% | $FR_4(C)$=16175<br>28,21% | $FR_5(C)$=12914<br>28,15% |
| | | | $FR_2(G)$=33062<br>28,83% | $FR_3(G)$=22101<br>28,91% | $FR_4(G)$=16564<br>28,89% | $FR_5(G)$=13110<br>28,58% |
| 2 | VACCG, Vaccinia virus Copenhagen, complete genome. 191737 bp. Accession M35027.1 | $F_1(A)$=63921<br>33,34% | $F_2(A)$=31832<br>33,20% | $F_3(A)$=21310<br>33,34% | $F_4(A)$=15932<br>33,24% | $F_5(A)$=12872<br>33,57% |
| | | $F_1(T)$=63776<br>33,26% | $F_2(T)$=31540<br>32,90% | $F_3(T)$=21513<br>33,66% | $F_4(T)$=15720<br>32,80% | $F_5(T)$=12796<br>33,37% |
| | | $F_1(C)$=32010<br>16,69% | $F_2(C)$=16257<br>16,96% | $F_3(C)$=10452<br>16,35% | $F_4(C)$=8137<br>16,98% | $F_5(C)$=6355<br>16,57% |
| | | $F_1(G)$=32030<br>16,71% | $F_2(G)$=16239<br>16,94% | $F_3(G)$=10637<br>16,64% | $F_4(G)$=8145<br>16,99% | $F_5(G)$=6324<br>16,49% |
| | | $\Sigma_1$ = 191737 | $\Sigma_2$ = 95868 | $\Sigma_3$ =63912 | $\Sigma_4$ =47934 | $\Sigma_5$ =38347 |
| | | | $FR_2$(A)=32088<br>33,47% | $FR_3$(A)=21337<br>33,38% | $FR_4$(A)=16024<br>33,43% | $FR_5$(A)=12711<br>33,15% |
| | | | $FR_2$(T)=32236<br>33,63% | $FR_3$(T)=21239<br>33,23% | $FR_4$(T)=16104<br>33,60% | $FR_5$(T)=12847<br>33,50% |
| | | | $FR_2$(C)=15753<br>16,43% | $FR_3$(C)=11007<br>17,22% | $FR_4$(C)=7871<br>16,42% | $FR_5$(C)=6307<br>16,45% |
| | | | $FR_2$(G)=15791<br>16,47% | $FR_3$(G)=10329<br>16,16% | $FR_4$(G)=7935<br>16,55% | $FR_5$(G)=6482<br>16,90% |
| 3 | MPOMTCG, Marchantia paleacea isolate A 18 mitochondrion, complete genome. 186609 bp. Accession M68929.1 | $F_1$(A)=53206<br>28,51% | $F_2$(A)=26575<br>28,48% | $F_3$(A)=17874<br>28,73% | $F_4$(A)=13285<br>28,48% | $F_5$(A)=10665<br>28,58% |
| | | $F_1$(T)=54264<br>29,08% | $F_2$(T)=27190<br>29,14% | $F_3$(T)=17831<br>28,67% | $F_4$(T)=13477<br>28,89% | $F_5$(T)=10782<br>28,89% |
| | | $F_1$(C)=39215<br>21,01% | $F_2$(C)=19603<br>21,01% | $F_3$(C)=12979<br>20,87% | $F_4$(C)=9882<br>21,18% | $F_5$(C)=7742<br>20,74% |
| | | $F_1$(G)=39924<br>21,39% | $F_2$(G)=19936<br>21,37% | $F_3$(G)=13519<br>21,73% | $F_4$(G)=10008<br>21,45% | $F_5$(G)=8132<br>21,79% |
| | | **$\Sigma_1$ = 186609** | **$\Sigma_2$ = 93304** | **$\Sigma_3$ =62203** | **$\Sigma_4$ =46652** | **$\Sigma_5$ =37321** |
| | | | $FR_2$(A)=26631<br>28,54% | $FR_3$(A)=17607<br>28,31% | $FR_4$(A)=13207<br>28,31% | $FR_5$(A)=10504<br>28,15% |
| | | | $FR_2$(T)=27073<br>29,02% | $FR_3$(T)=18339<br>29,48% | $FR_4$(T)=13600<br>29,15% | $FR_5$(T)=10860<br>29,10% |
| | | | $FR_2$(C)=19612<br>21,02% | $FR_3$(C)=13159<br>21,15% | $FR_4$(C)=9866<br>21,15% | $FR_5$(C)=8021<br>21,49% |
| | | | $FR_2$(G)=19988<br>21,42% | $FR_3$(G)=13098<br>21,06% | $FR_4$(G)=9979<br>21,39% | $FR_5$(G)=7936<br>21,26% |
| 4 | HS4B958RAJ, Epstein-Barr virus. 184113 bp. Accession M80517.1 | $F_1$(A)=36002<br>19,55% | $F_2$(A)=18098<br>19,66% | $F_3$(A)=11910<br>19,41% | $F_4$(A)=9044<br>19,65% | $F_5$(A)=7210<br>19,58% |
| | | $F_1$(T)=37665<br>20,46% | $F_2$(T)=18887<br>20,52% | $F_3$(T)=12160<br>19,81% | $F_4$(T)=9394<br>20,41% | $F_5$(T)=7571<br>20,56% |
| | | $F_1$(C)=55824<br>30,32% | $F_2$(C)=27823<br>30,22% | $F_3$(C)=19144<br>31,19% | $F_4$(C)=13990<br>30,39% | $F_5$(C)=11162<br>30,31% |
| | | $F_1$(G)=54622<br>29,67% | $F_2$(G)=27248<br>29,60% | $F_3$(G)=18157<br>29,59% | $F_4$(G)=13600<br>29,55% | $F_5$(G)=10879<br>29,54% |
| | | **$\Sigma_1$ = 184113** | **$\Sigma_2$ = 92056** | **$\Sigma_3$ =61371** | **$\Sigma_4$ =46028** | **$\Sigma_5$ =36822** |
| | | | $FR_2$(A)=17903<br>19,45% | $FR_3$(A)=12495<br>20,36% | $FR_4$(A)=9145<br>19,87% | $FR_5$(A)=7178<br>19,49% |
| | | | $FR_2$(T)=18778<br>20,40% | $FR_3$(T)=12940<br>21,08% | $FR_4$(T)=9499<br>20,64% | $FR_5$(T)=7583<br>20,59% |
| | | | $FR_2$(C)=28001<br>30,42% | $FR_3$(C)=18163<br>29,60% | $FR_4$(C)=13640<br>29,63% | $FR_5$(C)=11145<br>30,27% |
| | | | $FR_2$(G)=27374<br>29,74% | $FR_3$(G)=17773<br>28,96% | $FR_4$(G)=13744<br>29,86% | $FR_5$(G)=10916<br>29,65% |
| 5 | Nicotiana tabacum chloroplast genome DNA. 155943 bp. Accession Z00044.2 | $F_1$(A)=47860<br>30,69% | $F_2$(A)=23952<br>30,72% | $F_3$(A)=16025<br>30,83% | $F_4$(A)=12043<br>30,89% | $F_5$(A)=9661<br>30,98% |
| | | $F_1$(T)=49064<br>31,46% | $F_2$(T)=24497<br>31,42% | $F_3$(T)=16460<br>31,67% | $F_4$(T)=12313<br>31,58% | $F_5$(T)=9743<br>31,24% |
| | | $F_1$(C)=30014<br>19,25% | $F_2$(C)=15095<br>19,36% | $F_3$(C)=9974<br>19,19% | $F_4$(C)=7452<br>19,12% | $F_5$(C)=5997<br>19,23% |
| | | $F_1$(G)=29005<br>18,60% | $F_2$(G)=14427<br>18,50% | $F_3$(G)=9522<br>18,32% | $F_4$(G)=7177<br>18,41% | $F_5$(G)=5787<br>18,56% |
| | | **$\Sigma_1$ = 155943** | **$\Sigma_2$ = 77971** | **$\Sigma_3$ =51981** | **$\Sigma_4$ =38985** | **$\Sigma_5$ =31188** |
| | | | $FR_2$(A)=23907<br>30,66% | $FR_3$(A)=15602<br>30,01% | $FR_4$(A)=11867<br>30,44% | $FR_5$(A)=9490<br>30,43% |
| | | | $FR_2$(T)=24567<br>31,51% | $FR_3$(T)=16480<br>31,70% | $FR_4$(T)=12375<br>31,74% | $FR_5$(T)=9891<br>31,71% |
| | | | $FR_2$(C)=14919<br>19,13% | $FR_3$(C)=10077<br>19,39% | $FR_4$(C)=7433<br>19,07% | $FR_5$(C)=5970<br>19,14% |
| | | | $FR_2$(G)=14578<br>18,70% | $FR_3$(G)=9822<br>18,90% | $FR_4$(G)=7310<br>18,75% | $FR_5$(G)=5837<br>18,72% |
| 6 | Equine herpesvirus 1 | $F_1$(A)=32616<br>21,71% | $F_2$(A)=16273<br>21,66% | $F_3$(A)=11331<br>22,63% | $F_4$(A)=8082<br>21,52% | $F_5$(A)=6480<br>21,57% |
| | | $F_1$(T)=32482 | $F_2$(T)=16191 | $F_3$(T)=10871 | $F_4$(T)=8106 | $F_5$(T)=6514 |
| | strain Ab4, complete genome. 150224 bp . Accession AY665713.1 | 21,62% | 21,56% | 21,71% | 21,58% | 21,68% |
| | | $F_1$(C)=43173<br>28,74% | $F_2$(C)=21746<br>28,95% | $F_3$(C)=14287<br>28,53% | $F_4$(C)=10735<br>28,58% | $F_5$(C)=8646<br>28,78% |
| | | $F_1$(G)=41953<br>27,93% | $F_2$(G)=20902<br>27,83% | $F_3$(G)=13585<br>27,13% | $F_4$(G)=10633<br>28,31% | $F_5$(G)=8404<br>27,97% |
| | | **$\Sigma_1$** = 150224 | **$\Sigma_2$** = 75112 | **$\Sigma_3$** =50074 | **$\Sigma_4$** =37556 | **$\Sigma_5$** =30044 |
| | | | $FR_2$(A)=16343<br>21,76% | $FR_3$(A)=10938<br>21,84% | $FR_4$(A)=8222<br>21,89% | $FR_5$(A)=6542<br>21,77% |
| | | | $FR_2$(T)=16291<br>21,69% | $FR_3$(T)=10822<br>21,61% | $FR_4$(T)=8147<br>21,69% | $FR_5$(T)=6471<br>21,54% |
| | | | $FR_2$(C)=21427<br>28,53% | $FR_3$(C)=13997<br>27,95% | $FR_4$(C)=10724<br>28,55% | $FR_5$(C)=8723<br>29,03% |
| | | | $FR_2$(G)=21051<br>28,03% | $FR_3$(G)=14317<br>28,59% | $FR_4$(G)=10463<br>27,86% | $FR_5$(G)=8308<br>27,65% |
| 7 | Equine herpesvirus 1 strain Ab4, complete genome. 150224 bp. Accession AY665713.1 | $F_1$(A)=32616<br>21,71% | $F_2$(A)=16273<br>21,66% | $F_3$(A)=11331<br>22,63% | $F_4$(A)=8082<br>21,52% | $F_5$(A)=6480<br>21,57% |
| | | $F_1$(T)=32482<br>21,62% | $F_2$(T)=16191<br>21,56% | $F_3$(T)=10871<br>21,71% | $F_4$(T)=8106<br>21,58% | $F_5$(T)=6514<br>21,68% |
| | | $F_1$(C)=43173<br>28,74% | $F_2$(C)=21746<br>28,95% | $F_3$(C)=14287<br>28,53% | $F_4$(C)=10735<br>28,58% | $F_5$(C)=8646<br>28,78% |
| | | >G>=41953<br>27,93% | $F_2$(G)=20902<br>27,83% | $F_3$(G)=13585<br>27,13% | $F_4$(G)=10633<br>28,31% | $F_5$(G)=8404<br>27,97% |
| | | **$\Sigma_1$** = 150224 | **$\Sigma_2$** = 75112 | **$\Sigma_3$** =50074 | **$\Sigma_4$** =37556 | **$\Sigma_5$** =30044 |
| | | | $FR_2$(A)=16343<br>21,76% | $FR_3$(A)=10938<br>21,84% | $FR_4$(A)=8222<br>21,89% | $FR_5$(A)=6542<br>21,77% |
| | | | $FR_2$(T)=16291<br>21,69% | $FR_3$(T)=10822<br>21,61% | $FR_4$(T)=8147<br>21,69% | $FR_5$(T)=6471<br>21,54% |
| | | | $FR_2$(C)=21427<br>28,53% | $FR_3$(C)=13997<br>27,95% | $FR_4$(C)=10724<br>28,55% | $FR_5$(C)=8723<br>29,03% |
| | | | $FR_2$(G)=21051<br>28,03% | $FR_3$(G)=14317<br>28,59% | $FR_4$(G)=10463<br>27,86% | $FR_5$(G)=8308<br>27,65% |
| 8 | Oryza sativa cultivar TN1 chloroplast, complete genome. 134502 bp. Accession NC_031333.1 | $F_1$(A)=41231<br>30,65% | $F_2$(A)=20572<br>30,59% | $F_3$(A)=13637<br>30,42% | $F_4$(A)=10209<br>30,36% | $F_5$(A)=8302<br>30,86% |
| | | $F_1$(T)=40818<br>30,35% | $F_2$(T)=20391<br>30,32% | $F_3$(T)=13675<br>30,50% | $F_4$(T)=10159<br>30,21% | $F_5$(T)=8094<br>30,09% |
| | | $F_1$(C)=26129<br>19,43% | $F_2$(C)=13035<br>19,38% | $F_3$(C)=8815<br>19,66% | $F_4$(C)=6576<br>19,56% | $F_5$(C)=5341<br>19,86% |
| | | $F_1$(G)=26324<br>19,57% | $F_2$(G)=13253<br>19,71% | $F_3$(G)=8707<br>19,42% | $F_4$(G)=6681<br>19,87% | $F_5$(G)=5163<br>19,19% |
| | | **$\Sigma_1$** = 134502 | **$\Sigma_2$** = 67251 | **$\Sigma_3$** =44834 | **$\Sigma_4$** =33625 | **$\Sigma_5$** =26900 |
| | | | $FR_2$(A)=20659<br>30,72% | $FR_3$(A)=13722<br>30,61% | $FR_4$(A)=10286<br>30,59% | $FR_5$(A)=8197<br>30,47% |
| | | | $FR_2$(T)=20427<br>30,37% | $FR_3$(T)=13382<br>29,85% | $FR_4$(T)=10249<br>30,48% | $FR_5$(T)=8129<br>30,22% |
| | | | $FR_2$(C)=13094<br>19,47% | $FR_3$(C)=8704<br>19,41% | $FR_4$(C)=6596<br>19,62% | $FR_5$(C)=5307<br>19,73% |
| | | | $FR_2$(G)=13071<br>19,44% | $FR_3$(G)=9026<br>20,13% | $FR_4$(G)=6494<br>19,31% | $FR_5$(G)=5267<br>19,58% |
| 9 | IH1CG, Ictalurid herpesvirus 1 strain Auburn 1, complete genome. 134226 bp. Accession M75136.2 | $F_1$(A)=28727<br>21,40% | $F_2$(A)=14299<br>21,31% | $F_3$(A)=9855<br>22,03% | $F_4$(A)=7148<br>21,30% | $F_5$(A)=5795<br>21,59% |
| | | $F_1$(T)=30025<br>22,37% | $F_2$(T)=15047<br>22,42% | $F_3$(T)=10633<br>23,77% | $F_4$(T)=7535<br>22,46% | $F_5$(T)=5969<br>22,24% |
| | | $F_1$(C)=37767<br>28,14% | $F_2$(C)=18943<br>28,23% | $F_3$(C)=12066<br>26,97% | $F_4$(C)=9436<br>28,12% | $F_5$(C)=7480<br>27,86% |
| | | $F_1$(G)=37707<br>28,09% | $F_2$(G)=18824<br>28,05% | $F_3$(G)=12188<br>27,24% | $F_4$(G)=9437<br>28,12% | $F_5$(G)=7601<br>28,31% |
| | | **$\Sigma_1$** = 134226 | **$\Sigma_2$** = 67113 | **$\Sigma_3$** =44742 | **$\Sigma_4$** =33556 | **$\Sigma_5$** =26845 |
| | | | $FR_2$(A)=14428<br>21,50% | $FR_3$(A)=9954<br>22,25% | $FR_4$(A)=7233<br>21,56% | $FR_5$(A)=5756<br>21,44% |
| | | | $FR_2$(T)=14978<br>22,32% | $FR_3$(T)=9795<br>21,89% | $FR_4$(T)=7506<br>22,37% | $FR_5$(T)=5865<br>21,85% |
| | | | $FR_2$(C)=18824<br>28,05% | $FR_3$(C)=12106<br>27,06% | $FR_4$(C)=9413<br>28,05% | $FR_5$(C)=7682<br>28,62% |
| | | | $FR_2$(G)=18883<br>28,14% | $FR_3$(G)=12887<br>28,80% | $FR_4$(G)=9404<br>28,02% | $FR_5$(G)=7542<br>28,09% |
| 10 | Human herpesvirus 3 isolate 667/2005, complete genome. 124884 bp. Accession JN704693.1 | $F_1$(A)=33782<br>27,05% | $F_2$(A)=16995<br>27,22% | $F_3$(A)=11327<br>27,21% | $F_4$(A)=8467<br>27,12% | $F_5$(A)=6724<br>26,92% |
| | | $F_1$(T)=33623<br>26,92% | $F_2$(T)=16785<br>26,88% | $F_3$(T)=11586<br>27,83% | $F_4$(T)=8445<br>27,05% | $F_5$(T)=6767<br>27,09% |
| | | $F_1$(C)=29295<br>23,46% | $F_2$(C)=14590<br>23,37% | $F_3$(C)=9512<br>22,85% | $F_4$(C)=7336<br>23,50% | $F_5$(C)=5826<br>23,33% |
| | | $F_1$(G)=28184<br>22,57% | $F_2$(G)=14072<br>22,54% | $F_3$(G)=9203<br>22,11% | $F_4$(G)=6973<br>22,33% | $F_5$(G)=5659<br>22,66% |
| | | **$\Sigma_1$** = 124884 | **$\Sigma_2$** = 62442 | **$\Sigma_3$** =41628 | **$\Sigma_4$** =31221 | **$\Sigma_5$** =24976 |
| | | | $FR_2$(A)=16787<br>26,88% | $FR_3$(A)=11358<br>27,28% | $FR_4$(A)=8424<br>26,98% | $FR_5$(A)=6753<br>27,04% |
| | | | $FR_2$(T)=16838<br>26,97% | $FR_3$(T)=10995<br>26,41% | $FR_4$(T)=8390<br>26,87% | $FR_5$(T)=6807<br>27,25% |
| | | | $FR_2$(C)=14705<br>23,55% | $FR_3$(C)=9969<br>23,95% | $FR_4$(C)=7318<br>23,44% | $FR_5$(C)=5846<br>23,41% |
| | | | $FR_2$(G)=14112<br>22,60% | $FR_3$(G)=9306<br>22,36% | $FR_4$(G)=7089<br>22,71% | $FR_5$(G)=5570<br>22,30% |
| 11 | Marchantia paleacea chloroplast genome DNA. 121024 bp. Accession X04465.1 | $F_1$(A)=42896<br>35,44% | $F_2$(A)=21385<br>35,34% | $F_3$(A)=14349<br>35,57% | $F_4$(A)=10739<br>35,49% | $F_5$(A)=8517<br>35,19% |
| | | $F_1$(T)=43263<br>35,75% | $F_2$(T)=21703<br>35,87% | $F_3$(T)=14637<br>36,28% | $F_4$(T)=10870<br>35,93% | $F_5$(T)=8677<br>35,85% |
| | | $F_1$(C)=17309<br>14,30% | $F_2$(C)=8677<br>14,34% | $F_3$(C)=5687<br>14,10% | $F_4$(C)=4340<br>14,34% | $F_5$(C)=3448<br>14,25% |
| | | $F_1$(G)=17556<br>14,51% | $F_2$(G)=8747<br>14,45% | $F_3$(G)=5668<br>14,05% | $F_4$(G)=4307<br>14,24% | $F_5$(G)=3562<br>14,72% |
| | | **$\Sigma_1$** = 121024 | **$\Sigma_2$** = 60512 | **$\Sigma_3$** =40341 | **$\Sigma_4$** =30256 | **$\Sigma_5$** =24204 |
| | | | $FR_2$(A)=21511<br>35,55% | $FR_3$(A)=14273<br>35,38% | $FR_4$(A)=10729<br>35,46% | $FR_5$(A)=8552<br>35,33% |
| | | | $FR_2$(T)=21560<br>35,63% | $FR_3$(T)=13879<br>34,40% | $FR_4$(T)=10745<br>35,51% | $FR_5$(T)=8680<br>35,86% |
| | | | $FR_2$(C)=8632<br>14,26% | $FR_3$(C)=5965<br>14,79% | $FR_4$(C)=4344<br>14,36% | $FR_5$(C)=3436<br>14,20% |
| | | | $FR_2$(G)=8809<br>14,56% | $FR_3$(G)=6224<br>15,43% | $FR_4$(G)=4438<br>14,67% | $FR_5$(G)=3536<br>14,61% |
| 12 | Escherichia coli strain PSUO78 plasmid pPSUO78_1, complete sequence. 132464 bp. Accession CP012113.1 | $F_1$(A)=33030<br>24,94% | $F_2$(A)=16477<br>24,88% | $F_3$(A)=11106<br>25,15% | $F_4$(A)=8304<br>25,08% | $F_5$(A)=6592<br>24,88% |
| | | $F_1$(T)=33824<br>25,53% | $F_2$(T)=16910<br>25,53% | $F_3$(T)=11340<br>25,68% | $F_4$(T)=8526<br>25,75% | $F_5$(T)=6907<br>26,07% |
| | | $F_1$(C)=34062<br>25,71% | $F_2$(C)=17075<br>25,78% | $F_3$(C)=11321<br>25,64% | $F_4$(C)=8428<br>25,45% | $F_5$(C)=6813<br>25,72% |
| | | $F_1$(G)=31548<br>23,82% | $F_2$(G)=15770<br>23,81% | $F_3$(G)=10387<br>23,52% | $F_4$(G)=7858<br>23,73% | $F_5$(G)=6180<br>23,33% |
| | | **$\Sigma_1$** = 132464 | **$\Sigma_2$** = 66232 | **$\Sigma_3$** =44154 | **$\Sigma_4$** =33116 | **$\Sigma_5$** =26492 |
| | | | $FR_2$(A)=16553<br>24,99% | $FR_3$(A)=10821<br>24,51% | $FR_4$(A)=8357<br>25,24% | $FR_5$(A)=6595<br>24,89% |
| | | | $FR_2$(T)=16914<br>25,54% | $FR_3$(T)=11239<br>25,45% | $FR_4$(T)=8423<br>25,43% | $FR_5$(T)=6691<br>25,26% |
| | | | $FR_2$(C)=16987<br>25,65% | $FR_3$(C)=11592<br>26,25% | $FR_4$(C)=8509<br>25,69% | $FR_5$(C)=6805<br>25,69% |
| | | | $FR_2$(G)=15778<br>23,82% | $FR_3$(G)=10502<br>23,78% | $FR_4$(G)=7827<br>23,64% | $FR_5$(G)=6401<br>24,16% |
| 13 | HSIULR, Human herpesvirus 1 complete genome. 152261 bp. Accession X14112.1 | $F_1$(A)=24240<br>15,92% | $F_2$(A)=12192<br>16,01% | $F_3$(A)=7949<br>15,66% | $F_4$(A)=6121<br>16,08% | $F_5$(A)=4926<br>16,18% |
| | | $F_1$(T)=24050<br>15,80% | $F_2$(T)=11941<br>15,69% | $F_3$(T)=7848<br>15,46% | $F_4$(T)=5945<br>15,62% | $F_5$(T)=4862<br>15,97% |
| | | $F_1$(C)=51458<br>33,80% | $F_2$(C)=25758<br>33,83% | $F_3$(C)=17447<br>34,38% | $F_4$(C)=12906<br>33,91% | $F_5$(C)=10344<br>33,97% |
| | | $F_1$(G)=52513<br>34,49% | $F_2$(G)=26239<br>34,47% | $F_3$(G)=17509<br>34,50% | $F_4$(G)=13093<br>34,40% | $F_5$(G)=10320<br>33,89% |
| | | **$\Sigma_1$** = 152261 | **$\Sigma_2$** = 76130 | **$\Sigma_3$** =50753 | **$\Sigma_4$** =38065 | **$\Sigma_5$** =30452 |
| | | | $FR_2(A)=12048$ 15,83% | $FR_3(A)=8234$ 16,22% | $FR_4(A)=5969$ 15,68% | $FR_5(A)=4875$ 16,01% |
| | | | $FR_2(T)=12109$ 15,91% | $FR_3(T)=8365$ 16,48% | $FR_4(T)=6048$ 15,89% | $FR_5(T)=4802$ 15,77% |
| | | | $FR_2(C)=25699$ 33,76% | $FR_3(C)=16828$ 33,16% | $FR_4(C)=12818$ 33,67% | $FR_5(C)=10272$ 33,73% |
| | | | $FR_2(G)=26274$ 34,51% | $FR_3(G)=17326$ 34,14% | $FR_4(G)=13230$ 34,76% | $FR_5(G)=10503$ 34,49% |
| 14 | HUMNEUROF, Human oligodendrocyte myelin glycoprotein (OMG) exons 1-2; neurofibromatosis 1 (NF1) exons 28-49; ecotropic viral integration site 2B (EVI2B) exons 1-2; ecotropic viral integration site 2A (EVI2A) exons 1-2; adenylate kinase (AK3) exons 1-2. 100849 bp. Accession L05367.1 | $F_1(A)=30346$ 30,09% | $F_2(A)=15271$ 30,29% | $F_3(A)=10157$ 30,21% | $F_4(A)=7636$ 30,29% | $F_5(A)=5978$ 29,64% |
| | | $F_1(T)=32481$ 32,21% | $F_2(T)=16175$ 32,08% | $F_3(T)=10746$ 31,97% | $F_4(T)=8030$ 31,85% | $F_5(T)=6570$ 32,57% |
| | | $F_1(C)=18635$ 18,48% | $F_2(C)=9390$ 18,62% | $F_3(C)=6302$ 18,75% | $F_4(C)=4727$ 18,75% | $F_5(C)=3709$ 18,39% |
| | | $F_1(G)=19387$ 19,22% | $F_2(G)=9588$ 19,01% | $F_3(G)=6411$ 19,07% | $F_4(G)=4819$ 19,11% | $F_5(G)=3912$ 19,40% |
| | | $\Sigma_1 = 100849$ | $\Sigma_2 = 50424$ | $\Sigma_3 = 33616$ | $\Sigma_4 = 25212$ | $\Sigma_5 = 20169$ |
| | | | $FR_2(A)=15075$ 29,90% | $FR_3(A)=10134$ 30,15% | $FR_4(A)=7523$ 29,84% | $FR_5(A)=6125$ 30,37% |
| | | | $FR_2(T)=16306$ 32,34% | $FR_3(T)=10865$ 32,32% | $FR_4(T)=8159$ 32,36% | $FR_5(T)=6502$ 32,24% |
| | | | $FR_2(C)=9245$ 18,33% | $FR_3(C)=6167$ 18,35% | $FR_4(C)=4621$ 18,33% | $FR_5(C)=3630$ 18,00% |
| | | | $FR_2(G)=9798$ 19,43% | $FR_3(G)=6450$ 19,19% | $FR_4(G)=4909$ 19,47% | $FR_5(G)=3912$ 19,40% |
| 15 | Podospora anserina mitochondrion, complete genome. 100314 bp. Accession NC_001329.3 | $F_1(A)=35804$ 35,69% | $F_2(A)=17906$ 35,70% | $F_3(A)=11917$ 35,64% | $F_4(A)=8814$ 35,15% | $F_5(A)=7101$ 35,40% |
| | | $F_1(T)=34358$ 34,25% | $F_2(T)=17236$ 34,36% | $F_3(T)=11802$ 35,30% | $F_4(T)=8693$ 34,66% | $F_5(T)=6967$ 34,73% |
| | | $F_1(C)=13428$ 13,39% | $F_2(C)=6669$ 13,30% | $F_3(C)=4415$ 13,20% | $F_4(C)=3413$ 13,61% | $F_5(C)=2614$ 13,03% |
| | | $F_1(G)=16724$ 16,67% | $F_2(G)=8346$ 16,64% | $F_3(G)=5304$ 15,86% | $F_4(G)=4158$ 16,58% | $F_5(G)=3380$ 16,85% |
| | | $\Sigma_1 = 100314$ | $\Sigma_2 = 50157$ | $\Sigma_3 = 33438$ | $\Sigma_4 = 25078$ | $\Sigma_5 = 20062$ |
| | | | $FR_2(A)=17898$ 35,68% | $FR_3(A)=11726$ 35,07% | $FR_4(A)=8922$ 35,58% | $FR_5(A)=7261$ 36,19% |
| | | | $FR_2(T)=17122$ 34,14% | $FR_3(T)=11546$ 34,53% | $FR_4(T)=8572$ 34,18% | $FR_5(T)=6792$ 33,86% |
| | | | $FR_2(C)=6759$ 13,48% | $FR_3(C)=4565$ 13,65% | $FR_4(C)=3415$ 13,62% | $FR_5(C)=2700$ 13,46% |
| | | | $FR_2(G)=8378$ 16,70% | $FR_3(G)=5601$ 16,75% | $FR_4(G)=4169$ 16,62% | $FR_5(G)=3309$ 16,49% |
| 16 | HUMTCRADCV, Human T-cell receptor genes (Human Tcr-C-delta gene, exons 1-4; Tcr-V-delta gene, exons 1-2; | $F_1(A)=28063$ 28,74% | $F_2(A)=13947$ 28,57% | $F_3(A)=9319$ 28,64% | $F_4(A)=6869$ 28,14% | $F_5(A)=5662$ 29,00% |
| | | $F_1(T)=26383$ 27,02% | $F_2(T)=13185$ 27,01% | $F_3(T)=8723$ 26,80% | $F_4(T)=6622$ 27,13% | $F_5(T)=5189$ 26,57% |
| | | $F_1(C)=20949$ 21,46% | $F_2(C)=10479$ 21,47% | $F_3(C)=6987$ 21,47% | $F_4(C)=5254$ 21,53% | $F_5(C)=4242$ 21,72% |
| | | $F_1(G)=22235$ 22,77% | $F_2(G)=11204$ 22,95% | $F_3(G)=7514$ 23,09% | $F_4(G)=5662$ 23,20% | $F_5(G)=4433$ 22,70% |
| | | $\Sigma_1 = 97630$ | $\Sigma_2 = 48815$ | $\Sigma_3 = 32543$ | $\Sigma_4 = 24407$ | $\Sigma_5 = 19526$ |
| | T-cell receptor alpha (Tcr-alpha) gene, J1-J61 segments; and Tcr-C-alpha gene, exons 1-4). 97630 bp . Accession M94081.1 | | $FR_2(A)=14116$ 28,92% | $FR_3(A)=9304$ 28,59% | $FR_4(A)=7039$ 28,84% | $FR_5(A)=5593$ 28,64% |
| | | | $FR_2(T)=13198$ 28,92% | $FR_3(T)=8873$ 27,27% | $FR_4(T)=6568$ 26,91% | $FR_5(T)=5386$ 27,58% |
| | | | $FR_2(C)=10470$ 21,45% | $FR_3(C)=6970$ 21,42% | $FR_4(C)=5257$ 21,54% | $FR_5(C)=4163$ 21,32% |
| | | | $FR_2(G)=11031$ 22,60% | $FR_3(G)=7396$ 22,73% | $FR_4(G)=5543$ 22,71% | $FR_5(G)=4384$ 22,45% |
| 17 | MUSTCRA, Mouse T-cell receptor alpha/delta chain locus. 94647 bp. Accession M64239.1 | $F_1(A)=26359$ 27,85% | $F_2(A)=13093$ 27,67% | $F_3(A)=8879$ 28,14% | $F_4(A)=6527$ 27,59% | $F_5(A)=5303$ 28,02% |
| | | $F_1(T)=25769$ 27,23% | $F_2(T)=13008$ 27,49% | $F_3(T)=8621$ 27,33% | $F_4(T)=6476$ 27,37% | $F_5(T)=5124$ 27,07% |
| | | $F_1(C)=20790$ 21,97% | $F_2(C)=10284$ 27,49% | $F_3(C)=6911$ 21,91% | $F_4(C)=5166$ 21,83% | $F_5(C)=4108$ 21,70% |
| | | $F_1(G)=21729$ 22,96% | $F_2(G)=10938$ 23,11% | $F_3(G)=7138$ 22,63% | $F_4(G)=5492$ 23,21% | $F_5(G)=4394$ 23,21% |
| | | **$\Sigma_1$ = 94647** | **$\Sigma_2$ = 47323** | **$\Sigma_3$ =31549** | **$\Sigma_4$ =23661** | **$\Sigma_5$ =18929** |
| | | | $FR_2(A)=13265$ 28,03% | $FR_3(A)=8717$ 27,63% | $FR_4(A)=6633$ 28,03% | $FR_5(A)=5235$ 27,66% |
| | | | $FR_2(T)=12761$ 26,97% | $FR_3(T)=8527$ 27,03% | $FR_4(T)=6380$ 26,96% | $FR_5(T)=5164$ 27,28% |
| | | | $FR_2(C)=10506$ 22,20% | $FR_3(C)=7004$ 22,20% | $FR_4(C)=5259$ 22,23% | $FR_5(C)=4150$ 21,92% |
| | | | $FR_2(G)=10791$ 22,80% | $FR_3(G)=7301$ 23,14% | $FR_4(G)=5389$ 22,78% | $FR_5(G)=4380$ 23,14% |

Fig. 67. Collective frequences $F_n(A)$, $F_n(T)$, $F_n(C)$, $F_n(G)$, $FR_n(A)$, $FR_n(T)$, $FR_n(C)$ and $FR_n(G)$ of n-plets (n = 1, 2, 3, 4, 5) in 17 long nucleotide sequences. Percentages of each of collective frequencies are also shown (see explanations in the text). $\Sigma_1$, $\Sigma_2$, $\Sigma_3$, $\Sigma_4$ and $\Sigma_5$ mean quantities of mononucleotides, doublets, triplets, 4-plets and 5-plets correspondingly. Initial data of sequences are taken from Genbank on the basis of the list of sequences, which have more than 900000 nucleotides, from the article [Prahbu, 1993].

Our results (Fig. 67) of the comparative analysis of these 17 and some other sequences give initial evidences in favor of existence of the following new universal rules of equality in long nucleotide sequences of the single-stranded DNA (by some analogy with the universal character of the second Chargaff's rule of equality). More precisely, these rules concerns approximate equalities of collective frequencies of n-plets in two mentioned tetra-groups. Of course, these supposed rules should be tested on the basis of much more number of long nucleotide sequences (we are working to make such checks).

**The first rule** (the rule of an approximate equality of the following collective frequencies: $F_n(A) \approx F_n(T)$ and $F_n(C) \approx F_n(G)$):

- In long nucleotide sequences of single-stranded DNA, collective frequencies of all n-plets, which have the same first letter, are approximately equal to collective frequencies of all n-plets with the complementary first letter: $F_n(A) \approx F_n(T)$ and $F_n(C) \approx F_n(G)$.

For example, one can see from data of the sequence №1 in Fig. 67 the following:

- $F_2(A)=24800 \approx F_2(T)=24340$; $F_3(A)=16825 \approx F_3(T)=16601$; $F_4(A)=12423 \approx F_4(T)=12266$; $F_5(A)=9981 \approx F_5(T)=9660$;

- $F_2(C)=32407 \approx F_2(G)=33130$; $F_3(C)=21303 \approx F_3(G)=21722$;
  $F_4(C)=16132 \approx F_4(G)=16517$; $F_5(C)=12992 \approx F_5(G)=13237$.

We note that for n=1 this first rule is coincided with the second Chargaff's rule: $F_1(A) \approx F_1(T)$ and $F_1(C) \approx F_1(G)$).

We emphasize that this rule about equalities of collective frequencies is not-trivial since approximately equal collective frequencies $F_n(A)$ and $F_n(T)$ (as well as $F_n(C)$ and $F_n(G)$) can differ significantly in the values of individual frequencies in them. For example, in the same sequence №1 (Fig. 67), these frequencies are composed of the following individual frequencies:
- $F_2(A) = F(AA)+F(AC)+F(AG)+F(AT) = 6079+7448+6486+4787=24800$;
- $F_2(T) = F(TC)+F(TA)+F(TT) + F(TG) = 6992+4262+5857+7229 = 24340$;
- $F_2(C) = F(CC)+F(CA)+F(CT)+F(CG) = 7856+7279+6099+11173 = 32407$
- $F_2(G) = F(GC)+F(GA)+F(GT)+F(GG) = 10208+7055+7693+8174 = 33\ 130$.

**The second rule** (the rule of an approximate equality of the following collective frequencies: $FR_n(A) \approx FR_n(T)$ and $FR_n(C) \approx FR_n(G)$):

- In long nucleotide sequences of single-stranded DNA, collective frequencies of all n-plets, which have the same last letter, are approximately equal to collective frequencies of all n-plets with the complementary last letter: $FR_n(A) \approx FR_n(T)$ and $FR_n(C) \approx FR_n(G)$.

For example, one can see from data of the sequence №1 in Fig. 67 the following:

- $FR_2(A) = 24675 \approx FR_2(T) = 24436$; $FR_3(A) = 16648 \approx FR_3(T) = 16211$;
  $FR_4(A) = 12279 \approx FR_4(T) = 12320$; $FR_5(A) = 10057 \approx FR_5(T) = 9789$;
- $FR_2(C) = 32504 \approx FR_2(G) = 33062$; $FR_3(C) = 21491 \approx FR_3(G) = 22101$;
  $FR_4(C)=16175 \approx FR_4(G)=16564$; $FR_5(C)=12914 \approx FR_5(G)=13110$.

We emphasize that approximately equal collective frequencies $FR_n(A)$ and $FR_n(T)$ (as well as $FR_n(C)$ and $FR_n(G)$) can differ significantly by the values of individual frequencies in them. For example, in the same sequence №1 (Fig. 67), these frequencies are composed of the following individual frequencies:
- $FR_2(A) = F(AA)+F(CA)+F(GA)+(TA) = 6079+7279+7055+4262 = 24675$;
- $FR_2(T) = F(CT)+F(AT)+F(TT)+F(GT) = 6099+4787+5857+7693 = 24436$;
- $FR_2(C) = F(CC)+F(AC)+F(TC)+F(GC) = 7856+7448+6992+10208 = 32504$;
- $FR_2(G) = F(CG)+F(AG)+F(TG)+F(GG) = 11173+6486+7229+8174 = 33062$.

**The third rule** (the rule of an approximate equality of the following collective frequencies: $F_n(A) \approx FR_n(A)$, $F_n(T) \approx FR_n(T)$, $F_n(C) \approx FR_n(C)$, $F_n(G) \approx FR_n(G)$):

- In long nucleotide sequences of single-stranded DNA, collective frequencies of all n-plets, which have the same first letter, are approximately equal to collective frequencies of all n-plets with the complementary last letter: $F_n(A) \approx FR_n(A)$, $F_n(T) \approx FR_n(T)$, $F_n(C) \approx FR_n(C)$, $F_n(G) \approx FR_n(G)$.

For example, one can see from data of the sequence №1 in Fig. 67 the following:

- $F_2(A) =24800 \approx FR_2(A) =24675$, $F_3(A) =16825 \approx FR_3(A) =16648$,
  $F_4(A) =12423 \approx FR_4(A) =12279$, $F_5(A) =9981 \approx FR_5(A) =10057$;

- $F_2(T)=24340 \approx FR_2(T)=24436$, $F_3(T)=16601 \approx FR_3(T)=16648$, $F_4(T)=12266 \approx FR_4(T)=12320$, $F_5(T) =9660 \approx FR_5(T) =9789$;
- $F_2(C)=32407 \approx FR_2(C)=32504$, $F_3(C)=21303 \approx FR_3(C)=21491$, $F_4(C)=16132 \approx FR_4(C)=16175$, $F_5(A)=12992 \approx FR_5(C)=12914$;
- $F_2(G)=33130 \approx FR_2(G)=33062$, $F_3(G)=21722 \approx FR_3(G)=22101$, $F_4(G) =16517 \approx FR_4(G) =16564$, $F_5(G) \approx FR_5(G)$.

We emphasize that approximately equal collective frequencies $F_n(A) \approx FR_n(A)$ (as well as $F_n(T) \approx FR_n(T)$, $F_n(C) \approx FR_n(C)$, $F_n(G) \approx FR_n(G)$) can differ significantly by the values of individual frequencies in them.

One of many confirmations of these three rules is based on data about individual frequences of 64 triplets in the whole human genome, which contains 2.843.411.612 (about three billion) triplets. These initial data are taken from [Perez, 2010] and are reproduced in Fig. 68.

| triplet | triplet frequency | triplet | triplet frequency | triplet | triplet frequency | triplet | triplet frequency |
|---|---|---|---|---|---|---|---|
| AAA | 109143641 | CAA | 53776608 | GAA | 56018645 | TAA | 59167883 |
| AAC | 41380831 | CAC | 42634617 | GAC | 26820898 | TAC | 32272009 |
| AAG | 56701727 | CAG | 57544367 | GAG | 47821818 | TAG | 36718434 |
| AAT | 70880610 | CAT | 52236743 | GAT | 37990593 | TAT | 58718182 |
| ACA | 57234565 | CCA | 52352507 | GCA | 40907730 | TCA | 55697529 |
| ACC | 33024323 | CCC | 37290873 | GCC | 33788267 | TCC | 43850042 |
| ACG | 7117535 | CCG | 7815619 | GCG | 6744112 | TCG | 6265386 |
| ACT | 45731927 | CCT | 50494519 | GCT | 39746348 | TCT | 62964984 |
| AGA | 62837294 | CGA | 6251611 | GGA | 43853584 | TGA | 55709222 |
| AGC | 39724813 | CGC | 6737724 | GGC | 33774033 | TGC | 40949883 |
| AGG | 50430220 | CGG | 7815677 | GGG | 37333942 | TGG | 52453369 |
| AGT | 45794017 | CGT | 7137644 | GGT | 33071650 | TGT | 57468177 |
| ATA | 58649060 | CTA | 36671812 | GTA | 32292235 | TTA | 59263408 |
| ATC | 37952376 | CTC | 47838959 | GTC | 26866216 | TTC | 56120623 |
| ATG | 52222957 | CTG | 57598215 | GTG | 42755364 | TTG | 54004116 |
| ATT | 71001746 | CTT | 56828780 | GTT | 41557671 | TTT | 109591342 |

Fig. 68. Frequences of triplets in the whole human genome (data are taken from the work [Perez, 2010]).

A calculation from data in Fig. 68 gives the following collective frequencies for the tetra-group of triplets in the whole human genome:

| | |
|---|---|
| $F_3(A) = 839827642$ | $FR_3(A) = 839827334$ |
| $F_3(T) = 841214589$ | $FR_3(T) = 841214933$ |
| $F_3(C) = 581026275$ | $FR_3(C) = 581026487$ |
| $F_3(G) = 581343106$ | $FR_3(G) = 581342858$ |

Fig. 69. Collective frequencies in the tetra-group of triplets in the whole human genome.

These data about collective frequencies in the tetra-group of triplets in the whole human genome (Fig. 69) confirm - with a high precision - all three rules, which were formulated above:

- In accordance with the first rule, $F_3(A)=839827642 \approx F_3(T)=841214589$ (the difference in these values is 0,16%) and $F_3(C)=581026275 \approx F_3(G)=581343106$ (the difference in these values is 0,05%);
- In accordance with the second rule, $FR_3(A)=839827334 \approx FR_3(T)=841214933$ (the difference in these values is 0,16%) and $FR_3(C)=581026487 \approx FR_3(G)=581342858$ (the difference in these values is 0,05%);
- In accordance with the third rule, $F_3(A)=839827642 \approx FR_3(A)=839827334$ (the difference in these values is 0,00004%); $F_3(T)=841214589 \approx FR_3(T)=841214933$ (the difference in these values is 0,00004%); $F_3(C)=581026275 \approx FR_3(C)=581026487$ (the difference in these values is 0,00004%); $F_3(G)=581343106 \approx FR_3(G))=581342858$ (the difference in these values is 0,00004%). This third rule is fulfilled here with the highest accuracy equal to 0, 00004% (rounded) in all considered cases.

We note else the following additional fact from data in Fig. 68 concerning triplets with the same middle letter: in the whole human genome the collective frequence of triplets with the middle letter A (its value is 839827606) is approximately equal to the collective frequence of triplets with the middle letter T (its value is 841214880) and the collective frequence of triplets with the middle letter C (its value is 581026266) is approximately equal to the collective frequence of triplets with the middle letter G (its value is 581342860). These collective frequences - 839827606, 841214880, 581026266 and 581342860 - are approximately equal to appropriate collective frequencies of triplets $F_3(A)$, $F_3(T)$, $F_3(C)$ and $F_3(G)$ in Fig. 69.

This coincidence leads to the hypothesis that rules of approximate equality of collective frequencies exist not only for tetra-groups, where n-plets are combined in one subgroup by the criteria of the same first letter or the same last letter, but also for tetra-groups, where n-plets (with $n \geq 3$) are combined in one subgroup by the criterion of the same second letter, by the criterion of the same third letter, etc. Our preliminary tests of this hypothesis gave evidences in favor of existence of these extended rules.

For a simpler representation of results of these tests, we should improve denotations of collective frequences in cases of tetra-groups of n-plets, where n-plets are combined in one of 4 subgroups by the criteria of the same letter on the concrete position inside n-plets. We will use new symbols $F_n(A_k)$, $F_n(T_k)$, $F_n(C_k)$ and $F_n(G_k)$, where n = 1, 2, 3, 4,… and $k \leq n$. These symbols denote collective frequencies in long nucleotide sequences on the basis of a tetra-group of n-plets, where equivalent n-plets are combined in one of 4 subgroups by the criterion of the same letter on the position k. It is obvious that for each concrete value «n» there exist «n» kinds of such tetra-groups since n-plets have «n» positions inside them:

- if n=2, two tetra-groups exist, which define two appropriate sets of collective frequencies: firstly $F_2(A_1)$, $F_2(T_1)$, $F_2(C_1)$, $F_2(G_1)$ and secondly $F_2(A_2)$, $F_2(T_2)$, $F_2(C_2)$, $F_2(G_2)$;
- if n=3, three tetra-groups exist, which define three appropriate sets of collective frequences: firstly $F_3(A_1)$, $F_3(T_1)$, $F_3(C_1)$, $F_3(G_1)$, secondly $F_3(A_2)$, $F_3(T_2)$, $F_3(C_2)$, $F_2(G_2)$ and thirdly $F_3(A_3)$, $F_3(T_3)$, $F_3(C_3)$, $F_3(G_3)$;
- …..
- for arbitrary n-plets, «n» tetra-groups exist, which define «n» appropriate sets of collective frequencies: firstly $F_n(A_1)$, $F_n(T_1)$, $F_n(C_1)$, $F_n(G_1)$, secondly $F_n(A_2)$, $F_n(T_2)$, $F_n(C_2)$, $F_n(G_2)$, …, finally $F_n(A_n)$, $F_n(T_n)$, $F_n(C_n)$ and $F_n(G_n)$.

Fig. 70 shows results of the improved analysis of the sequence № 1 from Fig. 67 taking into account the extented list of the mentioned tetra-groups of n-plets and the extended list of collective frequences $F_n(A_k)$, $F_n(T_k)$, $F_n(C_k)$ and $F_n(G_k)$, where n = 1, 2, 3, 4, 5 and $k \le n$.

| **NUCLEOTIDES** | **DOUBLETS** | **TRIPLETS** | **4-PLETS** | **5-PLETS** |
|---|---|---|---|---|
| $\Sigma_1$ = 229354 | $\Sigma_2$ =114677 | $\Sigma_3$ =76451 | $\Sigma_4$ =57338 | $\Sigma_5$ =45870 |
| $F_1(A_1)$=49475<br>21,57% | $F_2(A_1)$=24800<br>21,63% | $F_3(A_1)$=16825<br>22,01% | $F_4(A_1)$=12423<br>21,67% | $F_5(A_1)$=9981<br>21,76% |
| $F_1(T_1)$=48776<br>21,27% | $F_2(T_1)$=24340<br>21,22% | $F_3(T_1)$=16601<br>21,71% | $F_4(T_1)$=12266<br>21,39% | $F_5(T_1)$=9660<br>21,06% |
| $F_1(C_1)$=64911<br>28,30% | $F_2(C_1)$=32407<br>28,26% | $F_3(C_1)$=21303<br>27,86% | $F_4(C_1)$=16132<br>28,13% | $F_5(C_1)$=12992<br>28,32% |
| $F_1(G_1)$=66192<br>28,86% | $F_2(G_1)$=33130<br>28,89% | $F_3(G_1)$=21722<br>28,41% | $F_4(G_1)$=16517<br>28,81% | $F_5(G_1)$=13237<br>28,86% |
| | $F_2(A_2)$=24675<br>21,52% | $F_3(A_2)$=16002<br>20,93% | $F_4(A_2)$= 12396<br>21,61% | $F_5(A_2)$= 9846<br>21,46% |
| | $F_2(T_2)$=24436<br>21,31% | $F_3(T_2)$=15964<br>20,88% | $F_4(T_2)$= 12116<br>21,13% | $F_5(T_2)$= 9707<br>21,16% |
| | $F_2(C_2)$=32504<br>28,34% | $F_3(C_2)$=22117<br>28,93% | $F_4(C_2)$= 16329<br>28,48% | $F_5(C_2)$= 13119<br>28,60% |
| | $F_2(G_2)$=33062<br>28,83% | $F_3(G_2)$=22368<br>29,26% | $F_4(G_2)$= 16497<br>28,77% | $F_5(G_2)$= 13198<br>28,77% |
| | | $F_3(A_3)$=16648<br>21,78% | $F_4(A_3)$= 12377<br>21,59% | $F_5(A_3)$= 9747<br>21,25% |
| | | $F_3(T_3)$=16211<br>21,20% | $F_4(T_3)$= 12074<br>21,06% | $F_5(T_3)$= 9842<br>21,46% |
| | | $F_3(C_3)$=21491<br>28,11% | $F_4(C_3)$= 16053<br>28,00% | $F_5(C_3)$= 12970<br>28,28% |
| | | $F_3(G_3)$=22101<br>28,91% | $F_4(G_3)$= 16612<br>28,97% | $F_5(G_3)$= 13311<br>29,02% |
| | | | $FR_4(A_4)$=12279<br>21,42% | $F_5(A_4)$= 9907<br>21,60% |
| | | | $FR_4(T_4)$=12320<br>21,49% | $F_5(T_4)$= 9733<br>21,22% |
| | | | $FR_4(C_4)$=16175<br>28,21% | $F_5(C_4)$= 12800<br>27,90% |
| | | | $FR_4(G_4)$=16564<br>28,89% | $F_5(G_4)$= 13333<br>29,07% |
| | | | | $FR_5(A_5)$=10057<br>21,93% |
| | | | | $FR_5(T_5)$=9789<br>21,34% |
| | | | | $FR_5(C_5)$=12914<br>28,15% |
| | | | | $FR_5(G_5)$=13110<br>28,58% |

Fig. 70. Collective frequencies $F_n(A_k)$, $F_n(T_k)$, $F_n(C_k)$ and $F_n(G_k)$ (n = 1, 2, 3, 4, 5 and $k \leq n$) for nucleotide sequences of n-plets, which have the same letter in their position k, in the case of the sequence Human cytomegalovirus strain AD169 complete genome, 229354 bp, GenBank, accession X17403.1 (compare with data about this sequence in Fig. 67).

Taking these new data into account, the first rule and the second rule mentioned above can be combained in the one joint rule of the following extended type for further checks:

- the extended first rule: in long nucleotide sequences of single-stranded DNA, collective frequencies $F_n(A_k)$, $F_n(T_k)$, $F_n(C_k)$ and $F_n(G_k)$ ($k \leq n$) of all n-plets, which have the same letter on their position k, are approximately equal to collective frequencies of all n-plets with the complementary letter on the same position k: $F_n(A_k) \approx F_n(T_k)$ and $F_n(C_k) \approx F_n(G_k)$.

Also these new data allow proposing the following extended version of the third rule for further checks:

- In long nucleotide sequences of single-stranded DNA, collective frequencies of all n-plets, which have the same letter on their position k, are approximately equal to collective frequencies of all n-plets with the same letter on their any another position j ($k, j \leq n$): $F_n(A_k) \approx F_n(A_j)$, $F_n(T_k) \approx F_n(T_j)$, $F_n(C_k) \approx F_n(C_j)$, $F_n(G_k) \approx F_n(G_j)$.

These extended rules give new opportunities to study connections of nucleotide sequences with cyclic permutations of positions in n-plets.

**The fourth rule** (the rule of approximate equality of the percentage of each of the considered collective frequencies, regardless of the length n of the considered n-plets):

- In long nucleotide sequences of single-stranded DNA, percentage values $(F_n(A_k)/\Sigma_n)*100\%$ ($k \leq n$) are approximately equal to each other for different values n. The same is true for percentage values $(F_n(T_k)/\Sigma_n)*100\%$, $(F_n(C_k)/\Sigma_n)*100\%$, $(F_n(G_k)/\Sigma_n)*100\%$.

In accordance with this rule, a set of percentage values in each of rows of the tables in Fig. 67 and in Fig. 70 contains only approximately equal values.

The introduced notion of tetra-groups of n-plets with the same letter on their position k allows a digitization of any genetic sequence of n-plets by a denotation of each of 4 members of the tetra-group by means of its individual number 0, 1, 2 or 3. For example, in the case of the tetra-group (13) of doublets with the same first letter, one can denote equivalent doublets with the first letter T by number 0, equivalent doublets with the second letter G – by number 1, equivalent doublets with the first letter A – by number 2, equivalent doublets with the first letter C – by number 3. In the case of such digitization any symbolic sequence of doublets is transformed into the corresponding digital sequence of four kinds of digits 0, 1, 2, 3. For example the symbolic sequence CA-GG-GT-CG-TA-AA-TG-AC-… is convoluted into a digital sequence 3-1-1-3-0-2-0-2-…, the length of which is shortened in half in comparison with the initial nucleotide sequence. We call this method "the method of tetra-groupic convolution".

The method of the Chaos Game Representation (CGR-method) for visualization of sequences of mononucleotides A, C, G, T, which are digitalized by means of a denotation of these 4 letters by digits 0, 1, 2 and 3, was proposed in [Jeffrey, 1990]. In this work an application of the CGR-method to visualizations of different genetic sequences of mononucleotides has revealed that these sequences are connected with different fractal patterns (CGR-fractals). We apply this known CGR-method to visualize sequences of n-plets, which are digitalized by our method of equaivalent digitization by tetra-groups. Fig. 71 and 72 show a few examples of CGR-fractals received by such approach. In these examples, numbers 0, 1, 2 and 3 symbolize doublets and triplets with letters T, G, A and C at appropriate positions correspondingly (if this correspondance is changed, pictures of CGR-patterns are also changed in a parallel way).

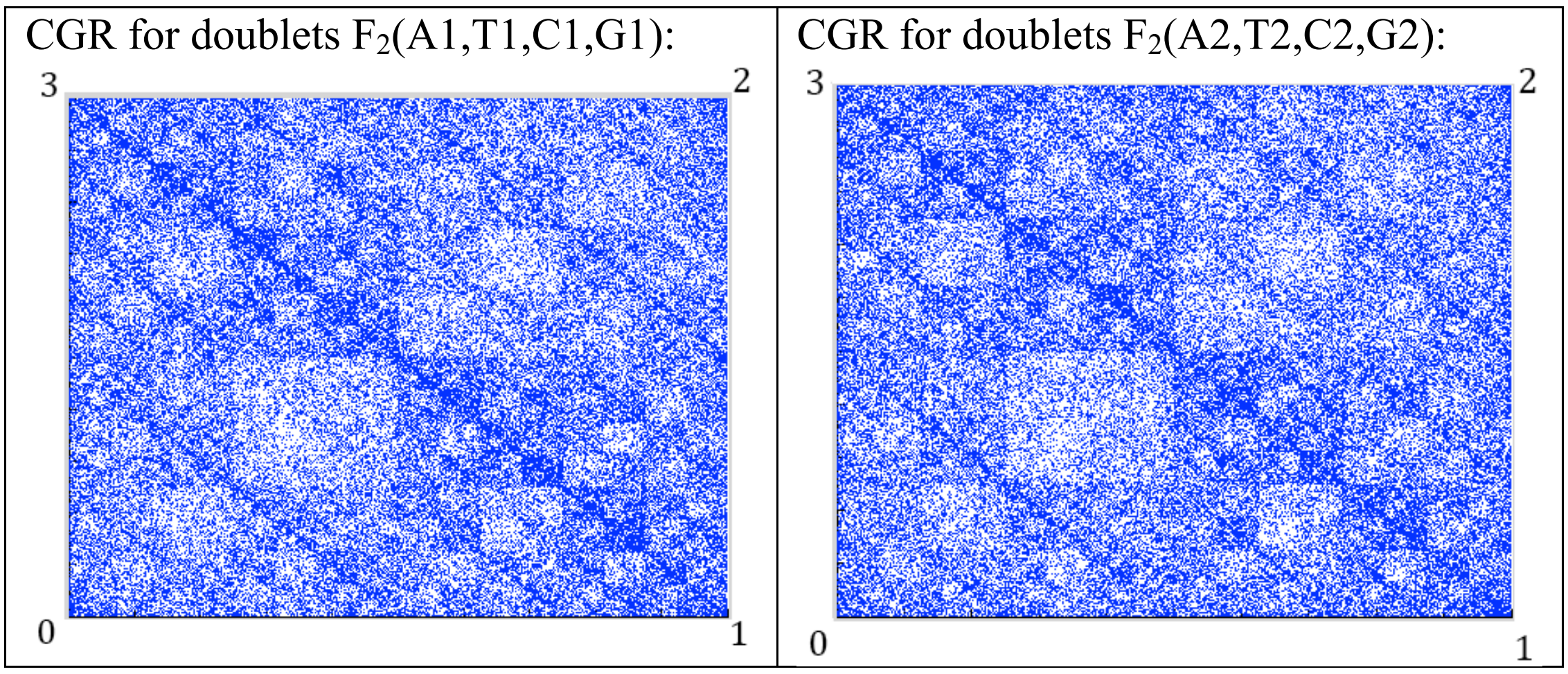

Fig. 71. The CGR-patterns of representation of the sequence of doublets of Human cytomegalovirus strain AD169 complete genome, 229354 bp, GenBank, accession X17403.1. These patterns were received by using the proposed method of equaivalent digitization by tetra-groups. Left: the case of the tetra-group of equivalency of doublets with the same letter on their first position. Right: the case of the tetra-group of equivalency of doublets with the same letter on their last position. In both cases here, digits 0, 1, 2, 3, symblolize doublets with the letters T, G, A, C on corresponding postions.

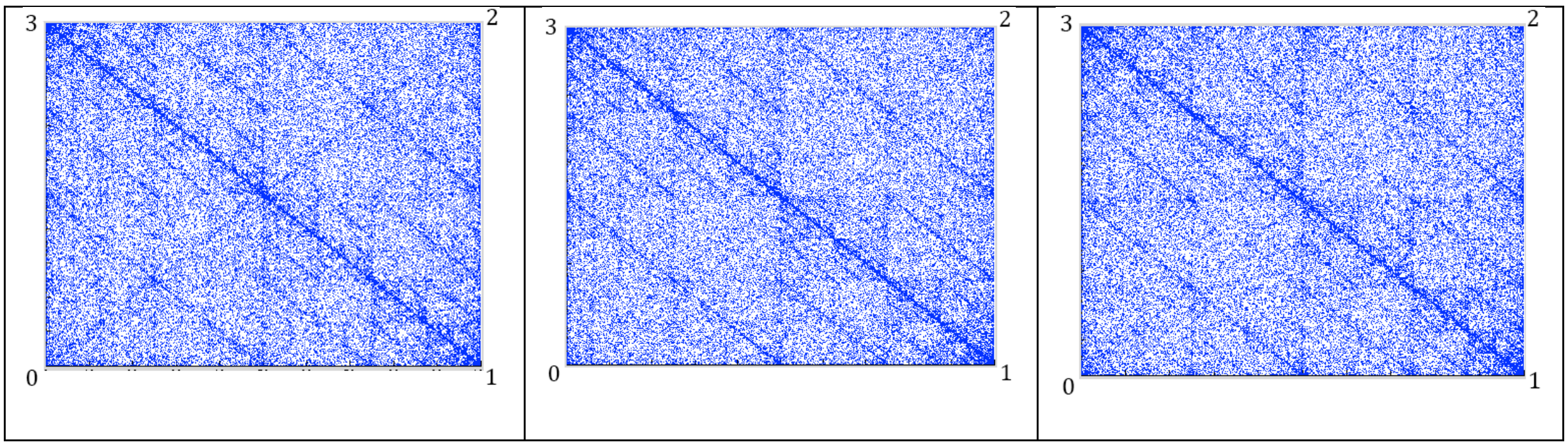

Fig. 72. The CGR-patterns of representation of the sequence of triplets of Human cytomegalovirus strain AD169 complete genome, 229354 bp, GenBank, accession X17403.1, which were received by using the proposed method of equaivalent digitization by tetra-groups. Left: the case of the tetra-group of equivalency of triplets with the same letter on the first position. Middle: the case of the tetra-group of equivalency of triplets with the same letter on the second position. Right: the case of the tetra-group of equivalency of triplets with the same letter on the third position. In all cases here, 0, 1, 2, 3, symblolize triples with the letters T, G, A, C on corresponding positions.

These examples (Fig. 71, 72) show that in all these different cases, the general form of the fractal CGR-patterns does not practically change.

This digitization of any long sequence of n-plets, which consist of 4 symbols A, C, G and T, - by means of the proposed method of tetra-groupic convolution - is transformed the symbolic sequence of n-plets into a digit sequence with 4 kinds of digital members (0, 1, 2 and 3). This digit sequence represents the first level of the convolution of the initial nucleotide sequence by the proposed method (for a sufficiently long nucleotide sequence, this method allows also obtaining the second and higher levels of its convolution). By analogy with the symbolic sequence of nucleotides, this digit sequence with 4 digital members can be considered as a sequence of "digital n-plets", which has its own 16 digital doublets, 64 digital triplets, etc. For example, the digital sequence 200132131101030221… can be considered as a sequence of digital doublets 20-01-32-13-11-01-03-02-21-… or as a sequence of digital triplets 200-132-131-101-030-221-…, etc. It is obvious that such digit sequences can be analogically analyzed from the standpoint of collective frequencies of members of their own tetra-groups, 4 members of which combine digital n-plets with the same digit on their corresponding positions. What one can say about possible rules for relevant collective frequencies $F_n(0_k)$, $F_n(1_k)$, $F_n(2_k)$ and $F_n(3_k)$ in such digital n-plets? Our preliminary results show that the similar rules of an approximate equality of appropriate collective frequencies are also fulfilled here and they give additional materials about fractal-like properties of long nucleotide sequences. Verification of this preliminary approval is continuing now. Fig. 73 represents one of examples of a calculation of collective frequencies in digital sequencies at the first level of the tetra-groupic convolution.

More precisely, we took the nucleotide sequence Human cytomegalovirus strain AD169 complete genome (some characteristics of which were already represented in Fig. 67,70-72) as a sequence of doublets. Then we digitized this sequence: each of doublets with the first letter T was replaced by digit 0, each of doublets with the first letter G was replaced by digit 1, each of doublets with the first letter A was replaced by digit 2, each of doublets with the first letter C was replaced by digit 3. In the result of this tetra-grouping convolution by the first letter, the digital sequence arised - 113301113011…, which was represented as the appropriate sequence of digital doublets 11-33-01-11-30-11-…. . At the final stage we calculate collective frequencies: $F_2(0_1)$ of doublets with the first digit 0; $F_2(1_1)$ of doublets with the first digit 1; $F_2(2_1)$ of doublets with the first digit 2; $F_2(3_1)$ of doublets with the first digit 3; $F_2(0_2)$ of doublets with the last digit 0; $F_2(1_1)$ of doublets with the last digit 1; $F_2(2_1)$ of doublets with the last digit 2; $F_2(3_1)$ of doublets with the last digit 3 (see received results in Fig. 73). We have also represented the same digital sequence 113301113011… as the sequence of digital triplets 113-301-113-011-… and have calculated collective frequencies of digital triplets $F_3(0_k)$, $F_3(1_k)$, $F_3(2_k)$ and $F_3(3_k)$, where k=1, 2, 3 shows the position of the digit in triplets (see received results in Fig. 73).

| **"DIGITAL NUCLEOTIDES"** | **DIGITAL DOUBLETS** | **DIGITAL TRIPLETS** |
|---|---|---|
| $\Sigma_1$ = 114677 | $\Sigma_2$ = 57338 | $\Sigma_3$ = 38225 |
| $F_1(2_1)$=24800<br>21,63% | $F_2(2_1)$= 12423<br>21,67% | $F_3(2_1)$= 8493<br>22,22% |
| $F_1(0_1)$=24340<br>21,22% | $F_2(0_1)$= 12266<br>21,39% | $F_3(0_1)$= 8291<br>21,69% |
| $F_1(3_1)$=32407<br>28,26% | $F_2(3_1)$= 16132<br>28,13% | $F_3(3_1)$= 10607<br>27,75 |
| $F_1(1_1)$=33130<br>28,89% | $F_2(1_1)$= 16517<br>28,81% | $F_3(1_1)$= 10834<br>28,34% |
| | $F_2(2_2)$= 12377<br>21,59% | $F_3(2_2)$= 8363<br>21,88% |
| | $F_2(0_2)$= 12074<br>21,06% | $F_3(0_2)$= 8101<br>21,19% |
| | $F_2(3_2)$= 16275<br>28,38% | $F_3(3_2)$= 10716<br>28,03% |
| | $F_2(1_2)$= 16612<br>28,97% | $F_3(1_2)$= 11045<br>28,89% |
| | | $F_3(2_3)$= 7943<br>20,78 |
| | | $F_3(0_3)$= 7948<br>20,79% |
| | | $F_3(3_3)$= 11084<br>29,00% |
| | | $F_3(1_3)$= 11250<br>29,43% |

Fig. 73. Collective frequencies $F_2(0_k)$, $F_2(1_k)$, $F_2(2_k)$, $F_3(3_k)$, $F_3(0_k)$, $F_3(1_k)$, $F_3(2_k)$ and $F_3(3_k)$ in the digital sequence, which arises in the result of the tetra-groupic convolution of the sequence of doublets (by their first letter) of Human cytomegalovirus strain AD169 complete genome, 229354 bp, GenBank, accession X17403.1 (see explanation in text). The percentage of each of collective frequencies is also shown.

Fig. 73 shows that collective frequencies of digital doublets and triplets at the first level of this tetra-convolution are approximately in 2 time less than collective frequencies of doublets and tripples of the considered nucleotide sequence in Fig. 70.

Our work [Petoukhov, Svirin, 2012] has introduced the notion «fractal genetic nets» with a description of special transformations of long nucleotide sequencies. This transformations generate new shorter sequences of nucleotides, to study of which the proposed method of tetra-groupic equivalency of oligonucleotides is also applied.

In the end of this Section, one should say the following. The Chargaff's rules have plaid an important role in development of bioinformatics. We hope that the rules, which are described in this Section, will be also useful for further development of bioinformatics. In particular, these new rules testify in favor of usage of principles of logical holography and fractal compression of informatics in mathematical modeling of genetic structures and phenomena.

Our results reveal new hidden connections of long nucleotide sequences with fractal structures, which are associated with strange attractors of the chaotic dynamics. As known, fractals are to chaos what geometry is to algebra. They are the geometric manifestation of the chaotic dynamics. They are called "the fingerprints of chaos" sometimes. The revealing new hidden connections of long nucleotide sequences with fractals show additional possibilities of usage of ideology and mathematical tools of non-linear dynamics or chaotic dynamical systems in genetics. In particular, modern knowledge about so called chaotic communication systems and a special type of computers, which work on principles of chaotic dynamical systems [Dmitriev, 1998], can be apllied to understand genetic phenomena. In such approach, a living organism is represented as a set of chaotic dynamical computers and chaotic communication systems. This approach can be useful for progress in both - genetic and technological - fields. Taking into account our received results about the described rules (on the basis of tetra-groups of equivalency of oligonucleotide), we propose the new notion of genetic and epigenetic «tetra-codes». Also we should note two following thoughts connected with these tetra-codes:

- Genetic and epigenetic systems are collective essences and their secrets are connected with interactions of whole ensembles of genetic elements such as the described tetra-groups of oligonucleotides. Correspondingly, collective characteristics of genetic ensembles should be studied to reveal new secrets of genetic systems (by analogy with the described collective frequences of oligonucleotides on the basis of tetra-groups);
- The described quantitative rules reflect that the genetic coding systems in the whole use tetra-codes, which are digital representations of the described tetra-groups of oligonucleotides. In our opinion, these tetra-codes are useful for understanding the geno-logical coding [Petoukhov, Petukhova, 2017]. Study of these tetra-codes should use binary representations of oligonucleotides (in particular, decimal numbers 0, 1, 2 and 3 can be represented as members of the 2-bits dyadic group of binary numbers 00, 01, 10 and 11 on the basis of their biochemical binary-oppositional traits.

To the theme of importance of tetra-groups and tetra-codes in genetics one can add the fact of existence of 4 histones, which serve as the basis in the well known hypothetical histone code. In eukaryote cells, filaments of DNA are coiled around nucleosomes, each of which is a shank consisting of the histones of the four types: H2A, H2B, H3 and H4. This set of four types is divided by nature into the pairs of one-specific histones. The histones H2A and H2B possess the important possibility to create the pair just one with another on the basis of their mutual revealing and mutual "attraction". Another pair consists of the histones H3 and H4, which possess the similar possibility to create the pairs just one with another on the analogical basis of their mutual revealing and mutual "attraction". It is an interesting open question about a correspondence between the described tetra-groups of oligonucleotides and the tetra-group of histones in epigenetic coding properties of DNA. We think that epigenetic coding uses a set of interrelated tetra-codes of different nature.

## SOME CONCLUDING REMARKS

As it was noted in the beginning of the article, living organism is a machine for coding and processing of information. For example, visual information about external objects is transmitted through the nerves from the eye retina to the brain already in a logarithmically encoded form. This article shows some evidences that oblique projection operators and their combinations, which are connected with matrix representations of the genetic coding system, can be a basis of adequate approaches to simulate ensembles of inherited biological phenomena. Only some of such phenomena were considered in this article. Some other phenomena will be described from the proposed point of view later. In addition, the author reminds about fractal genetic nets (FGN), which can be represented as a construction on a base of orthogonal projectors and which lead to new genetic rules in structures of long nucleotide sequences [Petoukhov, 2012; Petoukhov, Svirin, 2012].

The revealed genetic system of operators connected with oblique projectors allows modeling multi-dimensional phase spaces with many subspaces, processes in which can be selectively determined and controled. Speaking about importance of projectors in the nature, the following points should be noted (it seems that the nature «likes» projectors):

- Most people are familiar with the idea of projectors due to sun rays (and light rays in general) that are distributed in a straight line and provide shade from the subjects (from the ancient time sundials were constructed on the use of this). Light rays have the projection property;
- Electromagnetic vectors are the sum of their projections in the form of their electric and magnetic vectors;
- Evolution of living organisms is associated with the consumption of solar energy that is projected by means of sun rays to surfaces of living bodies (photosynthesis, which is for living matter one of its basic mechanisms, is the conversion of sunlight energy into biochemical energy for activity of organisms; circadian biorhythms are connected with external light cycles "day-night");
- Projection phenomena of birefringence in biological tissues and crystals exist;
- A great variety of living organisms has polarization eyesight;
- Our vision is based on the projection of images on the retina;
- Religious people may ask whether there is any indication in the Bible on this subject? Especially for them, it may be recalled the following. According to the Bible, God's creation of the world began with the creation of the light: "Let there be light." Many thinkers suggested previously for various reasons that, figuratively speaking, the body is woven from the light. This has some associations with our hypothesis that the body is woven from the projectors.

Now projectors and their combinations become interesting instruments to study and simulate genetic phenomena and inherited structures and processes in living matter. A new conceptual notion with appropriate mathematical formalisms are proposed about a multi-dimensional control space (or coding space) with subspaces of a selective control in each on basis of a participation of projection operators in such control. Here one can remember the statement: "*Profound study of nature is the most fertile source of mathematical discoveries*" (Fourier, 2006, Chapter 1, p. 7).

Using this ideology of projection operators, one can get many unexpected results and approaches. In author's opinion, one of many promising applications of projectors in mathematical biology and bioinformatics is the study of connections between genetic projectors and Boolean algebra. It is known that every family of commutative projectors generates a Boolean algebra of projectors. The Boolean algebra plays a great role in the modern science because of its connections with many scientific branches: mathematical logic, the problem of artificial intelligence, computer technologies, bases of theory of probability, etc. G.Boole was creating such algebra of logics (or logical operators), which would reflect inherited laws of human thought. One should note here that some of genetic projectors (which are not described in this article) form commutative pairs; this fact provokes thoughts about Boolean algebras in genetics and bioinformatics and also about genetic basis of logics of human thought. The connection of Boolean algebra with commutative projectors is essential also for development of our concept of geno-lohical coding [Petoukhov, 2016 ; Petoukhov, Petukhova, 2017].

Genetic molecules are subordinated to laws of quantum mechanics, which has begun from matrix mechanics by W. Heisenberg. Till this pioneer work by Heisenberg, matrices were not used in physics. It was very unexpected for scientific community that applying whole ensembles of numbers in a form of matrices can be useful and appropriate to describe natural phenomena and systems. Contemporary science uses matrices widely in many fields, and our work uses matrices to study molecular-genetic systems.

Materials of this article reinforce the author's point of view that living matter in its informational fundamentals is an algebraic essence. The author believes that a development of algebraic biology, elements of which are contained in this and other author's articles, is possible. By analogy with the known fact that molecular foundations of molecular genetics turned up unexpectedly very simple, perhaps algebraic foundations of living matter are also relative simple. In the infinite set of matrices, we find a small subset, which simulates the world of molecular genetic coding with many of its phenomenologic features; this discovery was possible due to studying the family of alphabets in the molecular-genetic system. The matrix-algebraic approach to structured genetic alphabets has led the author, in particular, to the rules of long nucleotide sequences, which are described and discussed in the Section 17.

The new theme of systems of $2^n$-block united-hypercomplex numbers (U-hypercomplex numbers), which are new mathematical tools for modeling different structures, was included in the 7th and 8th versions of this article (see Sections 15 and 16). These numeric systems generalize systems of complex numbers and hypercomplex numbers, which become a particular case of appropriate systems of $2^n$-block united-hypercomplex numbers when only one block in them differs from zero. Hypercomplex numbers, beginning from quaternions of Hamilton, were created as an extension of complex numbers by means of including new independent imaginary units in addition to a single imaginary unit of complex numbers. In our studies of genetic matrices we paid attention on the alternative way of extention of complex numbers: such extension is provided by means of including new independent real units (one or more real units, which we call local-real units) together with their particular complect of imaginary units (local-imaginary units). The world of sparse matrices gives a comfortable possibility of the creation of such U-hypercomplex numbers.

One can remind that the idea of multi-dimensional numbers and multi-dimensional spaces works intensively for a long time in theoretical physics and other fields of science for modeling the phenomena of our physical world. Our results add mathematical formalisms, first of all, into the fields of molecular genetics and bioinformatics. After the discovery of non-Euclidean geometries and of Hamilton quaternions, it is known that different natural systems can possess their own geometry and their own algebra (see about this [Kline, 1980]).

But what kinds of algebra are appropriate for living organisms in their bioinformational aspects? We believe that the algebraic organization of any of living organisms in its inherited bioinformational aspects is based (or can be modelled) on systems of $2^n$-block U-hypercomplex numbers. It seems that many difficulties of modern bioinformatics and mathematical biology are connected with utilizing - for biological structures - inadequate algebras, which were developed for completely other natural systems. Hamilton had similar difficulties in his attempts to describe transformations in a three-dimensional space by means of 3-dimensional numbers while this description needs 4-parametrical quaternions. The author simultaneously hopes that the proposed U-hypercomplex numbers will help to make progress not only in bioinformatics and mathematical biology, where algebraization of biology is under development now, but also in many fields of sciences and technologies. In physics, these U-hypercomplex numbers can lead to new unexpected concepts and theories by generalizing or combining existing concepts and theories. For example, the following question arises: is it possible to connect - in modelling aims - the "wave-particle" duality of quantum-mechanical objects (which are manifested in different measurement experiments or as a particle, or as a wave) with 2-block U-numbers, in which both blocks are mutually related, but in different experiments only one of two blocks manifests itself in its projection onto a measuring device? In other words, for quantum-mechanical objects in this approach different physical experiments manifest characteristics of different blocks of the same 2-block U-number.

Multiblock structure, which is genetically inherited, is one of the most characteristic features of a plurality of living bodies on different levels and branches of biological evolution. The author believes that systems of multiblock U-hypercomplex numbers represent a relevant mathematics for modeling such inherited structures. But non-living substances also consist of various blocks of different levels up to atoms atoms and elementary particles (one of the most important statements in the history of science has been done by Democritus approximately in the fourth century BC: "All bodies are composed of discrete units - atoms"). It is not excluded that the proposed systems of multiblock U-hypercomplex numbers will be also useful in theoretical physics to study block structure of substances.

"*Complexity of a civilization is reflected in complexity of numbers used by this civilization*" [Davis, 1967]. Pythagoras has formulated the idea: "*all things in the world are numbers*" or "*number rules the world*". B. Russell noted that he did not know other person who would exert such influence on thinking of people as Pythagoras. From this viewpoint, there is no more fundamental scientific idea in the world, than this idea about a basic meaning of numbers. Our researches in the field of matrix genetics have led to new systems of multidimensional numbers and have given new materials to the great idea by Pythagoras in its modernized formulating: "*All things are multi-dimensional numbers*".

This article proposes a new mathematical approach to study "*a partnership between genes and mathematics*" (see Section 1 above). In the author's opinion, the proposed kind of mathematics is beautiful and it can be used for further developing of algebraic biology, theoretical physics and informatics in accordance with the famous statement by P. Dirac, who taught that a creation of a physical theory must begin with the beautiful mathematical theory: "*If this theory is really beautiful, then it necessarily will appear as a fine model of important physical phenomena. It is necessary to search for these phenomena to develop applications of the beautiful mathematical theory and to interpret them as predictions of new laws of physics*" (this quotation is taken from [Arnold, 2007]). According to Dirac, all new physics, including relativistic and quantum, are developing in this way. One can suppose that this statement is also true for mathematical biology.

## APPENDIX 1. COMPLEX NUMBERS, CYCLIC GROUPS AND SUMS OF GENETIC PROJECTORS

This Appendix shows a connection between complex numbers and cyclic groups on the base of sums of (8*8)-projectors $s_0+s_2$, $s_0+s_3$, $s_1+s_2$, $s_1+s_3$, $s_4+s_6$, $s_4+s_7$, $s_5+s_6$, $s_5+s_7$ from Fig. 7 on the base of the Rademacher (8*8)-matrix $R_8$ (on Fig. 9 these sums were marked by green color and they corresponded to cyclic groups, if the weight coefficient $2^{-0.5}$ was used for them). Each of these 8 sums can be decomposed into two matrices $e_{2k}$ and $e_{2k+1}$ (k=0, 1, …7), a set of whose is closed relative to multiplication and has a multiplication table, which coincides with the multiplication table of basic elements of complex numbers (Fig. 74). It means that these matrices $e_{2k}$ and $e_{2k+1}$ represent basic elements of complex numbers in corresponding 2-dimensional planes of a 8-dimensional vector space. Sets of matrices $a_{2k}*e_{2k} + a_{2k+1}*e_{2k+1}$ (here $a_{2k}$ and $a_{2k+1}$ are real numbers; each of matrices $e_{2k}$ plays a role of unitary matrix inside the appropriate set $a_{2k}*e_{2k} + a_{2k+1}*e_{2k+1}$) represent complex numbers inside these 2-dimensional planes of the 8-dimensional space.

$$s_0+s_2=\begin{bmatrix}1&0&1&0&0&0&0&0\\1&0&1&0&0&0&0&0\\-1&0&1&0&0&0&0&0\\-1&0&1&0&0&0&0&0\\1&0&-1&0&0&0&0&0\\1&0&-1&0&0&0&0&0\\-1&0&-1&0&0&0&0&0\\-1&0&-1&0&0&0&0&0\end{bmatrix}=e_0+e_1=\begin{bmatrix}1&0&0&0&0&0&0&0\\1&0&0&0&0&0&0&0\\0&0&1&0&0&0&0&0\\0&0&1&0&0&0&0&0\\0&0&-1&0&0&0&0&0\\0&0&-1&0&0&0&0&0\\-1&0&0&0&0&0&0&0\\-1&0&0&0&0&0&0&0\end{bmatrix}+\begin{bmatrix}0&0&1&0&0&0&0&0\\0&0&1&0&0&0&0&0\\-1&0&0&0&0&0&0&0\\-1&0&0&0&0&0&0&0\\1&0&0&0&0&0&0&0\\1&0&0&0&0&0&0&0\\0&0&-1&0&0&0&0&0\\0&0&-1&0&0&0&0&0\end{bmatrix};$$

| | | |
|---|---|---|
| | $e_0$ | $e_1$ |
| $e_0$ | $e_0$ | $e_1$ |
| $e_1$ | $e_1$ | $-e_0$ |

$$s_0+s_3=\begin{bmatrix}1&0&0&1&0&0&0&0\\1&0&0&1&0&0&0&0\\-1&0&0&1&0&0&0&0\\-1&0&0&1&0&0&0&0\\1&0&0&-1&0&0&0&0\\1&0&0&-1&0&0&0&0\\-1&0&0&-1&0&0&0&0\\-1&0&0&-1&0&0&0&0\end{bmatrix}=e_2+e_3=\begin{bmatrix}1&0&0&0&0&0&0&0\\1&0&0&0&0&0&0&0\\0&0&0&1&0&0&0&0\\0&0&0&1&0&0&0&0\\0&0&0&-1&0&0&0&0\\0&0&0&-1&0&0&0&0\\-1&0&0&0&0&0&0&0\\-1&0&0&0&0&0&0&0\end{bmatrix}+\begin{bmatrix}0&0&0&1&0&0&0&0\\0&0&0&1&0&0&0&0\\-1&0&0&0&0&0&0&0\\-1&0&0&0&0&0&0&0\\1&0&0&0&0&0&0&0\\1&0&0&0&0&0&0&0\\0&0&0&-1&0&0&0&0\\0&0&0&-1&0&0&0&0\end{bmatrix};$$

| | | |
|---|---|---|
| | $e_2$ | $e_3$ |
| $e_2$ | $e_2$ | $e_3$ |
| $e_3$ | $e_3$ | $-e_2$ |

$$s_1+s_2=\begin{bmatrix}0&1&1&0&0&0&0&0\\0&1&1&0&0&0&0&0\\0&-1&1&0&0&0&0&0\\0&-1&1&0&0&0&0&0\\0&1&-1&0&0&0&0&0\\0&1&-1&0&0&0&0&0\\0&-1&-1&0&0&0&0&0\\0&-1&-1&0&0&0&0&0\end{bmatrix}=e_4+e_5=\begin{bmatrix}0&1&0&0&0&0&0&0\\0&1&0&0&0&0&0&0\\0&0&1&0&0&0&0&0\\0&0&1&0&0&0&0&0\\0&0&-1&0&0&0&0&0\\0&0&-1&0&0&0&0&0\\0&-1&0&0&0&0&0&0\\0&-1&0&0&0&0&0&0\end{bmatrix}+\begin{bmatrix}0&0&1&0&0&0&0&0\\0&0&1&0&0&0&0&0\\0&-1&0&0&0&0&0&0\\0&-1&0&0&0&0&0&0\\0&1&0&0&0&0&0&0\\0&1&0&0&0&0&0&0\\0&0&-1&0&0&0&0&0\\0&0&-1&0&0&0&0&0\end{bmatrix};$$

| | | |
|---|---|---|
| | $e_4$ | $e_5$ |
| $e_4$ | $e_4$ | $e_5$ |
| $e_5$ | $e_5$ | $-e_4$ |

$$s_1+s_3=\begin{bmatrix}0&1&0&1&0&0&0&0\\0&1&0&1&0&0&0&0\\0&-1&0&1&0&0&0&0\\0&-1&0&1&0&0&0&0\\0&1&0&-1&0&0&0&0\\0&1&0&-1&0&0&0&0\\0&-1&0&-1&0&0&0&0\\0&-1&0&-1&0&0&0&0\end{bmatrix}=e_6+e_7=\begin{bmatrix}0&1&0&0&0&0&0&0\\0&1&0&0&0&0&0&0\\0&0&0&1&0&0&0&0\\0&0&0&1&0&0&0&0\\0&0&0&-1&0&0&0&0\\0&0&0&-1&0&0&0&0\\0&-1&0&0&0&0&0&0\\0&-1&0&0&0&0&0&0\end{bmatrix}+\begin{bmatrix}0&0&0&1&0&0&0&0\\0&0&0&1&0&0&0&0\\0&-1&0&0&0&0&0&0\\0&-1&0&0&0&0&0&0\\0&1&0&0&0&0&0&0\\0&1&0&0&0&0&0&0\\0&0&0&-1&0&0&0&0\\0&0&0&-1&0&0&0&0\end{bmatrix};$$

| | | |
|---|---|---|
| | $e_6$ | $e_7$ |
| $e_6$ | $e_6$ | $e_7$ |
| $e_7$ | $e_7$ | $-e_6$ |

$$s_4+s_6=\begin{bmatrix}0&0&0&0&1&0&-1&0\\0&0&0&0&1&0&-1&0\\0&0&0&0&-1&0&-1&0\\0&0&0&0&-1&0&-1&0\\0&0&0&0&1&0&1&0\\0&0&0&0&1&0&1&0\\0&0&0&0&-1&0&1&0\\0&0&0&0&-1&0&1&0\end{bmatrix}=e_8+e_9=\begin{bmatrix}0&0&0&0&0&0&-1&0\\0&0&0&0&0&0&-1&0\\0&0&0&0&-1&0&0&0\\0&0&0&0&-1&0&0&0\\0&0&0&0&1&0&0&0\\0&0&0&0&1&0&0&0\\0&0&0&0&0&0&1&0\\0&0&0&0&0&0&1&0\end{bmatrix}+\begin{bmatrix}0&0&0&0&1&0&0&0\\0&0&0&0&1&0&0&0\\0&0&0&0&0&0&-1&0\\0&0&0&0&0&0&-1&0\\0&0&0&0&0&0&1&0\\0&0&0&0&0&0&1&0\\0&0&0&0&-1&0&0&0\\0&0&0&0&-1&0&0&0\end{bmatrix};$$

| | | |
|---|---|---|
| | $e_8$ | $e_9$ |
| $e_8$ | $e_8$ | $e_9$ |
| $e_9$ | $e_9$ | $-e_8$ |

$$s_4+s_7=\begin{pmatrix}0&0&0&0&1&0&0&-1\\0&0&0&0&1&0&0&-1\\0&0&0&0&-1&0&0&-1\\0&0&0&0&-1&0&0&-1\\0&0&0&0&1&0&0&1\\0&0&0&0&1&0&0&1\\0&0&0&0&-1&0&0&1\\0&0&0&0&-1&0&0&1\end{pmatrix}=e_{10}+e_{11}=\begin{pmatrix}0&0&0&0&0&0&0&-1\\0&0&0&0&0&0&0&-1\\0&0&0&0&-1&0&0&0\\0&0&0&0&-1&0&0&0\\0&0&0&0&1&0&0&0\\0&0&0&0&1&0&0&0\\0&0&0&0&0&0&0&1\\0&0&0&0&0&0&0&1\end{pmatrix}+\begin{pmatrix}0&0&0&0&1&0&0&0\\0&0&0&0&1&0&0&0\\0&0&0&0&0&0&0&-1\\0&0&0&0&0&0&0&-1\\0&0&0&0&0&0&0&1\\0&0&0&0&0&0&0&1\\0&0&0&0&-1&0&0&0\\0&0&0&0&-1&0&0&0\end{pmatrix};$$

| | $e_{10}$ | $e_{11}$ |
|---|---|---|
| $e_{10}$ | $e_{10}$ | $e_{11}$ |
| $e_{11}$ | $e_{11}$ | $-e_{10}$ |

$$s_5+s_6=\begin{pmatrix}0&0&0&0&0&1&-1&0\\0&0&0&0&0&1&-1&0\\0&0&0&0&0&-1&-1&0\\0&0&0&0&0&-1&-1&0\\0&0&0&0&0&1&1&0\\0&0&0&0&0&1&1&0\\0&0&0&0&0&-1&1&0\\0&0&0&0&0&-1&1&0\end{pmatrix}=e_{12}+e_{13}=\begin{pmatrix}0&0&0&0&0&0&-1&0\\0&0&0&0&0&0&-1&0\\0&0&0&0&0&-1&0&0\\0&0&0&0&0&-1&0&0\\0&0&0&0&0&1&0&0\\0&0&0&0&0&1&0&0\\0&0&0&0&0&0&1&0\\0&0&0&0&0&0&1&0\end{pmatrix}+\begin{pmatrix}0&0&0&0&0&1&0&0\\0&0&0&0&0&1&0&0\\0&0&0&0&0&0&-1&0\\0&0&0&0&0&0&-1&0\\0&0&0&0&0&0&1&0\\0&0&0&0&0&0&1&0\\0&0&0&0&0&-1&0&0\\0&0&0&0&0&-1&0&0\end{pmatrix};$$

| | $e_{12}$ | $e_{13}$ |
|---|---|---|
| $e_{12}$ | $e_{12}$ | $e_{13}$ |
| $e_{13}$ | $e_{13}$ | $-e_{12}$ |

$$s_5+s_7=\begin{pmatrix}0&0&0&0&0&1&0&-1\\0&0&0&0&0&1&0&-1\\0&0&0&0&0&-1&0&-1\\0&0&0&0&0&-1&0&-1\\0&0&0&0&0&1&0&1\\0&0&0&0&0&1&0&1\\0&0&0&0&0&-1&0&1\\0&0&0&0&0&-1&0&1\end{pmatrix}=e_{14}+e_{15}=\begin{pmatrix}0&0&0&0&0&0&0&-1\\0&0&0&0&0&0&0&-1\\0&0&0&0&0&-1&0&0\\0&0&0&0&0&-1&0&0\\0&0&0&0&0&1&0&0\\0&0&0&0&0&1&0&0\\0&0&0&0&0&0&0&1\\0&0&0&0&0&0&0&1\end{pmatrix}+\begin{pmatrix}0&0&0&0&0&1&0&0\\0&0&0&0&0&1&0&0\\0&0&0&0&0&0&0&-1\\0&0&0&0&0&0&0&-1\\0&0&0&0&0&0&0&1\\0&0&0&0&0&0&0&1\\0&0&0&0&0&-1&0&0\\0&0&0&0&0&-1&0&0\end{pmatrix};$$

| | $e_{14}$ | $e_{15}$ |
|---|---|---|
| $e_{14}$ | $e_{14}$ | $e_{15}$ |
| $e_{15}$ | $e_{15}$ | $-e_{14}$ |

Fig. 74. The decomposition of each of (8*8)-matrices $s_0+s_2$, $s_0+s_3$, $s_1+s_2$, $s_1+s_3$, $s_4+s_6$, $s_4+s_7$, $s_5+s_6$, $s_5+s_7$ from Fig. 7 into a set of two matrices $e_{2k}$ and $e_{2k+1}$ (k=0, 1, …7), a set of whose is closed relative to multiplication and gives the multiplication table of complex numbers (on the right)

In a general case, the described approach allows constructing selective operators of a $2^n$-dimensional vector space with a set of different 2-dimensional planes, each of which can contain a function of complex numbers (parameters of these functions in different planes can be independent or interrelated). Such selective operator allows simulating a combinatory behaviour of a multi-parametric system, which contains different 2-parametric subsystems, whose independent or interrelated behaviours can be simulated by means of functions of complex numbers. If these functions of complex numbers are cyclic, such selective operator describes a behaviour of a multi-parametric system, which contains an appropriate ensemble of 2-parametric subsystems with cyclic behaviours.

## APPENDIX 2. HYPERBOLIC NUMBERS AND SUMS OF GENETIC PROJECTORS

Let us show now a connection between hyperbolic numbers and sums of (8*8)-projectors $s_0+s_1$, $s_0+s_4$, $s_0+s_5$, $s_1+s_4$, $s_1+s_5$, $s_2+s_3$, $s_2+s_6$, $s_2+s_7$, $s_3+s_6$, $s_3+s_7$, $s_4+s_5$, $s_6+s_7$ from Fig. 7 on the base of the Rademacher (8*8)-matrix $R_8$ (on Fig. 9 these sums were marked by red color). Each of these 12 sums can be decomposed into two matrices $j_{2k}$ and $j_{2k+1}$ (k=0, 1, 2, …, 11), a set of whose is closed relative to multiplication and has a multiplication table, which coincides with the multiplication table of basic elements of hyperbolic numbers (Fig. 75). It means that these matrices $j_{2k}$ and $j_{2k+1}$ represent basic elements of hyperbolic numbers in corresponding 2-dimensional planes of a 8-dimensional vector space. Sets of matrices $a_{2k}*j_{2k} + a_{2k+1}*j_{2k+1}$ (here $a_{2k}$ and $a_{2k+1}$ are real numbers; each of matrices $j_{2k}$ plays a role of unitary matrix inside the appropriate set $a_{2k}*j_{2k} + a_{2k+1}*j_{2k+1}$) represent hyperbolic numbers inside these 2-dimensional planes of the 8-dimensional space.

$$s_0+s_1 = \begin{bmatrix} 1&1&0&0&0&0&0&0\\ 1&1&0&0&0&0&0&0\\ -1&-1&0&0&0&0&0&0\\ -1&-1&0&0&0&0&0&0\\ 1&1&0&0&0&0&0&0\\ 1&1&0&0&0&0&0&0\\ -1&-1&0&0&0&0&0&0\\ -1&-1&0&0&0&0&0&0 \end{bmatrix} = j_0+j_1= \begin{bmatrix} 1&0&0&0&0&0&0&0\\ 0&1&0&0&0&0&0&0\\ -1&0&0&0&0&0&0&0\\ 0&-1&0&0&0&0&0&0\\ 1&0&0&0&0&0&0&0\\ 0&1&0&0&0&0&0&0\\ -1&0&0&0&0&0&0&0\\ 0&-1&0&0&0&0&0&0 \end{bmatrix} + \begin{bmatrix} 0&1&0&0&0&0&0&0\\ 1&0&0&0&0&0&0&0\\ 0&-1&0&0&0&0&0&0\\ -1&0&0&0&0&0&0&0\\ 0&1&0&0&0&0&0&0\\ 1&0&0&0&0&0&0&0\\ 0&-1&0&0&0&0&0&0\\ -1&0&0&0&0&0&0&0 \end{bmatrix};$$

| | $j_0$ | $j_1$ |
|---|---|---|
| $j_0$ | $j_0$ | $j_1$ |
| $j_1$ | $j_1$ | $j_0$ |

$$s_0+s_4 = \begin{bmatrix} 1&0&0&0&1&0&0&0\\ 1&0&0&0&1&0&0&0\\ -1&0&0&0&-1&0&0&0\\ -1&0&0&0&-1&0&0&0\\ 1&0&0&0&1&0&0&0\\ 1&0&0&0&1&0&0&0\\ -1&0&0&0&-1&0&0&0\\ -1&0&0&0&-1&0&0&0 \end{bmatrix} = j_2+j_3= \begin{bmatrix} 1&0&0&0&0&0&0&0\\ 1&0&0&0&0&0&0&0\\ 0&0&0&0&-1&0&0&0\\ 0&0&0&0&-1&0&0&0\\ 0&0&0&0&1&0&0&0\\ 0&0&0&0&1&0&0&0\\ -1&0&0&0&0&0&0&0\\ -1&0&0&0&0&0&0&0 \end{bmatrix} + \begin{bmatrix} 0&0&0&0&1&0&0&0\\ 0&0&0&0&1&0&0&0\\ -1&0&0&0&0&0&0&0\\ -1&0&0&0&0&0&0&0\\ 1&0&0&0&0&0&0&0\\ 1&0&0&0&0&0&0&0\\ 0&0&0&0&-1&0&0&0\\ 0&0&0&0&-1&0&0&0 \end{bmatrix};$$

| | $j_2$ | $j_3$ |
|---|---|---|
| $j_2$ | $j_2$ | $j_3$ |
| $j_3$ | $j_3$ | $j_2$ |

$$s_0+s_5 = \begin{bmatrix} 1&0&0&0&0&1&0&0\\ 1&0&0&0&0&1&0&0\\ -1&0&0&0&0&-1&0&0\\ -1&0&0&0&0&-1&0&0\\ 1&0&0&0&0&1&0&0\\ 1&0&0&0&0&1&0&0\\ -1&0&0&0&0&-1&0&0\\ -1&0&0&0&0&-1&0&0 \end{bmatrix} = j_4+j_5= \begin{bmatrix} 1&0&0&0&0&0&0&0\\ 1&0&0&0&0&0&0&0\\ 0&0&0&0&0&-1&0&0\\ 0&0&0&0&0&-1&0&0\\ 0&0&0&0&0&1&0&0\\ 0&0&0&0&0&1&0&0\\ -1&0&0&0&0&0&0&0\\ -1&0&0&0&0&0&0&0 \end{bmatrix} + \begin{bmatrix} 0&0&0&0&0&1&0&0\\ 0&0&0&0&0&1&0&0\\ -1&0&0&0&0&0&0&0\\ -1&0&0&0&0&0&0&0\\ 1&0&0&0&0&0&0&0\\ 1&0&0&0&0&0&0&0\\ 0&0&0&0&0&-1&0&0\\ 0&0&0&0&0&-1&0&0 \end{bmatrix};$$

| | $j_4$ | $j_5$ |
|---|---|---|
| $j_4$ | $j_4$ | $j_5$ |
| $j_5$ | $j_5$ | $j_4$ |

$$s_1+s_4 = \begin{bmatrix} 0&1&0&0&1&0&0&0\\ 0&1&0&0&1&0&0&0\\ 0&-1&0&0&-1&0&0&0\\ 0&-1&0&0&-1&0&0&0 \end{bmatrix} = j_6+j_7= \begin{bmatrix} 0&1&0&0&0&0&0&0\\ 0&1&0&0&0&0&0&0\\ 0&0&0&0&-1&0&0&0\\ 0&0&0&0&-1&0&0&0 \end{bmatrix} + \begin{bmatrix} 0&0&0&0&1&0&0&0\\ 0&0&0&0&1&0&0&0\\ 0&-1&0&0&0&0&0&0\\ 0&-1&0&0&0&0&0&0 \end{bmatrix};$$

| | $j_6$ | $j_7$ |
|---|---|---|

$$\begin{array}{|cccccccc|}\hline 0&1&0&0&1&0&0&0\\0&1&0&0&1&0&0&0\\0&-1&0&0&-1&0&0&0\\0&-1&0&0&-1&0&0&0\\\hline\end{array}\qquad\begin{array}{|cccccccc|}\hline 0&0&0&0&1&0&0&0\\0&0&0&0&1&0&0&0\\0&-1&0&0&0&0&0&0\\0&-1&0&0&0&0&0&0\\\hline\end{array}\qquad\begin{array}{|cccccccc|}\hline 0&1&0&0&0&0&0&0\\0&1&0&0&0&0&0&0\\0&0&0&0&-1&0&0&0\\0&0&0&0&-1&0&0&0\\\hline\end{array}$$

| | | |
|---|---|---|
| $j_6$ | $j_6$ | $j_7$ |
| $j_7$ | $j_7$ | $j_6$ |

$$s_1+s_5=\begin{array}{|cccccccc|}\hline 0&1&0&0&0&1&0&0\\0&1&0&0&0&1&0&0\\0&-1&0&0&0&-1&0&0\\0&-1&0&0&0&-1&0&0\\0&1&0&0&0&1&0&0\\0&1&0&0&0&1&0&0\\0&-1&0&0&0&-1&0&0\\0&-1&0&0&0&-1&0&0\\\hline\end{array}=j_8+j_9=\begin{array}{|cccccccc|}\hline 0&1&0&0&0&0&0&0\\0&1&0&0&0&0&0&0\\0&0&0&0&0&-1&0&0\\0&0&0&0&0&-1&0&0\\0&0&0&0&0&1&0&0\\0&0&0&0&0&1&0&0\\0&-1&0&0&0&0&0&0\\0&-1&0&0&0&0&0&0\\\hline\end{array}+\begin{array}{|cccccccc|}\hline 0&0&0&0&0&1&0&0\\0&0&0&0&0&1&0&0\\0&-1&0&0&0&0&0&0\\0&-1&0&0&0&0&0&0\\0&1&0&0&0&0&0&0\\0&1&0&0&0&0&0&0\\0&0&0&0&0&-1&0&0\\0&0&0&0&0&-1&0&0\\\hline\end{array};$$

| | | |
|---|---|---|
| | $j_8$ | $j_9$ |
| $j_8$ | $j_8$ | $j_9$ |
| $j_9$ | $j_9$ | $j_8$ |

$$s_2+s_3=\begin{array}{|cccccccc|}\hline 0&0&1&1&0&0&0&0\\0&0&1&1&0&0&0&0\\0&0&1&1&0&0&0&0\\0&0&1&1&0&0&0&0\\0&0&-1&-1&0&0&0&0\\0&0&-1&-1&0&0&0&0\\0&0&-1&-1&0&0&0&0\\0&0&-1&-1&0&0&0&0\\\hline\end{array}=j_{10}+j_{11}=\begin{array}{|cccccccc|}\hline 0&0&1&0&0&0&0&0\\0&0&0&1&0&0&0&0\\0&0&1&0&0&0&0&0\\0&0&0&1&0&0&0&0\\0&0&-1&0&0&0&0&0\\0&0&0&-1&0&0&0&0\\0&0&-1&0&0&0&0&0\\0&0&0&-1&0&0&0&0\\\hline\end{array}+\begin{array}{|cccccccc|}\hline 0&0&0&1&0&0&0&0\\0&0&1&0&0&0&0&0\\0&0&0&1&0&0&0&0\\0&0&1&0&0&0&0&0\\0&0&0&-1&0&0&0&0\\0&0&-1&0&0&0&0&0\\0&0&0&-1&0&0&0&0\\0&0&-1&0&0&0&0&0\\\hline\end{array};$$

| | | |
|---|---|---|
| | $j_{10}$ | $j_{11}$ |
| $j_{10}$ | $j_{10}$ | $j_{11}$ |
| $j_{11}$ | $j_{11}$ | $j_{10}$ |

$$s_2+s_6=\begin{array}{|cccccccc|}\hline 0&0&1&0&0&0&-1&0\\0&0&1&0&0&0&-1&0\\0&0&1&0&0&0&-1&0\\0&0&1&0&0&0&-1&0\\0&0&-1&0&0&0&1&0\\0&0&-1&0&0&0&1&0\\0&0&-1&0&0&0&1&0\\0&0&-1&0&0&0&1&0\\\hline\end{array}=j_{12}+j_{13}=\begin{array}{|cccccccc|}\hline 0&0&1&0&0&0&0&0\\0&0&1&0&0&0&0&0\\0&0&0&0&0&0&-1&0\\0&0&0&0&0&0&-1&0\\0&0&0&0&0&0&1&0\\0&0&0&0&0&0&1&0\\0&0&-1&0&0&0&0&0\\0&0&-1&0&0&0&0&0\\\hline\end{array}+\begin{array}{|cccccccc|}\hline 0&0&0&0&0&0&-1&0\\0&0&0&0&0&0&-1&0\\0&0&1&0&0&0&0&0\\0&0&1&0&0&0&0&0\\0&0&-1&0&0&0&0&0\\0&0&-1&0&0&0&0&0\\0&0&0&0&0&0&1&0\\0&0&0&0&0&0&1&0\\\hline\end{array};$$

| | | |
|---|---|---|
| | $j_{12}$ | $j_{13}$ |
| $j_{12}$ | $j_{12}$ | $j_{13}$ |
| $j_{13}$ | $j_{13}$ | $j_{12}$ |

$$s_2+s_7=\begin{array}{|cccccccc|}\hline 0&0&1&0&0&0&0&-1\\0&0&1&0&0&0&0&-1\\0&0&1&0&0&0&0&-1\\0&0&1&0&0&0&0&-1\\0&0&-1&0&0&0&0&1\\0&0&-1&0&0&0&0&1\\0&0&-1&0&0&0&0&1\\0&0&-1&0&0&0&0&1\\\hline\end{array}=j_{14}+j_{15}=\begin{array}{|cccccccc|}\hline 0&0&1&0&0&0&0&0\\0&0&1&0&0&0&0&0\\0&0&0&0&0&0&0&-1\\0&0&0&0&0&0&0&-1\\0&0&0&0&0&0&0&1\\0&0&0&0&0&0&0&1\\0&0&-1&0&0&0&0&0\\0&0&-1&0&0&0&0&0\\\hline\end{array}+\begin{array}{|cccccccc|}\hline 0&0&0&0&0&0&0&-1\\0&0&0&0&0&0&0&-1\\0&0&1&0&0&0&0&0\\0&0&1&0&0&0&0&0\\0&0&-1&0&0&0&0&0\\0&0&-1&0&0&0&0&0\\0&0&0&0&0&0&0&1\\0&0&0&0&0&0&0&1\\\hline\end{array};$$

| | | |
|---|---|---|
| | $j_{14}$ | $j_{15}$ |
| $j_{14}$ | $j_{14}$ | $j_{15}$ |
| $j_{15}$ | $j_{15}$ | $j_{14}$ |

$$s_3+s_6=\begin{array}{|cccccccc|}\hline 0&0&0&1&0&0&-1&0\\0&0&0&1&0&0&-1&0\\0&0&0&1&0&0&-1&0\\0&0&0&1&0&0&-1&0\\\hline\end{array}=j_{16}+j_{17}=\begin{array}{|cccccccc|}\hline 0&0&0&1&0&0&0&0\\0&0&0&1&0&0&0&0\\0&0&0&0&0&0&-1&0\\0&0&0&0&0&0&-1&0\\\hline\end{array}+\begin{array}{|cccccccc|}\hline 0&0&0&0&0&0&-1&0\\0&0&0&0&0&0&-1&0\\0&0&0&1&0&0&0&0\\0&0&0&1&0&0&0&0\\\hline\end{array};$$

| | | |
|---|---|---|
| | $j_{16}$ | $j_{17}$ |

$$\begin{bmatrix} 0&0&0&-1&0&0&1&0\\ 0&0&0&-1&0&0&1&0\\ 0&0&0&-1&0&0&1&0\\ 0&0&0&-1&0&0&1&0 \end{bmatrix} \quad \begin{bmatrix} 0&0&0&0&0&0&1&0\\ 0&0&0&0&0&0&1&0\\ 0&0&0&-1&0&0&0&0\\ 0&0&0&-1&0&0&0&0 \end{bmatrix} \quad \begin{bmatrix} 0&0&0&-1&0&0&0&0\\ 0&0&0&-1&0&0&0&0\\ 0&0&0&0&0&0&1&0\\ 0&0&0&0&0&0&1&0 \end{bmatrix}$$

| | | |
|---|---|---|
| $j_{16}$ | $j_{16}$ | $j_{17}$ |
| $j_{17}$ | $j_{17}$ | $j_{16}$ |

$$s_3+s_7 = \begin{bmatrix} 0&0&0&1&0&0&0&-1\\ 0&0&0&1&0&0&0&-1\\ 0&0&0&1&0&0&0&-1\\ 0&0&0&1&0&0&0&-1\\ 0&0&0&-1&0&0&0&1\\ 0&0&0&-1&0&0&0&1\\ 0&0&0&-1&0&0&0&1\\ 0&0&0&-1&0&0&0&1 \end{bmatrix} = j_{18}+j_{19} = \begin{bmatrix} 0&0&0&1&0&0&0&0\\ 0&0&0&1&0&0&0&0\\ 0&0&0&0&0&0&0&-1\\ 0&0&0&0&0&0&0&-1\\ 0&0&0&0&0&0&0&1\\ 0&0&0&0&0&0&0&1\\ 0&0&0&-1&0&0&0&0\\ 0&0&0&-1&0&0&0&0 \end{bmatrix} + \begin{bmatrix} 0&0&0&0&0&0&0&-1\\ 0&0&0&0&0&0&0&-1\\ 0&0&0&1&0&0&0&0\\ 0&0&0&1&0&0&0&0\\ 0&0&0&-1&0&0&0&0\\ 0&0&0&-1&0&0&0&0\\ 0&0&0&0&0&0&0&1\\ 0&0&0&0&0&0&0&1 \end{bmatrix};$$

| | $j_{18}$ | $j_{19}$ |
|---|---|---|
| $j_{18}$ | $j_{18}$ | $j_{19}$ |
| $j_{19}$ | $j_{19}$ | $j_{18}$ |

$$s_4+s_5 = \begin{bmatrix} 0&0&0&0&1&1&0&0\\ 0&0&0&0&1&1&0&0\\ 0&0&0&0&-1&-1&0&0\\ 0&0&0&0&-1&-1&0&0\\ 0&0&0&0&1&1&0&0\\ 0&0&0&0&1&1&0&0\\ 0&0&0&0&-1&-1&0&0\\ 0&0&0&0&-1&-1&0&0 \end{bmatrix} = j_{20}+j_{21} = \begin{bmatrix} 0&0&0&0&1&0&0&0\\ 0&0&0&0&0&1&0&0\\ 0&0&0&0&-1&0&0&0\\ 0&0&0&0&0&-1&0&0\\ 0&0&0&0&1&0&0&0\\ 0&0&0&0&0&1&0&0\\ 0&0&0&0&-1&0&0&0\\ 0&0&0&0&0&-1&0&0 \end{bmatrix} + \begin{bmatrix} 0&0&0&0&0&1&0&0\\ 0&0&0&0&1&0&0&0\\ 0&0&0&0&0&-1&0&0\\ 0&0&0&0&-1&0&0&0\\ 0&0&0&0&0&1&0&0\\ 0&0&0&0&1&0&0&0\\ 0&0&0&0&0&-1&0&0\\ 0&0&0&0&-1&0&0&0 \end{bmatrix};$$

| | $j_{20}$ | $j_{21}$ |
|---|---|---|
| $j_{20}$ | $j_{20}$ | $j_{21}$ |
| $j_{21}$ | $j_{21}$ | $j_{20}$ |

$$s_6+s_7 = \begin{bmatrix} 0&0&0&0&0&0&-1&-1\\ 0&0&0&0&0&0&-1&-1\\ 0&0&0&0&0&0&-1&-1\\ 0&0&0&0&0&0&-1&-1\\ 0&0&0&0&0&0&1&1\\ 0&0&0&0&0&0&1&1\\ 0&0&0&0&0&0&1&1\\ 0&0&0&0&0&0&1&1 \end{bmatrix} = j_{22}+j_{23} = \begin{bmatrix} 0&0&0&0&0&0&-1&0\\ 0&0&0&0&0&0&0&-1\\ 0&0&0&0&0&0&-1&0\\ 0&0&0&0&0&0&0&-1\\ 0&0&0&0&0&0&1&0\\ 0&0&0&0&0&0&0&1\\ 0&0&0&0&0&0&1&0\\ 0&0&0&0&0&0&0&1 \end{bmatrix} + \begin{bmatrix} 0&0&0&0&0&0&0&-1\\ 0&0&0&0&0&0&-1&0\\ 0&0&0&0&0&0&0&-1\\ 0&0&0&0&0&0&-1&0\\ 0&0&0&0&0&0&0&1\\ 0&0&0&0&0&0&1&0\\ 0&0&0&0&0&0&0&1\\ 0&0&0&0&0&0&1&0 \end{bmatrix};$$

| | $j_{22}$ | $j_{23}$ |
|---|---|---|
| $j_{22}$ | $j_{22}$ | $j_{23}$ |
| $j_{23}$ | $j_{23}$ | $j_{22}$ |

Fig. 75. The decomposition of each of (8*8)-matrices $s_0+s_1$, $s_0+s_4$, $s_0+s_5$, $s_1+s_4$, $s_1+s_5$, $s_2+s_3$, $s_2+s_6$, $s_2+s_7$, $s_3+s_6$, $s_3+s_7$, $s_4+s_5$, $s_6+s_7$ from Fig. 7 into a set of two matrices $j_{2k}$ and $j_{2k+1}$ (k=0, 1, …, 11), a set of whose is closed relative to multiplication and gives the multiplication table of hyperbolic numbers (on the right)

## APPENDIX 3. ANOTHER TENSOR FAMILY OF GENETIC HADAMARD MATRICES

Hadamard matrices are well-known in noise-immunity coding, quantum mechanics, etc. Their rows are Walsh functions, which are widely used in radiocommunication for a code division in systems with many channels, etc., for example, in cellular standards such as IS-95, CDMA2000 or UMTS. Walsh functions and corresponding series and transforms find various applications in physics and engineereing, in particular, in digital signal processing. They are used in speech recognition, in medical and biological image processing, in digital holography, and other areas.

Above we have described the variant of the relation of the molecular-genetic system with the special tensor family of Hadamard matrices $H_4$, $H_8$, etc. (Fig. 1 and 38). This variant uses the phenomenological fact of existence of triplets with strong and weak roots; in other words this variant is based on a specifity of the degeneracy of the genetic code. This Appendix shows second variant of a relation of the genetic alphabets with another tensor family of Hadamard matrices. This new family of genetic Hadamard matrices is based only on properties of the genetic alphabet A, C, G, T and doesn't depend on the degeneracy of the genetic code. On the author's opinion, this new variant is more interesting and fundamental for further using in future genetic researches.

Fig. 76 shows the beginning of this new tensor family of Hadamard matrices $P^{(n)} = [1\ 1; -1\ 1]^{(n)}$ together with genetic matrices $[C\ T; A\ G]^{(m)}$ (m=1, 2, 3) of monoplets, duplets and triplets with their black-and-white mosaics which coincide with mosaics of the Hadamard matrices.

$P_2 =$

| | |
|---|---|
| 1 | -1 |
| 1 | 1 |

; $P_4 =$

| | | | |
|---|---|---|---|
| 1 | -1 | -1 | 1 |
| 1 | 1 | -1 | -1 |
| 1 | -1 | 1 | -1 |
| 1 | 1 | 1 | 1 |

; $P_8 =$

| | | | | | | | |
|---|---|---|---|---|---|---|---|
| 1 | -1 | -1 | 1 | -1 | 1 | 1 | -1 |
| 1 | 1 | -1 | -1 | -1 | -1 | 1 | 1 |
| 1 | -1 | 1 | -1 | -1 | 1 | -1 | 1 |
| 1 | 1 | 1 | 1 | -1 | -1 | -1 | -1 |
| 1 | -1 | -1 | 1 | 1 | -1 | -1 | 1 |
| 1 | 1 | -1 | -1 | 1 | 1 | -1 | -1 |
| 1 | -1 | 1 | -1 | 1 | -1 | 1 | -1 |
| 1 | 1 | 1 | 1 | 1 | 1 | 1 | 1 |

| | |
|---|---|
| C | T |
| A | G |

;

| | | | |
|---|---|---|---|
| CC | CT | TC | TT |
| CA | CG | TA | TG |
| AC | AT | GC | GT |
| AA | AG | GA | GG |

;

| | | | | | | | |
|---|---|---|---|---|---|---|---|
| CCC | CCT | CTC | CTT | TCC | TCT | TTC | TTT |
| CCA | CCG | CTA | CTG | TCA | TCG | TTA | TTG |
| CAC | CAT | CGC | CGT | TAC | TAT | TGC | TGT |
| CAA | CAG | CGA | CGG | TAA | TAG | TGA | TGG |
| ACC | ACT | ATC | ATT | GCC | GCT | GTC | GTT |
| ACA | ACG | ATA | ATG | GCA | GCG | GTA | GTG |
| AAC | AAT | AGC | AGT | GAC | GAT | GGC | GGT |
| AAA | AAG | AGA | AGG | GAA | GAG | GGA | GGG |

Fig. 76. Upper level: the beginning of the tensor family of Hadamard matrices $P^{(n)} = [1\ 1; -1\ 1]^{(n)}$, where $P_4 = [1\ 1; -1\ 1]^{(2)}$ and $P_8 = [1\ 1; -1\ 1]^{(3)}$. Bottom level: genetic matrices [C T; A G], $[C\ T; A\ G]^{(2)}$ and $[C\ T; A\ G]^{(3)}$ of monoplets, duplets and triplets with their black-and-white mosaics, which coincide with mosaics of the Hadamard matrices $P_2$, $P_4$, $P_8$.

On Fig. 76 the black-and-white matrices $P_2$, $P_4$, $P_8$ are Hadamard matrices because they satisfy the criterium $H_n*H_n^T=n*E$ [Ahmed, Rao,1975]. The genetic matriices [C T; A G], $[C\ T; A\ G]^{(2)}$ and $[C\ T; A\ G]^{(3)}$ algorithmically have the same black-and-white mosaics on the base of fundamental properties of the DNA alphabet (adenine A, cytosine C, guanine G and thymine T). These properties contrapose the letter T against three other letters of the DNA-alphabet by the following phenomenological facts:

- the thymine T is a single nitrogenous base in DNA which is replaced in RNA by another nitrogenous base U (uracil) for unknown reason;

- the thymine T is a single nitrogenous base in DNA without the amino-group $NH_2$ (Fig. 77), which plays an important role in molecular genetics. For instance, the amino-group in amino acids provides a function of recognition of the amino acids by ferments [Chapeville, Haenni, 1974]. A detachment of amino-groups in nitrogenous bases A and C in RNA under action of nitrous acid $HNO_2$ determines a property of amino-mutating of these bases, which was used to divide the set of 64 triplets into eight natural subsets with 8 triplets in each [Wittmann, 1961].

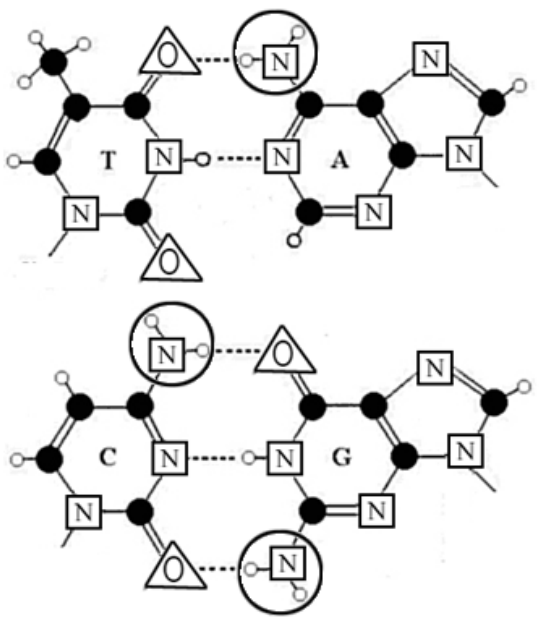

Fig. 77. The complementary pairs of the four nitrogenous bases in DNA. A-T (adenine and thymine), C-G (cytosine and guanine). Amino-groups $NH_2$ are marked by big circles. Black circles are atoms of carbon; small white circles are atoms of hydrogen; squares with the letter N are atoms of nitrogen; triangles with the letter O are atoms of oxygen.

From the point of view of these two facts, the letters A, C, G are identical to each other and the letter T is opposite to them. Correspondingly this binary-oppositional division inside the DNA-alphabet can be reflected by the symbol "+1" for each of the letters A, C, G and by the opposite symbol "-1" for the letter T. Concerning the genetic matrices $[C\ T;\ A\ G]^{(n)}$ this approach leads to a simple algorithm for assigning a sign «+1» or «-1» to each of multiplets (monoplets, duplets, triplets, etc.): each multiplet is considered as a product of the signs «+1» or «-1» of its letters (A=C=G=1, T=-1). For example, the triplet CTG has the sign «-1» because 1*(-1)*1=-1; the triplet CTT has the sign «+1» because 1*(-1)*(-1)=+1, etc. Inside of the genetic matrices $[C\ T;\ A\ G]^{(n)}$ on Fig. 76, all multiplets with the symbol «+1» are denoted by black color and all multiplets with the symbol «-1» are denoted by white color. In the result we have the connection of the genetic matrices $[C\ T;\ A\ G]^{(n)}$ with Hadamard matrices $P^{(n)} = [1\ \ 1;\ -1\ \ 1]^{(n)}$ on the base of fundamental molecular-genetic properties, which can be used in genetic computers of living organisms.

The Hadamard matrices $P_4$ and $P_8$ (Fig. 76) can be correspondingly decomposed into 4 and 8 sparse matrices by analogy with the «column» decompositions of the Hadamard matrices $H_4$ and $H_8$ on Fig. 10 and 14. Results of such decompositions are shown on Fig. 78 and 79.

$$P_4 = K_0+K_1+K_2+K_3 = \begin{vmatrix} 1&0&0&0\\ 1&0&0&0\\ 1&0&0&0\\ 1&0&0&0 \end{vmatrix} + \begin{vmatrix} 0&-1&0&0\\ 0&1&0&0\\ 0&-1&0&0\\ 0&1&0&0 \end{vmatrix} + \begin{vmatrix} 0&0&-1&0\\ 0&0&-1&0\\ 0&0&1&0\\ 0&0&1&0 \end{vmatrix} + \begin{vmatrix} 0&0&0&1\\ 0&0&0&-1\\ 0&0&0&-1\\ 0&0&0&1 \end{vmatrix}$$

Fig. 78. The «column» decomposition of the Hadamard matrix $P_4$ from Fig. 48 into 4 sparse matrices $K_0$, $K_1$, $K_2$ and $K_3$, each of which is a projector.

$$P_8=J_0+J_1+J_2+J_3+J_4+J_5+J_6+J_7 = \begin{vmatrix} 1&0&0&0&0&0&0&0\\ 1&0&0&0&0&0&0&0\\ 1&0&0&0&0&0&0&0\\ 1&0&0&0&0&0&0&0\\ 1&0&0&0&0&0&0&0\\ 1&0&0&0&0&0&0&0\\ 1&0&0&0&0&0&0&0\\ 1&0&0&0&0&0&0&0 \end{vmatrix} + \begin{vmatrix} 0&-1&0&0&0&0&0&0\\ 0&1&0&0&0&0&0&0\\ 0&-1&0&0&0&0&0&0\\ 0&1&0&0&0&0&0&0\\ 0&-1&0&0&0&0&0&0\\ 0&1&0&0&0&0&0&0\\ 0&-1&0&0&0&0&0&0\\ 0&1&0&0&0&0&0&0 \end{vmatrix} +$$

$$\begin{vmatrix} 0&0&-1&0&0&0&0&0\\ 0&0&-1&0&0&0&0&0\\ 0&0&1&0&0&0&0&0\\ 0&0&1&0&0&0&0&0\\ 0&0&-1&0&0&0&0&0\\ 0&0&-1&0&0&0&0&0\\ 0&0&1&0&0&0&0&0\\ 0&0&1&0&0&0&0&0 \end{vmatrix} + \begin{vmatrix} 0&0&0&1&0&0&0&0\\ 0&0&0&-1&0&0&0&0\\ 0&0&0&-1&0&0&0&0\\ 0&0&0&1&0&0&0&0\\ 0&0&0&1&0&0&0&0\\ 0&0&0&-1&0&0&0&0\\ 0&0&0&-1&0&0&0&0\\ 0&0&0&1&0&0&0&0 \end{vmatrix} + \begin{vmatrix} 0&0&0&0&-1&0&0&0\\ 0&0&0&0&-1&0&0&0\\ 0&0&0&0&-1&0&0&0\\ 0&0&0&0&-1&0&0&0\\ 0&0&0&0&1&0&0&0\\ 0&0&0&0&1&0&0&0\\ 0&0&0&0&1&0&0&0\\ 0&0&0&0&1&0&0&0 \end{vmatrix} +$$

$$\begin{vmatrix} 0&0&0&0&0&1&0&0\\ 0&0&0&0&0&-1&0&0\\ 0&0&0&0&0&1&0&0\\ 0&0&0&0&0&-1&0&0\\ 0&0&0&0&0&-1&0&0\\ 0&0&0&0&0&1&0&0\\ 0&0&0&0&0&-1&0&0\\ 0&0&0&0&0&1&0&0 \end{vmatrix} + \begin{vmatrix} 0&0&0&0&0&0&1&0\\ 0&0&0&0&0&0&1&0\\ 0&0&0&0&0&0&-1&0\\ 0&0&0&0&0&0&-1&0\\ 0&0&0&0&0&0&-1&0\\ 0&0&0&0&0&0&-1&0\\ 0&0&0&0&0&0&1&0\\ 0&0&0&0&0&0&1&0 \end{vmatrix} + \begin{vmatrix} 0&0&0&0&0&0&0&-1\\ 0&0&0&0&0&0&0&1\\ 0&0&0&0&0&0&0&1\\ 0&0&0&0&0&0&0&-1\\ 0&0&0&0&0&0&0&1\\ 0&0&0&0&0&0&0&-1\\ 0&0&0&0&0&0&0&-1\\ 0&0&0&0&0&0&0&1 \end{vmatrix}$$

Fig. 79. The «column» decomposition of the Hadamard matrix $P_8$ from Fig. 76 into 8 sparse matrices $J_0$, $J_1$, $J_2$, $J_3$, $J_4$, $J_5$, $J_6$ and $J_7$, each of which is a projector.

Each of these sparse matrices $K_0$, $K_1$, $K_2$, $K_3$ and $J_0$, $J_1$, .., $J_7$ is a projector. Now one can analyze properties of sums of pairs of these new projectors in relation to their exponentiation: $(K_0+K_1)^n$, $(K_0+K_2)^n$, …, $(K_2+K_3)^n$, $(J_0+J_1)^n$, $(J_0+J_2)^n$, …, $(J_6+J_7)^n$. By analogy with the similar analysis of sums of projectors from the first variant of the Hadamard matrices $H_4$ and $H_8$ (Fig. 1, 10-14, 16 and 38), we have got results of such analysis shown in tabular forms on Fig. 80 in the case of the Hadamard matrices $P_4$ and $P_8$.

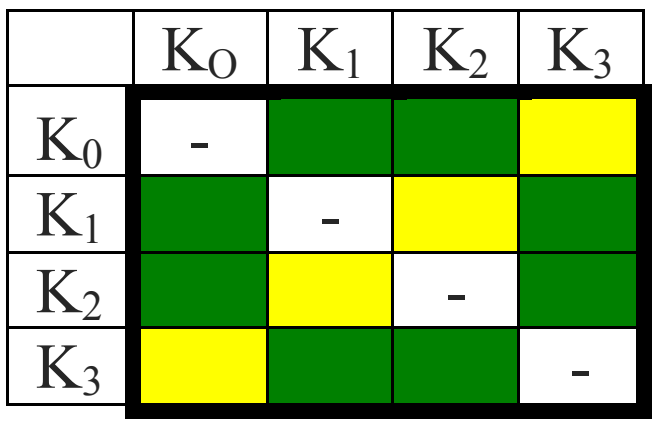

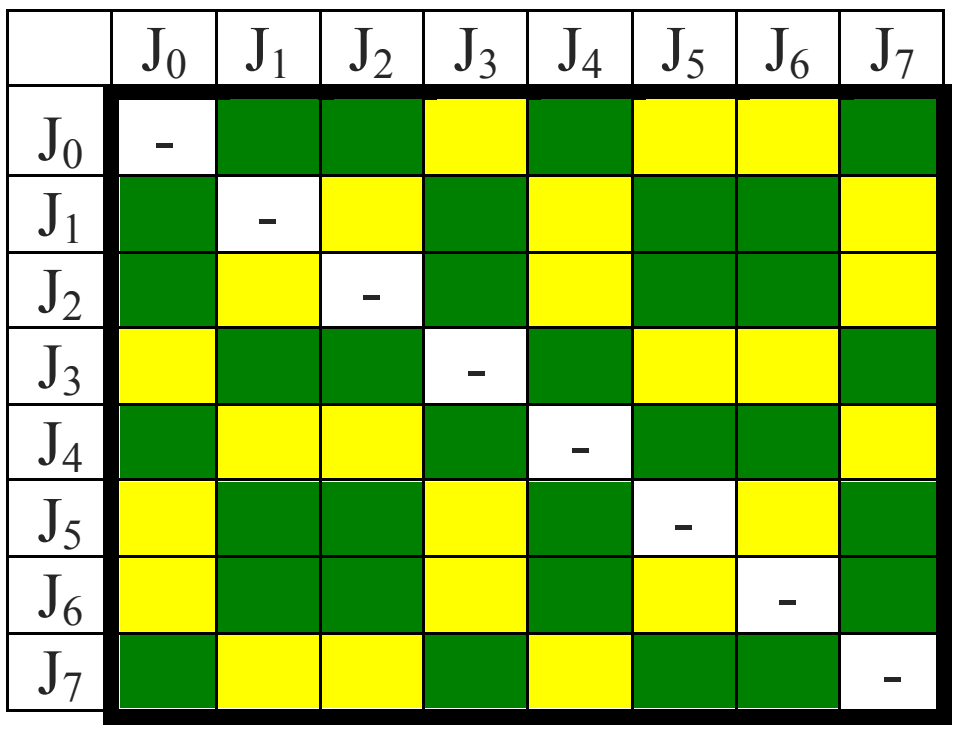

Fig. 80. Bi-symmetrical tables of features of sums of pairs of projectors $K_0$, $K_1$, $K_2$, $K_3$ and $J_0$, $J_1$, .., $J_7$ (shown on Fig. 50 and 51) from the Hadamard matrices $P_4$ and $P_8$ in relation to their exponentiation: $(K_0+K_1)^n$, $(K_0+K_2)^n$, ..., $(K_2+K_3)^n$, $(J_0+J_1)^n$, $(J_0+J_2)^n$, ..., $(J_6+J_7)^n$.

By analogy with Fig. 4, 9, 11, 16 and 19, tabular cells with green color on Fig. 80 correspond to those matrices, exponentiations of which generate cyclic groups with a period 8 in the case of using the normalizing factor $2^{-0.5}$. For example, $(2^{-0.5}*(K_0+K_1))^n = (2^{-0.5}*(K_0+K_1))^{n+8}$. One can show that the matrices in cells with green color represent complex numbers with unit coordinates inside appropriate 2-dimensional planes of 4-dimensional or 8-dimensional spaces correspondingly. Tabular cells with yellow color on Fig. 80 correspond to matrices with the «quadruplet property»: for example, $((K_0+K_3)^2)^n = 4^{n-1}*(K_0+K_3)^2$ and $((J_0+J_3)^2)^n = 4^{n-1}*(J_0+J_3)^2$, where n = 1, 2, 3… Similar tabular results can be also obtained for the case of «row» decompositions of Hadamard matrices $P_4$ and $P_8$.

By analogy with the equation (2) the following equation holds true to receive Hadamard $(2^n*2^n)$-matrices $P_K$ (where $K=2^n$, n = 4, 5, 6,…): $P_4 \otimes [1\ -1;\ 1\ 1]^{(n-2)} = P_K$.

## APPENDIX 4. ABOUT SOME APPLICATIONS IN ROBOTICS

Turning once more to tensornumbers (first of all, to tensorcomplex numbers), which were described above, one can note their possible application in some tasks of robotics when movement control of a group of robots is needed. In this case a special class of multi-parametric system of a tensornumber organization is under consideration. Control and encoding in these systems can be organized so that each of the representative set of subsystems may be selectively controlled and coded independently from other subsystems. An example of such a multi-parameter system is a group of robots, each of which moves along a certain trajectory plane in accordance with the program from a matrix operator, whose components are functions of time; the entire set of these individual operators incorporated into a single matrix operator, whose multiplication with another matrix operator of a similar structure generates a new matrix operator, endowed with the same property independent motion control of each robot in a new regime (This is the problem of collective motion control of a set of robots, each of which can move quite independently of the others due to the fact that the management of its movement is carried by its "personal" sub-operator from a general matrix operator, allowing collective restructuring of all sub-operators by means of simple multiplication of the general operator with a matrix operator of a similar structure).

Consider an example of collective management of an 8-parametric system, which has a tensorcomplex type of its organization and which consists of four 2-parametric subsystems (four robots), the status of each of which may change over time regardless of the status of the other three subsystems. This management can be carried out using an (8 * 8)-matrix M(t), which is an operator of the tensorcomplex type ; all its 8 components $a_0$, $a_1$, ..., $a_7$ are functions of time t (Figure 81). A state of the whole system inside its configuration space during time can be characterized by an 8-dimensional vector $[x_0, x_1, x_2, ..., x_7]$, which is determined by the operator M (t).

M(t) =

| | | | | | | | |
|---|---|---|---|---|---|---|---|
| $a_0(t)$ | $-a_1(t)$ | $a_2(t)$ | $-a_3(t)$ | $-a_4(t)$ | $a_5(t)$ | $a_6(t)$ | $-a_7(t)$ |
| $a_0(t)$ | $a_1(t)$ | $a_2(t)$ | $a_3(t)$ | $-a_4(t)$ | $-a_5(t)$ | $a_6(t)$ | $a_7(t)$ |
| $-a_0(t)$ | $a_1(t)$ | $a_2(t)$ | $-a_3(t)$ | $a_4(t)$ | $-a_5(t)$ | $a_6(t)$ | $-a_7(t)$ |
| $-a_0(t)$ | $-a_1(t)$ | $a_2(t)$ | $a_3(t)$ | $a_4(t)$ | $a_5(t)$ | $a_6(t)$ | $a_7(t)$ |
| $a_4(t)$ | $-a_5(t)$ | $-a_6(t)$ | $a_7(t)$ | $a_0(t)$ | $-a_1(t)$ | $a_2(t)$ | $-a_3(t)$ |
| $a_4(t)$ | $a_5(t)$ | $-a_6(t)$ | $-a_7(t)$ | $a_0(t)$ | $a_1(t)$ | $a_2(t)$ | $a_3(t)$ |
| $-a_4(t)$ | $a_5(t)$ | $-a_6(t)$ | $a_7(t)$ | $-a_0(t)$ | $a_1(t)$ | $a_2(t)$ | $-a_3(t)$ |
| $-a_4(t)$ | $-a_5(t)$ | $-a_6(t)$ | $-a_7(t)$ | $-a_0(t)$ | $-a_1(t)$ | $a_2(t)$ | $a_3(t)$ |

Fig. 81. The example of (8*8)-matrix as a general operator for movement control of 4 robots in a case when they form an 8-parametric system (explanation in text).

In this example, behavior of each of the four 2-parametric subsystems is graphically depicted by means of movement of a point inside one of four coordinate planes ($x_0$, $x_4$), ($x_1$, $x_5$), ($x_2$, $x_6$), ($x_3$, $x_7$) of the configuration space of the system; this behavior can determined entirely independently from behavior of the other three subsystems. To specify the path and pace of movement of each point along its individual trajectory, which is defined parametrically, it is only necessary to set the functions $a_n(t)$ of the operator M (t).

For instance, independent movements of the points in the planes ($x_0$, $x_4$), ($x_1$, $x_5$), ($x_2$, $x_6$), ($x_3$, $x_7$) occur along known types of curves - "cardoid", "petal clover", " 5-petal rose" and "logarithmic spiral " (Figure 82) - in case, when the functions $a_n(t)$ are the following:

$a_0(t)=\cos(t)*(1+\cos(t))$; $a_4(t)=\sin(t)*(1+\cos(t))$;

$a_1(t)=\cos(t)+\cos(t)*\cos(3*t)+\cos(t)*\sin^2(3*t)$;

$a_5(t)=\sin(t)+\sin(t)*\cos(3*t)+\sin(t)*\sin^2(3*t)$; $a_2(t)=2*\cos(t)*\sin(5/3*t)$;

$a_6(t)=\sin(t)*\sin(5/3*t)$; $a_3(t)=\cos(t)*(13/12)^t$; $a_7(t)=\sin(t)*(13/12)^t$.

| | |
|---|---|
| In the plane ($x_0$, $x_4$), motion occurs along a cardoid: 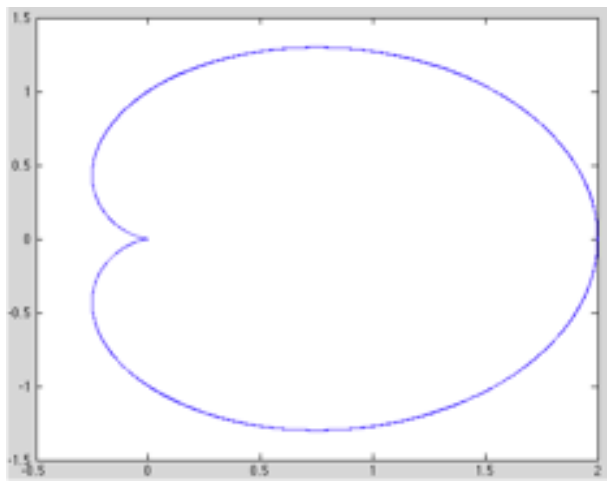 | In the plane ($x_1$, $x_5$), motion occurs along a "petal clover": 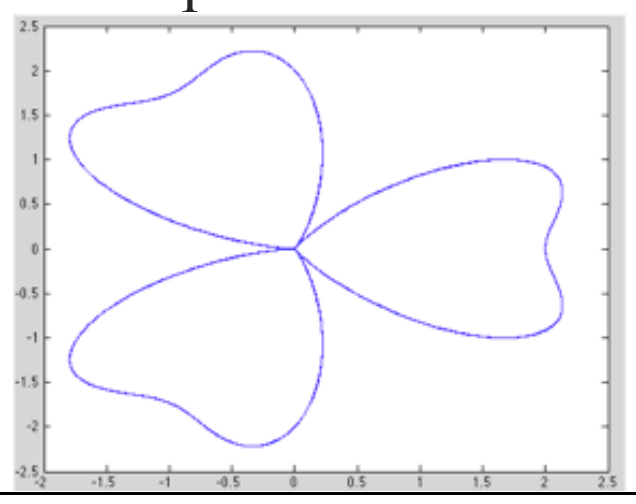 |
| In the plane ($x_2$, $x_6$), motion occurs along a "5-petal rose": 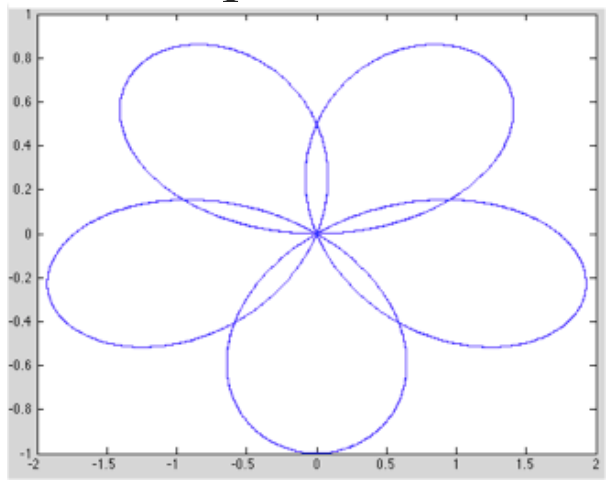 | In the plane ($x_3$, $x_7$), motion occurs along a logarithmic spiral: 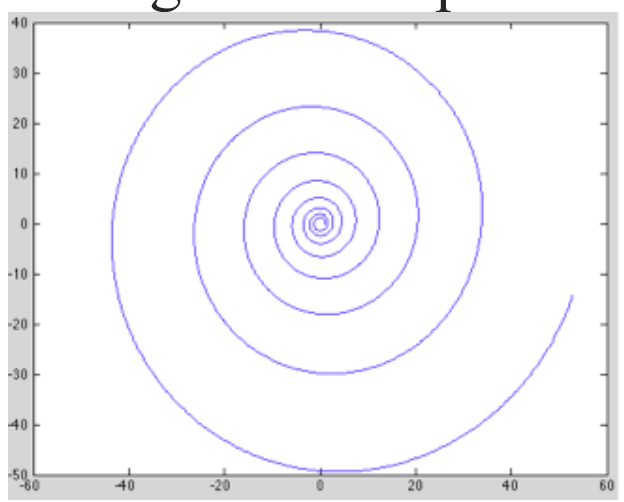 |

Fig. 82. The example of motion trajectories for 4 sub-systems of the 8-parametric system (explanation in text).